\documentclass[a4paper,12pt]{article}
\usepackage[T1]{fontenc}
\usepackage{amsmath,amssymb,amsthm}
\usepackage[ruled,vlined]{algorithm2e}
\usepackage[mathscr]{euscript}
\usepackage{mathrsfs}
\usepackage{enumitem}
\usepackage{color}
\usepackage{graphicx}
\usepackage{epsfig}
\usepackage{pdfpages}
\usepackage{float}
\usepackage{array}
\usepackage{subcaption}
\usepackage[font=small]{caption}
\usepackage[hang,flushmargin]{footmisc} 
\usepackage{booktabs}
\usepackage{rotating,tabularx}
\usepackage{tikz}
\usetikzlibrary{positioning}
\usepackage{natbib}
\usepackage{setspace}
\usepackage{longtable}
\usepackage{mathabx}
\usepackage[section]{placeins}
\usepackage[left=3cm,right=3cm,bottom=3cm,top=2.5cm]{geometry}
\numberwithin{equation}{section}
\numberwithin{figure}{section}
\numberwithin{table}{section}
\renewcommand{\baselinestretch}{1.2}\normalsize

\allowdisplaybreaks

\usepackage{hyperref}
\hypersetup{
hidelinks,
}

\usepackage{ifthen}
\newcommand{\comments}{0} 
\ifthenelse{\comments=1}{\newcommand{\KOM}[1]{{\color{red}#1}}}{\newcommand{\KOM}[1]{}}

\newcommand{\reals}{\mathbb{R}}

\newcommand{\pr}{\mathbb{P}} 
\newcommand{\ex}{\mathbb{E}} 
\newcommand{\var}{\textnormal{Var}} 
\newcommand{\normal}{N} 
\newcommand{\ind}{1} 
\newcommand{\bs}[1]{\boldsymbol{#1}}

\newcommand{\eig}{{\color{black}{\psi}}} 
\newcommand{\Eig}{{\color{black}{\Psi}}} 
\newcommand{\pen}{{\color{black}{\lambda}}} 
\newcommand{\pennw}{{\color{black}{\kappa}}} 
\newcommand{\resnw}{{\color{black}{\Delta}}} 
\newcommand{\betanw}{{\color{black}{\theta}}} 
\newcommand{\summable}{{\color{black}{a}}} 
\newcommand{\Summable}{{\color{black}{A}}} 
\newcommand{\moments}{{\color{black}{\nu}}} 
\newcommand{\scaling}{\mathcal{N}}

\DeclareMathOperator*{\argmin}{arg\,min}

\newcommand{\convd}{\stackrel{d}{\longrightarrow}} 
\newcommand{\convp}{\stackrel{p}{\longrightarrow}} 

\theoremstyle{plain}
\newtheorem{theorem}{Theorem}[section]

\newtheorem{definition}{Definition}[section]
\newtheorem{lemmaA}{Lemma}
\newtheorem{propA}{Proposition}

\newtheorem{lemmaAprime}{Lemma}
\newtheorem{propAprime}{Proposition}

\newtheorem{lemmaB}{Lemma}
\newtheorem{propB}{Proposition}

\newtheorem{lemmaBprime}{Lemma}
\newtheorem{propBprime}{Proposition}

\newtheorem{lemmaC}{Lemma}

\theoremstyle{remark}
\newtheorem{remark}{Remark}[section]

\makeatletter
\newcommand{\lefteqno}{\let\veqno\@@leqno}
\makeatother

\makeatletter
\newsavebox\myboxA
\newsavebox\myboxB
\newlength\mylenA

\newcommand*\xoverline[2][0.75]{%
    \sbox{\myboxA}{$\m@th#2$}%
    \setbox\myboxB\null
    \ht\myboxB=\ht\myboxA%
    \dp\myboxB=\dp\myboxA%
    \wd\myboxB=#1\wd\myboxA
    \sbox\myboxB{$\m@th\overline{\copy\myboxB}$}
    \setlength\mylenA{\the\wd\myboxA}
    \addtolength\mylenA{-\the\wd\myboxB}%
    \ifdim\wd\myboxB<\wd\myboxA%
       \rlap{\hskip 0.5\mylenA\usebox\myboxB}{\usebox\myboxA}%
    \else
        \hskip -0.5\mylenA\rlap{\usebox\myboxA}{\hskip 0.5\mylenA\usebox\myboxB}%
    \fi}
\makeatother

\newcommand{\heading}[3]
{  \setcounter{page}{1}
   \begin{center}

   \phantom{Distance to upper boundary}
   \vspace{0.5cm}

   {\LARGE \textbf{#1}}\\[0.35cm]
   {\LARGE \textbf{#2}}\\[0.3cm]
   {\LARGE \textbf{#3}}
   \end{center}
}

\newcommand{\headingSupp}[4]
{  \setcounter{page}{1}
   \begin{center}

   \phantom{Distance to upper boundary}
   \vspace{0.5cm}

   {\LARGE \textbf{#1}}\\[0.3cm]
   {\LARGE \textbf{#2}}\\[0.35cm]
   {\LARGE \textbf{#3}}\\[0.3cm]
   {\LARGE \textbf{#4}}
   \end{center}
}

\newcommand{\authors}[4]
{  \begin{center}
      \begin{minipage}[c][2cm][c]{5.5cm}
      \begin{center} 
      {\large #1} 
      
      #2 
      \end{center}
      \end{minipage}
      \begin{minipage}[c][2cm][c]{5.5cm}
      \begin{center} 
      {\large #3}

      #4 
      \end{center}
      \end{minipage}
   \end{center}
}

\begin{document}

\heading{Estimation and Inference in}{High-Dimensional Panel Data Models}{with Interactive Fixed Effects}

\authors
{Maximilian R\"ucker\renewcommand{\thefootnote}{1}\footnotemark[1]}{Ulm University}
{Michael Vogt\renewcommand{\thefootnote}{2}\footnotemark[2]}{Ulm University}
\vspace{-1.25cm}

\authors
{Oliver Linton\renewcommand{\thefootnote}{3}\footnotemark[3]}{University of Cambridge}
{Christopher Walsh\renewcommand{\thefootnote}{4}\footnotemark[4]}{Newcastle University}

\footnotetext[1]{Address: Institute of Statistics, Department of Mathematics and Economics, Ulm University, Helmholtzstrasse 20, 89081 Ulm, Germany. Email: \texttt{maximilian.ruecker@uni-ulm.de}.}
\renewcommand{\thefootnote}{2}
\footnotetext[2]{Corresponding author. Address: Institute of Statistics, Department of Mathematics and Economics, Ulm University, Helmholtzstrasse 20, 89081 Ulm, Germany. Email: \texttt{m.vogt@uni-ulm.de}.}
\renewcommand{\thefootnote}{3}
\footnotetext[3]{Address: Faculty of Economics, University of Cambridge, Austin Robinson Building, Sidgwick Avenue, Cambridge, CB3 9DD, UK. Email: \texttt{obl20@cam.ac.uk}.}
\renewcommand{\thefootnote}{4}
\footnotetext[4]{Address: Newcastle University Business School, 5 Barrack Road, Newcastle upon Tyne, NE1 4SE, UK. Email: \texttt{chris.walsh@newcastle.ac.uk}.}
\renewcommand{\thefootnote}{\arabic{footnote}}
\setcounter{footnote}{4}

\vspace{-1cm}

\renewcommand{\baselinestretch}{1.0}\normalsize
\renewcommand{\abstractname}{}
\begin{abstract}
\noindent We develop new econometric methods for estimation and inference in high-dimensional panel data models with interactive fixed effects. Our approach can be regarded as a non-trivial extension of the very popular common correlated effects (CCE) approach. Roughly speaking, we proceed as follows: We first construct a projection device to eliminate the unobserved factors from the model by applying a dimensionality reduction transform to the matrix of cross-sectionally averaged covariates. The unknown parameters are then estimated by applying lasso techniques to the projected model. For inference purposes, we derive a desparsified version of our lasso-type estimator.  While the original CCE approach is restricted to the low-dimensional case where the number of regressors is small and fixed, our methods can deal with both low- and high-dimensional situ\-ations where the number of regressors is large and may even exceed the overall sample size. We derive theory for our estimation and inference methods both in the large-$T$-case, where the time series length $T$ tends to infinity, and in the small-$T$-case, where $T$ is a fixed natural number. Specifically, we derive the convergence rate of our estimator and show that its desparsified version is asymptotically normal under suitable regularity conditions. The theoretical analysis of the paper is complemented by a simulation study and an empirical application to characteristic based asset pricing.
\end{abstract}

\renewcommand{\baselinestretch}{1.2}\normalsize



\section{Introduction}

Nowadays, economic panel data sets are often ``high-dimensional'': they contain a wide variety of time-varying characteristics and controls whose number is quite substantial in comparison to the sample size and may even exceed it. 
For example, in the case of low-frequency financial panel data, there is a rapidly evolving literature on the so-called ``factor zoo'', which involves a large number of observed firm-specific characteristics that have been proposed as potential drivers of stock risk premia. \cite{Harvey-et-al2015} document $382$ such factors published in top journals and also point to the ongoing exponential growth in their number. 
A microeconomic example where the covariate dimensionality is pressing is presented in \cite{Belloni2016} who attempt to determine the social costs of gun ownership. Their analysis involves estimating the effect of gun prevalence on crime rates in a fixed-effect panel model. Their data set comprises information on $988$ explanatory variables, while the overall sample size is $nT=3705$ (with $n=195$ being the cross-section dimension and $T=19$ the time series length).
Another example from macroeconomics concerns the determinants of economic growth. \cite{LuSu2016} estimate the effect of various possible determinants on the GDP growth rate in a panel with $n=108$ countries over $T=36$ years. As they point out, the survey by \cite{Durlauf2005} lists $145$ potential determinants of economic growth. Additionally allowing for interaction terms and nonlinear transformations of these variables, one easily arrives at a situation where the number of available covariates is comparable to or even exceeds the sample size.

Estimating high-dimensional panel models where the number of explanatory variables $p$ is large relative to the sample size $nT$ is far from trivial. Standard methods and theory from high-dimensional statistics are mostly tailored to a simple cross-sectional i.i.d.\ data structure. Panel data, in contrast, comprise complicated dependence structures that need to be taken into account: they usually exhibit non-negligible correlation in the time (and cross-sectional) dimension. Moreover, in order to account for unobserved heterogeneity, models with intricate error structures involving fixed-effect or factor-type components are considered rather than models with simple i.i.d.\ errors. These complicated data structures are presumably the reason why the literature on econometric methods for high-dimensional panels is quite limited. Below, we give a brief overview of the existing literature and of how our contribution fits into it.

The main aim of this paper is to develop estimation and inference methods for the high-dimensional panel data model with interactive fixed effects: 
\begin{equation}
Y_{it}=\beta ^{\top }X_{it}+\gamma _{i}^{\top }F_{t}+\varepsilon _{it}
\label{model-intro}
\end{equation}%
for $1\leq t\leq T$ and $1\leq i\leq n$, where $i$ is the cross-section index and $t$ the time series index, $Y_{it}$ is a real-valued response variable, $X_{it}$ is a vector of $p$ regressors and $\beta $ is the unknown parameter vector of length $p$. We allow the number of potential regressors $p$ to be very large but impose sparsity on $\beta$ in the sense that the number of non-zero elements of $\beta$ is relatively small compared to $p$. The error structure of the model comprises two components: a standard idiosyncratic error term $\varepsilon _{it}$ and the interactive fixed effects component $\gamma _{i}^{\top }F_{t}$, where $F_{t}$ is a vector of unobserved factors and $\gamma _{i}$ is a vector of unobserved factor loadings. The regressors $X_{it}$ are allowed to be correlated with the factor structure, which induces endogeneity issues in model \eqref{model-intro}. The interactive fixed effects in \eqref{model-intro} allow to model unobserved heterogeneity in a quite flexible manner, in particular, much more flexibly than standard fixed effects $a_{i}$ and $b_{t}$ in a model of the form $Y_{it}=\beta^{\top}X_{it}+a_{i}+b_{t}+\varepsilon _{it}$.

In the traditional low-dimensional case where the number of regressors $p$ is small and fixed, model \eqref{model-intro} has been analyzed extensively in the literature. The most popular estimator of $\beta$ in this traditional setting is the common correlated effects (CCE) estimator of \cite{Pesaran2006}. In this paper, we develop an estimator that can be regarded as an extension of the CCE method to the case where there are many potential covariates or controls. 
As the original CCE method, our approach is based on the following strategy:
we ``project away'' the unobserved factors, i.e., we (approximately) eliminate them by applying a particular projection device to the response and the covariates. To estimate $\beta$, we then apply $\ell_1$-penalized least squares methods, i.e., lasso methods to the projected regression. We call our estimator a high-dimensional CCE (HD-CCE) estimator.


As detailed in Section \ref{sec:est}, the original CCE approach breaks down completely as soon as $p>T$ (and
performs very poorly already for $p$ somewhat smaller than $T$), thus
imposing very strong restrictions on the number of explanatory variables $p$. 
Our approach, in contrast, works for $p$ in a very wide range: we can deal
with the standard \textquotedblleft low-dimensional\textquotedblright\ case
where $p$ is small and fixed, the \textquotedblleft moderately
high-dimensional\textquotedblright\ case where $p$ is fairly large but still
smaller than the sample size $nT$ and the \textquotedblleft truly
high-dimensional\textquotedblright\ case where $p$ exceeds the sample size $%
nT$. Precise conditions on the size of $p$ in comparison to $n$ and $T$ are
provided in the context of our theoretical results in Section \ref%
{sec:theory}. For our estimator to work with $p$ in such a wide range, we
require a projection device that approximately eliminates the factors no
matter whether $p$ is small or big. To construct such a device, we make use
of methods from high-dimensional factor analysis \citep[see e.g.][]{Fan2013}
that are based on singular value decompositions of high-dimensional
covariance matrices and principal components thresholding. Related principal
components based methods have been used in the CCE context before 
\citep[see
e.g.][]{Juodis2021b}, however, for very different purposes and only in the
low-dimensional case with $p$ small and fixed. Notably, the ability of our
estimator to deal with both low- and high-dimensional situations does not
come without cost: in contrast to the original CCE approach, we require an
estimate of the number of factors. We propose a simple estimation procedure
which can be regarded as a formalization of scree plots that are very common
in applied factor analysis.

In the theoretical part of the paper, we derive the convergence rate of our HD-CCE
estimator. We further establish an inference procedure
for scalar parameters of interest. As usual in high-dimensional statistics,
we need to desparsify or debias our lasso-type estimator in order to perform
inference. We show that the desparsified version of our procedure is
asymptotically normal and provide consistent standard errors that can be used for confidence intervals or hypothesis tests. We note that in
contrast to most of the literature on panel models with interactive fixed
effects, our methods and theory are not restricted to the large-$T$-case
where both $n$ and $T$ tend to infinity but we also cover the small-$T$-case
where $n$ tends to infinity and $T$ is a fixed natural number. This makes
our methods applicable in a very wide range of application contexts. We
provide numerical evidence on the performance of our methods by Monte Carlo experiments and illustrate the usefulness of our methods by 
an application to financial panel data.

\subsubsection*{Literature review}

In the low-dimensional case with $p$ small and fixed, panel data models with interactive fixed effects are well understood. Since its introduction, the CCE estimator has become a standard tool for their analysis, giving rise to a whole
new strand of the literature with numerous extensions such as \cite%
{Kapetanios2011}, \cite{ChudikPesaranTosetti2011}, \cite{PesaranTosetti2011}%
, \cite{ChudikPesaran2015}, \cite{Westerlund2018}, \cite{Westerlund2019}, 
\cite{Juodis2021a}, \cite{BrownSchmidtWooldridge2021} and \cite{Juodis2021b} to name just a few. There are
several alternatives to the CCE estimator which can be used to estimate the
parameter vector $\beta$ in the low-dimensional case. The most important
one is a least squares approach 
originally studied in \cite{Bai2009} and theoretically further explored in 
\cite{MoonWeidner2015} among others. The philosophy behind this approach is
quite different from that of the CCE method: whereas the CCE approach
eliminates the factors and the corresponding loadings by a suitable
transformation of the model, the least squares approach treats them as
additional parameters to be estimated. One disadvantage of the least squares
approach is that the criterion function to be minimized is not convex.
Hence, to compute the estimator, one needs to solve a non-convex
optimization problem. Recently, least squares estimation with nuclear norm
penalization has been proposed to overcome this problem. The resulting
estimator minimizes a convex criterion function and can thus be efficiently
computed by standard methods from convex optimization. It has, however, the
disadvantage that its convergence rate is fairly slow in general. Recent
studies on nuclear norm penalized estimators for panel data models with
interactive fixed effects include \cite{ChernozhukovHansenLiaoZhu2018}, \cite%
{BeyhumGautier2019} and \cite{MoonWeidner2019}. A state-of-the-art review of methods for fixed effects panels, including interactive fixed effects and other variants, can be found in \cite{Bonhomme2024}.

Whereas panel models with interactive fixed effects are well studied in the
low-dimensional case, they are largely unexplored in high dimensions.
Indeed, the literature on high-dimensional
panels is rather limited in general. High-dimensional panel models with
random and fixed effects have been considered in \cite{Kock2013, Kock2016}, 
\cite{Belloni2016} and \cite{KockTang2019}: \cite{Kock2013} derives theory
for bridge estimators in both random and fixed effects models, while \cite{Kock2016} analyzes a model with a hybrid error structure that is in-between
random and fixed effects. \cite{Belloni2016} introduce the so-called
cluster-lasso to estimate the unknown parameters in a model with an
individual fixed effect. Econometric methods for high-dimensional panel
models with interactive fixed effects have been developed in \cite{LuSu2016}
and \cite{BelloniChenPadillaWang2019}: \cite{LuSu2016} extend the least
squares method of \cite{Bai2009} to a high-dimensional dynamic panel model
by adding a group-lasso penalty. However, they only consider a situation
where $p$ grows fairly slowly with the sample size. \cite{BelloniChenPadillaWang2019} develop nuclear norm penalized estimation
methods for high-dimensional quantile panel regression. A high-dimensional
version of the panel partial factor model, which is closely related to panel
models with interactive fixed effects, is investigated in \cite{HansenLiao2019}. 
\cite{Cheng-et-al2024} study another closely related model framework, specifically, a high-dimensional panel regression model for financial data with interactive fixed effects where the factor loadings are driven nonparametrically by observed stock-specific characteristics or covariates. In their model, the covariates are assumed to be weakly dependent across both cross section and time series, which is incompatible with the type of factor structure in the covariates that we assume and exploit in this paper. To the best of our knowledge, CCE-type approaches suited to high dimensions have not been developed at all in the literature so far.

\subsubsection*{Structure of the paper}

The model framework which underlies our
theoretical analysis is introduced in detail in Section \ref{sec:model}, while
identification issues are discussed in Section \ref{sec:ident}. The HD-CCE
estimator and its desparsified version are derived step by step in Section \ref{sec:est}. Section \ref{sec:est:tuning} is dedicated to the practical implementation of our
estimators, in particular, to the choice of the involved tuning parameters.
The main theoretical results are laid out in Section \ref{sec:theory}. We
provide a simulation study in the supplementary material and illustrate our methods by an analysis of the ``factor zoo'' in Section \ref{sec:application}. A brief overview of the simulation study can be found in Section \ref{sec:sim-overview}.

\subsubsection*{\texttt{R} code}

Our methods are implemented in the \texttt{R} package \texttt{hdcce} which can be downloaded from \texttt{https://github.com/RueckerM/hdcce}. Moreover, replication files are available at \linebreak \texttt{https://github.com/RueckerM/hdcce-ReplicationFiles}.

\subsubsection*{Notation}

Matrices are denoted by bold letters,
whereas scalars and vectors are printed in normal font. For a vector $v =
(v_1,\ldots,v_q)^\top \in \reals^q$ and a set $S \subseteq \{1,\ldots,q\}$,
we let $v_S = (v_i: i \in S)$ be the vector which consists of the entries $%
v_i$ with $i \in S$ only. In addition, we sometimes write $v_{-i}$ to denote
the vector $v$ without the $i$-th component. We let $\|v\| = (\sum_i
v_i^2)^{1/2}$ denote the Euclidean norm of $v$, $\|v\|_1 = \sum_i|v_1|$ its $%
\ell_1$-norm, and $\|v\|_{\infty} = \max_i|v_i|$ its $\ell_\infty$-norm. For
a generic matrix $\bs{A}^{T \times p}$, we denote the row vectors by $A_t$
and the column vectors by $A_{(j)}$, that is, $\bs{A} = (A_1 \ldots
A_T)^\top = (A_{(1)} \ldots A_{(p)})$. Moreover, the matrix $\bs{A}$ without
the $t$-th row is denoted by $\bs{A}_{-t}$ and that without the $j$-th
column by $\bs{A}_{(-j)}$. The symbols $\eig_{\min}(\bs{A})$ and $\eig%
_{\max}(\bs{A})$ are used to denote the minimal and maximal eigenvalue of a
square matrix $\bs{A} \in \reals^{q \times q}$. In addition, we sometimes
write $\eig_1(\bs{A}) \ge \eig_2(\bs{A}) \ge \ldots \ge \eig_q(\bs{A})$ to
denote the eigenvalues of $\bs{A}$ (in decreasing order). For a general (not
necessarily square) matrix $\bs{A} = (a_{ij})$, $\| \bs{A} \|$, $\|\bs{A}%
\|_1 $, $\|\bs{A}\|_\infty$ and $\|\bs{A}\|_{\text{max}}$ are its spectral
norm, $\ell_1$-norm, $\ell_\infty$-norm and elementwise norm, respectively.
In particular, $\| \bs{A} \| = \eig_{\max}^{1/2}(\bs{A}^\top \bs{A})$, $\|%
\bs{A}\|_1 = \max_j \sum_i |a_{ij}|$, $\|\bs{A}\|_\infty = \max_i \sum_j
|a_{ij}|$ and $\|\bs{A}\|_{\max} = \max_{ij} |a_{ij}|$. The symbol $\bs{A}%
^{-}$ stands for the generalized inverse of a matrix $\bs{A}$ and the symbol 
$\bs{I}_q$ for the $q \times q$ identity matrix. Sometimes, we also write $%
\bs{I}$ instead of $\bs{I}_q$ for short. Finally, the indicator function is
denoted by $\ind(\cdot)$ and the cardinality of a set $S$ by $|S|$.

\section{Model framework}\label{sec:model}

We observe a sample of panel data $\{ (Y_{it}, X_{it}): 1 \le t \le T, \, 1 \le i \le n \}$ with real-valued random variables $Y_{it}$ and $\reals^p$-valued random vectors $X_{it} = (X_{it,1},\ldots,X_{it,p})^\top$, where $n$ is the cross-section dimension and $T$ the time series length. We consider the following two scenarios: 
\begin{enumerate}[label=(\roman*),leftmargin=1.4cm,parsep=0pt,itemsep=0pt]
\item the large-$T$-case where both $n \to \infty$ and $T \to \infty$ 
\item the small-$T$-case where $n \to \infty$ but $T$ is a fixed natural number. 
\end{enumerate}
We regard both $T$ and $p$ as a function of $n$, that is, $T=T(n)$ and $p=p(n)$. Hence, asymptotic statements are to be understood in the sense that $n \to \infty$ (and $T=T(n) \to \infty$ in the large-$T$-case). 
The dimension $p$ of the random vector $X_{it}$ is allowed to be large, potentially much larger than $n$ and $T$. Put differently, we allow $p$ to grow with $n$ (and $T$). The only restriction is that $p$ does not grow too quickly compared to $n$ (and $T$). Hence, the methods and theory of this paper are valid for any $p$ which is not too large compared to $n$ (and $T$). In particular, we cover both the traditional low-dimensional case where $p$ is small and fixed and the high-dimensional case where $p$ grows potentially much faster than $n$ (and $T$). Precise conditions on the size of $p$ compared to $n$ and $T$ are provided in Section \ref{sec:theory}.

We consider a high-dimensional version of the linear panel data model with interactive fixed effects analyzed in \cite{Pesaran2006}. The model has the form 
\begin{equation}\label{eq:model-CCE}
Y_i = \bs{X}_i \beta + \bs{F} \gamma_i + \varepsilon_i \qquad (1 \le i \le n)
\end{equation}
for each cross-sectional unit $i$, where $Y_i = (Y_{i1},\ldots,Y_{iT})^\top \in \reals^T$ is the response vector, $\beta = (\beta_1,\ldots,\beta_p)^\top$ is the unknown parameter vector, $\bs{X}_i = (X_{i1} \ldots X_{iT})^\top \in \reals^{T \times p}$ is the regressor matrix, $\varepsilon_i = (\varepsilon_{i1},\ldots,\varepsilon_{iT})^\top \in \reals^T$ is the vector of idiosyncratic errors with $\ex[\varepsilon_{it}] = 0$ for all $i$ and $t$, and $\bs{F} \gamma_i$ is the interactive fixed effects part of the error. More specifically, $\bs{F} = (F_1 \ldots F_T)^\top \in \reals^{T \times K}$ with $F_t = (F_{t,1},\ldots,F_{t,K})^\top$ is a matrix of unobserved factors and $\gamma_i = (\gamma_{i,1},\ldots,\gamma_{i,K})^\top$ is a vector of (unknown) individual-specific factor loadings. Throughout the paper, we treat the factors $F_t$ as non-random parameters. Put differently, we implicitly condition on the factors $F_1,\ldots,F_T$ in our theoretical analysis as is common in the literature \citep[see e.g.][]{MoonWeidner2015}. The factor loadings $\gamma_i$, in contrast, are considered to be random. The regressors in \eqref{eq:model-CCE} are supposed to have the structure 
\begin{equation}\label{eq:model-CCE-reg}
\bs{X}_i = \bs{F} \bs{\Gamma}_i^\top + \bs{Z}_i \qquad (1 \le i \le n), 
\end{equation}
where $\bs{\Gamma}_i \in \reals^{p \times K}$ is a matrix of individual-specific factor loadings and $\bs{Z}_i = (Z_{i1} \ldots$ $\ldots Z_{iT})^\top \in \reals^{T \times p}$ represents the idiosyncratic part of the regressors with $\ex[Z_{it}] = 0$ for all $i$ and $t$. This structure implies that the regressors $\bs{X}_i$ are in general correlated with the unobserved part of equation \eqref{eq:model-CCE}, $e_i = \bs{F}\gamma_i + \varepsilon_i$, via the interactive fixed effects.

The main difference of model \eqref{eq:model-CCE}--\eqref{eq:model-CCE-reg} from Pesaran's original model is that we allow the dimension $p$ of the regressors $X_{it} = (X_{it,1},\ldots,X_{it,p})^\top$ to be large, possibly much larger than the overall sample size $nT$. Without structural constraints on the parameter vector $\beta$, model \eqref{eq:model-CCE}--\eqref{eq:model-CCE-reg} is not estimable in general. As usual in the literature on high-dimensional statistics, we impose a sparsity constraint on $\beta$. In particular, we assume that the set $S=\{j: \beta_j \ne 0\}$ of non-zero components of $\beta$ has cardinality $s := |S|$ considerably smaller than the sample size $nT$. Hence, only a small subset of regressors is active, that is, enters the model with a non-zero coefficient. Precise conditions on the size of the sparsity index $s$ are provided in Section \ref{sec:theory}. 
Notably, there have been some attempts to perform estimation and inference in high-dimensional models with non-sparse structures in recent years \citep{ZhuBradic2018,SilinFan2022}. However, even though the assumption of sparsity is not harmless \citep{Kolesar2025}, we here follow the main bulk of the literature on high-dimensional statistics and work under a sparsity constraint.  
In contrast to the number of regressors $p$, the number of unknown factors $K$ is assumed to be fixed in magnitude as in Pesaran's model. Assuming that $K$ is comparably small makes sense as $K$ plays a role analogous to the number of active regressors $s$ rather than the total number of regressors $p$. We do not assume that $K$ is known a priori, and determine it from the data.\footnote{Following \cite{Pesaran2006}, one may additionally include observed factors in model \eqref{eq:model-CCE}--\eqref{eq:model-CCE-reg} and allow for heterogeneous parameter vectors $\beta_i = \beta + \eta_i$ with i.i.d.\ disturbances $\eta_i$. In particular, as long as the random disturbances $\eta_i$ produce sparse parameter vectors $\beta_i$ and we are in the large-$T$-case, it is possible to estimate the individual $\beta_i$'s. However, if the $\beta_i$'s are non-sparse, it will in general only be possible to estimate the (sparse) mean vector $\beta$.}

If our interest focuses on point estimation of $\beta$, the above model description is fully sufficient. If the aim is to perform inference, in contrast, we need additional structure. Suppose in particular we want to compute confidence bands for the coefficient $\beta_j$ of the $j$-th regressor. To be able to do so, we additionally impose the nodewise regression equation
\begin{equation}\label{eq:model-CCE-nodewise}
X_{i(j)} = \bs{X}_{i(-j)}\betanw + \bs{F}\nu_{i} + u_i \qquad (1 \le i \le n),
\end{equation}
where $X_{i(j)}$ is the $j$-th column of the matrix $\bs{X}_i$, $\bs{X}_{i(-j)}$ is the matrix $\bs{X}_i$ without the $j$-th column, $\betanw$ is a sparse parameter vector, $u_i$ denotes the idiosyncratic error term with $\ex[u_i] = 0$ and $\nu_i \in \reals^K$ is a vector of factor loadings. The quantities $\betanw$, $\nu_i$ and $u_i$ depend on $j$. For convenience, however, we suppress this dependence in the notation. According to \eqref{eq:model-CCE-nodewise}, the $j$-th regressor can be represented as a sparse linear function of the other regressors plus an interactive fixed effects error structure. Such a nodewise regression equation is very common in high-dimensional inference, both when desparsified lasso techniques \citep[cp.][]{vandeGeer2014} and double selection techniques \citep[cp.][]{Belloni2014} are used.
Notably, it is no problem to satisfy both the regressor equation \eqref{eq:model-CCE-reg} and the nodewise equation \eqref{eq:model-CCE-nodewise} for the $j$-th regressor in our framework. In particular, if the $j$-th regressor is modelled via the nodewise equation \eqref{eq:model-CCE-nodewise}, it also fulfills \eqref{eq:model-CCE-reg}: $X_{i(j)} = \bs{F} \Gamma_{i,j} + Z_{i(j)}$ with $\Gamma_{i,j} := \bs{\Gamma}_{i,-j}^\top \theta + \nu_i$ and $Z_{i(j)} := \bs{Z}_{i(-j)} \theta + u_i$, which follows immediately upon plugging equation \eqref{eq:model-CCE-reg} for all but the $j$-th regressor, i.e., the equation $\bs{X}_{i(-j)} = \bs{F} \bs{\Gamma}_{i,-j}^\top + \bs{Z}_{i(-j)}$ into \eqref{eq:model-CCE-nodewise}.

A list of the technical conditions that we impose on the model components in equations \eqref{eq:model-CCE}, \eqref{eq:model-CCE-reg} and \eqref{eq:model-CCE-nodewise} to derive our theoretical results can be found in Section \ref{sec:theory}.

\section{Identification}\label{sec:ident}

Model \eqref{eq:model-CCE}--\eqref{eq:model-CCE-reg} contains the following unobserved components: the parameter vector $\beta$, the factor structure $\Theta_{\text{fac}} = \{ \bs{F}, \{ \bs{\Gamma}_i, \gamma_i \}_{i=1}^n \}$ consisting of the factors and their loadings, and the idiosyncratic structure $\Theta_{\text{idio}} =\{ \bs{Z}_i, \varepsilon_i\}_{i=1}^n$ consisting of the idiosyncratic part of the regressors and the idiosyncratic errors. Importantly, the parameter vector $\beta$, the factor structure $\Theta_{\text{fac}}$ and the idiosyncratic structure $\Theta_{\text{idio}}$ are in general not identified. Put differently, the parameter vector $\beta$, the factor structure $\Theta_{\text{fac}}$ and the idiosyncratic structure $\Theta_{\text{idio}}$ which satisfy model equations \eqref{eq:model-CCE}--\eqref{eq:model-CCE-reg} and the technical assumptions of Section \ref{sec:theory} are in general not unique.

In what follows, we show that the parameter vector $\beta$ and the number of factors $K$ are identified if certain additional constraints are imposed. That the factor structure $\Theta_{\text{fac}}$ (apart from $K$) and the idiosyncratic structure $\Theta_{\text{idio}}$ remain unidentified is no problem at all for our methods and theory. For our theoretical arguments to work, it suffices to consider some factor structure $\Theta_{\text{fac}}$ and some idiosyncratic structure $\Theta_{\text{idio}}$ such that the model equations and the technical conditions are fulfilled. Which version is considered does not matter.

We start with identification of $K$. As usual, we normalize the factors $F_t$ to be orthonormal: 
\begin{enumerate}[label=(ID\arabic*), leftmargin=1.15cm]
\item \label{C:id1} It holds that $(\bs{F}^\top\bs{F})/T = \bs{I}_{K\times K}$. 
\end{enumerate}
Given this normalization, we impose the following assumption on the mean loading matrix $\bs{\Gamma} = \ex[\bs{\Gamma}_i] \in \reals^{p \times K}$:
\begin{enumerate}[label=(ID\arabic*),leftmargin=1.15cm]
\setcounter{enumi}{1}
\item \label{C:id2} 
The minimal and the maximal eigenvalue $\eig_{\min}(\bs{\Gamma}^\top \bs{\Gamma}/p)$ and $\eig_{\max}(\bs{\Gamma}^\top \bs{\Gamma}/p)$ of the matrix $\bs{\Gamma}^\top \bs{\Gamma}/p$ are such that $0 < c_{\min} \le \eig_{\min} (\bs{\Gamma}^\top \bs{\Gamma}/p) \le \eig_{\max} (\bs{\Gamma}^\top \bs{\Gamma}/p) \le c_{\max} < \infty$ for some fixed constants $c_{\min}$ and $c_{\max}$. 
\end{enumerate}
\ref{C:id2} is a standard condition in the literature on high-dimensional approximate factor models; see e.g.\ \cite{Fan2013} and \cite{BaiLiao2016}. By imposing it, we focus on the case of strong factors.\footnote{It is in principle possible to weaken \ref{C:id2}. In particular, our theory does not require all eigenvalues of $\bs{\Gamma}^\top \bs{\Gamma}$ to be of order $p$ as assumed in \ref{C:id2}. Instead, we could allow the eigenvalues to be of different order as long as their orders are large enough, in particular, larger than $p \sqrt{\log p} / \sqrt{n}$. Such a generalization of condition \ref{C:id2} would however influence the convergences rates of our HD-CCE estimator.} 
Under \ref{C:id2}, the eigenvalues of $\bs{\Gamma}^\top \bs{\Gamma} / p$ are strictly positive for all $p$, which implies that the matrix $\bs{\Gamma}$ has full rank $K$ for all $p$. An analogous full-rank condition is required in the original CCE approach of \cite{Pesaran2006}.\footnote{Note that \cite{Pesaran2006} also treats the rank-deficient case. However, as shown in \cite{WesterlundUrbain2013}, his results only hold if $\gamma_i$ and $\bs{\Gamma}_i$ are uncorrelated. Hence, the full-rank condition on $\bs{\Gamma}$ is indeed required in the original CCE approach unless one is willing to make the very strong assumption that $\gamma_i$ and $\bs{\Gamma}_i$ are uncorrelated.}
Under \ref{C:id1} and \ref{C:id2}, we can prove the following identification result. 
\begin{theorem}\label{lemma:id-K-large}
Let the technical conditions \ref{C:F}--\ref{C:mixing} and \ref{C:nTp-large}--\ref{C:K-large} from Section \ref{sec:theory} be satisfied in the large-$T$-case and \ref{C:F}--\ref{C:Z} together with \ref{C:nTp-small}--\ref{C:K-small} in the small-$T$-case. If \ref{C:id1}--\ref{C:id2} are fulfilled, then the number of factors $K$ is unique for sufficiently large $n$.
\end{theorem}
The proof of this as well as the subsequent results on identification can be found in the supplementary material.

We next turn to identification of $\beta$. In the high-dimensional case with $p$ potentially larger than the full sample size $nT$ itself, there is of course no way to identify $\beta$ in general. However, we can get identification if we restrict attention to para\-meter vectors $\beta$ with certain properties. Specifically, we focus on vectors $\beta$ which are $s$-sparse, that is, which have at most $s$ non-zero components. As we will see, under certain constraints, there is a unique $s$-sparse parameter vector $\beta$ which satisfies model \eqref{eq:model-CCE}. In order to formulate the precise identification result, we introduce some notation: Let $\mathcal{L}_{\bs{F}} = \{\bs{F} v: v \in \reals^K\}$ be the column space of the matrix $\bs{F}$ and $\bs{\Pi} = \bs{I} - \bs{F} (\bs{F}^\top \bs{F})^{-1} \bs{F}^\top$ the projection matrix onto the orthogonal complement of $\mathcal{L}_{\bs{F}}$. Since $\bs{\Pi} \bs{F} = \bs{0}$ by construction, applying $\bs{\Pi}$ to the model equation in \eqref{eq:model-CCE} yields $\bs{\Pi} Y_i = \bs{\Pi} \bs{X}_i \beta + \bs{\Pi} \varepsilon_i$. Stacking the projected model equations $\bs{\Pi} Y_i = \bs{\Pi} \bs{X}_i \beta + \bs{\Pi} \varepsilon_i$ for all $i$, we obtain the model 
\begin{equation}\label{eq:full-model-projected}
Y^\perp = \bs{X}^\perp \beta + \varepsilon^\perp, 
\end{equation} 
where
\[ Y^\perp = \begin{pmatrix} \bs{\Pi} Y_1 \\ \vdots \\ \bs{\Pi} Y_n \end{pmatrix}, \ \ \bs{X}^\perp = \begin{pmatrix} \bs{\Pi} \bs{X}_1 \\ \vdots \\ \bs{\Pi} \bs{X}_n \end{pmatrix}, \ \ \varepsilon^\perp = \begin{pmatrix} \bs{\Pi} \varepsilon_1 \\ \vdots \\ \bs{\Pi} \varepsilon_n \end{pmatrix}. \]
In order to identify the $s$-sparse parameter vector $\beta$, we impose a restricted eigenvalue (or compatibility) condition on the design matrix $\bs{X}^\perp$ in model \eqref{eq:full-model-projected}. Such a condition is very common in high-dimensional statistics \citep[see e.g.][]{BuehlmannvandeGeer2011} and can be formulated as follows.    
\begin{definition}\label{def:RE-condition}
A matrix $\bs{A} \in \reals^{nT \times p}$ fulfills the restricted eigenvalue condition $\textnormal{RE}(I,\varphi)$ for some index set $I \subseteq \{1,\ldots,p\}$ and a constant $\varphi > 0$ if  
\[ \| b_{I} \|_1^2 \le \frac{\| \bs{A} b \|^2}{nT} \frac{|I|}{\varphi^2} \qquad \text{for all } b \text{ with } 3 \| b_I \|_1 \ge \| b_{I^c} \|_1. \] 
\end{definition}
\noindent We assume that with probability tending to $1$, the design matrix $\bs{X}^\perp$ satisfies the $\textnormal{RE}(I,\varphi)$ condition for all $I \subseteq \{1,\ldots,p\}$ with $|I| \le 2s$. More formally:
\begin{enumerate}[label=(ID\arabic*),leftmargin=1.15cm]
\setcounter{enumi}{2}
\item \label{C:id3} It holds that 
\[ \pr\Big( \bs{X}^\perp \text{ fulfills } \textnormal{RE}(I,\varphi) \text { for all } I \subseteq \{1,\ldots,p\} \text{ with } |I| \le 2s\Big) \ge 1 - c_{n,T}, \]
where $\varphi > 0$ is a fixed constant and $\{c_{n,T}\}$ is a sequence of non-negative numbers with $c_{n,T} \to 0$. 
\end{enumerate}
Under \ref{C:id3}, the parameter vector $\beta$ is identified in the following sense.
\begin{theorem}\label{lemma:id-beta-large}
Let the technical conditions \ref{C:F}--\ref{C:mixing} and \ref{C:nTp-large}--\ref{C:K-large} from Section \ref{sec:theory} be satisfied in the large-$T$-case and \ref{C:F}--\ref{C:Z} together with \ref{C:nTp-small}--\ref{C:K-small} in the small-$T$-case. If \ref{C:id1}--\ref{C:id3} are fulfilled, then the $s$-sparse parameter vector $\beta$ is unique for sufficiently large $n$. 
\end{theorem}

How reasonable are the restricted eigenvalue conditions on $\bs{X}^\perp$ in \ref{C:id3}? It can be shown that \ref{C:id3} is implied by an analogous assumption on the idiosyncratic matrix $\bs{Z} = (\bs{Z}_1^\top \ldots \bs{Z}_n^\top)^\top$ from equation \eqref{eq:model-CCE-reg}. Specifically, Lemmas \ref{lemma:RE-Z-large-T} and \ref{lemma:RE-Z-small-T} in the supplementary material show that \ref{C:id3} is implied by the following condition:
\begin{enumerate}[label=(ID\arabic*'),leftmargin=1.3cm]
\setcounter{enumi}{2}
\item \label{C:id4} It holds that 
\[ \pr\Big( \bs{Z} \text{ fulfills } \textnormal{RE}(I,\varphi) \text { for all } I \subseteq \{1,\ldots,p\} \text{ with } |I| \le 2s\Big) \ge 1 - c_{n,T}, \]
where $\varphi > 0$ is a fixed constant and $\{c_{n,T}\}$ is a sequence of non-negative numbers with $c_{n,T} \to 0$. 
\end{enumerate}
As the matrix $\bs{Z}$ does not depend on the factors $\bs{F}$, it has a completely standard structure and can be regarded as an ``ordinary'' design matrix in a setting with sample size $nT$ and dimension $p$. Hence, imposing a restricted eigenvalue condition on $\bs{Z}$ is as restrictive or unrestrictive as imposing such a condition on the design matrix in a plain vanilla high-dimensional linear model. Notably, it is possible to verify that $\bs{Z}$ fulfills \ref{C:id4} under certain distributional assumptions. Theorem 1 in \cite{RaskuttiWainwrightYu2010}, for example, shows that \ref{C:id4} is satisfied if the random vectors $Z_{it}$ are independent across $i$ and $t$ and $Z_{it} \sim \normal(0,\bs{\Lambda})$ with $\eig_{\min}(\bs{\Lambda}) \ge c > 0$ and $\max_{1 \le j \le p} \bs{\Lambda}_{jj} \le C < \infty$. This result remains to hold true when the variables $Z_{it}$ are non-Gaussian with sufficiently light tails; see e.g.\ Theorem 7 in \cite{JavanmardMontanari2014}.

\section{Estimation and inference}\label{sec:est}

A very popular technique to estimate the parameter vector $\beta$ in the low-dimen\-sional case is the common correlated effects (CCE) approach of \cite{Pesaran2006}. In the high-dimensional case, however, this estimation technique breaks down and straightforward extensions are not possible. In this section, we construct a novel estimator which does work in high dimensions. As it is similar in spirit to the CCE approach, we call it a high-dimensional CCE estimator, or HD-CCE estimator for short. The section is structured as follows: First, we outline the general strategy to estimate $\beta$ which underlies both our and the CCE approach. We then explain why the CCE estimator collapses in high dimensions. Next, we introduce our estimation approach and give some heuristic discussion why it works. Finally, we explain how to perform inference based on the HD-CCE estimator.

\subsection{A general estimation strategy}

A general strategy to estimate $\beta$ in the panel data model \eqref{eq:model-CCE}--\eqref{eq:model-CCE-reg} with interactive fixed effects is to eliminate or ``project away'' the unknown factors from the model equation by a suitable transformation and then to apply regression techniques to the transformed data.

To formalize this idea, we first consider the oracle case where the factors $\bs{F}$ are observed. In this case, the model equation $Y_i = \bs{X}_i \beta + \bs{F} \gamma_i + \varepsilon_i$ can be regarded as a partitioned regression model, where the factors $\bs{F}$ are additional regressors and the design matrix is given by $(\bs{X}_i \ \bs{F})$. The factors can be eliminated as follows: As already defined above, let $\mathcal{L}_{\bs{F}} = \{ \bs{F} v: v \in \reals^K \}$ be the column space of the factor matrix $\bs{F}$ and 
\[ \bs{\Pi} = \bs{I} - \bs{F} (\bs{F}^\top \bs{F})^{-1} \bs{F}^\top \]
the projection matrix onto the orthogonal complement of $\mathcal{L}_{\bs{F}}$. Since $\bs{\Pi} \bs{F} = \bs{0}$ by construction, we can pre-multiply the model equation by $\bs{\Pi}$ to get that
$\bs{\Pi} Y_i 
  = \bs{\Pi} \bs{X}_i \beta + \bs{\Pi} \bs{F} \gamma_i + \bs{\Pi} \varepsilon_i 
  = \bs{\Pi} \bs{X}_i \beta + \bs{\Pi} \varepsilon_i$, 
thus ``projecting away'' the factors $\bs{F}$. An estimator of $\beta$ can be obtained by applying regression techniques for high-dimensional linear models to the transformed data $\{ (\bs{\Pi} Y_i, \bs{\Pi} \bs{X}_i): 1 \le i \le n \}$. Specifically, running a lasso regression on the transformed data leads to the estimator 
\[ \widehat{\beta}_\pen^{\text{oracle}} \in \underset{b \in \reals^p}{\text{argmin}} \bigg\{ \frac{1}{nT} \sum_{i=1}^n \big\| \bs{\Pi} Y_i - \bs{\Pi} \bs{X}_i b \big\|^2 + \pen \|b\|_1 \bigg\}, \]
where $\pen > 0$ is the penalty constant of the lasso. If $p$ is much smaller than the sample size $nT$ (in particular, in the low-dimensional case with fixed $p$), there is of course no need to work with the lasso. One may rather set the penalty constant $\pen$ to $0$ and use the least squares estimator $\widehat{\beta}_0^{\text{oracle}}$.

Obviously, the oracle estimator $\widehat{\beta}_\pen^{\text{oracle}}$ is not feasible in practice: since the factors $\bs{F}$ are not observed, the projection matrix $\bs{\Pi} = \bs{I} - \bs{F} (\bs{F}^\top \bs{F})^{-1} \bs{F}^\top$ and thus the estimator $\widehat{\beta}_\pen^{\text{oracle}}$ cannot be computed. To obtain a feasible estimator of $\beta$, we need to replace the unknown matrix $\bs{\Pi}$ by a proxy. The construction of such a proxy in high dimensions turns out to be quite intricate. This is the main technical challenge we need to deal with.

\subsection{Breakdown of the CCE estimator in high dimensions}

Before we construct a proxy of $\bs{\Pi}$ in high dimensions, we review the traditional low-dimensional case where (i) the number of regressors $p$ is a fixed natural number, (ii) $p$ is small in the sense that $p < T$, and (iii) the number of factors $K$ is not larger than $p$, that is, $K \le p$.

The CCE approach of \cite{Pesaran2006} provides an elegant way to proxy $\bs{\Pi}$ in this low-dimensional case. For simplicity, we only use the regressors $X_{it}$ for the construction (and thus ignore the responses $Y_{it}$). This gives a clearer picture of the approach and does not affect our argumentation. For a generic random variable $R_{it}$, let $\overline{R}_t = n^{-1} \sum_{i=1}^n R_{it}$ be its cross-sectional average. The CCE approach proxies the projection matrix $\bs{\Pi} = \bs{I} - \bs{F} (\bs{F}^\top \bs{F})^{-1} \bs{F}^\top$ by
\[ \overline{\bs{\Pi}} = \bs{I} - \overline{\bs{X}} (\overline{\bs{X}}^\top \overline{\bs{X}})^{-} \overline{\bs{X}}^\top, \]
where $\overline{\bs{X}} = (\overline{X}_1 \ldots \overline{X}_T)^\top$ is the matrix containing the cross-sectional averages $\overline{X}_t = (\overline{X}_{t,1},\ldots,\overline{X}_{t,p})^\top$ of the regressor variables.
Under suitable regularity conditions, it can be shown that $\overline{\bs{\Pi}} Y_i \approx \overline{\bs{\Pi}} \bs{X}_i \beta + \overline{\bs{\Pi}} \varepsilon_i$ in the low-dimensional case. Hence, pre-multiplying the model equation by $\overline{\bs{\Pi}}$ approximately eliminates the factors. We may thus use $\overline{\bs{\Pi}}$ as an observable proxy of $\bs{\Pi}$ and estimate $\beta$ by applying least squares methods to the sample of transformed data $\{ (\overline{\bs{\Pi}} Y_i, \overline{\bs{\Pi}} \bs{X}_i): 1 \le i \le n \}$.

Why does the CCE approach not work in the high-dimensional case where $p$ is large? In particular, why not simply estimate $\beta$ by applying lasso rather than least squares techniques to the sample of transformed data $\{ (\overline{\bs{\Pi}} Y_i, \overline{\bs{\Pi}} \bs{X}_i): 1 \le i \le n \}$? The problem is that the CCE proxy $\overline{\bs{\Pi}}$ breaks down completely in high dimensions. To see this, consider the following situation: 
\begin{enumerate}[label=(\roman*),leftmargin=1.4cm,parsep=0pt,itemsep=0pt]
\item the number of regressors $p$ is at least as large as $T$, that is, $p \ge T$ 
\item the matrix $\overline{\bs{X}} \in \reals^{T \times p}$  has full rank, that is, $\text{rank}(\overline{\bs{X}}) = T$.  
\end{enumerate}
In this situation, the column space of $\overline{\bs{X}}$ is con\-si\-der\-ably larger than the column space of $\bs{F}$. In particular, the columns of $\overline{\bs{X}}$ span the whole space $\reals^T$. As a consequence, $\overline{\bs{\Pi}} = \bs{I} - \overline{\bs{X}} (\overline{\bs{X}}^\top \overline{\bs{X}})^{-} \overline{\bs{X}}^\top$ is the projection matrix onto the orthogonal complement of $\reals^T$, which is the linear space consisting of the null vector only. Put differently, $\overline{\bs{\Pi}}$ is the null matrix (that is, the matrix with the entry $0$ everywhere), which is obviously an extremely poor proxy of the projection matrix $\bs{\Pi}$. 
These observations point to a general shortcoming of the CCE approach which is well-known in the literature \citep[see][]{Karabiyik2017}: If $p$ is comparably large, the column space of $\overline{\bs{X}}$ tends to be much larger than the column space of $\bs{F}$, implying that $\overline{\bs{\Pi}}$ is a poor proxy of $\bs{\Pi}$. In the worst case scenario, the columns of $\overline{\bs{X}}$ span the whole space $\reals^T$, which means that $\overline{\bs{\Pi}} = \bs{0}$. This worst case occurs whenever $\overline{\bs{X}} \in \reals^{T \times p}$ has full rank $T$. Importantly, this may already happen when $p \ge T$. Hence, the CCE approach runs into trouble not only in the high-dimensional case where $p$ is much larger than $n$ and $T$, but already when $p$ has size comparable to $T$. The larger $p$, the more likely it is that the matrix $\overline{\bs{X}}$ has rank $T$. Hence, in high dimensions, the proxy $\overline{\bs{\Pi}}$ of the CCE approach is not reliable and can be expected to break down frequently.

\subsection{Definition of the HD-CCE estimator}\label{sec:est:def-HDCCE}

We now construct a proxy of the unknown projection matrix $\bs{\Pi}$ which does work in high dimensions and build an estimator of $\beta$ based on it. 

\subsubsection*{Step 1: Estimation of the unknown number of factors $\bs{K}$}

Compute the $p \times p$ matrix 
$\widehat{\bs{\Sigma}} = T^{-1} \sum_{t=1}^T \overline{X}_t \overline{X}_t^\top$
from the cross-sectional averages $\overline{X}_t$ and perform an eigendecomposition of $\widehat{\bs{\Sigma}}$, which yields the eigenvalues $\widehat{\eig}_1 \ge \widehat{\eig}_2 \ge \ldots \ge \widehat{\eig}_p \ge 0$ and the corresponding orthonormal eigenvectors $\widehat{U}_1,\ldots,\widehat{U}_p$. Estimate the unknown number of factors $K$ by 
\[ \widehat{K} = \sum_{j=1}^p \ind\big(\widehat{\eig}_j \ge \tau\big), \]
where $\tau = \tau_{n,T}$ is a threshold parameter that is of slightly smaller order than $p$. Precise technical conditions on $\tau$ can be found in Section \ref{sec:theory} and rules for selecting $\tau$ in practice are discussed in Section \ref{sec:est:tuning}.

\subsubsection*{Step 2: Approximation of the unknown projection matrix $\bs{\Pi}$}

Let $\widehat{\bs{U}} = (\widehat{U}_1 \ldots \widehat{U}_{\widehat{K}})$ be the matrix of eigenvectors of $\widehat{\bs{\Sigma}}$ that correspond to the $\widehat{K}$ largest eigenvalues $\widehat{\eig}_1 \ge \ldots \ge \widehat{\eig}_{\widehat{K}}$ and define $\widehat{\bs{W}} = \overline{\bs{X}} \widehat{\bs{U}}$. Approximate $\bs{\Pi}$ by 
\[ \widehat{\bs{\Pi}} = \bs{I} - \widehat{\bs{W}} (\widehat{\bs{W}}^\top \widehat{\bs{W}})^{-} \widehat{\bs{W}}^\top. \]

\subsubsection*{Step 3: Estimation of $\bs{\beta}$} 

Run a lasso regression on the transformed data sample $\{ (\widehat{Y}_i, \widehat{\bs{X}}_i): 1 \le i \le n \}$, where $\widehat{Y}_i = \widehat{\bs{\Pi}} Y_i$ and $\widehat{\bs{X}}_i = \widehat{\bs{\Pi}} \bs{X}_i$. Specifically, define the lasso estimator of $\beta$ by  
\[ \widehat{\beta}_\pen \in \underset{b \in \reals^p}{\text{argmin}} \bigg\{ \frac{1}{nT} \sum_{i=1}^n \big\| \widehat{Y}_i - \widehat{\bs{X}}_i b \big\|^2 + \pen \|b\|_1 \bigg\}, \]
where $\pen > 0$ is the penalty constant of the lasso.

\subsection{Heuristic idea behind the HD-CCE estimator}

We now give some heuristic arguments why our estimation approach works in high dimensions. We in particular explain why the matrix $\widehat{\bs{\Pi}}$ defined in Step 2 of the algorithm provides a good approximation to the unknown projection matrix $\bs{\Pi}$ even when $p$ is very large. Since the heuristics are essentially the same for large and small $T$, we restrict attention to the large-$T$-case.

Our estimation algorithm is based on the following observation: The cross-sectional averages $\overline{X}_t = n^{-1} \sum_{i=1}^n X_{it}$ satisfy a high-dimensional approximate factor model of the form 
\begin{equation}\label{eq:approx-factor-model-Xbar}
\overline{X}_t = \bs{\Gamma} F_t + u_t \qquad \text{with} \qquad u_t = (\overline{\bs{\Gamma}} - \bs{\Gamma}) F_t + \overline{Z}_t. 
\end{equation}
The error terms $u_t = (u_{t,1},\ldots,u_{t,p})^\top$ in this model are negligible in the sense that $u_{t,j} = o_p(1)$ for any $t$ and $j$ as $n \to \infty$. This directly follows from the fact that under our regularity conditions, $\overline{\Gamma}_j = \Gamma_j + o_p(1)$ and $\overline{Z}_{t,j} = o_p(1)$ for any $t$ and $j$ as $n \to \infty$, where $\Gamma_j$ and $\overline{\Gamma}_j$ denote the $j$-th row of $\bs{\Gamma}$ and $\overline{\bs{\Gamma}}$, respectively. Hence, it holds that $\overline{X}_t \approx \bs{\Gamma} F_t$, or put differently, $\overline{\bs{X}} \approx \bs{F} \bs{\Gamma}^\top$, which means that the variables $\overline{X}_t$ approximately follow a factor model.

In Step 1 of the estimation algorithm, we exploit this observation as follows: As $\overline{X}_t$ satisfies \eqref{eq:approx-factor-model-Xbar}, the matrix $\overline{\bs{\Sigma}} = \ex[T^{-1} \sum_{t=1}^T \overline{X}_t \overline{X}_t^\top]$ is closely related to a high-dimensional covariance matrix in an approximate factor model. Such covariance matrices tend to have spiked eigenvalues as observed and exploited e.g.\ in \cite{Fan2013}. 
We thus expect the eigenvalues of $\overline{\bs{\Sigma}}$ to be spiked as well. More formally, we can show that under our assumptions, the first $K$ eigenvalues of $\overline{\bs{\Sigma}}$ are (at least) of order $p$ (in the sense of being bounded from below by $cp$ for some positive constant $c$ and sufficiently large $n$), whereas the others are of (much) smaller order (in the sense of being $o(p)$).
The eigenvalues $\widehat{\eig}_1 \ge \ldots \ge \widehat{\eig}_p$ of the estimator $\widehat{\bs{\Sigma}} = T^{-1} \sum_{t=1}^T \overline{X}_t \overline{X}_t^\top$ can be shown to behave similarly: whereas the $K$ largest eigenvalues are of order $p$, the others are of considerably smaller order. This suggests to estimate $K$ by thresholding the eigenvalues of $\widehat{\bs{\Sigma}}$. In particular, we may work with the estimator $\widehat{K} = \sum_{j=1}^p \ind(\widehat{\eig}_j \ge \tau)$ introduced in Step 1 of the algorithm.

In Step 2 of the algorithm, we exploit the observation that $\overline{X}_t$ satisfies an approximate factor model as follows: Let $\widehat{\bs{U}} = (\widehat{U}_1 \ldots \widehat{U}_{\widehat{K}})$ be the matrix of eigenvectors of $\widehat{\bs{\Sigma}}$ that correspond to the $\widehat{K}$ largest eigenvalues $\widehat{\eig}_1 \ge \ldots \ge \widehat{\eig}_{\widehat{K}}$. Since $\overline{\bs{X}} \approx \bs{F} \bs{\Gamma}^\top$, it holds that 
\[ \widehat{\bs{\Sigma}} = \frac{\overline{\bs{X}}^\top \overline{\bs{X}}}{T} \approx \bs{\Gamma} \Big(\frac{\bs{F}^\top \bs{F}}{T} \Big) \bs{\Gamma}^\top = \bs{\Gamma} \bs{\Gamma}^\top, \]
where we have used that $\bs{F}^\top \bs{F}/T =  \bs{I}_K$ by \ref{C:id1}. Let $\bs{\Gamma} = \bs{U} \bs{D} \bs{V}^\top$ be the singular value decomposition of $\bs{\Gamma}$, where the matrices $\bs{U} \in \reals^{p \times K}$ and $\bs{V} \in \reals^{K \times K}$ have orthonormal columns and $\bs{D}$ is a diagonal matrix which contains the singular values on its main diagonal. With this decomposition, we further obtain that 
\[ \widehat{\bs{\Sigma}} \approx \bs{\Gamma} \bs{\Gamma}^\top = \bs{U} \bs{D}^2 \bs{U}^\top. \]  
This suggests that the matrix $\widehat{\bs{U}}$ of the first $\widehat{K}$ eigenvectors of $\widehat{\bs{\Sigma}}$ can be regarded as an estimator of the matrix $\bs{U}$ whose columns are the first $K$ eigenvectors of $\bs{\Gamma} \bs{\Gamma}^\top$. So far, we have seen that $\widehat{\bs{U}} \approx \bs{U}$ and $\overline{\bs{X}} \approx \bs{F} \bs{\Gamma}^\top$, which taken together yields that 
\begin{equation}\label{eq:W-What}
\overline{\bs{X}} \widehat{\bs{U}} \approx \bs{F} \bs{\Gamma}^\top \bs{U} = \bs{F} \bs{V} \bs{D}.  
\end{equation}
Since $\bs{V} \bs{D}$ is invertible under the full-rank condition on $\bs{\Gamma}$ in \ref{C:id2}, the $K$ columns of the matrix $\bs{W} := \bs{F} \bs{V} \bs{D}$ span the same linear space as those of $\bs{F}$. Consequently,
\begin{align*}
\bs{\Pi} & = \bs{I} - \bs{F} (\bs{F}^\top \bs{F})^{-1} \bs{F}^\top = \bs{I} - \bs{W} (\bs{W}^\top \bs{W})^{-1} \bs{W}^\top. 
\end{align*}
Moreover, since $\bs{W} \approx \widehat{\bs{W}} := \overline{\bs{X}} \widehat{\bs{U}}$ by \eqref{eq:W-What}, a good proxy of the projection matrix $\bs{\Pi}$ should be given by 
\[ \widehat{\bs{\Pi}} = \bs{I} - \widehat{\bs{W}} (\widehat{\bs{W}}^\top \widehat{\bs{W}})^{-} \widehat{\bs{W}}^\top, \]
which is the proxy defined in Step 2 of the algorithm.

From the heuristic discussion so far, it follows that $\widehat{\bs{\Pi}} \bs{F} \approx \bs{\Pi} \bs{F} = \bs{0}$. Hence, applying the matrix $\widehat{\bs{\Pi}}$ to the model equation $Y_i = \bs{X}_i \beta + \bs{F} \gamma_i + \varepsilon_i$ leads to the transformed (approximate) model equation $\widehat{\bs{\Pi}} Y_i \approx \widehat{\bs{\Pi}} \bs{X}_i \beta + \widehat{\bs{\Pi}} \varepsilon_i$ for each $i$. Stacking these equations for all $i$, we obtain the (approximate) high-dimensional linear panel regression model 
\[ \widehat{Y} \approx \widehat{\bs{X}} \beta + \widehat{\varepsilon} \quad \text{with} \quad \widehat{Y} = \begin{pmatrix} \widehat{\bs{\Pi}} Y_1 \\ \vdots \\ \widehat{\bs{\Pi}} Y_n \end{pmatrix}, \ \widehat{\bs{X}} = \begin{pmatrix} \widehat{\bs{\Pi}} \bs{X}_1 \\ \vdots \\ \widehat{\bs{\Pi}} \bs{X}_n \end{pmatrix} \text{ and } \ \widehat{\varepsilon} = \begin{pmatrix} \widehat{\bs{\Pi}} \varepsilon_1 \\ \vdots \\ \widehat{\bs{\Pi}} \varepsilon_n \end{pmatrix}, \]
which does not have any interactive fixed effects in the errors. To obtain an estimator of $\beta$, we apply standard techniques from high-dimensional linear regression to this transformed model. Specifically, we work with lasso techniques, which leads to the estimator $\widehat{\beta}_\pen$ defined in Step 3 of the algorithm.

\subsection{A least squares version of the HD-CCE estimator}

So far, our discussion has concentrated on the high-dimensional case where $p$ is large and may even exceed the sample size $nT$. However, the CCE approach does not only break down in this high-dimensional setting. It rather becomes unreliable as soon as $p \ge T$. This is particularly problematic when the time series length $T$ is fairly short as often happens in microeconomic applications. In this case, the number of available regressors $p$ easily exceeds $T$, which means that we are faced with the following situation: 
\vspace{-0.5cm}

\begin{equation}\label{eq:micro-situation}
T \text{ is relatively small and } T \le p \ll nT,  
\end{equation}
\vspace{-0.6cm}

\noindent where the symbol $a \ll b$ is here used informally to express that $a$ is considerably smaller than $b$.

In the situation given by \eqref{eq:micro-situation}, the CCE method is essentially inapplicable. Our estimator, in contrast, works perfectly fine. It is also possible to replace it by a least squares version since there is no need to use the lasso when $p \ll nT$. This is done as follows: We construct $\widehat{K}$ and $\widehat{\bs{\Pi}}$ exactly as described in the first two steps of the estimation algorithm. However, instead of using the lasso in the third step, we apply least squares to the transformed data $\{ (\widehat{Y}_i, \widehat{\bs{X}}_i): 1 \le i \le n \}$ with $\widehat{Y}_i = \widehat{\bs{\Pi}} Y_i$ and $\widehat{\bs{X}}_i = \widehat{\bs{\Pi}} \bs{X}_i$. This yields the least-squares-type estimator
\[ \widehat{\beta}_{\text{LS}} \in \underset{b \in \reals^p}{\text{argmin}} \bigg\{ \frac{1}{nT} \sum_{i=1}^n \big\| \widehat{Y}_i - \widehat{\bs{X}}_i b \big\|^2 \bigg\}, \]
which is nothing else than the lasso $\widehat{\beta}_\pen$ with $\pen = 0$.

It depends of course on the specific sizes of $n$, $T$ and $p$ whether it makes more sense to use the lasso $\widehat{\beta}_\pen$ (with some $\pen > 0$) or the least squares version $\widehat{\beta}_{\text{LS}}$. If $p$ is only slightly larger than $T$ in the situation given by \eqref{eq:micro-situation}, there is only a small number of regressors in the model and one may prefer to use the least squares estimator $\widehat{\beta}_{\text{LS}}$. This in particular has the advantage that we do not have to select the penalty parameter $\pen$. If $p$ is substantially larger than $T$, that is, if there is a comparably large number of regressors in the model, one may prefer to use the lasso instead for the following reasons: The least squares estimator can be expected to be outperformed by penalized least squares methods such as the lasso. Moreover, since the lasso performs not only estimation but also variable selection, it produces results that are easier to interpret.

\subsection{The desparsified HD-CCE estimator}\label{sec:inference}

As is well known, the lasso -- and thus in particular our HD-CCE estimator -- has a very complicated limiting distribution which is hardly tractable. For this reason, it cannot be used for statistical inference in practice. A common way to circumvent this issue is to desparsify or debias the lasso; see \cite{vandeGeer2014}, \cite{JavanmardMontanari2014}, \cite{ZhangZhang2014} and \cite{Belloni2014}. In what follows, we demonstrate how desparsified lasso techniques can be applied to our HD-CCE estimator. To do so, we focus on the following inference problem: we want to compute (asymptotic) confidence bands for the coefficient $\beta_j$ of the $j$-th regressor. Our approach to solve this inference problem is as follows.


\subsubsection*{Step 1: Estimation of the projection matrix}

For inference purposes, we need to construct the proxy of the projection matrix $\bs{\Pi}$ slightly differently than we did for estimation purposes. In particular, we need to replace the matrix $\overline{\bs{X}} = n^{-1} \sum_{i=1}^n \bs{X}_i$ by $\overline{\bs{X}}_{(-j)}$ which results from eliminating the $j$-th column of $\overline{\bs{X}}$. This helps us to control certain bias terms in the asymptotic theory. Once this replacement is done, the construction proceeds as before: (i) Compute the $(p-1) \times (p-1)$ matrix $\widetilde{\bs{\Sigma}} = \overline{\bs{X}}_{(-j)}^\top \overline{\bs{X}}_{(-j)}/T $ with eigenvalues $\widetilde{\eig}_1 \geq \ldots \geq  \widetilde{\eig}_{p-1} \geq 0$ and corresponding eigenvectors $\widetilde{U}_{(1)}, \ldots, \widetilde{U}_{(p-1)}$. 
(ii) Estimate $\bs{\Pi}$ by
\begin{align*}
\widetilde{\bs{\Pi}} = \bs{I} - \widetilde{\bs{W}}(\widetilde{\bs{W}}^\top \widetilde{\bs{W}})^{-}\widetilde{\bs{W}}^\top,
\end{align*}
where $\widetilde{\bs{W}} = \overline{\bs{X}}_{(-j)} \widetilde{\bs{U}}$ and $\widetilde{\bs{U}} = (\widetilde{U}_{1} \ldots \widetilde{U}_{\widehat{K}})$ with $\widehat{K}$ as defined before.\footnote{It is possible to replace the estimator $\widehat{K}$ by $\widetilde{K} = \sum_{\ell=1}^{p-1} \ind\big(\widetilde{\eig}_\ell \ge \tau\big)$ with an appropriately chosen threshold sequence $\tau = \tau_{n,T}$. However, there is no need to do so from a theoretical point of view.} 
Notably, the matrix $\widetilde{\bs{\Sigma}}$ depends on $j$. The same holds for the quantities based on it such as $\widetilde{\bs{\Pi}}$, $\widetilde{\bs{W}}$ and $\widetilde{\bs{U}}$ as well as further expressions defined in the subsequent steps. For simplicity of notation, we however suppress the dependence on $j$ throughout. 

\subsubsection*{Step 2: Estimation of the parameter vectors $\bs{\beta}$ and $\bs{\betanw}$}

We estimate $\beta$ as before by applying lasso techniques to the projected sample of data $\{ (\widetilde{Y}_i, \widetilde{\bs{X}}_i): 1 \le i \le n \}$ with $\widetilde{Y}_i = \widetilde{\bs{\Pi}} Y_i$ and $\widetilde{\bs{X}}_i = \widetilde{\bs{\Pi}} \bs{X}_i$. This yields the estimator
\begin{align*}
\widetilde{\beta}_\pen \in \argmin_{b \in\reals^{p}} \left\{\frac{1}{nT}\sum_{i=1}^n \big\|\widetilde{Y}_i - \widetilde{\bs{X}}_i b \big\|^2 + \pen\|b \|_1\right\}.
\end{align*}
Analogously, we estimate the parameter vector $\betanw$ in the nodewise equation by
\begin{align*}
\widetilde{\betanw}_\pennw \in \argmin_{\vartheta \in\reals^{p-1}}\left\{\frac{1}{nT}\sum_{i=1}^n \big\|\widetilde{X}_{i(j)} - \widetilde{\bs{X}}_{i(-j)}\vartheta \big\|^2 + \pennw\|\vartheta \|_1\right\}
\end{align*}
where $\widetilde{X}_{i(j)} = \widetilde{\bs{\Pi}} X_{i(j)}$, $\widetilde{\bs{X}}_{i(-j)} = \widetilde{\bs{\Pi}} \bs{X}_{i(-j)}$ and $\pennw$ is the penalty constant of the lasso.

\subsubsection*{Step 3: Desparsifying the HD-CCE estimator}

Let $\widetilde{\resnw}_i = \widetilde{X}_{i(j)} - \widetilde{\bs{X}}_{i(-j)} \widetilde{\betanw}_{\pennw}$ be the residual vector from the nodewise lasso regression and write $\widetilde{\resnw} = (\widetilde{\resnw}_1^\top,\ldots,\widetilde{\resnw}_n^\top)^\top$. Following the strategy in \cite{vandeGeer2014}, we define the desparsified HD-CCE estimator of $\beta_j$ by
\begin{align*}
\widetilde{b}_j = \widetilde{\beta}_{\pen,j} + \frac{\widetilde{\resnw}^\top(\widetilde{Y}-\widetilde{\bs{X}}\widetilde{\beta}_\pen)}{\widetilde{\resnw}^\top \widetilde{X}_{(j)}}
\end{align*}
with $\widetilde{Y} = (\widetilde{Y}_1^\top, \dots, \widetilde{Y}_n^\top)^\top$, $\widetilde{\bs{X}} = (\widetilde{\bs{X}}_1^\top \dots \widetilde{\bs{X}}_n^\top)^\top$ and $\widetilde{X}_{(j)} = ( \widetilde{X}_{1(j)}^\top, \dots, \widetilde{X}_{n(j)}^\top)^\top$.   
Notably, this estimator is closely related (but not identical) to the double lasso \citep[see][p.106/7]{ChernozhukovHansen2022}, which amounts to a least squares regression of lasso residuals from the main equation on lasso residuals from the nodewise equation.

\subsubsection*{Step 4: Inference with the desparsified HD-CCE estimator}

To perform inference with the desparsified HD-CCE estimator, we consider the statistic
\[ \mathbb{T}_j = \frac{\widetilde{\resnw}^\top \widetilde{X}_{(j)}}{\scaling}(\widetilde{b}_{j} - \beta_j) \quad \text{with} \quad \scaling = \sqrt{\sum_{t,t'=1}^T \bigg\{ \sum_{i=1}^n \widetilde{\resnw}_{it} \widetilde{\resnw}_{it'} \ex[ \varepsilon_{it} \varepsilon_{it'} ] \bigg\}}. \]
This statistic can be treated as approximately standard normal, which is formally justified in Section \ref{sec:theory}. 
Hence, 
\begin{equation} \label{eq:confidence_interval} 
\mathbb{C}_{j,\alpha} = \left[ \widetilde{b}_j + \frac{\scaling \, q_{\frac{\alpha}{2}}}{\widetilde{\resnw}^\top \widetilde{X}_{(j)}}, \ \widetilde{b}_j + \frac{\scaling \, q_{1-\frac{\alpha}{2}}}{\widetilde{\resnw}^\top \widetilde{X}_{(j)}} \right]
\end{equation} 
with $q_\kappa$ the $\kappa$-quantile of the standard normal distribution is an approximate confidence band of level $(1-\alpha)$ for $\beta_j$, that is, $\pr( \beta_j \in \mathbb{C}_{j,\alpha}) \approx 1 - \alpha$.
As the normalization term $\scaling$ in the definition of $\mathbb{T}_j$ is not available in practice, we need to replace it by an estimator $\widetilde{\scaling}$. There are different ways to do so:
\begin{enumerate}[label=(\roman*),leftmargin=0.75cm]

\item If we assume the errors $\varepsilon_{it}$ to be i.i.d.\ not only across $i$ but also across $t$, then $\scaling$ simplifies to $\scaling = \sigma_\varepsilon \| \widetilde{\resnw} \|$, where $\sigma_\varepsilon^2 = \ex[\varepsilon_{it}^2]$ denotes the idiosyncratic error variance. In this case, $\scaling$ can simply be estimated by 
\[ \widetilde{\scaling}^{\hspace{1pt} \text{IID}} = \widetilde{\sigma}_\varepsilon \| \widetilde{\resnw} \| \quad \text{with} \quad \widetilde{\sigma}_\varepsilon^2 = \bigg(\frac{T}{T-\widehat{K}}\bigg) \frac{\| \widetilde{\epsilon} \|^2}{nT}, \]
where $\widetilde{\epsilon} = (\widetilde{\epsilon}_1^\top,\ldots,\widetilde{\epsilon}_n^\top)^\top$ with $\widetilde{\epsilon}_i = \widetilde{\bs{\Pi}} Y_i - \widetilde{\bs{\Pi}} \bs{X}_i \widetilde{\beta}_\pen$ is the vector of lasso residuals from the main model equation and $\widetilde{\sigma}_\varepsilon^2$ can be shown to be a consistent estimator of the error variance $\sigma_\varepsilon^2$ under suitable regularity conditions. 
\item If we allow the errors $\varepsilon_{it}$ to be heteroskedastic across $i$ (in the sense that the error variance $\sigma_{\varepsilon,i}^2 = \ex[\varepsilon_{it}^2]$ may vary across $i$), then $\scaling = \{ \sum_i \sigma_{\varepsilon,i}^2 \|\widetilde{\resnw}_i\|^2\}^{1/2}$. Similarly to case (i), we may estimate $\scaling$ by
\[ \widetilde{\scaling}^{\hspace{1pt} \text{HET}} = \sqrt{\sum_{i=1}^n \widetilde{\sigma}_{\varepsilon,i}^2 \|\widetilde{\resnw}_i\|^2} \quad \text{with} \quad \widetilde{\sigma}_{\varepsilon,i}^2 = \bigg(\frac{T}{T-\widehat{K}}\bigg) \frac{\| \widetilde{\epsilon}_i \|^2}{T}. \]
\item  If we allow the errors $\varepsilon_{it}$ to be both heteroskedastic across $i$ and serially dependent across $t$, we can estimate $\scaling$ by 
the HAC-type estimator 
\[ \widetilde{\scaling}^{\hspace{1pt} \textnormal{HAC}} = \sqrt{\sum_{\substack{1 \le t,t' \le T, \\ |t-t'| \le h_T}} \bigg\{ \sum_{i=1}^n \widetilde{\resnw}_{it} \widetilde{\resnw}_{it'} \widetilde{\epsilon}_{it} \widetilde{\epsilon}_{it'} \bigg\}}, \]
where $h_T$ is a bandwidth parameter (depending only on $T$). Notably, we set $h_T = T$, i.e., we do not regularize at all. This (quite unusual) choice should work well in our framework, at least as long as $T$ is not too large in comparison to $n$. The heuristic reason is as follows: 
If the errors $\varepsilon_{it}$ are independent from the regressors $X_{it}$, we can write $\scaling^2 / (nT)  = \ex_{\bs{X}} [(nT)^{-1} \sum_i (\widetilde{\resnw}_i^\top \varepsilon_i)^2] = \ex_{\bs{X}} [(nT)^{-1} \sum_i (\widetilde{\resnw}_i^\top \widetilde{\varepsilon}_i)^2] = T^{-1} \sum_{t,t'} \ex_{\bs{X}} [n^{-1} \sum_i \widetilde{\resnw}_{it} \widetilde{\resnw}_{it'}$ $\widetilde{\varepsilon}_{it} \widetilde{\varepsilon}_{it'}]$, where $\widetilde{\varepsilon}_i = \widetilde{\bs{\Pi}} \varepsilon_i$ and $\ex_{\bs{X}}$ denotes the conditional expectation given $\bs{X}$. For any fixed pair of time points $(t,t')$, a law-of-large-numbers-type argument suggests that $n^{-1} \sum_i \widetilde{\resnw}_{it} \widetilde{\resnw}_{it'} \linebreak  \widetilde{\varepsilon}_{it} \widetilde{\varepsilon}_{it'} \approx \ex_{\bs{X}}[n^{-1} \sum_i \widetilde{\resnw}_{it} \widetilde{\resnw}_{it'} \widetilde{\varepsilon}_{it} \widetilde{\varepsilon}_{it'}]$. Moreover, if the projected lasso residuals $\widetilde{\epsilon}_{i}$ are close to the projected errors $\widetilde{\varepsilon}_{i}$, we should get that $n^{-1} \sum_i \widetilde{\resnw}_{it} \widetilde{\resnw}_{it'} \widetilde{\epsilon}_{it} \widetilde{\epsilon}_{it'} \approx n^{-1} \sum_i \widetilde{\resnw}_{it} \widetilde{\resnw}_{it'} \widetilde{\varepsilon}_{it} \widetilde{\varepsilon}_{it'}$. Hence, as long as the number of terms in the double sum $\sum_{t,t'}$ is not too large, we expect that 
\[\frac{\scaling^2}{nT} = \frac{1}{T} \sum_{t,t'} \ex_{\bs{X}} \bigg[ \frac{1}{n} \sum_i \widetilde{\resnw}_{it} \widetilde{\resnw}_{it'} \widetilde{\varepsilon}_{it} \widetilde{\varepsilon}_{it'} \bigg] \approx \frac{1}{T} \sum_{t,t'} \bigg\{ \frac{1}{n} \sum_i \widetilde{\resnw}_{it} \widetilde{\resnw}_{it'} \widetilde{\epsilon}_{it} \widetilde{\epsilon}_{it'} \bigg\} = \frac{(\widetilde{\scaling}^{\hspace{1pt} \textnormal{HAC}})^2}{nT} \]
with $h_T = T$. 
\end{enumerate}
In our \texttt{R} package, the normalizations $\widetilde{\scaling}^{\hspace{1pt} \text{IID}}$, $\widetilde{\scaling}^{\hspace{1pt} \text{HET}}$ and $\widetilde{\scaling}^{\hspace{1pt} \text{HAC}}$ are available as different options. In the supplement, we run a number of simulation exercises to explore the performance of the statistic $\mathbb{T}_j$ with these normalizations. In the theoretical analysis of Section \ref{sec:theory}, we focus on the i.i.d.\ case and thus on the normalization $\widetilde{\scaling}^{\hspace{1pt} \text{IID}}$.

\section{Implementation}\label{sec:est:tuning}

The HD-CCE estimator $\widehat{\beta}_\pen$ depends on two tuning parameters: the threshold parameter $\tau$ for the estimation of $K$ and the penalty parameter $\pen$ of the lasso. The desparsified HD-CCE estimator $\widetilde{b}_j$ additionally involves the penalty parameter $\pennw$ from the nodewise lasso regression. We now discuss how to select these tuning parameters in practice.

\subsection{Choice of $\bs{\tau}$}\label{sec:est:tuning:tau}

Our estimator of $K$ is defined as $\widehat{K} = \sum_{k=1}^p \ind(\widehat{\eig}_k \ge \tau)$, where $\widehat{\eig}_1 \ge \ldots \ge \widehat{\eig}_p$ are the eigenvalues of $\widehat{\bs{\Sigma}}$ in descending order. It can be shown formally that the eigenvalues $\widehat{\eig}_k$ are of order $p$ for $k \le K$ but of much smaller order for $k > K$. Hence, to ensure that $\widehat{K}$ is a consistent estimator of $K$, we need to choose $\tau$ such that it separates the ``large'' eigenvalues of order $p$ (that is, those with $k \le K$) from the ``small'' ones (that is, those with $k > K$). As a practical rule-of-thumb, we regard an eigenvalue $\widehat{\eig}_k$ as ``small'' if $\widehat{\eig}_k/\widehat{\eig}_1 < \alpha$ with some small $\alpha$ (such as $\alpha = 0.05$ or $\alpha = 0.01$). Put differently, we regard $\widehat{\eig}_k$ as ``small'' if it is less than $100\cdot\alpha\%$ of the largest eigenvalue $\widehat{\eig}_1$ in size. This rule-of-thumb results in the choice $\tau = \alpha \widehat{\eig}_1$.\footnote{From a theoretical point of view, we need to let $\alpha = \alpha_{n,T}$ slowly go to $0$ with increasing sample size to make sure that $\tau = \alpha \widehat{\eig}_1$ is of somewhat smaller order than $p$ and thus produces a consistent estimator $\widehat{K}$ of $K$. In practice, however, the sample size is fixed, implying that $\alpha$ is a fixed number as well. We thus do not reflect the dependence of $\alpha$ on $n$ and $T$ in the notation.}

The estimator $\widehat{K}$ is closely related to a simple graphical tool that is frequently used in factor analysis: a scree plot which depicts the eigenvalues $\widehat{\eig}_1 \ge \ldots \ge \widehat{\eig}_p$ in descending order. Typically, a large gap or elbow becomes visible in such a plot which allows to distinguish the large eigenvalues from the small ones. The estimator $\widehat{K}$ formalizes this graphical tool by thresholding the eigenvalues. 
There are many alternatives to the estimator $\widehat{K}$. Determining the number of factors is a well-understood problem in factor analysis. See for example \cite{Kapetanios2010} and \cite{Onatski2010} as well as Chapter 6 in \cite{Jolliffe2002} for an overview of common approaches. 

\subsection{Choice of $\bs{\pen}$ for the HD-CCE estimator}\label{sec:est:tuning:lambda}

We choose the penalty parameter $\lambda$ by a version of cross-validation, the details of which are explained below. Another possibility is to adapt selection methods that are based on the effective noise of the lasso \citep[][]{LedererVogt2021} to the setting at hand. Yet another possibility is to adapt the method of \cite{Belloni2016}. This would, however, require to estimate the factors $F_t$ and the loadings $\gamma_i$, which goes a bit against the philosophy of our approach to eliminate or ``project away'' the factors rather than estimate them. 
Generally speaking, it is highly non-trivial to derive theory for data-driven selection of the lasso's tuning parameter already in a plain-vanilla linear model with i.i.d.\ cross-sectional data; see \cite{ChetverikovLiaoChernozhukov2021} for cross-validated lasso and \cite{LedererVogt2021} for effective noise based methods.
We thus take a pragmatic approach to the problem of selecting $\pen$ in this paper: As in most other theoretical treatments of the lasso in the literature, we regard the penalty parameter $\pen$ as a deterministic quantity that converges to $0$ at an appropriate rate when deriving our theory. In the empirical part of the paper, we choose $\pen$ by the following version of $L$-fold cross-validation: Divide the sample $\{(\widehat{Y}_i,\widehat{X}_i): i=1,\ldots,n\}$ of the projected data into $L$ folds $\mathcal{F}_1,\ldots,\mathcal{F}_{L}$, where $\mathcal{F}_{\ell} = \{(\widehat{Y}_i,\widehat{X}_i): i \in \mathcal{I}_\ell \}$ with $\mathcal{I}_\ell = \{ (\ell-1) \lfloor n/L \rfloor + 1,\ldots, \ell \lfloor n/L \rfloor\}$ for $\ell=1,\ldots,L-1$ and $\mathcal{I}_L = \{ (L-1) \lfloor n/L \rfloor + 1,\ldots, n\}$. Then run standard $L$-fold cross-validation over a grid of $\lambda$-values.\footnote{In our \texttt{R} package, we use the grid chosen by the cross-validation function of the \texttt{glmnet} package.}

\subsection{Choice of $\bs{\pen}$ and $\bs{\pennw}$ for the desparsified HD-CCE estimator}\label{sec:est:tuning:node}

We follow the selection strategy advocated in \cite{dezeure2015nodewisechoice}. Specifically, the penalty parameter $\pen$ is chosen by cross-validation as before and the nodewise penalty constant $\pennw$ is selected by the following procedure \citep[see p.554 in][]{dezeure2015nodewisechoice}: 
\begin{enumerate}[label=(\roman*),leftmargin=0.7cm]
\item Run the nodewise lasso regression of $\widetilde{X}_{i(j)}$ on $\widetilde{\bs{X}}_{i(-j)}$ with cross-validated $\pennw$ (where cross-validation is implemented in the same way as for $\pen$) and denote the resulting residual vector by $\widetilde{\resnw}$.
\item Compute $\widetilde{\scaling}^2 /(\widetilde{\resnw}^\top \widetilde{X}_{(j)})^2$ (with $\widetilde{\scaling} = \widetilde{\scaling}^{\hspace{1pt} \text{IID}}$, $\widetilde{\scaling} = \widetilde{\scaling}^{\hspace{1pt} \text{HET}}$ or $\widetilde{\scaling} = \widetilde{\scaling}^{\hspace{1pt} \text{HAC}}$), which estimates the asymptotic variance of the desparsified HD-CCE estimator $\widetilde{b}_j$.
\item Increase the variance by 25\%, i.e., set $V_j = 1.25 \, \widetilde{\scaling}^2 /(\widetilde{\resnw}^\top \widetilde{X}_{(j)})^2$.
\item Let $\widetilde{\resnw}(\pennw)$ be the residual from the nodewise lasso regression carried out with penalty parameter $\pennw$ and let $\widetilde{\scaling}(\pennw)$ be the normalization term computed with $\widetilde{\resnw}(\pennw)$. Select the smallest $\pennw$ such that \[\frac{\widetilde{\scaling}^2(\pennw)}{(\widetilde{\resnw}(\pennw)^\top \widetilde{X}_{(j)})^2} \leq V_j.\]
\end{enumerate}

\subsection{Data normalization}

As the standard lasso, our HD-CCE estimator is not invariant to the scaling of the regressors. In the literature on the lasso, it is common practice to normalize the regressors prior to estimation to have empirically zero mean and unit variance and then to return the estimated coefficients on the original scale. This is, for example, the baseline procedure when fitting the lasso with the very popular \texttt{R} package \texttt{glmnet}. We essentially follow this convention in our \texttt{R} package \texttt{hdcce}: we compute the HD-CCE estimator (and the nodewise estimator for its desparsified variant) from a normalized version of the projected regressors $\widehat{\bs{\Pi}} \bs{X}_i$, but we output the coefficient estimates on the original scale. Specifically, we normalize the projected regressors to have empirical variance $1$ (but we do not centre them as the projection approximately centres them anyway). For simplicity, this normalization step is not reflected in our theory, but it is possible to adjust the theory accordingly.

\section{Theoretical results}\label{sec:theory}

\KOM{[Notation: In some places, double indices $\{c_{n,T}\}$ are used for sequences, in others single indices $\{c_n\}$ are used. Should make the notation more consistent.]}

\subsection{Assumptions for the analysis of the HD-CCE estimator}

The components of model \eqref{eq:model-CCE}--\eqref{eq:model-CCE-reg} are assumed to satisfy the following regularity conditions: 
\begin{enumerate}[label=(M\arabic*),leftmargin=1.1cm]
\item \label{C:F} The factors $F_t$ are deterministic parameters with the property that for some $\moments > 8$,
$\max_{1 \le k \le K} \{ T^{-1} \sum_{t=1}^T |F_{t,k}|^\moments \} \le C < \infty$.
\item \label{C:loadings} The factor loadings $\gamma_i$ and $\bs{\Gamma}_i$ are independent from $Z_{i^\prime t}$ and $\varepsilon_{i^\prime t}$ for all $i$, $i^\prime$ and $t$. Moreover, they are independent across $i$ with means $\gamma = \ex[\gamma_i]$ and $\bs{\Gamma} = \ex[\bs{\Gamma}_i]$. Finally, $\max_{i,j,k} \ex[|\Gamma_{i,jk}|^\moments]\leq C < \infty$ for some $\moments > 8$. 
\item \label{C:eps} The idiosyncratic errors $\varepsilon_{it}$ are independent from $Z_{i^\prime t^\prime}$ for all $i$, $i^\prime$, $t$ and $t^\prime$. Moreover, they are independent across $i$. For all $i$ and $t$, it holds that $\ex[\varepsilon_{it}] = 0$ and $\ex|\varepsilon_{it}|^\moments \le C < \infty$ for some $\moments > 8$. 
\item \label{C:Z} The variables $Z_{it}$ are independent across $i$. For all $i$, $j$ and $t$, it holds that $\ex[Z_{it,j}] = 0$ and $\ex|Z_{it,j}|^{\moments} \le C < \infty$ for some $\moments > 8$. 
\end{enumerate}
In the large-$T$-case, we additionally assume that the model variables form weakly dependent time series processes that satisfy the following mixing conditions:
\begin{enumerate}[label=(M\arabic*),leftmargin=1.1cm]
\setcounter{enumi}{4}
\item \label{C:mixing} \KOM{\color{red} [Need to slightly strengthen this assumption.]} Let $\alpha(m)$ be non-negative real numbers which decay exponentially fast to $0$ as $m \to \infty$, in particular, $\alpha(m) \le C a^m$ for some $0 \le a < 1$ and $C > 0$. 
\begin{enumerate}[label=(\alph*),leftmargin=0.7cm,parsep=0pt,itemsep=0pt,topsep=0pt]
\item For each $i$, the time series $\mathcal{E}_{i,T} = \{ \varepsilon_{it}: 1 \le t \le T \}$ is strongly mixing with mixing coefficients $\alpha_{i,T}^{\varepsilon}(m) \le \alpha(m)$. 
\item For each $i$ and $j$, the time series $\mathcal{Z}_{ij,T} = \{ Z_{it,j}: 1 \le t \le T \}$ is strongly mixing with mixing coefficients $\alpha_{ij,T}^{Z}(m) \le \alpha(m)$. 
\end{enumerate}
\end{enumerate}
\ref{C:F}--\ref{C:mixing} are very similar to the assumptions in \cite{Pesaran2006}. However, unlike there, we do not impose any linearity or stationarity assumptions on the involved time series. 
Notably, the final requirement in \ref{C:F} according to which $\max_{1 \le k \le K} \{ T^{-1} \sum_{t=1}^T |F_{t,k}|^\nu \} \le C < \infty$ is rather mild. If $\{F_{t,k}: t =1,\ldots,T\}$ were a time series of weakly dependent random variables with sufficiently many moments, then standard concentration bounds would imply that $T^{-1} \sum_{t=1}^T |F_{t,k}|^\nu \le C < \infty$ with probability approaching $1$. Hence, if we think of our factors as realizations of such time series, the final requirement in \ref{C:F} will be fulfilled with high probability.  
Combined with \ref{C:id2}, the moment conditions on the entries of the loading matrices $\bs{\Gamma}_i$ in \ref{C:loadings} imply that the factors are pervasive (i.e., each factor must drive a sufficiently large number of regressors), which is a common assumption in the literature \citep[see e.g.][]{Onatski2012}.  
It is in principle possible to drop \ref{C:mixing} in the large-$T$-case and to do without any conditions on the time series dependence of the model variables as in the small-$T$-case. However, then we could not fully account for the time series information in the data. As a consequence, we would obtain a slower convergence rate for our estimator of $\beta$. For simplicity, the mixing coefficients in \ref{C:mixing} are assumed to decay to zero exponentially fast. It is possible though to allow for sufficiently fast polynomial decay instead.

Besides the conditions \ref{C:F}--\ref{C:mixing} on the model components, we need some restrictions on the dimension $p$ and the sparsity index $s$. In the large-$T$-case, we impose the following conditions on the dimension para\-meters $n$, $T$, $p$, $s$ and $K$: 
\begin{enumerate}[label=(D$_{\ell}$\arabic*), leftmargin=1.15cm]
\item \label{C:nTp-large} The dimensions $n$, $T$ and $p$ are such that $n^{(\moments/2)-1}/T \gg p$ and $T^{(\moments/2)-1}/n \gg p$, where $a_{n,p,T} \gg b_{n,p,T}$ means that $b_{n,p,T}/a_{n,p,T} \le C (npT)^{-\xi}$ for some small $\xi > 0$ and $\moments$ is specified in \ref{C:F}--\ref{C:Z}.
\item \label{C:s-large} The set $S=\{j: \beta_j \ne 0\}$ of non-zero components of $\beta$ has cardinality $s := |S|$ with $s = o(\min\{n,T\} / \log(npT))$. 
\item \label{C:K-large} The number of factors $K$ is a fixed natural number with $K < T$ and $K \le p$.
\end{enumerate}
\ref{C:nTp-large} essentially says that $p$ is not allowed to grow too quickly in comparison to $n$ and $T$. To better understand the restrictions on $p$, let us consider the special case $n=T$. In this case, the two restrictions of \ref{C:nTp-large} simplify to $(nT)^{(\moments/4)-1} \gg p$. Hence, how fast $p$ can grow in comparison to the sample size $nT$ depends on how many moments $\moments$ the model variables have. If all moments exist, $\moments$ can be chosen as large as desired and $p$ can grow as any polynomial of $nT$. If $\moments$ is quite small in contrast, say $\moments = 8 + \delta$ for some small $\delta > 0$, then $p$ can only grow slightly faster than the sample size $nT$. 
\ref{C:s-large} imposes constraints on the growth of the sparsity index $s$, that is, on the number of non-zero components of $\beta$. As one can see, $s$ is restricted to grow slightly more slowly than $\min\{n,T\}$. In the special case $n=T$, in particular, $s$ can only grow slightly more slowly than $\sqrt{nT}$. 
In the small-$T$-case, our conditions on the dimension parameters $n$, $T$, $p$, $s$ and $K$ are as follows: 
\begin{enumerate}[label=(D$_s$\arabic*), leftmargin=1.15cm]
\item \label{C:nTp-small} The dimensions $n$ and $p$ are such that $n^{(\moments/4)-1} \gg p$, where $a_{n,p} \gg b_{n,p}$ means that $b_{n,p}/a_{n,p} \le C (np)^{-\xi}$ for some small $\xi > 0$ and $\moments$ is specified in \ref{C:F}--\ref{C:Z}.
\item \label{C:s-small} The set $S=\{j: \beta_j \ne 0\}$ of non-zero components of $\beta$ has cardinality $s := |S|$ with $s = o((np)^{-2/\moments} \sqrt{n/\log p})$. 
\item \label{C:K-small} The number of factors $K$ is a fixed natural number with $K < T$ and $K \le p$.
\end{enumerate}
\ref{C:nTp-small} puts restrictions on the growth of $p$. Analogously to the large-$T$-case, the more moments $\moments$ exist, the faster $p$ is allowed to grow in comparison to $n$. In particular, if all moments exist, then $p$ can grow as any polynomial of $n$. \ref{C:s-small} imposes constraints on the growth of the sparsity index $s$. As can be seen, the more moments $\moments$ exist, the faster $s$ is allowed to increase. In particular, if all moments exist, then $s$ can grow almost as fast as $\sqrt{n}$. In contrast, if only a few moments exist, say $\moments = 8 + \delta$ for some small $\delta > 0$, then $s$ must grow considerably more slowly than $\sqrt{n}$.

All in all, the above conditions on the dimensions $n$, $T$ and $p$ allow us to deal with a wide range of scenarios (as long as the tails of the model variables are not too thick, i.e., as long as sufficiently many moments $\moments$ exist). In particular, we can deal with ``standard low-dimensional'' scenarios where $p$ is small and fixed, with ``moderately high-dimensional'' scenarios where $p$ is fairly large but still smaller than the sample size $nT$ and with ``truly high-dimensional'' scenarios where $p$ exceeds the sample size $nT$. 

\subsection{Convergence rate of the HD-CCE estimator}

We now derive the convergence rate of the HD-CCE estimator $\widehat{\beta}_\pen$. To formulate the result, we let $\{h_n\}$ be any sequence of positive real numbers which slowly diverges to infinity. For instance, we may choose $h_n = C \log \log n$ with some constant $C > 0$.   
\begin{theorem}[Convergence rate of the HD-CCE estimator]\label{theo:rate}
\hfill  
\begin{enumerate}[label=(\alph*),leftmargin=0.7cm]  
\item Consider the large-$T$-case. Assume that \ref{C:F}--\ref{C:mixing}, \ref{C:nTp-large}--\ref{C:K-large}, \ref{C:id1}--\ref{C:id2} and \ref{C:id4} are satis\-fied.
Let the penalty parameter $\pen$ be equal to $\pen = h_n \log(npT) /$ $\min\{n,\sqrt{nT}\}$ and choose the threshold parameter $\tau$ such that $\tau = o(p)$ and $\{p \sqrt{\log p}/\sqrt{n}\}  / \tau = o(1)$. Then 
\[ \| \widehat{\beta}_\pen - \beta \|_1 = O_p \Big( s \, \frac{h_n \log(npT)}{\min\{n,\sqrt{nT}\}} \Big). \] 
\item Consider the small-$T$-case. Assume that \ref{C:F}--\ref{C:Z}, \ref{C:nTp-small}--\ref{C:K-small} and \ref{C:id1}--\ref{C:id3} are satisfied. Let $\pen = h_n (n^2p)^{1/\moments} \sqrt{\log p / n}$ and choose $\tau$ such that $\tau = o(p)$ and $\{p\sqrt{\log p}/\sqrt{n}\} / \tau = o(1)$. Then,
\[ \| \widehat{\beta}_\pen - \beta \|_1 = O_p\Big(s \frac{h_n (n^2p)^{1/\moments}\sqrt{\log p}}{\sqrt{n}} \Big). \] 
\end{enumerate}
\end{theorem}
\noindent The proof of Theorem \ref{theo:rate} is provided in the technical appendices. We briefly give some remarks on the derived convergence rates.

\begin{remark} 
If we replace \ref{C:id4} by \ref{C:id3} in Theorem \ref{theo:rate}(a), we get the same convergence rate but can weaken the restrictions on the sparsity index $s$ a bit. In particular, we can replace the restriction $s = o(\min\{n,T\} / \log(npT))$ in \ref{C:s-large} by $s = o(\min\{n,\sqrt{nT}\} / h_n \log(npT))$. \KOM{[Maybe formulate the result directly in terms of \ref{C:id3} to save space.]} 
\end{remark}

\begin{remark} 
To get some intuition on the rate in the large-$T$-case, it is instructive to consider the special case where $n=T$ and the sparsity index $s$ is a fixed number which does not grow with $n=T$. In this case, the best rate we can hope for is the parametric rate $1/\sqrt{nT}$. According to Theorem \ref{theo:rate}(a), it holds that 
\[ \| \widehat{\beta}_\pen - \beta \|_1 = O_p \Big( \frac{h_n \log(npT)}{\sqrt{nT}} \Big). \] 
Hence, up to the log-factor $h_n \log(npT)$ (where we can e.g.\ choose  $h_n = C \log \log n$), the estimator $\widehat{\beta}_\pen$ attains the parametric rate $1/\sqrt{nT}$. The additional log-factor stems from the fact that the set $S = \{j: \beta_j \ne 0 \}$ of non-zero components of $\beta$ is unknown. If the sparsity index $s = |S|$ grows with $n=T$, it becomes visible in the rate as a multiplicative factor. In particular, the rate changes to $O_p( s  h_n \log(npT)/ \sqrt{nT})$. Both the additional log-factor and the appearance of $s$ as a multiplicative factor in the rate are completely in line with standard theory for the lasso. 
\end{remark}

\begin{remark} Interestingly, the convergence rate in the large-$T$-case is not symmetric in $n$ and $T$: If $n = o(T)$, the rate is $s h_n \log(npT)/ n$. If $T = o(n)$, it is $s h_n \log(npT)/ \sqrt{nT}$ in contrast (rather than $s h_n \log(npT)/ T$). The reason is that the time series and the cross-section direction do not play the same role in the construction of the estimator $\widehat{\beta}_\pen$. In particular, the construction of the projection matrix $\widehat{\bs{\Pi}}$ involves computing cross-sectional averages $\overline{X}_t$ of the regressors, whereas time series averages do not come into play. This gets reflected by an asymmetric dependence of the convergence rate on $n$ and $T$. 
Notably, there is a simple intuition why we should get the rate $s h_n \log(npT)/ \sqrt{nT}$ in the case with $T =o(n)$ (rather than the rate $s h_n \log(npT)/ T$): In the small-$T$-case where $T$ is a fixed natural number, the best rate we can hope for is the standard parametric rate $1/\sqrt{n}$ (neglecting log-factors and the sparsity index $s$). In the large-$T$-case where $T \to \infty$, in contrast, we obtain more and more time series information that we can exploit. Intuitively, this additional information should get reflected in a better rate. Hence, we should be able to obtain a faster rate than $1/\sqrt{n}$ in the large-$T$-case even if $T$ grows very slowly in comparison to $n$. This intuition is indeed correct: Even if $T$ is of much smaller order than $n$, Theorem \ref{theo:rate}(a) yields the rate $1/ \sqrt{nT}$ (neglecting the log-factor $h_n \log(npT)$ and the multiplicative factor $s$), which is faster than $1/\sqrt{n}$.
\end{remark}

\begin{remark}
Unlike in the large-$T$-case, the convergence rate in the small-$T$-case depends on how many moments $\moments$ the model variables have. In particular, the more moments $\moments$ exist, the faster the rate. In the extreme case where the model variables have all moments and $\moments$ can thus be chosen as large as desired, Theorem \ref{theo:rate}(b) yields the rate 
\[ \| \widehat{\beta}_\pen - \beta \|_1 = O_p\Big(s \frac{h_n n^\delta \sqrt{\log p}}{\sqrt{n}} \Big), \]
where $\delta > 0$ is an arbitrarily small constant. The estimator $\widehat{\beta}_\pen$ thus converges to $\beta$ at the fast parametric rate $1 / \sqrt{n}$ (up to the slowly diverging factor $h_n n^\delta \sqrt{\log p}$ and the multiplicative factor $s$). If only a small number of moments $\moments$ exist, in contrast, the rate is significantly slowed down by the multiplicative factor $(n^2p)^{1/\moments}$. 
\end{remark}

\begin{remark}
Why does the convergence rate in the small-$T$-case depend on the number of moments $\moments$? 
In the proof of Theorem \ref{theo:rate}, we need to analyze statistics of the form 
\[ \mathcal{S} := \max_{1 \le i \le n} \Big| \frac{1}{\sqrt{T}} \sum_{t=1}^T V_{it} \Big|, \]
where the random variables $V_{it}$ are independent across $i$ (and weakly dependent across $t$ in the large-$T$-case). In Lemmas \ref{lemma:aux1} and \ref{lemmaprime:aux1}, for instance, we consider such a statistic with $V_{it} = F_{t,k} \varepsilon_{it}$. The behaviour of the statistic $\mathcal{S}$ is very different depending on whether $T$ is bounded or tends to infinity: 
In the large-$T$-case ($T \to \infty$), we can invoke a suitable central limit theorem to show that for each $i$, the statistic $\mathcal{S}_i := T^{-1/2} \sum_{t=1}^T  V_{it}$ is (asymptotically) normally distributed. Hence, $\mathcal{S} = \max_{1 \le i \le n} |\mathcal{S}_i|$ is (approximately) the maximum over (absolute values of) normally distributed random variables, which suggests that $\mathcal{S} = O_p(\sqrt{\log n})$. 
In the small-$T$-case ($T$ fixed), in contrast, the distribution of the statistics $\mathcal{S}_i$ strongly depends on the distribution of the variables $V_{it}$ (and may be far from normal). Moreover, the behaviour of the maximum statistic $\mathcal{S} = \max_{1 \le i \le n} |\mathcal{S}_i|$ strongly depends on the distribution of the variables $\mathcal{S}_i$, in particular, on how many moments they have (or put differently, on how thick their tails are). This gets reflected in the convergence rate of $\widehat{\beta}_\pen$ in the small-$T$-case. 
\end{remark}


\begin{remark}
Theorem \ref{theo:rate} focuses on the $\ell_1$-rate of the HD-CCE estimator. However, because of the norm inequality $\|\widehat{\beta}_\lambda - \beta\|_2 \leq \|\widehat{\beta} - \beta\|_1$, it also yields a bound on the $\ell_2$-rate. This bound on the $\ell_2$-rate can be improved if we replace the restricted eigenvalue condition from Definition \ref{def:RE-condition} by the more restrictive version
\begin{align*}
\| b \|_2^2 \le \frac{1}{\varphi^2}\frac{\| \bs{A} b \|^2}{ nT}  \qquad \text{for all } b \text{ with } 3 \| b_I \|_1 \ge \| b_{I^c} \|_1,
\end{align*}
where we use the same notation as in Definition \ref{def:RE-condition}. Under this stronger condition, a slightly modified version of the proof of Theorem \ref{theo:rate} yields the sharper $\ell_2$-bound
\begin{align*}
\|\widehat{\beta}_\lambda - \beta\|_2 = O_p\left(\sqrt{s} \pen \right)
\end{align*}
with $\pen$ chosen as in Theorem \ref{theo:rate}. 
\end{remark}

\subsection{Assumptions for the analysis of the desparsified HD-CCE estimator}

In addition to \ref{C:F}--\ref{C:mixing}, we impose the following conditions on the components in model equations \eqref{eq:model-CCE}, \eqref{eq:model-CCE-reg} and \eqref{eq:model-CCE-nodewise}: 
\begin{enumerate}[label=(M\arabic*),leftmargin=1.1cm]
\setcounter{enumi}{5}
\item \label{C:eps-stronger} The error variables $\varepsilon_{it}$ are independent and identically distributed across $i$ and $t$. The error variance is denoted by $\sigma_\varepsilon^2 = \ex[\varepsilon_{it}^2]$.
\item \label{C:nodewise-error} The nodewise error terms $u_{it}$ are independent across $i$ and $t$ with $\ex[u_{it}] = 0$ and $\ex[|u_{it}|^\moments]\leq C< \infty$ for some $\moments > 8$. Moreover, the error vector $u_i$ is independent of $\bs{\Gamma}_{i'},\gamma_{i'},\varepsilon_{i'}, \bs{Z}_{i'(-j)}$ and $\nu_{i'}$ for all $i$ and $i'$. 
\item \label{C:nodewise-loadings} The factor loadings $\nu_i$ are independent across $i$. Moreover, it holds that $\max_{1 \le k \le K} \ex[|\nu_{i,k}|^{\moments}]\leq C <\infty$ for some $C>0$ and $\moments>8$.
\KOM{ \newline [Double-check that this is sufficient for \\
$\sum_{n=1}^\infty \pr\left(\frac{1}{n}\sum_{i=1}^n \| \nu_i\|_2^2 > C\right)< \infty$ \\
$\sum_{n=1}^\infty \pr\left(\frac{1}{n}\sum_{i=1}^n \| \nu_i\|_2 > C\right)< \infty$ \\ 
$\sum_{n=1}^\infty\pr(\max_{i,k} |\nu_{ik}| > C(npT)^{\frac{2+\xi}{\moments}})<\infty$ with $\xi>0$ small.]} 
\end{enumerate}
Assumptions \ref{C:eps-stronger} and \ref{C:nodewise-error} require the error terms $\varepsilon_{it}$ and $u_{it}$ to be independent both across $i$ and $t$. We impose these rather strong independence conditions to avoid certain complications in the extremely technical derivation of the limit distribution of the desparsified HD-CCE estimator. However, we conjecture that it is possible to weaken these conditions (in particular, to allow for weak dependence across $t$ and non-identical distributions across $i$). In the simulations, we assess the performance of our inference methods in situations where the errors are not i.i.d. The assumptions on the factor loadings $\nu_i$ in \ref{C:nodewise-loadings} are rather mild.

Conditions \ref{C:nTp-large}--\ref{C:K-large} and \ref{C:nTp-small}--\ref{C:K-small} on the dimension para\-meters $n$, $T$, $p$, $s$ and $K$ are sufficient to guarantee a good convergence behaviour of the HD-CCE estimator. For inference purposes, however, we require additional constraints on them. Specifically, in the large-$T$-case, we assume the following: 
\begin{enumerate}[label=(D$_{\ell}$\arabic*), leftmargin=1.15cm]
\setcounter{enumi}{3}  
\item \label{C:nTp-large-INF} 
Let $\pennw = C_{\pennw} \sqrt{\log(np^2)} \log(npT) [ (nT)^{-1/2} + n^{-1} ]$ with a sufficiently large constant $C_{\pennw}$, $\{h_n\}$ any sequence of positive real numbers which slowly diverges to infinity, and $\xi>0$ an arbitrarily small but fixed constant. It holds that 
\begin{enumerate}[label=(\alph*),leftmargin=0.7cm]
\item \label{C:nTp-large-INF-a} $\displaystyle{\frac{\sqrt{T}\log(pT)}{\sqrt{n}} = o(1)}$ and $\displaystyle{\frac{T}{n^{\moments/2 - 1}} (npT)^{2+\xi}} = o(1)$, 
\item \label{C:nTp-large-INF-b} $\displaystyle{\| \betanw \|_0 = o\bigg( \min \Big\{\frac{1}{\pennw}, \frac{1}{\pennw^2 T \log(pT)}\Big\} \bigg)}$ and $\displaystyle{s = o \bigg( \frac{1}{h_n \log(npT) \pennw} \bigg)}$.
\end{enumerate}
\end{enumerate}
The first part of \ref{C:nTp-large-INF-a} requires that $T$ diverges somewhat more slowly than $n$. Hence, for our inference theory, we need to restrict the growth of $T$ more strongly than for the convergence analysis of the HD-CCE estimator. The second part of \ref{C:nTp-large-INF-a} is fulfilled, e.g., if $p$ grows at most polynomially in $nT$ and the number of moments $\moments$ of the model variables is sufficiently large. Given \ref{C:nTp-large-INF-a}, the growth conditions on the sparsity indices $s$ and $\|\betanw\|_0$ in \ref{C:nTp-large-INF-b} are satisfied if $s$ and $\| \betanw \|_0$ diverge a bit more slowly than $\sqrt{nT}$.     
In the small-$T$-case, we impose additional constraints similar to those in \ref{C:nTp-large-INF}:
\begin{enumerate}[label=(D$_s$\arabic*), leftmargin=1.15cm]
\setcounter{enumi}{3}
\item \label{C:np-small-INF} Let $\pennw = C_{\pennw} \sqrt{\log(p)/n}(np)^{(4+2\xi)/\moments}$ with $C_{\pennw} > 0$ sufficiently large and $\xi>0$ an arbitrarily small but fixed constant. It holds that 
\begin{enumerate}[label=(\alph*),leftmargin=0.7cm]
\item \label{C:np-small-INF-a} $\displaystyle{\{n^{3+\xi}p^{2+\xi}\}/n^{\moments/2} = o(1)}$,
\item \label{C:np-small-INF-b} $\displaystyle{\|\betanw\|_0  = o(\pennw^{-1})}$ and $\displaystyle{s = o \bigg( \frac{\sqrt{n}}{h_n \log(p) (n^2p)^{1/\moments} (np)^{(4+2\xi)/\moments}} \bigg).}$
\end{enumerate}
\end{enumerate}
Condition \ref{C:np-small-INF-a} is satisfied if $p$ grows at most polynomially in $n$ and the number of moments $\moments$ is large enough. Moreover, if $\moments$ can be chosen as large as desired (meaning that all moments of the model variables exist), the sparsity indices $s$ and $\|\betanw\|_0$ can grow almost as fast as $\sqrt{n}$ under condition \ref{C:np-small-INF-b}.

We finally need to strengthen conditions \ref{C:id2} and \ref{C:id3} a bit. We in particular impose the following constraints additional to them:
\begin{enumerate}[label=(INF\arabic*), leftmargin=1.4cm]
\item \label{C:Gamma-INF} The matrix of factor loadings $\bs{\Gamma}_{-j} = \ex\left[ \bs{\Gamma}_{i,-j}\right] \in \reals^{(p-1)\times K}$ has full rank $K$ and its eigenvalues have the property that 
\[ 0 < c_{\textnormal{min}} \leq \lambda_{\textnormal{min}}\left(\frac{\bs{\Gamma}_{-j}^\top\bs{\Gamma}_{-j}}{p} \right) \leq \lambda_{\textnormal{max}}\left(\frac{\bs{\Gamma}_{-j}^\top\bs{\Gamma}_{-j}}{p} \right) < c_{\textnormal{max}} <\infty. \]
\item\label{C:RE-INF} Let $\widetilde{\mathcal{T}}_{\textnormal{RE}}$ be the event that $\widetilde{\bs{X}}$ satisfies the restricted eigenvalue condition $\textnormal{RE}(S,\phi)$ with some $\phi > 0$. There exists a sequence $\{c_{n}\}$  of non-negative real numbers with $\sum_{n=1}^\infty c_{n} < \infty$ such that $\pr(\widetilde{\mathcal{T}}_{\textnormal{RE}}) \ge 1 - c_{n}$ for all $n$. 
\item\label{C:RE-nodewise-INF} Let $\widetilde{\mathcal{T}}_{\textnormal{RE}}^{\textnormal{node}}$ be the event that $\widetilde{\bs{X}}_{(-j)}$ satisfies the restricted eigenvalue condition $\textnormal{RE}(\textnormal{supp}(\betanw), \phi)$ with some $\phi > 0$, where $\textnormal{supp}(\betanw) = \{j: \betanw_j \ne 0 \}$. There exists a sequence $\{c_{n}\}$ of non-negative real numbers with $\sum_{n=1}^\infty c_{n} < \infty$ such that $\pr(\widetilde{\mathcal{T}}_{\textnormal{RE}}^{\textnormal{node}}) \ge 1 - c_{n}$ for all $n$.
\end{enumerate}
As in the analysis of the HD-CCE estimator, we could impose the restricted eigenvalue conditions on the matrices $\bs{X}^\perp$ and $\bs{X}_{(-j)}^\perp$ (or on the idiosyncratic parts $\bs{Z}$ and $\bs{Z}_{(-j)}$) rather than directly on $\widetilde{\bs{X}}$ and $\widetilde{\bs{X}}_{(-j)}$ at the cost of additional technical arguments. However, as these technical arguments are completely analogous to those in the analysis of the HD-CCE estimator, we work with assumptions \ref{C:RE-INF} and \ref{C:RE-nodewise-INF} for the sake of simplicity. 

\subsection{Limit distribution of the desparsified HD-CCE estimator}

We now show that the desparsified HD-CCE estimator is asymptotically normal both in the large-$T$ and the small-$T$-case. As before, we let $\{h_n\}$ be any sequence of positive real numbers which slowly diverges to infinity (e.g., $h_n = C \log \log n$ with some constant $C > 0$).

\begin{theorem}[Asymptotic normality of the desparsified HD-CCE estimator]\label{theo:normality}
\hfill  
\begin{enumerate}[label=(\alph*),leftmargin=0.7cm]  
\item Consider the large-$T$-case and let \ref{C:F}--\ref{C:nodewise-loadings}, \ref{C:nTp-large}--\ref{C:nTp-large-INF}, \ref{C:id1}--\ref{C:id3} as well as \ref{C:Gamma-INF}--\ref{C:RE-nodewise-INF} be satisfied. Let the penalty parameters $\pen$ and $\pennw$ be equal to $\pen =  h_n \log(npT)/\min\{n,\sqrt{nT}\}$ and $\pennw = C_{\pennw} \sqrt{\log(np^2)} \log(npT) [ (nT)^{-1/2} + n^{-1} ]$. Moreover, choose the threshold parameter $\tau$  such that $\tau = o(p)$ and $\{ p \sqrt{\log p}/\sqrt{n} \} / \tau = o(1)$.   
\item Consider the small-$T$-case and let \ref{C:F}--\ref{C:Z}, \ref{C:eps-stronger}--\ref{C:nodewise-loadings}, \ref{C:nTp-small}--\ref{C:np-small-INF}, \ref{C:id1}--\ref{C:id3} and \ref{C:Gamma-INF}--\ref{C:RE-nodewise-INF} be satisfied. Moreover, let $\pen =  h_n (n^2p)^{1/\moments} \sqrt{\log p / n}$, $\pennw = C_{\pennw} \sqrt{\log(p)/n}(np)^{(4+2\xi)/\moments}$ with some arbitrarily small but fixed $\xi > 0$, and choose $\tau$ such that $\tau = o(p)$ and $\{ p \sqrt{\log p}/\sqrt{n} \} / \tau = o(1)$.
\end{enumerate}
In both cases (a) and (b), it holds that 
\begin{align*}
\frac{\widetilde{\resnw}^\top \widetilde{X}_{(j)}}{\widetilde{\scaling}^{\hspace{1pt} \textnormal{IID}}}(\widetilde{b}_{j} - \beta_j) \convd \normal(0,1)
\end{align*}
with $\widetilde{\scaling}^{\hspace{1pt} \textnormal{IID}} = \widetilde{\sigma}_\varepsilon \lVert\widetilde{\resnw}\rVert$.
\end{theorem}
\noindent The proof of Theorem \ref{theo:normality} is given in the technical appendices.

\section{Simulations}\label{sec:sim-overview}

In the supplementary material, we evaluate our estimation and inference methods by Monte Carlo experiments. 
In the first part of the simulation study, we examine the estimation performance of the HD-CCE estimator (and its least squares version) in various low- and high-dimensional scenarios. The Monte Carlo experiments show that the estimator performs well, being almost as accurate as certain oracle methods which presuppose knowledge of the true projection matrix $\bs{\Pi}$ and/or the true active set $S = \{j: \beta_j \ne 0 \}$. 
In the second part, we evaluate the size and power properties of the inference procedures that are based on the desparsified HD-CCE estimator. The simulation results demonstrate accurate size control as well as good power numbers in both low and high dimensions. 
Finally, in the third part, we run various robustness checks: we demonstrate that our methods are robust to moderate overestimation of $K$, we investigate what happens when some factors are much stronger than others, and we demonstrate that our inference methods are robust to heteroskedasticity and serial correlation of the idiosyncratic errors.

\section{Empirical study}\label{sec:application}

\begin{table}[b!]
\caption{Ticker symbols and company names of stocks used in the application.}
\label{tab:Stock_list_Application} 
\hspace{0.5cm}
\scriptsize{
\begin{tabular}{p{1.2cm}p{5.25cm}p{1.2cm}p{5.25cm}}
\textbf{Ticker}  &  \textbf{Company} & \textbf{Ticker}  &  \textbf{Company} \\[0.1cm]
\textit{AAPL}  & Apple Inc.  & \textit{INTC}  & Intel Corporation  \\
\textit{AMGN}  & Amgen Inc.  & \textit{JNJ}  & Johnson \& Johnson  \\
\textit{AMZN}  & Amazon.com, Inc. & \textit{JPM}   & JPMorgan Chase \& Co.  \\
\textit{AXP}   & American Express Company & \textit{KO}  & The Coca-Cola Company    \\
\textit{BA}  & The Boeing Company  & \textit{MCD}  & McDonald's Corporation  \\
\textit{CAT}  & Caterpillar Inc.  & \textit{MMM}  & 3M Company  \\
\textit{CRM}   & Salesforce, Inc. & \textit{MSFT}  & Microsoft Corporation   \\
\textit{CSCO}  & Cisco Systems, Inc. & \textit{NKE}   & Nike, Inc.  \\
\textit{CVX}  & Chevron Corporation  & \textit{PFE}  & Pfizer Inc.  \\
\textit{DIS}  & The Walt Disney Company  & \textit{PG}   & The Procter \& Gamble Company  \\
\textit{GE}  & General Electric Company & \textit{TRV}   & The Travelers Companies, Inc.  \\
\textit{GS}   & The Goldman Sachs Group, Inc. &  \textit{UNH} & UnitedHealth Group Incorporated 
\\ \textit{HD}  & The Home Depot, Inc. & \textit{VZ}  & Verizon Communications Inc.  \\
\textit{HON}  & Honeywell International Inc. & \textit{WMT}  & Walmart Inc.  \\	  
\textit{IBM}   & Internat.\ Business Machines Corp. & & \\[0.1cm]
\end{tabular}}
\end{table}


We apply our methods to a financial dataset that has been the subject of much previous work. The goal is to identify firm characteristics that are informative about the firm's stock return. We suppose that the monthly excess stock return $R_{it}$ of firm $i$ at time $t$ satisfies the model equation
\begin{equation}\label{eq:model-app}
R_{it}=\mu_i + \beta^{\top} C_{i,{t-1}}+\sum_{k=1}^{K}\gamma _{i,k}F_{t,k}+\varepsilon_{it} 
\end{equation}
for $1 \le i \le n$ and $1 \le t \le T$, where the latent factors $F_{t,k}$ capture the comovement of returns, $\mu_i$ is a stock-specific mean and {$C_{i,{t-1}}$} is a vector of ``slowly moving'' firm characteristics observed at time point $t-1$. Using the notation $\gamma_{i,0} := \mu_i$ and $F_{t,0} := 1$ for all $t$, model \eqref{eq:model-app} can be reformulated as 
\begin{equation}\label{eq:model-app1}
R_{it} = \beta^\top C_{i,{t-1}} + \sum_{k=0}^{K}\gamma _{i,k}F_{t,k}+\varepsilon_{it}
\end{equation}
and thus corresponds to our main model equation \eqref{eq:model-CCE} with $Y_{it}=R_{it}$ and $X_{it}=C_{i,{t-1}}$. \cite{Daniel1997} called \eqref{eq:model-app} the characteristic-based pricing model and interpret $\mu_{it}:=\mu_i + \beta^{\top}C_{i,{t-1}}$ as the expected return, although they specified only a scalar attribute $C_{i,{t-1}}$ at a time and work with observed Fama French factors. \cite{Chen2023} work with a similar model and consider a large number of firm-specific characteristics, but their methodology is portfolio sorting based. Our work is also closely related to \cite{Green2017} whose data we use. Model \eqref{eq:model-app} is in the spirit of much recent ``machine learning'' work in finance, see e.g.\ Gu et al.\ (2020, eq.\ 22 and 23) \nocite{Gu2020} and Nagel (2021, eq.\ 3.1 and 3.2) \nocite{Nagel2021} and the discussion therein.

\begin{table}[b!]
\scriptsize
\caption{Firm characteristics used in the application. The table contains the acronym and a brief description of the considered characteristics. More details on the variable definitions and constructions can be found in \cite{Green2017}. }\label{tab:Var_list_Application} 
\begin{tabular}{@{} p{2.1cm} p{4.95cm}}
\textbf{Acronym} & \textbf{Description} \\[0.1cm]
\textit{mve}  & size  \\
\textit{lev}  & leverage \\
\textit{saleinv}  & sales to inventory  \\
\textit{depr}  & depreciation to PP \& E \\ 
\textit{mve\_ia}  & size (ind adj)  \\
\textit{baspread}   & bid-ask spread \\ 
\textit{std\_turn}  & turnover volatility  \\
\textit{grltnoa}  & growth in long term net-op.\ assets \\
\textit{beta}   & beta \\
\textit{pchsale\_pchxsga}  & \% change in sales growth \\ & - \% change in overheads \\
\textit{rd\_sale}  & R\&{}D to sales  \\
\textit{sfe}  & scaled earnings forecast  \\
\textit{zerotrade}  & (weighted) days with zero trades  \\
\textit{divo}  & dividend omission \\
\textit{chinv}   & change in inventory \\  
\textit{roeq}  & return on equity  \\
\textit{std\_dolvol} & volume volatility  \\
\textit{ps}   & financial health score  \\
\textit{tang}   & asset tangibility  \\
\textit{lgr}  & growth in liabilities \\
\textit{absacc}  & absolute accruals \\ 
\textit{acc}   & working capital accruals \\ 
\textit{aeavol}   & abnormal earnings announcement vol.\ \\  
\textit{age}  & years since first Compustat coverage \\
\textit{agr}   & asset growth \\
\textit{betasq}   & beta squared \\
\textit{bm}   & book-to-market \\ 
\textit{bm\_ia}  & book-to-market (ind adj) \\ 
\textit{cash}  & cash to assets \\ 
\textit{cashdebt}   & cash flow to debt \\ 
\textit{cashpr}  & cash productivity \\ 
\textit{cfp}  & operating cash flow to price \\  
\textit{cfp\_ia}   & operating cash flow to price (ind adj) \\ 
\textit{chatoia}   & change in asset turnover (ind adj) \\  
\textit{chcsho}  & change in outstanding shares \\  
\textit{chempia}  & change in employees (ind adj) \\ 
\textit{chfeps}  & change in EPS forecast \\ 
\textit{chnanalyst}  & change in analyst coverage \\ 
\textit{chpmia} & change in profit margin (ind adj) \\ 
\textit{chtx}   & change in taxes \\
\textit{cinvest}   & corporate investment \\ 
\textit{convind}   & convertible debt indicator \\  
\textit{currat}   & current ratio \\ 
\textit{disp}   & EPS forecast dispersion \\
\textit{dolvol}  & dollar volume \\
\end{tabular}
\begin{tabular}{p{2.1cm} p{4.95cm}}
\textbf{Acronym} & \textbf{Description} \\[0.1cm]
\textit{dy}   & dividend yield \\
\textit{egr}   & growth in book value of equity \\
\textit{ep}  & earning to price \\ 
\textit{fgr5yr}  & 5yr EPS growth forecast \\
\textit{gma}  & gross profits to assets \\
\textit{grcapx}   & growth in capital expenditure \\ 
\textit{herf}  & industry sales concentration \\
\textit{hire}  & employment growth \\
\textit{idiovol}    & idiosyncratic return volatility \\
\textit{indmom}  & industry momentum \\ 
\textit{invest}  & capital expenditures and inventory \\ 
\textit{ms} & financial performance score \\  
\textit{nanalyst}   & analyst coverage  \\
\textit{nincr}  & length of earnings run   \\
\textit{operprof}   & operating profits / book equity   \\
\textit{orgcap}  & organizational capital  \\
\textit{pchcapx\_ia}   & \% change in capital exp. (ind adj)   \\
\textit{pchcurrat}  & \% change in current ratio  \\
\textit{pchdepr}   & \% change in depreciation to PP\&{}E   \\
\textit{pchgm\_pchsale}  &  \% change in sales \\
\textit{pchquick} &  \% change in quick ratio \\
\textit{pchsale\_pchinvt}  &    \% change in inventory   \\
\textit{pchsale\_pchrect} &    \% change in receivables    \\
\textit{pchsaleinv} & \% change in sales-to-inventory   \\
\textit{pctacc}  & percent accruals  \\
\textit{pricedelay}  &  price delay \\
\textit{quick}   & quick ratio  \\
\textit{rd}  & large R\&{}D increase  \\
\textit{rd\_mve}  & R\&{}D to size   \\
\textit{realestate}   & real estate holdings  \\
\textit{roaq}   & return on assets  \\
\textit{roavol}  & earnings volatility \\
\textit{roic}   & return on invested capital  \\
\textit{rsup}  & revenue surprise  \\
\textit{salecash}   & sales to cash  \\
\textit{salerec}   & sales to receivables  \\
\textit{secured}   & secured debt  \\
\textit{securedind}  & secured debt indicator \\
\textit{sgr}   & sales growth  \\
\textit{sp}  & sales to price  \\
\textit{stdacc}  & accrual volatility \\
\textit{stdcf}   & cash flow volatility  \\
\textit{sue}  & surprise earnings  \\
\textit{tb} & tax to book income  \\
\textit{turn}  & turnover  \\
 & \\
\end{tabular} 
\end{table}

We apply model \eqref{eq:model-app} to a sample of large cap stocks {($n=29$)} from April 2017 to March 2022 {($T=60$)} that are or recently were constituent stocks of the Dow Jones Industrial Average. The names and ticker symbols of the stocks are given in Table \ref{tab:Stock_list_Application}. 
The firm characteristics we use in the application including a brief description are listed in Table \ref{tab:Var_list_Application}. They comprise a subset of the characteristics collected by \cite{Green2017} for the period 1980--2014, which have been extended to 2022 by Shaoran Li, who kindly shared the data with us. We refer to the appendix of \cite{Green2017} for details on their precise definitions. In comparison to the full set of $102$ characteristics collected by \cite{Green2017}, we drop two characteristics that are constant over the sample period, nine characteristics that are calculated from past stock returns, and one characteristic that is zero for most firms, leaving us with a total of $p=90$ characteristics.\footnote{The two constant characteristics were \textit{ipo}, indicating an initial stock issue, and \textit{sin}, indicating stocks of firms that operate in a sinful industry. Our sample contained neither initial stock issues nor firms from a sinful industry.
The nine characteristics that are calculated from past returns include so-called momentum measures that correspond to the cumulative past returns over a given period (\textit{mom1m}, \textit{mom6m}, \textit{mom12m}, \textit{mom36m}), the change in certain momentum measures (\textit{chmom}), past returns at specific time points (\textit{ear}), the maximum daily return (\textit{maxret}) and the variation of daily returns (\textit{retvol}) over the past month, as well as an illiquidity measure (\textit{ill}). Finally, dividend initiation (\textit{divi}) is zero for nearly all firms not allowing to run the nodewise regressions to calculate the desparsified estimate for this characteristic.}
The data  are sourced from CRSP, Compustat and I/B/E/S, missing values are imputed using \cite{Freyberger2024}, the industry adjusted size measure \textit{mve\_ia} is scaled by $10^{-4}$ and the weighted number of days without a trade (\textit{zerotrade}) is scaled by $10^{5}$ to bring their values more in line with the other characteristics. \cite{Green2017} do Fama Macbeth regressions (averaging across time the OLS coefficients of cross-sectional regressions on all characteristics) using data on all common stocks on the NYSE, AMEX, and NASDAQ with available Compustat accounting data over the period 1980--2014 and the subperiods 1980--2003 and 2003--2014, that is, they had both large $n$ and large $T$ and $p$ smaller than either dimension. They found that over the full period only 12 of the considered characteristics consistently predicted returns for non-microcap stocks. However, return predictability sharply declined in 2003, reducing independent predictors to just two characteristics after that year. We consider only large cap stocks and only a five year period to mitigate nonstationarity issues, which is consistent with a lot of standard practice in this literature.

In what follows, we estimate the parameter vector $\beta$ in model \eqref{eq:model-app1} by our methods and compute pointwise confidence bands for the coefficients $\beta_j$. Notably, the firm characteristics $C_{i,t-1}$ are lagged by one time period, which requires some slight modifications of our methodology. We first present and interpret the estimation results and then provide the details on how to modify and implement our methods.

\subsubsection*{Estimation results}

\begin{figure}[t!]
    \centering
    \includegraphics[width=0.9\textwidth]{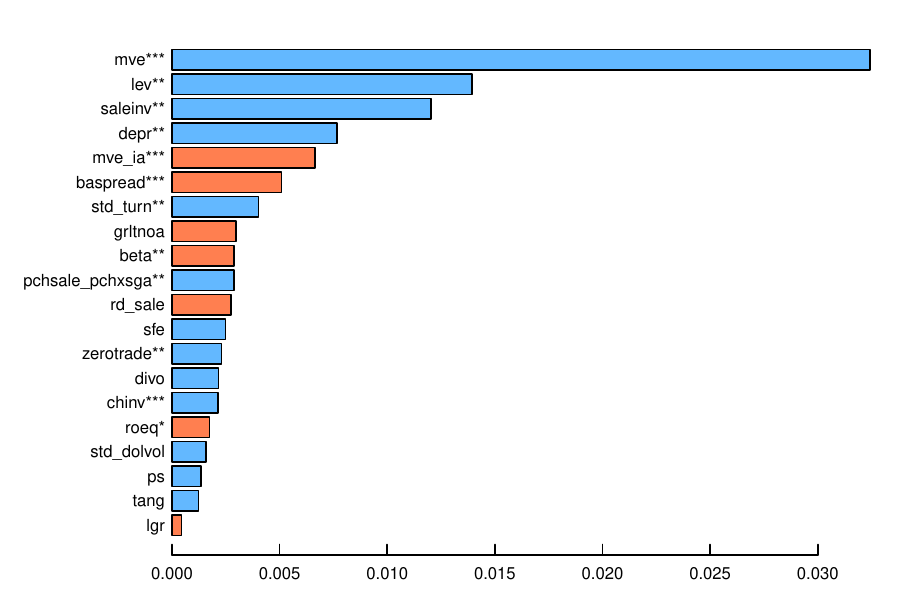}
    \caption{
    Scaled coefficient estimates $\widehat{\beta}_{\lambda,j}^{\text{sc}}$ ordered by absolute size. The stars $^*$, $^{**}$ and $^{***}$ after the variable names indicate significance at the $10\%$ , $5\%$ and $1\%$ level, respectively. The orange color corresponds to a positive sign of $\widehat{\beta}_{\lambda,j}^{\text{sc}}$, whereas the blue color indicates a negative sign.
    \label{fig:estimatedcoefficientsimportance}}
\end{figure}

\begin{figure}[t!]
    \centering
    \includegraphics[width=\linewidth]{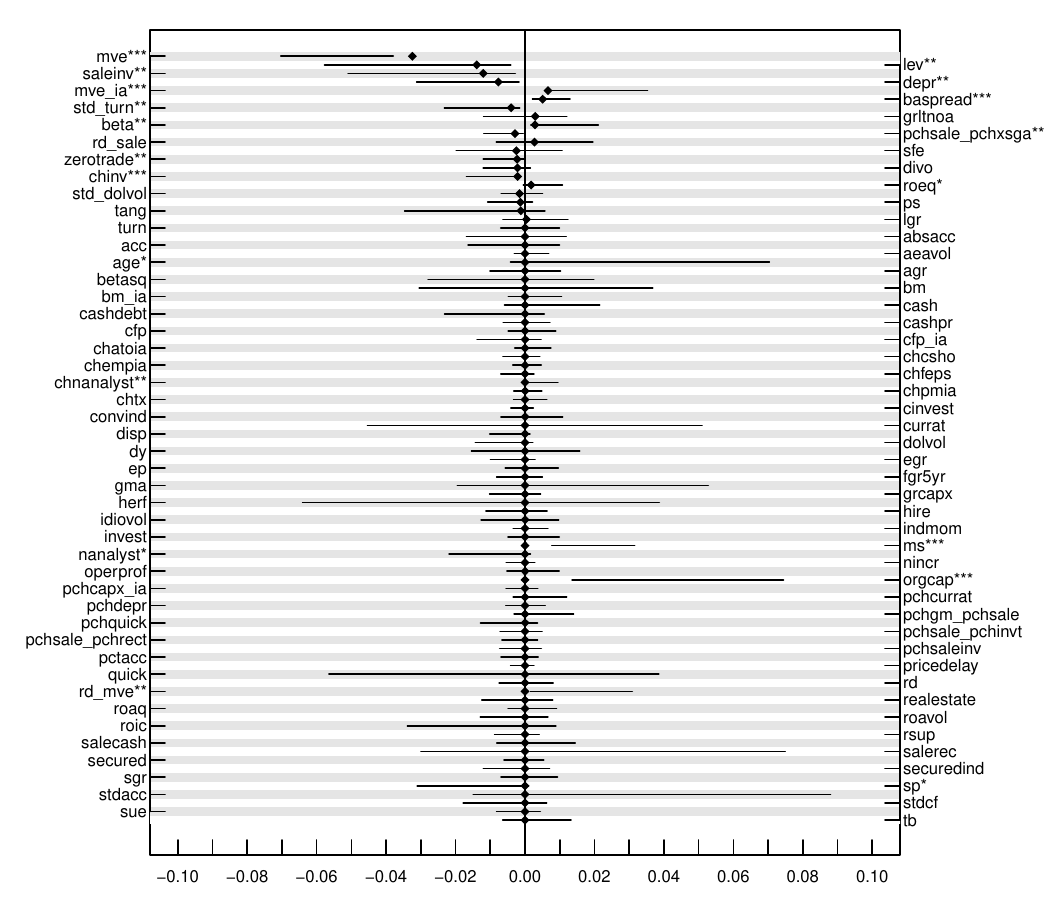}
    \caption{Scaled HD-CCE estimates $\widehat{\beta}_{\lambda,j}^{\text{sc}}$ (black dots) and 95\% confidence bands (black lines) for all firm characteristics. The stars $^*$, $^{**}$ and $^{***}$ after the variable names indicate significance at the $10\%$, $5\%$ and $1\%$ level, respectively.}
    \label{fig:application:inferenceplot}
\end{figure}

Our estimates of the coefficients $\beta_j$ and their confidence bands are presented in Figures \ref{fig:estimatedcoefficientsimportance} and \ref{fig:application:inferenceplot}. For better interpretability, we report the scaled coefficients $\widehat{\beta}_{\lambda,j}^{\text{sc}} := \widehat{\beta}_{\lambda,j} \cdot \widehat{s}_j$, where $\widehat{s}_j^2$ is the empirical variance of the $j$-th firm characteristic. We thus normalize the firm characteristics to have empirical variance $1$ and report the coefficients on the resulting scale. Figure \ref{fig:estimatedcoefficientsimportance} shows the non-zero coefficients $\widehat{\beta}_{\lambda,j}^{\text{sc}}$ ordered by their absolute size, while Figure \ref{fig:application:inferenceplot} shows all estimated coefficients $\widehat{\beta}_{\lambda,j}^{\text{sc}}$ including the zero ones along with their 95\% confidence bands.\footnote{In Figure \ref{fig:application:inferenceplot}, it is possible that a coefficient $\widehat{\beta}_{\lambda,j}^{\text{sc}}$ is not included in its confidence band. The reason is as follows: the confidence band is computed from the desparsified, i.e., the debiased version of the HD-CCE estimator and the bias of the HD-CCE estimator may be so strong that it lies outside the band. In particular, a coefficient $\widehat{\beta}_{\lambda,j}^{\text{sc}}$ may be equal to zero, while the corresponding confidence interval does not include zero. In Figure \ref{fig:application:inferenceplot}, this happens in four cases: \textit{chnanalyst}, \textit{ms}, \textit{orgcap} and \textit{rd\_mve}.}

We share some findings with \cite{Green2017}. Namely, we find only a small number of characteristics $j$ with a significant non-zero coefficient $\widehat{\beta}_{\lambda,j}^{\text{sc}}$. However, the characteristics we find hardly overlap with those identified by \cite{Green2017}, albeit they used quite different methodology and their data covered an earlier time period. Nevertheless, the signs of our signi\-fi\-cant non-zero coefficients are in line with prior literature in most cases. The size effect ($mve$) is strongly negative \citep{Banz1981} with a positive offset from the industry-adjusted size variable $mve\_ia$; both effects are statistically significant at the $1\%$ level and are amongst the variables with the largest coefficients. Leverage ($lev$) has a negative and statistically significant effect on excess returns at the $5\%$ level, while illiquidity ($baspread$) has a positive effect and is significant at the $1\%$ level, consistent with much earlier work on liquidity premiums \citep{Amihud1986}. Finally, $beta$ has a positive effect on excess returns, being significant at the $5\%$ level.

\subsubsection*{Implementation details}

We observe the data sample $\{ (Y_{it},C_{it}): 1 \le t \le T, \, 1 \le i \le n \}$ and consider the model consisting of the two equations
$R_{it} = \beta^\top C_{i,t-1} + \sum_{k=0}^K \gamma_{i,k} F_{t,k} + \varepsilon_{it}$ and $C_{i,t-1} = \bs{\Gamma}_i F_{t-1} + Z_{i,t-1}$
for $2 \le t \le T$ and $1 \le i \le n$, where $F_t = (F_{t,0}, F_{t,1},\ldots,F_{t,K})^\top$ and $F_{t,0} := 1$ for all $t$ is a time-constant factor. These equations can equivalently be expressed as 
\begin{align}
R_{i,2:T} & = \bs{C}_{i,1:(T-1)} \beta + \bs{F}_{2:T} \, \gamma_i + \varepsilon_{i,2:T} \label{eq:model-app-1} \\
\bs{C}_{i,1:(T-1)} & = \bs{F}_{1:(T-1)} \bs{\Gamma}_i^\top + \bs{Z}_{i,1:(T-1)} \label{eq:model-app-2}
\end{align}
for $1 \le i \le n$, where $\bs{F} \in \reals^{T \times (K+1)}$ is the factor matrix with $t$-th row $F_t$, $\bs{C}_i = (C_{i1} \ldots C_{iT})^\top \in \reals^{T \times p}$ is the matrix of firm characteristics and for a general matrix  $\bs{A} \in \reals^{T \times q}$, we let $\bs{A}_{k:\ell}$ be the matrix which results from $\bs{A}$ by deleting the rows $1,\ldots,k-1$ and $\ell+1,\ldots,T$. Notably, the error structure in \eqref{eq:model-app-1} depends on the factor matrix $\bs{F}_{2:T}$, whereas the firm characteristics in \eqref{eq:model-app-2} depend on the factor matrix $\bs{F}_{1:(T-1)}$ lagged by one time period. To take this into account, we slightly modify the projection matrix $\widehat{\bs{\Pi}}$. Specifically, we define $\widehat{\bs{\Pi}}$ to be the projection onto the orthogonal complement of the linear space $\mathcal{L} = \{ (\bs{1}, \widehat{\bs{W}}_{1:(T-1)}, \widehat{\bs{W}}_{2:T}) \, v: v \in \reals^{2\widehat{K}+1} \}$ which is spanned by the constant factor $\bs{1} = (1,\ldots,1)^\top \in \reals^{T-1}$ and the columns of $\widehat{\bs{W}}_{1:(T-1)}$ and $\widehat{\bs{W}}_{2:T}$. Here, $\widehat{\bs{W}}$ is defined exactly as in Section \ref{sec:est:def-HDCCE} with $\overline{\bs{X}} = \overline{\bs{C}}$ and $\overline{\bs{C}} = n^{-1} \sum_{i=1}^n \bs{C}_i$ the cross-sectional average of the matrices $\bs{C}_i$. With this choice of the projection matrix $\widehat{\bs{\Pi}}$, we compute the HD-CCE estimator exactly as described in Section \ref{sec:est:def-HDCCE}. The desparsified HD-CCE estimator is modified analogously. Our theory can be easily adapted to these modifications.

We implement our estimators as follows: The tuning parameter $\tau$ is chosen as suggested in Section \ref{sec:est:tuning:tau}, in particular, $\tau=\alpha\widehat{\psi}_1$ with $\alpha=0.01$. This results in the estimate $\widehat{K}=2$. 
The penalty constant $\pen$ of the HD-CCE estimator $\widehat{\beta}_\lambda$ is chosen by a leave-one-firm-out version of the cross-validation procedure outlined in Section \ref{sec:est:tuning:lambda} (i.e., $L = n$). Moreover, for each $j$, the penalty constant $\kappa$ of the nodewise lasso $\widetilde{\theta}_\kappa$ is chosen as detailed in Section \ref{sec:est:tuning:node} (with leave-one-firm-out cross-validation). The confidence bands depicted in Figure \ref{fig:application:inferenceplot} are computed according to \eqref{eq:confidence_interval}, where we estimate $\scaling$ by $\widetilde{\scaling}^{\hspace{1pt} \text{HET}}$. We use the heteroskedasticity robust normalization $\widetilde{\scaling}^{\hspace{1pt} \text{HET}}$ rather than $\widetilde{\scaling}^{\hspace{1pt} \text{HAC}}$ because the idiosyncratic errors $\varepsilon_{it}$ in equation \eqref{eq:model-app-1} can be interpreted as idiosyncratic return shocks that are uncorrelated over time. The stars $^*$, $^{**}$ and $^{***}$ after the variable names in Figures \ref{fig:estimatedcoefficientsimportance} and \ref{fig:application:inferenceplot} indicate whether a variable is significant at the $10\%$ , $5\%$ and $1\%$ level, respectively, i.e., whether the corresponding confidence band excludes zero. 
In order to make the estimated coefficients (and the corresponding confidence bands) better interpretable, we scale them by the empirical variances of the firm characteristics, i.e., we report the scaled versions $\widehat{\beta}_{\lambda,j}^{\text{sc}}$ and the correspondingly scaled bands.

\section{Concluding remarks}

In this paper, we have developed new estimation and inference methods for high-dimensional panel data models with interactive fixed effects. Our methods rely on the following general idea: rather than estimating the unobserved factor structure, we eliminate the factors from the model by a projection. Our methods can thus be regarded as a high-dimensional analogue of the CCE approach which is frequently used in the standard low-dimensional case. The projection device of the CCE approach breaks down completely in high dimensions and a simple fix is not possible. One of the main contributions of the paper is to come up with a projection device which works in both low and high dimensions. This device can be combined with high-dimensional regression techniques to obtain an estimator of the unknown parameter vector. We have focused on (desparsified) lasso techniques, but it is in principle possible to work with other techniques (Dantzig selector, SCAD, etc.). In our theoretical analysis, we have derived the convergence rate of our HD-CCE estimator and the asymptotic distribution of its desparsified version. Of course, this is only a first step towards a comprehensive theory. In what follows, we briefly outline some avenues for future research.

\subsubsection*{Variable selection} 
Another interesting issue besides parameter estimation and inference is variable selection. As the original lasso itself, our HD-CCE estimator will be selection consistent only under extremely strong conditions. To get better variable selection properties, it should be possible to combine our HD-CCE approach with adaptive lasso techniques.

\subsubsection*{Overestimating the number of factors $\boldsymbol{K}$} 
In our theory, we have assumed that we can consistently estimate the unknown number of factors $K$ (and we have provided a consistent estimator $\widehat{K}$). We conjecture that the theory can be extended to estimators $\widehat{K}$ with the property that $K \le \widehat{K} \le K_{\max}$ with probability tending to $1$, where $K_{\max}$ is a given upper bound on the number of factors. Such an extension would be formal proof that (moderate) overestimation of the number of factors is indeed unproblematic, as suggested by the simulation evidence in the supplementary material.

\subsubsection*{Dealing with nonlinear transformations}
Suppose we observe a sample of panel data $\{(Y_{it}, X_{it}^{\text{raw}}): 1 \le t \le T, \, 1 \le i \le n \}$, where $X_{it}^{\text{raw}} = (X_{it,1}^{\text{raw}}, \ldots, X_{it,p_0}^{\text{raw}})^\top$ is a vector of $p_0$ directly observed variables. Rather than only using the raw variables as regressors in the model, we would also like to include interactions $X_{it,j}^{\text{raw}} \cdot X_{it,k}^{\text{raw}}$ and nonlinear transformations such as polynomials $(X_{it,j}^{\text{raw}})^q$. Collecting all of the resulting regressors -- the raw, the interacted and the transformed variables -- in a long vector $X_{it} = (X_{it,1},\ldots,X_{it,p})^\top$, we consider the high-dimensional model 
$Y_{it} = \beta^\top X_{it} + \gamma_i^\top F_t + \varepsilon_{it}$.
As before, we assume that the observed variables $X_{it}^{\text{raw}}$ satisfy an approximate factor model of the form \eqref{eq:model-CCE-reg}, i.e., 
$X_{it}^{\text{raw}} = \bs{\Gamma}_i F_t +  Z_{it}$.
However, this factor structure is in general not preserved when the variables $X_{it}^{\text{raw}}$ are transformed nonlinearly. Hence, we cannot assume that all covariates $X_{it,j}$ satisfy an approximate factor model and thus have to modify our HD-CCE estimation algorithm from Section \ref{sec:est:def-HDCCE}.

We suggest to proceed as follows: we run Steps 1 and 2 of the algorithm on the raw variables $X_{it}^{\text{raw}}$ only, i.e., we construct $\widehat{K}$ and the projection matrix $\widehat{\bs{\Pi}}$ on the basis of $X_{it}^{\text{raw}}$. We then apply Step 3 with the thus constructed projection matrix. In this way, we can easily accommodate interactions and nonlinear transformations. Our theoretical results on the HD-CCE estimator from Theorem \ref{theo:rate} should still be valid after this modification, provided that the projected design matrix $\bs{X}^\perp$ satisfies a restricted eigenvalue (RE) condition as detailed in Section \ref{sec:ident}. However, showing that $\bs{X}^\perp$ indeed satisfies such an RE condition is extremely difficult, in particular, much more difficult than in the case analyzed in this paper. Moreover, our desparsification strategy needs to be adapted properly. A natural way would be to partition the covariates $X_{it,j}$ for $j=1,\ldots,p$ into groups. In particular, neglecting interactions, we may put all transformations of a given variable $X_{it,\ell}^{\textnormal{raw}}$ into one group and make inference on the resulting groups of coefficients. It is, however, not straightforward to extend our desparsified HD-CCE procedure to do so. The main issue is this: even if 
$X_{it,\ell}^{\textnormal{raw}}$ satisfies a nodewise regression equation with an interactive fixed effects error structure similar to \eqref{eq:model-CCE-nodewise}, nonlinear transformations of $X_{it,\ell}^{\textnormal{raw}}$ will in general not do so.

\section*{Acknowledgements}
We thank the Editor and three referees for helpful comments that greatly improved the paper. We thank Alexei Onatski and Hashem Pesaran for insightful discussions and Shaoran Li for supplying the data used in Section \ref{sec:application}.
Financial support by the DFG (German Research Foundation) -- project number 501082519 -- is gratefully acknow\-ledged.
Computing time granted on the supercomputer CLAIX at RWTH Aachen as part of the NHR4CES infrastructure is thankfully acknowledged as well.

\bibliographystyle{ims}
{\small
\setlength{\bibsep}{0.1em}
\bibliography{bibliography}}

\newpage
\allowdisplaybreaks[3]

\begin{center}
{\LARGE \textbf{Technical Appendices}}
\end{center}
\vspace{7pt}

\noindent Throughout the appendices, we let $c$ and $C$ denote generic positive constants that may take a different value on each occurrence. The symbols $c_j$ and $C_j$ with subscript $j$ (which may be either a natural number or a letter) are specific constants that are defined in the course of the appendices. Unless stated differently, the constants $c$, $C$, $c_j$ and $C_j$ depend neither on the dimensions $n$, $T$, $p$ nor on the sparsity index $s$. To emphasize that they do not depend on any of these parameters, we sometimes refer to them as absolute constants.

\def\theequation{A.\arabic{equation}}
\setcounter{equation}{0}
\section*{Appendix A: Proof of Theorem \ref{theo:rate}(a)}

In what follows, we prove Theorem \ref{theo:rate}. As the proof for small $T$ (part (b) of the theorem) is very similar to that for large $T$ (part (a) of the theorem), we concentrate on the large-$T$-case and provide a brief overview of the proof for the small-$T$-case in the supplementary material. We first lay out the main proof strategy (Steps 1--3) and then fill in the proofs of some intermediate propositions (Step 4). Throughout the appendix, we take implicitly for granted (e.g.\ when formulating the intermediate propositions) that the conditions of Theorem \ref{theo:rate}(a) are satisfied. Under these conditions, the matrix  $\widehat{\bs{W}}^\top \widehat{\bs{W}}$ is invertible with probability tending to $1$. Hence, we can replace the generalized inverse in the definition of the projection matrix $\widehat{\bs{\Pi}}$ by the proper inverse. More precisely speaking, we can write $\widehat{\bs{\Pi}} = \bs{I} - \widehat{\bs{W}} (\widehat{\bs{W}}^\top \widehat{\bs{W}})^{-1} \widehat{\bs{W}}^\top$ with probability tending to $1$. In what follows, we make use of this formulation but often suppress the specifier ``with probability tending to $1$'' for simplicity.

\subsection*{Step 1: Analysis of the eigenstructure of $\bs{\widehat{\Sigma}}$}

We first derive some rough bounds on the distances between the matrices $\bs{\Sigma}$, $\overline{\bs{\Sigma}}$ and $\widehat{\bs{\Sigma}}$, which are given by $\widehat{\bs{\Sigma}} = T^{-1} \bs{X}^\top \bs{X} = T^{-1} \sum_{t=1}^T \overline{X}_t \overline{X}_t^\top$, $\overline{\bs{\Sigma}} = \ex[T^{-1} \bs{X}^\top \bs{X}] = \ex[ T^{-1} \sum_{t=1}^T \overline{X}_t \overline{X}_t^\top ]$ and $\bs{\Sigma} = \bs{\Gamma} \bs{\Gamma}^\top$.

\begin{propA}\label{lemma1:eigenstructure}
It holds that 
\begin{enumerate}[label=(\roman*),leftmargin=0.975cm,topsep=0.5cm]
\item $\displaystyle{ \| \widehat{\bs{\Sigma}} - \overline{\bs{\Sigma}} \| = O_p\Big( p \sqrt{\frac{\log p}{n}} \Big) }$. 
\item $\displaystyle{ \| \overline{\bs{\Sigma}} - \bs{\Sigma} \| = O\Big(  \frac{p}{n} \Big) }$. 
\end{enumerate}
\end{propA}

\noindent With these bounds at hand, we have a closer look at the eigenstructure of the matrices $\bs{\Sigma}$, $\overline{\bs{\Sigma}}$ and $\widehat{\bs{\Sigma}}$. The first result shows that the matrix $\bs{\Sigma}$ has spiked eigenvalues: its first $K$ eigenvalues are extremely large (in particular, of order $p$) whereas the others are equal to $0$.

\begin{propA}\label{lemma2:eigenstructure}
The eigenvalues $\eig_1 \ge \ldots \ge \eig_p \ge 0$ of the matrix $\bs{\Sigma}$ have the following property: there exists an absolute constant $c_0 > 0$ such that 
\[ \eig_k \ge c_0 \, p \quad \text{for all } k \le K, \]
whereas $\eig_k = 0$ for all $k > K$. 
\end{propA}

\begin{proof}[Proof of Proposition \ref{lemma2:eigenstructure}.]
The claim follows upon considering the singular value decomposition $\bs{\Gamma} = \bs{U} \bs{D} \bs{V}^\top$, where the matrices $\bs{U} \in \reals^{p \times K}$ and $\bs{V} \in \reals^{K \times K}$ have orthonormal columns and $\bs{D} = \text{diag}(d_1,\ldots,d_K)$ is a diagonal matrix with $d_1 \ge \ldots \ge d_K \ge 0$. With this decomposition, we get that 
\[ \bs{\Sigma}/p = \bs{\Gamma} \bs{\Gamma}^\top/p = \bs{U} (\bs{D}^2/p) \bs{U}^\top, \]
which implies that the first $K$ eigenvalues of $\bs{\Sigma}/p = \bs{\Gamma} \bs{\Gamma}^\top/p$ are $d_1^2/p \ge \ldots \ge d_K^2/p$, while the others are equal to $0$. Since the first $K$ eigenvalues of $\bs{\Sigma}/p = \bs{\Gamma} \bs{\Gamma}^\top/p$ are identical to those of $\bs{\Gamma}^\top \bs{\Gamma}/p$, \ref{C:id2} yields that $0 < c_{\min} \le d_K^2/p \le \ldots \le d_1^2/p \le c_{\max} < \infty$. From this, the statement of the proposition follows immediately.
\end{proof}

\noindent The next proposition shows that the matrix $\overline{\bs{\Sigma}}$ has spiked eigenvalues as well. More specifically, the first $K$ eigenvalues are of the order $p$ whereas the others are of substantially smaller order.

\begin{propA}\label{lemma3:eigenstructure}
The eigenvalues $\overline{\eig}_1 \ge \ldots \ge \overline{\eig}_p \ge 0$ of $\overline{\bs{\Sigma}}$ have the following property: there exist an absolute constant $c_0 > 0$ and a natural number $n_0$ such that 
\[ \overline{\eig}_k \ge c_0 \, p \quad \text{for all } k \le K \text{ and } n \ge n_0, \]
whereas $\overline{\eig}_k = O( p/n)$ for all $k > K$. 
\end{propA}

\begin{proof}[Proof of Proposition \ref{lemma3:eigenstructure}.]
By Proposition \ref{lemma1:eigenstructure}, $\| \overline{\bs{\Sigma}} - \bs{\Sigma} \| = O(p/n)$. Hence, Weyl's theorem yields that 
\begin{equation}\label{eq:lemma3:eigenstructure:Weyl}
| \overline{\eig}_k - \eig_k | \le \| \overline{\bs{\Sigma}} - \bs{\Sigma} \| = O\Big( \frac{p}{n}\Big)
\end{equation}
for any $k$. Since $\eig_k \ge c_0 p$ for $k \le K$ and $\eig_k = 0$ for $k > K$ by Proposition \ref{lemma2:eigenstructure}, the result follows immediately from \eqref{eq:lemma3:eigenstructure:Weyl}. 
\end{proof}

\noindent We finally verify that the sample autocovariance matrix $\widehat{\bs{\Sigma}}$ has spiked eigenvalues similar to $\bs{\Sigma}$ and $\overline{\bs{\Sigma}}$.

\begin{propA}\label{lemma4:eigenstructure}
The eigenvalues $\widehat{\eig}_1 \ge \ldots \ge \widehat{\eig}_p \ge 0$ of $\widehat{\bs{\Sigma}}$ have the following property: there exists an absolute constant $c_0 > 0$ such that with probability tending to $1$,
\[ \widehat{\eig}_k \ge c_0 \, p \quad \text{for } k \le K, \]
whereas $\widehat{\eig}_k = O_p( p \sqrt{\log p} /\sqrt{n}) = o_p(p)$ for all $k > K$. 
\end{propA}

\begin{proof}[Proof of Proposition \ref{lemma4:eigenstructure}.]
Using that $\|\widehat{\bs{\Sigma}} - \overline{\bs{\Sigma}} \| = O_p( p \sqrt{\log p} / \sqrt{n})$, we can argue analogously as in the proof of Proposition \ref{lemma3:eigenstructure}. 
\end{proof}

\noindent An immediate consequence of the above propositions is the following.

\begin{propA}\label{lemma5:eigenstructure} 
It holds that $\widehat{K} \convp K$.
\end{propA} 
\noindent Put differently, $\widehat{K} = K$ with probability tending to $1$.

\subsection*{Step 2: Analysis of the projection matrix $\bs{\widehat{\Pi}}$}

In this step, we link the proxy $\bs{\widehat{\Pi}}$ of our method to the unknown projection matrix $\bs{\Pi}$. With $\widehat{\bs{W}} = \overline{\bs{X}} \widehat{\bs{U}} = \bs{F} \overline{\bs{\Gamma}}^\top \widehat{\bs{U}} + \overline{\bs{Z}} \widehat{\bs{U}}$, we can write $\widehat{\bs{\Pi}}$ as
\begin{align*} 
\widehat{\bs{\Pi}} 
 & = \bs{I} - \widehat{\bs{W}} (\widehat{\bs{W}}^\top \widehat{\bs{W}})^{-1} \widehat{\bs{W}}^\top \\
 & = \bigg\{ \bs{I} - \frac{1}{T} (\bs{F} \overline{\bs{\Gamma}}^\top \widehat{\bs{U}}) \Big[ \frac{1}{T} (\bs{F} \overline{\bs{\Gamma}}^\top \widehat{\bs{U}})^\top (\bs{F} \overline{\bs{\Gamma}}^\top \widehat{\bs{U}}) \Big]^{-1} (\bs{F} \overline{\bs{\Gamma}}^\top \widehat{\bs{U}})^\top \bigg\} - \widehat{\bs{R}},
\end{align*}
where 
\begin{align*} 
\widehat{\bs{R}} & = \frac{1}{T} (\bs{F} \overline{\bs{\Gamma}}^\top \widehat{\bs{U}}) \bigg\{ \widehat{\bs{\Eig}}^{-1} - \Big[ \frac{1}{T} (\bs{F} \overline{\bs{\Gamma}}^\top \widehat{\bs{U}})^\top (\bs{F} \overline{\bs{\Gamma}}^\top \widehat{\bs{U}}) \Big]^{-1} \bigg\} (\bs{F} \overline{\bs{\Gamma}}^\top \widehat{\bs{U}})^\top \\
 & \quad + \frac{1}{T} (\bs{F} \overline{\bs{\Gamma}}^\top \widehat{\bs{U}}) \widehat{\bs{\Eig}}^{-1} (\overline{\bs{Z}} \widehat{\bs{U}})^\top \\
 & \quad + \frac{1}{T} (\overline{\bs{Z}} \widehat{\bs{U}}) \widehat{\bs{\Eig}}^{-1} (\bs{F} \overline{\bs{\Gamma}}^\top \widehat{\bs{U}})^\top \\
 & \quad + \frac{1}{T} (\overline{\bs{Z}} \widehat{\bs{U}}) \widehat{\bs{\Eig}}^{-1} (\overline{\bs{Z}} \widehat{\bs{U}})^\top 
\end{align*} 
and 
\[ \frac{\widehat{\bs{W}}^\top \widehat{\bs{W}}}{T} = \widehat{\bs{U}}^\top \Big( \frac{\overline{\bs{X}}^\top \overline{\bs{X}}}{T} \Big) \widehat{\bs{U}} = \widehat{\bs{U}}^\top \widehat{\bs{\Sigma}} \widehat{\bs{U}} = \text{diag}(\widehat{\eig}_1,\ldots,\widehat{\eig}_{\widehat{K}}) =: \widehat{\bs{\Eig}}. \]
With this notation at hand, we can show that the two projection matrices $\bs{\widehat{\Pi}}$ and $\bs{\Pi}$ are related to each other as follows.

\begin{propA}\label{lemma2:projection}
With probability tending to $1$, $\widehat{\bs{\Pi}} = \bs{\Pi} - \widehat{\bs{R}}$.
\end{propA}

\noindent Proposition \ref{lemma2:projection} allows us to decompose the observed projection matrix $\widehat{\bs{\Pi}}$ into the ``oracle'' projection matrix $\bs{\Pi}$, which presupposes knowledge of the factors $\bs{F}$, and a remainder term $\widehat{\bs{R}}$. In order to exploit this decomposition, we need to make sure that the approximation error produced by the remainder $\widehat{\bs{R}}$ is asymptotically negligible. To do so, we examine the behaviour of the various components that show up in $\widehat{\bs{R}}$. This is done in Lemma \ref{lemma3:projection} and in the proofs of Propositions \ref{lemma2:Lasso} and \ref{lemma3:Lasso} below.

\subsection*{Step 3: Analysis of the lasso $\bs{\widehat{\beta}_\pen}$}

The lasso $\widehat{\beta}_\pen$ can be formulated as 
\[ \widehat{\beta}_\pen \in \underset{b \in \reals^p}{\text{argmin}} \bigg\{ \frac{1}{nT} \big\| \widehat{Y} - \widehat{\bs{X}} b \big\|^2 + \pen \|b\|_1 \bigg\} \]
with $\widehat{Y} = (\widehat{Y}_1^\top,\ldots,\widehat{Y}_n^\top)^\top$ and $\widehat{\bs{X}} = (\widehat{\bs{X}}_1^\top \ldots \widehat{\bs{X}}_n^\top)^\top$. Let $\mathcal{T}_{\text{RE}}$ be the event that the design matrix $\widehat{\bs{X}}$ fulfills the $\text{RE}(S,\phi)$ condition with some constant $\phi > 0$ and define the event $\mathcal{T}_\pen$ as 
\[ \mathcal{T}_\pen = \Big\{ \frac{4 \| \widehat{\bs{X}}^\top e \|_\infty}{nT} \le \pen \Big\}, \]
where $e = (e_1^\top,\ldots,e_n^\top)^\top$ with $e_i = \bs{F} \gamma_i + \varepsilon_i$. We first show that the lasso is well-behaved on the event $\mathcal{T}_\pen \cap \mathcal{T}_{\text{RE}}$ in the following sense.
\begin{propA}\label{lemma1:Lasso}
On the event $\mathcal{T}_\pen \cap \mathcal{T}_{\textnormal{RE}}$, it holds that 
\[ \| \widehat{\beta}_\pen - \beta \|_1 \le \frac{4}{\phi^2} \pen s. \]
\end{propA}
\noindent Proposition \ref{lemma1:Lasso} follows from standard finite-sample theory for the lasso. A proof is provided below for completeness.

We next have a closer look at the events $\mathcal{T}_\pen$ and $\mathcal{T}_{\text{RE}}$ that show up in Proposition \ref{lemma1:Lasso}. If we can prove that these two events occur with probability tending to $1$ for sufficiently small values of $\pen$, Theorem \ref{theo:rate}(a) is an immediate consequence of Proposition \ref{lemma1:Lasso}. In order to deal with the event $\mathcal{T}_\pen$, we derive the convergence rate of $\| \widehat{\bs{X}}^\top e \|_\infty / (nT)$, which is stated in the following result. 
\begin{propA}\label{lemma2:Lasso} 
It holds that 
\[ \frac{\| \widehat{\bs{X}}^\top e \|_\infty}{nT} = O_p\Big( \frac{\log pT}{n} + \sqrt{\frac{\log(npT) \log(np)}{nT}} \Big). \]
\end{propA}
\noindent Roughly speaking, the strategy to prove Proposition \ref{lemma2:Lasso} is as follows: 
Let $X_{i(j)}$ be the $j$-th column of $\bs{X}_i$, $Z_{i(j)}$ the $j$-th column of $\bs{Z}_i$ and $\Gamma_{i,j}$ the $j$-th row of $\bs{\Gamma}_i$. We write  
\[ \frac{\| \widehat{\bs{X}}^\top e \|_\infty}{nT} = \frac{1}{nT} \max_{1 \le j \le p} \Big| \sum_{i=1}^n \widehat{X}_{i(j)}^\top e_i \Big| \]
along with 
\[ \sum_{i=1}^n \widehat{X}_{i(j)}^\top e_i = \sum_{i=1}^n \big\{ \widehat{\bs{\Pi}} X_{i(j)} \big\}^\top \big\{ \widehat{\bs{\Pi}} e_i \big\} = \sum_{i=1}^n \big\{ \widehat{\bs{\Pi}} (\bs{F} \Gamma_{i,j} + Z_{i(j)}) \big\}^\top \big\{ \widehat{\bs{\Pi}} (\bs{F} \gamma_i + \varepsilon_i) \big\} \]
and exploit the main result from Step 2 in these formulas, according to which $\widehat{\bs{\Pi}} = \bs{\Pi} - \widehat{\bs{R}}$ with probability tending to $1$. A detailed proof is provided below. From Proposition \ref{lemma2:Lasso}, it immediately follows that 
\begin{equation}\label{eq:T-lambda}
\pr(\mathcal{T}_\pen) \to 1 \quad \text{for any choice} \quad \pen = h_n \frac{\log(npT)}{\min\{n,\sqrt{nT}\}}, 
\end{equation}
where $h_n$ slowly diverges to infinity. Hence, $\mathcal{T}_\pen$ occurs with probability tending to $1$ if $\pen$ is chosen of slightly larger order than $1/\min\{n,\sqrt{nT}\}$.

In order to cope with the event $\mathcal{T}_{\text{RE}}$, we first show that the covariance matrix $\widehat{\bs{X}}^\top \widehat{\bs{X}}/(nT)$ is close to $\bs{Z}^\top \bs{Z}/(nT)$ in the following sense. 
\begin{propA}\label{lemma3:Lasso}
It holds that
\[ \Big\| \frac{\widehat{\bs{X}}^\top \widehat{\bs{X}}}{nT} - \frac{\bs{Z}^\top \bs{Z}}{nT} \Big\|_{\max} = O_p \Big( \frac{\log(npT)}{\min\{n,T\}} \Big). \]
\end{propA}
\noindent The proof strategy is similar to that for Proposition \ref{lemma2:Lasso}. In particular, we rewrite the term of interest in a suitable way and then make heavy use of the fact that $\widehat{\bs{\Pi}} = \bs{\Pi} - \widehat{\bs{R}}$ with probability tending to $1$. The details are provided below. Since $s = o(\min\{n,T\} / \log(npT))$ by \ref{C:s-large}, Proposition \ref{lemma3:Lasso} implies that 
\begin{equation}\label{eq:Corollary6.8-vdGB} 
\frac{32 s}{\varphi^2} \Big\| \frac{\widehat{\bs{X}}^\top \widehat{\bs{X}}}{nT} - \frac{\bs{Z}^\top \bs{Z}}{nT} \Big\|_{\max} \le 1 
\end{equation}
with probability tending to $1$ for any given constant $\varphi > 0$. We can now use Corollary 6.8 in \cite{BuehlmannvandeGeer2011}, which says the following when applied to our context: Whenever $\bs{Z}$ fulfills the $\text{RE}(S,\varphi)$ condition and \eqref{eq:Corollary6.8-vdGB} is fulfilled, $\widehat{\bs{X}}$ satisfies the $\text{RE}(S,\phi)$ condition with $\phi = \varphi/\sqrt{2}$. Since $\bs{Z}$ obeys the $\text{RE}(S,\varphi)$ condition with probability tending to $1$ by assumption, we can infer that $\widehat{\bs{X}}$ must satisfy the $\text{RE}(S,\phi)$ condition with probability tending to $1$, that is,  
\begin{equation}\label{eq:T-RE}
\pr(\mathcal{T}_{\text{RE}}) \to 1. 
\end{equation}

Combining Proposition \ref{lemma1:Lasso} with \eqref{eq:T-lambda} and \eqref{eq:T-RE}, we finally arrive at the following statement: 
\[ \| \widehat{\beta}_\pen - \beta \|_1 \le \frac{4}{\phi^2} \pen s \]
for any $\pen = h_n \log(npT) /  \min\{n,\sqrt{nT}\}$ with probability tending to $1$, which implies Theorem \ref{theo:rate}(a).

\subsection*{Step 4: Proof of intermediate results}

It remains to prove Propositions \ref{lemma1:eigenstructure} and \ref{lemma2:projection}--\ref{lemma3:Lasso}. To do so, we need a series of auxiliary lemmas which are formulated below. The proofs of the propositions and auxiliary lemmas can be found in the supplementary material. In the proofs, we repeatedly make use of the following two facts: $\eig_{\max}(\bs{A}) \le p \|\bs{A}\|_{\max}$ for square matrices $\bs{A} \in \reals^{p \times p}$ and $\| \bs{B} \| = \eig_{\max}^{1/2}(\bs{B}^\top \bs{B}) \le \{ p \| \bs{B}^\top \bs{B} \|_{\max} \}^{1/2}$ for general (not necessarily square) matrices $\bs{B} \in \reals^{q \times p}$.

\begin{lemmaA}\label{lemma:aux1}
It holds that 
\begin{enumerate}[label=(\roman*),leftmargin=0.975cm,topsep=0.5cm]
\item \label{lemma:aux1:Z} $\displaystyle{\max_{1 \le j \le p} \max_{1 \le t \le T} \Big| \frac{1}{n} \sum_{i=1}^n Z_{it,j} \Big| = O_p\Big(\sqrt{\frac{\log(pT)}{n}}\Big)}$.
\item \label{lemma:aux1:Feps} $\displaystyle{\max_{1 \le k \le K} \max_{1 \le i \le n} \Big| \frac{1}{T} \sum_{t=1}^T F_{t,k} \varepsilon_{it} \Big| = O_p\Big(\sqrt{\frac{\log n}{T}}\Big)}$.  \KOM{[Double-check whether the conditions on the factors in \ref{C:F} are enough to get this rate.]}
\item \label{lemma:aux1:FZ} $\displaystyle{\max_{1 \le k \le K} \max_{1 \le i \le n} \max_{1 \le j \le p} \Big| \frac{1}{T} \sum_{t=1}^T F_{t,k} Z_{it,j} \Big| = O_p\Big(\sqrt{\frac{\log(np)}{T}}\Big)}$. \KOM{ [Double-check whether the conditions on the factors in \ref{C:F} are enough to get this rate.]}
\end{enumerate}
\end{lemmaA}

\begin{lemmaA}\label{lemma:aux2}
It holds that 
\begin{enumerate}[label=(\roman*),leftmargin=0.975cm,topsep=0.5cm]
\item \label{lemma:aux2:ZF} $\displaystyle{\max_{1 \le k \le K} \max_{1 \le j \le p} \Big| \frac{1}{T} \sum_{t=1}^T \Big\{ \frac{1}{n} \sum_{i=1}^n Z_{it,j} \Big\} F_{t,k} \Big| = O_p\Big( \sqrt{\frac{\log(npT)\log p}{nT}}\Big)}$.  \KOM{[Double-check whether the conditions on the factors in \ref{C:F} are enough to get this rate.]}
\item \label{lemma:aux2:Zeps} $\displaystyle{\max_{1 \le i \le n} \max_{1 \le j \le p} \Big| \frac{1}{T} \sum_{t=1}^T \Big\{ \frac{1}{n} \sum_{i^\prime=1}^n Z_{i^\prime t,j} \Big\} \varepsilon_{it} \Big| = O_p\Big( \sqrt{\frac{\log(npT)\log(np)}{nT}}\Big)}$.
\item \label{lemma:aux2:ZZ} $\displaystyle{\max_{1 \le i \le n} \max_{1 \le j \le p} \max_{1 \le j^\prime \le p} \Big| \frac{1}{T} \sum_{t=1}^T \Big\{ \frac{1}{n} \sum_{i^\prime=1}^n Z_{i^\prime t,j^\prime} \Big\} Z_{it,j} \Big| = O_p\Big( \sqrt{\frac{\log(npT)\log(np^2)}{nT}} + \frac{1}{n}\Big)}$.
\end{enumerate}
\end{lemmaA}

\begin{lemmaA}
\label{lemma:aux5}
It holds that 
\begin{enumerate}[label=(\roman*),leftmargin=0.975cm,topsep=0.5cm]
\item \label{lemma:aux5:1} $\displaystyle{\big\| \overline{\bs{\Gamma}} - \bs{\Gamma} \big\| = O_p \Big( \sqrt{\frac{p \log p}{n}} \Big)}$.
\item \label{lemma:aux5:2} $\displaystyle{\big\| \overline{\bs{\Gamma}} \overline{\bs{\Gamma}}^\top - \ex \, \overline{\bs{\Gamma}}  \overline{\bs{\Gamma}}^\top \big\| = O_p \Big( p \sqrt{\frac{\log p}{n}} \Big)}$.
\end{enumerate}
\end{lemmaA}

\begin{lemmaA}\label{lemma:aux3}
It holds that
\begin{enumerate}[label=(\roman*),leftmargin=0.975cm,topsep=0.5cm]
\item \label{lemma:aux3:Feps} $\displaystyle{\max_{1 \le i \le n} \Big\| \frac{\bs{F}^\top \varepsilon_i}{T} \Big\| = O_p\Big( \sqrt{\frac{\log n}{T}} \Big)}$.
\item \label{lemma:aux3:FZ} $\displaystyle{\max_{1 \le i \le n} \max_{1 \le j \le p} \Big\| \frac{\bs{F}^\top Z_{i(j)}}{T} \Big\| = O_p\Big(\sqrt{\frac{\log(np)}{T}}\Big)}$.
\end{enumerate}
\end{lemmaA}

\begin{lemmaA}\label{lemma:aux4}
It holds that 
\begin{enumerate}[label=(\roman*),leftmargin=0.975cm,topsep=0.5cm]
\item \label{lemma:aux4:barZ} $\displaystyle{\big\| \overline{\bs{Z}} \big\| = O_p\Big(\sqrt{\frac{pT\log(pT)}{n}}\Big)}$.
\item \label{lemma:aux4:barZF} $\displaystyle{\Big\| \frac{\overline{\bs{Z}}^\top \bs{F}}{T} \Big\| = O_p\Big( \sqrt{\frac{p \log(npT) \log p}{nT}}\Big)}$.
\item \label{lemma:aux4:barZeps} $\displaystyle{\max_{1 \le i \le n} \Big\| \frac{\overline{\bs{Z}}^\top \varepsilon_i}{T} \Big\| = O_p\Big( \sqrt{\frac{p \log(npT)\log(np)}{nT}}\Big)}$.
\item \label{lemma:aux4:barZZ} $\displaystyle{\max_{1 \le i \le n} \max_{1 \le j \le p} \Big\| \frac{\overline{\bs{Z}}^\top Z_{i(j)}}{T} \Big\| = O_p\Big( \sqrt{\frac{p \log(npT)\log(np^2)}{nT}} + \frac{\sqrt{p}}{n}\Big)}$. 
\end{enumerate}
\end{lemmaA}

\begin{lemmaA}\label{lemma3:projection}
It holds that 
\begin{enumerate}[label=(\roman*),leftmargin=0.975cm,topsep=0.5cm]
\item \label{lemma3:projection:i}
$\displaystyle{\Big\| \widehat{\bs{\Eig}} - \frac{1}{T} (\bs{F} \overline{\bs{\Gamma}}^\top \widehat{\bs{U}})^\top (\bs{F} \overline{\bs{\Gamma}}^\top \widehat{\bs{U}}) \Big\| = O_p\bigg( p \Big \{ \frac{\log(pT)}{n} + \sqrt{\frac{\log(npT) \log p}{nT}} \Big\} \bigg)}$.
\item \label{lemma3:projection:ii}
$\displaystyle{\Big\| \widehat{\bs{\Eig}}^{-1} - \Big[ \frac{1}{T} (\bs{F} \overline{\bs{\Gamma}}^\top \widehat{\bs{U}})^\top (\bs{F} \overline{\bs{\Gamma}}^\top \widehat{\bs{U}}) \Big]^{-1} \Big\|  = O_p\bigg( \frac{1}{p} \Big \{ \frac{\log(pT)}{n} + \sqrt{\frac{\log(npT) \log p}{nT}} \Big\} \bigg)}$. 
\end{enumerate}
\end{lemmaA}

\def\theequation{B.\arabic{equation}}
\setcounter{equation}{0}
\section*{Appendix B: Proof of Theorem \ref{theo:normality}(a)}

In this appendix, we prove Theorem \ref{theo:normality} for the large-$T$-case, that is, part (a) of the theorem. The proof for the small-$T$-case, that is, for part (b) is very similar and thus relegated to the supplementary material. We proceed analogously as in Appendix A: we first lay out the main proof strategy (Steps 1--7) and then fill in the proofs of some intermediate results (Step 8). We use the following notation, which parallels that from Appendix A: 
\[ \widetilde{\bs{\Sigma}} = \frac{\overline{\bs{X}}_{(-j)}^\top\overline{\bs{X}}_{(-j)}}{T}, \quad
\overline{\bs{\Sigma}}^{[-j]} = \frac{\ex[\overline{\bs{X}}_{(-j)}^\top\overline{\bs{X}}_{(-j)}]}{T} \quad \text{and} \quad 
\bs{\Sigma}^{[-j]} = \bs{\Gamma}_{-j}\bs{\Gamma}_{-j}^\top. \]
Moreover, we frequently make use of the shorthands $\widetilde{\bs{X}}_i = \widetilde{\bs{\Pi}} \bs{X}_i$, $\widetilde{\varepsilon}_i = \widetilde{\bs{\Pi}} \varepsilon_i$, $\widetilde{u}_i = \widetilde{\bs{\Pi}} u_i$, etc. Finally, we let $\{ \summable_n \}$ be a generic sequence which is non-negative and summable, i.e., $a_n \ge 0$ for all $n$ and $\sum_{n=1}^\infty \summable_n < \infty$. Whereas $\{ \summable_n \}$ may denote a different sequence on each occurrence, we let $\{ \summable_n^{(\ell)} \}$ with $\ell =1,2,3,\ldots$ denote specific non-negative and summable sequences. 
Throughout the appendix, we assume for simplicity that the number of factors $K$ is known. This is essentially without loss of generality: the arguments can be easily adapted to unknown $K$ since the estimator $\widehat{K}$ has the property that 
$\pr(\widehat{K} \neq K) \leq a_n$ 
(which immediately follows from the proof of Proposition \ref{lemma4:eigenstructure} and the fact that Lemma \ref{lemmaC:aux2}\ref{lemmaC:aux2:sigmahatsigmabar}
also holds for $\overline{\bs{\Sigma}}$ and $\widehat{\bs{\Sigma}}$).

\subsection*{Step 1: Analysis of $\bs{\widetilde{\bs{\Pi}}}$}

The results on $\bs{\widehat{\bs{\Pi}}}$ from Appendix A directly carry over to $\bs{\widetilde{\bs{\Pi}}}$. However, these results do not suffice for the present proof but need to be refined. Analogously as in Appendix A, we consider the decomposition
\begin{align*} 
\widetilde{\bs{\Pi}} 
 & = \bigg\{ \bs{I} - \frac{1}{T} (\bs{F} \overline{\bs{\Gamma}}_{-j}^\top \widetilde{\bs{U}}) \Big[ \frac{1}{T} (\bs{F} \overline{\bs{\Gamma}}_{-j}^\top \widetilde{\bs{U}})^\top (\bs{F} \overline{\bs{\Gamma}}_{-j}^\top \widetilde{\bs{U}}) \Big]^{-1} (\bs{F} \overline{\bs{\Gamma}}_{-j}^\top \widetilde{\bs{U}})^\top \bigg\} - \widetilde{\bs{R}},
\end{align*}
where
\begin{align*} 
\widetilde{\bs{R}} 
 & = \frac{1}{T} (\bs{F} \overline{\bs{\Gamma}}_{-j}^\top \widetilde{\bs{U}}) \bigg\{ \widetilde{\bs{\Eig}}^{-1} - \Big[ \frac{1}{T} (\bs{F} \overline{\bs{\Gamma}}_{-j}^\top \widetilde{\bs{U}})^\top (\bs{F} \overline{\bs{\Gamma}}_{-j}^\top \widetilde{\bs{U}}) \Big]^{-1} \bigg\} (\bs{F} \overline{\bs{\Gamma}}_{-j}^\top \widetilde{\bs{U}})^\top \\
 & \quad + \frac{1}{T} (\bs{F} \overline{\bs{\Gamma}}_{-j}^\top \widetilde{\bs{U}}) \widetilde{\bs{\Eig}}^{-1} (\overline{\bs{Z}}_{(-j)} \widetilde{\bs{U}})^\top \\
 & \quad + \frac{1}{T} (\overline{\bs{Z}}_{(-j)} \widetilde{\bs{U}}) \widetilde{\bs{\Eig}}^{-1} (\bs{F} \overline{\bs{\Gamma}}_{-j}^\top \widetilde{\bs{U}})^\top \\
 & \quad + \frac{1}{T} (\overline{\bs{Z}}_{(-j)} \widetilde{\bs{U}}) \widetilde{\bs{\Eig}}^{-1} (\overline{\bs{Z}}_{(-j)} \widetilde{\bs{U}})^\top 
\end{align*} 
and $\widetilde{\bs{\Eig}} = \textnormal{diag}(\widetilde{\eig}_1,\ldots,\widetilde{\eig}_{K})$. With this notation at hand, we verify that $\bs{\widetilde{\bs{\Pi}}}$ has the following properties.

\begin{propB}\label{propC:tilde-pi-equals-pi-r}
There exists a non-negative summable sequence $\{\summable_n^{(1)}\}$ such that
\begin{align*}
\pr\left( \widetilde{\bs{\Pi}} \neq \bs{\Pi} - \bs{\widetilde{R}} \right) \leq \summable_n^{(1)}.
\end{align*}
\end{propB}

\begin{propB}\label{propC:max-tildeR-u}
There exists a non-negative summable sequence $\{\summable_n^{(2)}\}$ with the following properties:
\begin{enumerate}[label=(\roman*),leftmargin=0.975cm,topsep=0.5cm]
\item \label{propC:max-tildeR-u-i} For sufficiently large $\Summable_2 > 0$,
\begin{align*}
\pr\left(\max_{i}\| \widetilde{\bs{R}}u_i\| \geq \Summable_2 \frac{\sqrt{\log(npT)\log(np)}}{\sqrt{n}}  \right) \leq \summable_n^{(2)}.
\end{align*}    
\item \label{propC:max-tildeR-u-ii} For any $\varepsilon>0$,
\begin{align*}
\pr\left(  \left|\frac{1}{nT}\sum_{i=1}^n \| \widetilde{\bs{\Pi}}u_i\|^2 - \frac{1}{nT}\sum_{i=1}^n \| \bs{\Pi}u_i\|^2 \right| >\varepsilon   \right) \leq \summable_n^{(2)}.
\end{align*}
\end{enumerate}
\end{propB}

\begin{propB}\label{propC:norm-tilde-pi-F}
There exist a positive constant $\Summable_3$ and a non-negative summable sequence $\{\summable_n^{(3)}\}$ such that
\begin{align*}
\pr\left( \|\widetilde{\bs{\Pi}}\bs{F}\| > \Summable_3 \sqrt{\frac{T\log(pT)}{n}} \right) \leq \summable_n^{(3)}.
\end{align*}
\end{propB}

\subsection*{Step 2: Analysis of the lasso}

The lasso $\widetilde{\beta}_\pen$ can be analyzed fully analogously to $\widehat{\beta}_\pen$. Let $\widetilde{\mathcal{T}}_{\textnormal{RE}}$ be the event that the design matrix $\widetilde{\bs{X}}$ fulfills the $\textnormal{RE}(S,\phi)$ condition with some $\phi > 0$ and define the event $\widetilde{\mathcal{T}}_\pen$ as
\[ \widetilde{\mathcal{T}}_\pen = \bigg\{ \frac{4\|\widetilde{\bs{X}}^\top e\|_\infty }{nT} \le \pen \bigg\}. \]
The same arguments as for Proposition \ref{lemma2:Lasso} in Appendix A yield that $\pr(\widetilde{\mathcal{T}}_\pen) \to 1$. Since $\pr(\widetilde{\mathcal{T}}_{\textnormal{RE}}) \to 1$ by \ref{C:RE-INF}, we obtain that
\begin{equation}\label{eq:conv-Tlambda-Tre}
\pr(\widetilde{\mathcal{T}}_\pen \cap \widetilde{\mathcal{T}}_{\textnormal{RE}}) \to 1.
\end{equation}
Repeating the arguments for Proposition \ref{lemma1:Lasso} further shows that the lasso $\widetilde{\beta}_\pen$ has the following properties on $\widetilde{\mathcal{T}}_\pen \cap \widetilde{\mathcal{T}}_{\textnormal{RE}}$. 
\begin{propB}\label{propC:Lasso}
On the event $\widetilde{\mathcal{T}}_\pen \cap \widetilde{\mathcal{T}}_{\textnormal{RE}}$,
\begin{align*}
\| \widetilde{\beta}_\pen - \beta \|_1 & \le \frac{4}{\phi^2} \pen \|\beta\|_0 \\
\frac{1}{nT} \| \widetilde{\bs{X}} (\widetilde{\beta}_\pen - \beta) \|^2 & \le \frac{4}{\phi^2} \pen^2 \|\beta\|_0.    
\end{align*}  
\end{propB}

\subsection*{Step 3: Analysis of the nodewise lasso}

We next turn to the nodewise lasso $\widetilde{\betanw}_\pennw$ which requires a more detailed analysis than the lasso $\widetilde{\beta}_\pen$. Similarly as before, let $\widetilde{\mathcal{T}}_{\textnormal{RE}}^{\textnormal{node}}$ be the event that the design matrix $\widetilde{\bs{X}}_{(-j)}$ fulfills the $\textnormal{RE}(\textnormal{supp}(\betanw), \phi)$ condition and define the event $\widetilde{\mathcal{T}}_\pennw^{\textnormal{node}}$ as
\[ \widetilde{\mathcal{T}}_\pennw^{\textnormal{node}} = \bigg\{ \frac{4\|\widetilde{\bs{X}}_{(-j)}^\top w\|_\infty }{nT} \le \pennw \bigg\}, \]
where $w = (w_1^\top,\ldots,w_n^\top)^\top$ with $w_i = \bs{F} \nu_i + u_i$ is the error vector in the nodewise lasso regression. The first result gives a probabilistic bound on the effective noise term $4\|\widetilde{\bs{X}}_{(-j)}^\top w\|_\infty / (nT)$ in the event $\widetilde{\mathcal{T}}_\pennw^{\textnormal{node}}$.  
\begin{propB}\label{propC:eff-noise-nodewise}
There exist a positive constant $C_\pennw$ and a non-negative summable sequence $\{\summable_n\}$ such that
\begin{align*}
\pr\left(\frac{4\|\widetilde{\bs{X}}_{(-j)}^\top w\|_\infty }{nT} > C_\pennw \sqrt{\log(np^2)} \log(npT) \left[ \frac{1}{\sqrt{nT}} + \frac{1}{n} \right] \right) \leq \summable_n
\end{align*}
for sufficiently large $n$.
\end{propB}
\noindent Choosing
\[ \pennw =  C_\pennw \sqrt{\log(np^2)} \log(npT) \left[ \frac{1}{\sqrt{nT}} + \frac{1}{n} \right], \]
this immediately implies that $\pr(\widetilde{\mathcal{T}}_\pennw^{\textnormal{node}}) \ge 1 - \summable_n$ with $\sum_{n=1}^\infty \summable_n < \infty$. By \ref{C:RE-nodewise-INF}, $\widetilde{\mathcal{T}}_{\textnormal{RE}}^{\textnormal{node}}$ is assumed to have an analogous property: $\pr(\widetilde{\mathcal{T}}_{\textnormal{RE}}^{\textnormal{node}}) \ge 1 - \summable_n$ with some non-negative summable sequence $\{ \summable_n \}$. We can thus infer that
\begin{equation}\label{eq:conv-Tlambda-Tre-nodewise}
\pr(\widetilde{\mathcal{T}}_\pen^{\textnormal{node}} \cap \widetilde{\mathcal{T}}_{\textnormal{RE}}^{\textnormal{node}}) \ge 1 - \summable_n^{(4)}
\end{equation}
with some non-negative summable sequence $\{ \summable_n^{(4)} \}$. As above, we can prove (by repeating the arguments for Proposition \ref{lemma2:Lasso}) that the nodewise lasso has the following properties on $\widetilde{\mathcal{T}}_\pen^{\textnormal{node}} \cap \widetilde{\mathcal{T}}_{\textnormal{RE}}^{\textnormal{node}}$.
\begin{propB}\label{propC:nodewise-prediction-l1}
On the event $\widetilde{\mathcal{T}}_\pennw^{\textnormal{node}} \cap \widetilde{\mathcal{T}}_{\textnormal{RE}}^{\textnormal{node}}$,
\begin{align*}
\| \widetilde{\betanw}_\pennw - \betanw \|_1 & \le \frac{4}{\phi^2} \pennw \|\betanw\|_0 \\
\frac{1}{nT} \| \widetilde{\bs{X}}_{(-j)} (\widetilde{\betanw}_\pennw - \betanw) \|^2 & \le \frac{4}{\phi^2} \pennw^2 \|\betanw\|_0.    
\end{align*}  
\end{propB}
\noindent We now analyze the nodewise lasso residuals $\widetilde{\resnw} = (\widetilde{\resnw}_1^\top,\ldots,\widetilde{\resnw}_n^\top)^\top$ with $\widetilde{\resnw}_i = \widetilde{X}_{i(j)} - \widetilde{\bs{X}}_{i(-j)} \widetilde{\betanw}_{\pennw}$ for $1 \le i \le n$. The following two results show that they can be bounded from above and from below (in a certain probabilistic sense). 
\begin{propB}\label{propC:residuals-unif-lower-bounded}
The event 
\begin{align*}
\mathcal{E}_n^{>} = \left\{ \frac{\| \widetilde{\resnw} \|^2}{nT} > c_\resnw \right\}
\end{align*}
with $c_\resnw > 0$ sufficiently small has the property that
\begin{align*}
\pr\left(\bigcup_{m=1}^\infty \bigcap_{n\geq m} \mathcal{E}_n^{>} \right) = 1.
\end{align*}
\end{propB}
\begin{propB}\label{propC:residuals-growth}
For $C_\resnw$ sufficiently large and $\xi > 0$ arbitrarily small but fixed, the event 
\begin{align*}
\mathcal{E}_n^{\le} = \left\{ \| \widetilde{\resnw} \|_\infty \leq C_\resnw \sqrt{T} (npT)^{\frac{2+\xi}{\moments}} \right\}
\end{align*}
has the property that 
\begin{align*}
\pr\left(\bigcup_{m=1}^\infty \bigcap_{n\geq m} \mathcal{E}_n^{\le} \right) =1.
\end{align*}
\end{propB}

\subsection*{Step 4: Decomposition of the desparsified lasso $\bs{\widetilde{b}_j}$}

It holds that
\begin{align*}
\frac{\widetilde{\resnw}^\top \widetilde{X}_{(j)}}{\|\widetilde{\resnw}\|}(\widetilde{b}_{j} - \beta_j)
 &= \frac{\widetilde{\resnw}^\top\widetilde{X}_{(j)}}{\|\widetilde{\resnw}\|}(\widetilde{\beta}_{\pen,j} - \beta_j) + \frac{\widetilde{\resnw}^\top(\widetilde{Y}-\widetilde{\bs{X}}\widetilde{\beta}_\pen)}{\|\widetilde{\resnw}\|}  \\
 &= \frac{\widetilde{\resnw}^\top\widetilde{X}_{(j)}}{\|\widetilde{\resnw}\|}(\widetilde{\beta}_{\pen,j} - \beta_j) + \frac{\widetilde{\resnw}^\top\widetilde{\bs{X}}(\beta-\widetilde{\beta}_\pen)}{\|\widetilde{\resnw}\|} + \frac{\widetilde{\resnw}^\top\widetilde{e}}{\|\widetilde{\resnw}\|} \\
 & = \frac{\widetilde{\resnw}^\top\widetilde{\bs{X}}_{(-j)}(\beta_{-j}-\widetilde{\beta}_{\pen,-j})}{\|\widetilde{\resnw}\|} + \frac{\widetilde{\resnw}^\top\widetilde{e}}{\|\widetilde{\resnw}\|},
\end{align*}
where $\widetilde{e} = (\widetilde{e}_1^\top,\ldots,\widetilde{e}_n^\top)^\top$ with $\widetilde{e}_i = \widetilde{\bs{\Pi}}e_i = \widetilde{\bs{\Pi}}(\bs{F}\gamma_i+ \varepsilon_i)$. Moreover,
\begin{align*}
\frac{\widetilde{\resnw}^\top\widetilde{e}}{\|\widetilde{\resnw}\|} 
   = \frac{1}{\| \widetilde{\resnw} \|} \sum_{i=1}^n \widetilde{\resnw}_i^\top  \widetilde{e}_i 
 & = \frac{1}{\| \widetilde{\resnw} \|} \sum_{i=1}^n \widetilde{\resnw}_i^\top  \widetilde{\bs{\Pi}}\bs{F}\gamma_i  + \frac{1}{\| \widetilde{\resnw} \|} \sum_{i=1}^n \widetilde{\resnw}_i^\top  \widetilde{\bs{\Pi}} \varepsilon_i.
\end{align*}
Taken together, these calculations yield that
\begin{align*}
\frac{\widetilde{\resnw}^\top \widetilde{X}_{(j)}}{\|\widetilde{\resnw}\|}(\widetilde{b}_{j} - \beta_j) = \widetilde{\Upsilon}_A + \widetilde{\Upsilon}_B + \widetilde{\Upsilon}_C
\end{align*}
with
\begin{align*}
\widetilde{\Upsilon}_A &= \frac{\widetilde{\resnw}^\top\widetilde{\bs{X}}_{(-j)}(\beta_{-j}-\widetilde{\beta}_{\pen,-j})}{\|\widetilde{\resnw}\|} \\
\widetilde{\Upsilon}_B &= \frac{1}{\| \widetilde{\resnw} \|} \sum_{i=1}^n \widetilde{\resnw}_i^\top  \widetilde{\bs{\Pi}}\bs{F}\gamma_i \\
\widetilde{\Upsilon}_C &= \frac{1}{\| \widetilde{\resnw} \|} \sum_{i=1}^n \widetilde{\resnw}_i^\top  \widetilde{\bs{\Pi}} \varepsilon_i.
\end{align*}

\subsection*{Step 5: Analysis of $\bs{\widetilde{\Upsilon}_A}$ and $\bs{\widetilde{\Upsilon}_B}$}

We now show that $\widetilde{\Upsilon}_A$ and $\widetilde{\Upsilon}_B$ are asymptotically negligible in the sense of being $o_p(1)$.

\begin{propB}\label{propC:DeltaA}
It holds that $\widetilde{\Upsilon}_A = o_p(1)$.
\end{propB}

\begin{proof}
By the KKT conditions, any solution $\widetilde{\betanw}_{\pennw} \in \reals^{p-1}$ to 
\begin{align*}
\argmin_{\vartheta \in \reals^{p-1}} \bigg\{ \frac{1}{nT}\|\widetilde{X}_{(j)} - \widetilde{\bs{X}}_{(-j)}\vartheta\|^2 + \pennw \|\vartheta\|_1 \bigg\}
\end{align*}
is uniquely characterized by
\begin{align*}
\frac{2}{nT} \widetilde{\bs{X}}_{(-j)}^\top \widetilde{\resnw} = \pennw v
\end{align*}
for some vector $v$ with $\| v\|_\infty\leq 1$. From this, it follows that 
\begin{align*}
|\widetilde{\Upsilon}_A|
 & \leq \|\beta-\widetilde{\beta}_\pen\|_1  \frac{\|\widetilde{\bs{X}}_{(-j)}^\top \widetilde{\resnw}\|_\infty}{\|\widetilde{\resnw}\|} \leq \|\beta-\widetilde{\beta}_\pen\|_1 \frac{\pennw \, nT}{2\|\widetilde{\resnw}\|}.
\end{align*}
We now use the following:
\begin{enumerate}[label=(\roman*)]
\item From \eqref{eq:conv-Tlambda-Tre} and Proposition \ref{propC:Lasso}, it follows that 
\begin{align*}
\|\beta-\widetilde{\beta}_\pen\|_1 =O_p\left( s \, \frac{h_n \log(npT)}{\min\{n,\sqrt{nT}\}} \right).
\end{align*}
\item The tuning parameter of the nodewise lasso is $\pennw = C_\pennw \sqrt{\log(np^2)}  \log(npT)$ \linebreak $[ (nT)^{-1/2} + n^{-1} ]$ with some sufficiently large constant $C_\pennw$.
\item Since $\|\widetilde{\resnw}\|^2/(nT) > c_\resnw$ with probability tending to $1$ by Proposition \ref{propC:residuals-unif-lower-bounded}, 
\[ \frac{\pennw \, nT}{2\|\widetilde{\resnw}\|} = \frac{\pennw \sqrt{nT}}{2 \sqrt{\frac{1}{nT} \| \widetilde{\resnw} \|^2}} \leq \frac{\pennw \sqrt{nT}}{2 \sqrt{c_\resnw}} \]
with probability tending to $1$. 
\end{enumerate}
Combining these facts leads to
\begin{align*}
|\widetilde{\Upsilon}_A|
 & = O_p \bigg(s \, \frac{h_n \log(npT)}{\min\{n,\sqrt{nT}\}} \pennw \sqrt{nT} \bigg) = o_p(1),
\end{align*}
where the last equality uses assumption \ref{C:nTp-large-INF}. 
\end{proof}

\begin{propB}\label{lemmaC: deltaB small}
It holds that $\widetilde{\Upsilon}_B = o_p(1)$.
\end{propB}

\begin{proof}
Decomposing $\widetilde{\Upsilon}_B$ leads to  
\begin{align}
\widetilde{\Upsilon}_B = \frac{1}{\| \widetilde{\resnw} \|} \sum_{i=1}^n \widetilde{\resnw}_{i}^\top \widetilde{\bs{\Pi}}\bs{F}\gamma_i 
 & = \frac{1}{\| \widetilde{\resnw} \|} \sum_{i=1}^n (\widetilde{\bs{X}}_{i(-j)}(\betanw - \widetilde{\betanw}_\pennw))^\top \widetilde{\bs{\Pi}}\bs{F}\gamma_i \tag{I} \\
 & \quad + \frac{1}{\| \widetilde{\resnw} \|} \sum_{i=1}^n (\widetilde{\bs{\Pi}}\bs{F}\nu_i)^\top \widetilde{\bs{\Pi}}\bs{F}\gamma_i \tag{II}\\
 & \quad + \frac{1}{\| \widetilde{\resnw} \|} \sum_{i=1}^n (\widetilde{\bs{\Pi}}u_i)^\top \widetilde{\bs{\Pi}}\bs{F}\gamma_i. \tag{III} 
\end{align}
\begin{enumerate}[label=(\Roman*):,leftmargin=1.1cm]

\item With Propositions \ref{propC:norm-tilde-pi-F}, \ref{propC:nodewise-prediction-l1} and \ref{propC:residuals-unif-lower-bounded}, we obtain that 
\begin{align*}
(\textnormal{I})
 & \leq \| \widetilde{\bs{\Pi}}\bs{F} \| \frac{1}{\| \widetilde{\resnw} \|} \sqrt{\sum_{i=1}^n \| \widetilde{\bs{X}}_{i(-j)}(\betanw - \widetilde{\betanw}_\pennw)\|^2 } \sqrt{\sum_{i=1}^n \| \gamma_i \|^2} \\
 & = O_p\left(\sqrt{\frac{T\log(pT)}{n}}\right)\frac{1}{\frac{1}{\sqrt{nT}}\| \widetilde{\resnw} \|} \sqrt{\frac{1}{nT}\sum_{i=1}^n \| \widetilde{\bs{X}}_{i(-j)}(\betanw - \widetilde{\betanw}_\pennw)\|^2 } \ O_p(\sqrt{n}) \\
 & =  O_p\left(\sqrt{\frac{T\log(pT)}{n}}\right) O_p(1) O_p(\pennw \sqrt{\| \betanw \|_0}) O_p(\sqrt{n}) = o_p(1),  
 \end{align*}
where the last equality follows from assumption \ref{C:nTp-large-INF}.

\item We have
\begin{align*}
(\textnormal{II})
 & \leq \| \widetilde{\bs{\Pi}}\bs{F} \|^2 \frac{1}{\frac{1}{\sqrt{nT}} \| \widetilde{\resnw} \|} \frac{1}{\sqrt{nT}} \sum_{i=1}^n \| \nu_i \| \|\gamma_i \| \\
 & = O_p\left(\frac{T\log(pT)}{n} \right) O_p(1) \, O_p\left(\sqrt{\frac{n}{T}}\right) = O_p\left(\frac{\sqrt{T}\log(pT)}{\sqrt{n}} \right) = o_p(1),
\end{align*}
where we have used Propositions  \ref{propC:norm-tilde-pi-F} and \ref{propC:residuals-unif-lower-bounded} as well as assumptions \ref{C:nodewise-loadings} and \ref{C:nTp-large-INF}.

\item By Proposition \ref{propC:norm-tilde-pi-F}, $\pr(\| \widetilde{\bs{\Pi}}\bs{F} \| > \Summable_3 \sqrt{T\log(pT) / n}) \le \summable_n^{(3)}$ with a summable sequence $\{\summable_n^{(3)}\}$. Inspecting the proof of Proposition \ref{propC:norm-tilde-pi-F} in detail reveals that $\{\summable_n^{(3)}\}$ converges to $0$ so fast that $T^2 \summable_n^{(3)} = o(\{ \log(pT)T / n\}^2)$. Using this and the fact that $\| \bs{F} \| = \sqrt{T}$,
we obtain that 
\begin{align*}
\| \bs{F} \|^4 \, \pr\left(\| \widetilde{\bs{\Pi}}\bs{F} \| > \Summable_3 \sqrt{\frac{T\log(pT)}{n}}  \right) = o\left(\frac{\log(pT)^2T^2}{n^2}\right).
\end{align*}
Hence,
\begin{align*}
\ex\left[\| \widetilde{\bs{\Pi}}\bs{F} \|^4  \right]  
 & \leq \ex\left[\| \widetilde{\bs{\Pi}}\bs{F} \|^4 \ \ind\Big\{\| \widetilde{\bs{\Pi}}\bs{F} \| \leq \Summable_3 \sqrt{\frac{T\log(pT)}{n}} \Big\} \right] \\
 & \quad + \ex\left[\| \widetilde{\bs{\Pi}}\bs{F} \|^4 \ \ind\Big\{\| \widetilde{\bs{\Pi}}\bs{F} \| > \Summable_3 \sqrt{\frac{T\log(pT)}{n}} \Big\} \right] \\
 & \leq \frac{\Summable_3^4 \log(pT)^2T^2}{n^2} + \| \bs{F} \|^4 \, \pr\left(\| \widetilde{\bs{\Pi}}\bs{F} \| > \Summable_3 \sqrt{\frac{T\log(pT)}{n}}  \right) \\
 &= O\Big(\frac{\log(pT)^2T^2}{n^2} \Big). 
\end{align*}
From this, it follows that
\begin{align*}
\ex \bigg( \sum_{i=1}^n u_i^\top \widetilde{\bs{\bs{\Pi}}}\bs{F}\gamma_i \bigg)^2 
 & = \sum_{i,l =1}^n \ex\left[ \gamma_l^\top (\widetilde{\bs{\Pi}}\bs{F})^\top u_l u_i^\top (\widetilde{\bs{\Pi}}\bs{F})\gamma_i \right] \\
 & = \sum_{i,l =1}^n \ex\left[\gamma_l^\top (\widetilde{\bs{\Pi}}\bs{F})^\top \ex\big[ u_l u_i^\top \big| \widetilde{\bs{\Pi}},\gamma_i,\gamma_l \big] (\widetilde{\bs{\Pi}}\bs{F})\gamma_i \right] \\
 & = \sum_{i,l =1}^n \ex\left[\gamma_l^\top (\widetilde{\bs{\Pi}}\bs{F})^\top \ex\big[u_l u_i^\top\big] (\widetilde{\bs{\Pi}}\bs{F})\gamma_i \right] \\
 & = \sigma_u^2 \sum_{i =1}^n \ex\left[\gamma_i^\top (\widetilde{\bs{\Pi}}\bs{F})^\top (\widetilde{\bs{\Pi}}\bs{F})\gamma_i \right] \\
 & \leq \sigma_u^2 \sum_{i =1}^n \ex\left[\| \widetilde{\bs{\Pi}}\bs{F} \|^2 \| \gamma_i \|^2 \right] \\
 & \leq n \sigma_u^2 \sqrt{\max_i\ex[\| \gamma_i \|^4 ]} \sqrt{\ex\left[\| \widetilde{\bs{\Pi}}\bs{F} \|^4  \right]} 
   = O(\log(pT)T), 
\end{align*}
where we have used that $u_{i}u_l$ is independent of $\widetilde{\bs{\Pi}}$ and $\gamma_i, \gamma_l$ by \ref{C:nodewise-error}. By Markov's inequality, this implies that
$\sum_{i=1}^n u_i^\top \widetilde{\bs{\Pi}}\bs{F}\gamma_i = O_p(\sqrt{\log(pT)T})$
and hence
\begin{align*}
(\textnormal{III})
 & = \frac{1}{\frac{1}{\sqrt{nT} }\| \widetilde{\resnw} \|} \frac{1}{\sqrt{nT}} \sum_{i=1}^n u_i^\top \widetilde{\bs{\Pi}}\bs{F}\gamma_i 
   = O_p\bigg(\sqrt{\frac{\log(pT)}{n}} \bigg) = o_p(1),
\end{align*}
where we have used Propositions \ref{propC:norm-tilde-pi-F} and \ref{propC:residuals-unif-lower-bounded}. \qedhere

\end{enumerate}
\end{proof}

\subsection*{Step 6: Derivation of the limit distribution of $\bs{\widetilde{\Upsilon}_C}$}

It remains to show that $\widetilde{\Upsilon}_C$ converges to a normal distribution. We in particular prove the following.

\begin{propB}\label{lemma:inference-delta}
It holds that $\widetilde{\Upsilon}_C \convd \normal(0,\sigma_\varepsilon^2)$.
\end{propB}

\begin{proof}
To avoid additional notation, we let $\sigma_\varepsilon^2 = 1$ without loss of generality. For the proof, we use the following two observations:
\begin{enumerate}[label=(\alph*),leftmargin=0.7cm]  
\item By assumption, the random vector $\varepsilon$ is independent from $\bs{X}$. Since the residuals $\widetilde{\resnw}$ are constructed solely from $\bs{X}$, this implies that $\varepsilon$ is independent from $\widetilde{\resnw}$ as well.
\item By Propositions \ref{propC:residuals-unif-lower-bounded} and \ref{propC:residuals-growth}, there exists an event $\mathcal{E}$ with $\pr(\mathcal{E}) = 1$ such that for each $\omega \in \mathcal{E}$, $\| \widetilde{\resnw}(\omega) \|_\infty \leq C_\resnw \sqrt{T} (npT)^{\frac{2+\xi}{\moments}} $ and  $(nT)^{-1} \| \widetilde{\resnw}(\omega) \|^2 > c_\resnw$ for all $n \ge N(\omega)$. 
\end{enumerate}
In what follows, we show that for each $\omega \in \mathcal{E}$, 
\begin{align}\label{eq:cond-weak-conv-delta}
\pr\big(\widetilde{\Upsilon}_C \leq z \, | \, \widetilde{\resnw} = \widetilde{\resnw}(\omega)\big) \to \Phi(z)
\end{align}
for all $z \in \reals$, where $\Phi$ is the distribution function of a standard normal random variable. By Lebesgue's dominated convergence theorem, this implies that
\begin{align*}
\pr(\widetilde{\Upsilon}_C \leq z) = \ex\left[\pr(\widetilde{\Upsilon}_C \leq z \, | \, \widetilde{\resnw})\right] \to \Phi(z)
\end{align*}
for all $z \in \reals$, which means that $\widetilde{\Upsilon}_C \stackrel{d}{\to} \normal(0,1)$. It thus remains to prove (\ref{eq:cond-weak-conv-delta}). In order to do so, we write 
\begin{align*}
\widetilde{\Upsilon}_C = \sum_{i=1}^n \sum_{t=1}^T \xi_{it} \qquad \text{with} \qquad \xi_{it} = \frac{\widetilde{\resnw}_{it}\varepsilon_{it}}{\|\widetilde{\resnw}\|}
\end{align*}
which is independent across $i$ and $t$ conditioned on $\widetilde{\resnw}$ by the conditions in \ref{C:loadings}, \ref{C:eps} and \ref{C:eps-stronger}. 
Using observation (a) and the fact that $\ex[\varepsilon_i\varepsilon_i^\top]=\bs{I}$, we have that for any $\omega \in \mathcal{E}$, $\ex[\xi_{it} \, | \, \widetilde{\resnw} = \widetilde{\resnw}(\omega)] = 0$ and 
\begin{align*}
\ex\big[\xi_{it}^2 \, | \, \widetilde{\resnw} = \widetilde{\resnw}(\omega)\big] = \frac{\widetilde{\resnw}_{it}^2(\omega) \ex[\varepsilon_{it}^2]}{\|\widetilde{\resnw}(\omega)\|^2} = \frac{\widetilde{\resnw}_{it}(\omega)^2}{\|\widetilde{\resnw}(\omega)\|^2}, 
\end{align*}
which in particular implies that $\sum_{i=1}^n \sum_{t=1}^T \ex[\xi_{it}^2 \, | \, \widetilde{\resnw} = \widetilde{\resnw}(\omega)] = 1$. We now verify that the Lindeberg condition is satisfied for any $\omega \in \mathcal{E}$ (and $n \ge N(\omega)$):  
\begin{align*}
 & \sum_{i=1}^n \sum_{t=1}^T \ex\left[\xi_{it}^2 \ind\{|\xi_{it}| \geq c\} \, \bigg| \, \widetilde{\resnw} = \widetilde{\resnw}(\omega)\right] \\
 & = \sum_{i=1}^n \sum_{t=1}^T \frac{\widetilde{\resnw}_{it}^2(\omega)}{\|\widetilde{\resnw}(\omega)\|^2}\ex\left[\varepsilon_{it}^2 \ind\left\{ \frac{|\widetilde{\resnw}_{it}\varepsilon_{it}|}{\|\widetilde{\resnw}\|} \geq c \right\} \bigg| \widetilde{\resnw} = \widetilde{\resnw}(\omega)\right] \\
 & \leq \sum_{i=1}^n \sum_{t=1}^T \frac{\max_{i,t}|\widetilde{\resnw}_{it}(\omega)|^2}{\|\widetilde{\resnw}(\omega)\|^2} \ex\left[|\varepsilon_{it}|^2 \ind\left\{\frac{|\varepsilon_{it}|\max_{i,t}|\widetilde{\resnw}_{it}| }{\|\widetilde{\resnw}\|}  \geq c\right\}\bigg|\widetilde{\resnw} = \widetilde{\resnw}(\omega)\right]  \\
 &\leq \sum_{i=1}^n \sum_{t=1}^T \frac{\max_{i,t}|\widetilde{\resnw}_{it}(\omega)|^\moments}{\|\widetilde{\resnw}(\omega)\|^\moments c^{\moments-2}} \ex\left[|\varepsilon_{it}|^\moments \bigg|\widetilde{\resnw} = \widetilde{\resnw}(\omega)\right]  \\
 &\leq C (nT)/c^{\moments-2} \left(\frac{\sqrt{T} (npT)^{\frac{2+\xi}{\moments}}}{\sqrt{nT}}\right)^\moments = o(1)
\end{align*}
for any $c > 0$, where we have used observation (b) together with assumption \ref{C:nTp-large-INF} and \ref{C:eps}. Applying Lindeberg's central limit theorem now yields (\ref{eq:cond-weak-conv-delta}). 
\end{proof}

\subsection*{Step 7: Consistent estimation of $\bs{\sigma_\varepsilon^2}$}

We finally show that $\widetilde{\sigma}_\varepsilon^2$ is a consistent estimator of the unknown error variance $\sigma_\varepsilon^2$.

\begin{propB}\label{propC:error-variance}
It holds that 
\[\widetilde{\sigma}_\varepsilon^2 = \frac{1}{n(T-K)}\sum_{i=1}^n \|\widetilde{\bs{\Pi}}Y_i - \widetilde{\bs{\Pi}}\bs{X}_i \widetilde{\beta}_\pen \|^2 \convp \sigma_\varepsilon^2.\]
\end{propB}

\subsection*{Step 8: Proof of intermediate results}

It remains to complete the proofs of Propositions \ref{propC:tilde-pi-equals-pi-r}--\ref{propC:error-variance}. To do so, we require a couple of auxiliary lemmas which are formulated below. The remaining proofs of the propositions and auxiliary lemmas can be found in the supplementary material.


\begin{lemmaB}\label{lemmaC:aux1}
There exist positive constants $\Summable_\ell$ and non-negative summable sequences $\{\summable_n^{(\ell)}\}$ for $5 \le \ell \le 10$ such that for sufficiently large $n$,
\begin{enumerate}[label=(\roman*),leftmargin=0.975cm,topsep=0.5cm]
\item \label{lemmaC:aux1:normbarZ-j} 
$\displaystyle{ \pr\left( \left\| \overline{\bs{Z}}_{(-j)} \right\| > \Summable_5 \sqrt{\frac{pT \log(pT)}{n}} \right) \leq \summable_n^{(5)}}$
\item \label{lemmaC:aux1:maxFui}
$\displaystyle{\pr\left(\max_{1 \le i \le n} \left\| \frac{\bs{F}^\top u_i}{T} \right\| > \Summable_6 \sqrt{\frac{\log n}{T}} \right) \leq \summable_n^{(6)}}$
\item \label{lemmaC:aux1:maxnormZijF}
$\displaystyle{ \pr\Bigg( \max_{\substack{1 \le i \le n \\ 1 \le j' \le n, j'\neq j}} \bigg\| \frac{Z_{i(j')}^\top \bs{F}}{T} \bigg\| > \Summable_7 \sqrt{\frac{\log(np)}{T}} \Bigg) \leq \summable_n^{(7)}}$
\item \label{lemmaC:aux1:normbarZF}
$\displaystyle{\pr\left( \bigg\| \frac{\bs{F}^\top \overline{\bs{Z}}_{(-j)}}{T} \bigg\| > \Summable_8 \sqrt{\frac{p \log(npT) \log p}{nT}}\right) \leq \summable_n^{(8)}}$ 
\item \label{lemmaC:aux1:normbarZui}
$\displaystyle{\pr\left(\max_{1 \le i \le n} \bigg\| \frac{\overline{\bs{Z}}_{(-j)}^\top u_i}{T} \bigg\| > \Summable_9 \sqrt{\frac{p\log(npT)\log(np)}{nT}} \right) \leq \summable_n^{(9)}}$
\item \label{lemmaC:aux1:maxnormZijbarZ}
$\displaystyle{\pr\Bigg( \max_{\substack{1 \le i \le n \\ 1 \le j' \le n, j'\neq j}} \bigg\|\frac{Z_{i(j')}^\top \overline{\bs{Z}}_{(-j)}}{T} \bigg\| > \Summable_{10} \sqrt{p} \bigg[ \sqrt{\frac{\log(npT)\log(np^2)}{nT}} + \frac{1}{n}\bigg] \Bigg) \leq \summable_n^{(10)}}$.            
\end{enumerate}
\end{lemmaB}

\begin{lemmaB}\label{lemmaC:aux2}
There exist positive constants $\Summable_\ell$ and non-negative summable sequences $\{\summable_n^{(\ell)}\}$ for $11 \le \ell \le 16$ such that for sufficiently large $n$,
\begin{enumerate}[label=(\roman*),leftmargin=0.975cm,topsep=0.5cm]
\item \label{lemmaC:aux2:normbargamma}
$\displaystyle{\pr\left(\|\overline{\bs{\Gamma}}_{-j}\| > \Summable_{11} \sqrt{p} \right) \leq \summable_n^{(11)}}$
\item \label{lemmaC:aux2:sigmasigmabar}
$\displaystyle{\|\bs{\Sigma}^{[-j]}-\overline{\bs{\Sigma}}^{[-j]}\| \le \frac{\Summable_{12} p}{n}}$
\item \label{lemmaC:aux2:sigmahatsigmabar}
$\displaystyle{\pr\left(\|\overline{\bs{\Sigma}}^{[-j]}-\widetilde{\bs{\Sigma}}\| > \Summable_{13} p\sqrt{\frac{\log p}{n}} \right) \leq \summable_n^{(13)}}$. 
\item \label{lemmaC:aux2:normpsi-1}
$\displaystyle{\pr\left(\big\| \widetilde{\bs{\Eig}}^{-1} \big\| > \frac{\Summable_{14}}{p} \right) \leq \summable_n^{(14)}}$   
\item \label{lemmaC:aux2:psibar1gammaUF1}
$\displaystyle{\pr\Bigg(\bigg\| \widetilde{\bs{\Eig}} - \left(\frac{1}{T}(\bs{F}\overline{\bs{\Gamma}}_{-j}^\top\widetilde{\bs{U}})^\top\bs{F}\overline{\bs{\Gamma}}_{-j}^\top\widetilde{\bs{U}}\right)\bigg\|}$ \\*[0.2cm]
$\displaystyle{\qquad > \Summable_{15} p \bigg\{ \sqrt{\frac{\log(p)\log(npT)}{nT}} + \frac{\log(pT)}{n} \bigg\} \Bigg) \leq \summable_n^{(15)}}$
\item \label{lemmaC:aux2:psibar-1gammaUF-1}
$\displaystyle{\pr\Bigg(\bigg\| \widetilde{\bs{\Eig}}^{ -1} - \left(\frac{1}{T}(\bs{F}\overline{\bs{\Gamma}}_{-j}^\top\widetilde{\bs{U}})^\top \bs{F}\overline{\bs{\Gamma}}_{-j}^\top\widetilde{\bs{U}}\right)^{-1} \bigg\|}$ \\*[0.2cm]
$\displaystyle{\qquad > \frac{\Summable_{16}}{p} \bigg\{ \sqrt{\frac{\log(p)\log(npT)}{nT}} + \frac{\log(pT)}{n} \bigg\} \Bigg) \leq \summable_n^{(16)}}$.
\end{enumerate}
\end{lemmaB}


\begin{lemmaB}\label{lemmaC:propertiesofPiu}
\hfill
\begin{enumerate}[label=(\roman*),leftmargin=0.975cm]
\item \label{lemmaC:propertiesofPiu_itemi}
It holds that $\ex\left[ \|\bs{\Pi} u_i \|^2 \right] = (T - K)\ex[u_{11}^2] $.
\item \label{lemmaC:propertiesofPiu-itemii} 
For any $\varepsilon>0$, there exists a non-negative summable sequence $\{\summable_n^{(17)}\}$ such that 
\begin{align*}
\pr\left(  \left|\frac{1}{nT}\sum_{i=1}^n (\| \bs{\Pi}u_i \|^2 -\ex\left[ \|\bs{\Pi} u_i \|^2 \right])   \right| >\varepsilon   \right) \leq \summable_n^{(17)}.
\end{align*}
\end{enumerate}
\end{lemmaB}

\newpage

\headingSupp{Supplement to}
{"Estimation and Inference in}
{High-Dimensional Panel Data Models}
{with Interactive Fixed Effects"}

\authors
{Maximilian R\"ucker}{Ulm University}
{Michael Vogt}{Ulm University}
\vspace{-1.25cm}

\authors
{Oliver Linton}{University of Cambridge}
{Christopher Walsh}{Newcastle University}
\vspace{-1cm}

\thispagestyle{empty}

\renewcommand{\baselinestretch}{1.0}\normalsize
\renewcommand{\abstractname}{}
\begin{abstract}
\noindent The supplementary material comprises two main parts. In the first part, we provide the simulation exercises summarized in Section \ref{sec:sim-overview} of the paper. In the second part, we give the proofs and technical details that are omitted in the paper.
\end{abstract}

\renewcommand{\baselinestretch}{1.2}\normalsize

\newpage

\setcounter{section}{0}
\def\thesection{S.\arabic{section}}
\def\thetable{S.\arabic{table}}
\def\thefigure{S.\arabic{figure}}
\def\theequation{S.\arabic{equation}}
\setcounter{subsection}{0}
\setcounter{equation}{0}
\setcounter{page}{1}

\begin{center}
{\LARGE \textbf{Supplement: Simulation Study}}
\end{center}

\section{Simulation design}\label{sec:sim-design}
\setcounter{table}{0}
\setcounter{figure}{0}

We simulate data from the model $Y_{it} = \beta^\top X_{it} + \gamma_i^\top F_t + \varepsilon_{it}$ with $K=3$ unobserved factors. The model components are generated as follows:
\begin{itemize}[leftmargin=0.45cm]

\item The error terms $\varepsilon _{it}$ are standard normal draws independent across $i$ and $t$.

\item The unobserved factors $F_{t}=(F_{t,1},F_{t,2},F_{t,3})^{\top }$ are generated as stationary AR(1) processes with zero means and unit variances. Specifically, for each $k\in \{1,2,3\}$, we let $F_{t,k}=0.5F_{t-1,k}+w_{t,k}$, where the innovations $w_{t,k}$ are $N(0,0.75)$-distributed and independent across $t$ and $k$. By construction, the factors are orthonormal in the sense that $\ex[F_tF_t^\top]=\boldsymbol{I}_{K}$. 

\item The $p=1+3d$ covariates $X_{it}$ are constructed as follows: The first covariate is generated according to the nodewise regression structure 
\begin{equation}\label{eq:nodewise-sim}
X_{it,1} = X_{it,-1} \theta + F_t^\top \nu_i + u_{it}, 
\end{equation}
where $\theta = (\theta_1,0,\ldots,0)^\top$ is a sparse parameter vector whose only non-zero entry is the first element $\theta_1$ (the value of which is chosen below). Moreover, the variables $u_{it}$ are standard normal draws independent across $i$ and $t$, and we set $\nu_i = 0$ for all $i$ for simplicity. The remaining $3d$ regressors are generated using the factor structure 
\[ X_{it,-1} = \bs{\Gamma}_{i,-1}F_{t}+Z_{it,-1}, \]
where the random vectors $Z_{it,-1}\in \reals^{3d}$ are drawn independently across $i$ and $t$ from a multivariate standard normal distribution $\normal(0,\bs{I})$ and 
\[ \bs{\Gamma}_{i,-1}=\begin{pmatrix} \Gamma^{(1)}_i & 0 & 0 \\ 0 & \Gamma^{(2)}_i & 0 \\ 0 & 0 & \Gamma^{(3)}_i \end{pmatrix} \in \reals ^{3d\times 3}\]
with random vectors $\Gamma _{i}^{(1)}=(\Gamma _{i,1},\dots ,\Gamma _{i,d})^{\top }$, $\Gamma_{i}^{(2)}=(\Gamma _{i,d+1},\dots ,\Gamma _{i,2d})^{\top }$ and $\Gamma_{i}^{(3)}=(\Gamma _{i,2d+1},\dots ,\Gamma _{i,3d})^{\top }$ of length $d$ (which are specified below). 

\item We collect the factor loadings from the outcome and the regressor equations (except the $\nu_i$'s which are equal to $0$) in a large vector $G_{i}=(\gamma _{i}^{\top}$, $\{\Gamma _{i}^{(1)}\}^{\top }$, $\{\Gamma
_{i}^{(2)}\}^{\top }$, $\{\Gamma _{i}^{(3)}\}^{\top })^{\top }$ and draw the random vectors $G_{i}$ independently from a multivariate normal distribution $\normal(\mu ,\bs{\Omega})$. Here, $\mu$ is a vector of ones and 
\[
\boldsymbol{\Omega }=
\begin{pmatrix}
1 & \rho & \cdots & \rho\\
\rho & \ddots & \ddots & \vdots\\
\vdots & \ddots & \ddots & \rho\\
\rho & \cdots & \rho & 1
\end{pmatrix},
\]
so that $\rho$ governs the pairwise correlation between the factor loadings. We set $\rho = 0.25$ throughout. 

\item We pick $\theta_1$ such that the covariates $X_{it,j}$ have the same mean and variance for all $j$. The resulting choice is $\theta_1 = \sqrt{2/3}$. (Obviously, $\ex[X_{it,j}]=0$ for all $j$. Moreover, straightforward calculations yield that  $\ex[X_{it,j}^2]=3$ for all $j > 1$ and $\ex[X_{it,1}^2] = 3 \theta_1^2 + 1$. Setting $\theta_1 = \sqrt{2/3}$, we thus get that $\ex[X_{it,1}^2] = 3$.)

\item For the analysis of the HD-CCE estimator, we consider a parameter vector $\beta$ of the form 
\[ \beta = (c^*, \beta^{(1)}, \beta^{(2)}, \beta^{(3)})^\top, \]
where $c^*$ is specified below and $\beta^{(\ell)} = (c^*,c^*,c^*,0,\ldots,0) \in \reals^d$ for $\ell \in \{1,2,3\}$. (In the special case that $d \le 3$, $\beta^{(\ell)}$ reduces to a vector with all entries equal to $c^*$.) The constant $c^*$ is chosen such that the signal-to-noise ratio $\textnormal{SNR} := \var(X_{it}^\top \beta) / \var(F_t^\top \gamma_i + \varepsilon_{it})$ is approximately equal to $1$. The resulting choice is $c^* = 0.4$. For the analysis of the desparsified HD-CCE estimator, we let $\beta$ have the form 
\[ \beta = (c^{**}, \beta^{(1)}, \beta^{(2)}, \beta^{(3)})^\top, \]
where the subvectors $\beta^{(\ell)}$ are chosen as above and $c^{**}$ takes different values which are specified below. 

\begin{table}[b!]
\centering
{\small
\caption{Simulation settings}\label{table:sim:settings} 
\begin{tabular}{
p{3cm}
>{\centering}p{3cm}
>{\centering}p{3cm}
>{\centering\arraybackslash}p{3cm}
}
\toprule \\[-0.4cm]
                & Scenario A & Scenario B & Scenario C \\
                & $p<T$& $T\leq p < nT$ & $nT < p$ \\[0.1cm]
\hline\\[-0.4cm]
$(n,T)=(50,15)$ & $p=\phantom{0}7,10,13$ & $p=31,151,301$ & $p=\phantom{0}901$ \\
$(n,T)=(50,50)$ & $p=16,31,46$ & $p=91, 451, 901$ & $p=3001$ \\[0.1cm]
\bottomrule
\end{tabular}}
\end{table}

\item We consider different choices for $(n,T,p=3d+1)$ which are summarized in Table \ref{table:sim:settings}. Scenario A ($p < T$) covers ``low-dimensional'' cases where $p$ is rather small, Scenario B ($T \le p < nT$) \textquotedblleft moderately
high-dimensional\textquotedblright\ cases where $p$ is fairly large but still smaller than the sample size $nT$, and Scenario C ($nT < p$) \textquotedblleft truly high-dimensional\textquotedblright\ cases where $p$ exceeds the sample size $nT$. Notably, the CCE estimator is only available in Scenario A because it breaks down as soon as $p \ge T$.

\end{itemize}
All Monte Carlo experiments are based on $1000$ simulation runs. The tuning parameters $\tau$, $\pen$ and $\pennw$ of our methods are chosen as recommended in Section \ref{sec:est:tuning}. In particular, we set $\tau = \alpha \widehat{\eig}_1$ with $\alpha = 0.01$ and use $L=10$ folds for cross-validation.

\section{Parameter estimation}\label{sec:sim-est}
\setcounter{table}{1}
\setcounter{figure}{0}

\begin{table}[b]
\caption{Overview of estimators}\label{table:estimators}
\centering
{\small\begin{tabular}{@{\extracolsep{5pt}} llll} 
\toprule\\[-0.4cm]
estimator & label & description & availability \\[0.1cm]
\hline\\[-0.4cm]
$\widehat{\beta}_\lambda$ & L & $\bs{\widehat{\Pi}}$ + lasso & Scenario A, B, C \\[0.05cm]
$\widehat{\beta}_{\lambda}^{\text{oracle}}$ & L-O & $\bs{\Pi}$ + lasso & Scenario A, B, C \\[0.05cm]
$\widehat{\beta}_{\textnormal{LS}}$ & LS & $\bs{\widehat{\Pi}}$ + least squares & Scenario A, B \\[0.05cm]
$\widehat{\beta}_{\textnormal{LS}}^{\text{oracle}}$ & LS-O & $\bs{\Pi}$ + least squares & Scenario A, B \\[0.05cm]
$\widehat{\beta}_{\textnormal{LS}}^{\text{double-oracle}}$ & LS-O$^2$  & $\bs{\Pi}$ + least squares on support $S$ & Scenario A, B, C \\[0.05cm]
$\widehat{\beta}_{\textnormal{CCE}}$ & CCE & CCE from \cite{Pesaran2006} & Scenario A \\[0.1cm]
\bottomrule
\end{tabular}}
\end{table}

In the first part of the simulation study, we evaluate the finite sample performance of our HD-CCE estimator $\widehat{\beta}_\lambda$ and its least squares version $\widehat{\beta}_{\textnormal{LS}}$. To do so, we compare them with oracle versions $\widehat{\beta}_\lambda^{\text{oracle}}$ and $\widehat{\beta}_{\text{LS}}^{\text{oracle}}$ which are computed in exactly the same way except that the proxy $\widehat{\bs{\Pi}}$ is replaced by the ``oracle'' matrix $\bs{\Pi}$. Moreover, we consider a ``double oracle'' estimator $\widehat{\beta}_{\text{LS}}^{\text{double-oracle}}$ which makes use of the true projection matrix $\bs{\Pi}$ and the true support $S = \{j: \beta_j \ne 0\}$ of the parameter vector $\beta$. This ``double oracle''  estimator is constructed exactly as $\widehat{\beta}_{\text{LS}}^{\text{oracle}}$ except that only the regressors $j \in S$ are included in the estimation. We further compute the original CCE estimator of \cite{Pesaran2006},  in particular, the CCEP version from equation (65) therein with the weights $\theta_i = w_i = 1/N$. Some of the estimators cannot be computed in all Scenarios A, B and C. The CCE estimator, for instance, is only available in Scenario A. Table \ref{table:estimators} gives an overview of the considered estimators and indicates in which scenarios they are available.

In our simulation design, there are four different groups of covariates: (a) the first covariate defined via the nodewise regression equation \eqref{eq:nodewise-sim}, (b) the covariates $j \in \{2,\ldots,d+1\}$ which are influenced only by the first factor, (c) the covariates $j \in \{d+2,\ldots,2d+1\}$ which are influenced only by the second factor, and (d) the covariates $j \in \{2d+2,\ldots,3d+1\}$ which are influenced only by the third factor. Moreover, there are three different types of covariates in group (b): the first covariate in this group has a non-zero coefficient in both $\beta$ and the nodewise parameter vector $\theta$, the second and third have a non-zero coefficient in $\beta$ but a zero coefficient in $\theta$, and the remaining ones have zero coefficients in both $\beta$ and $\theta$. Similarly, there are two different types of covariates in groups (c) and (d): the first three covariates in each group have a non-zero coefficient in $\beta$, while the others all have a zero coefficient in $\beta$. This makes eight different types of regressors in total. As the model is completely symmetric in the regressors of each type, it suffices to report the simulation results for one representative regressor per type. We in particular pick the regressors $j=1,2,3,5,d+2,d+5,2d+2,2d+5$ as the representatives of the eight types. (Note that in Scenario A with $p \in \{7,10\}$, there are only five types of regressors as the parameter vector $\beta$ does not include any zeros.)

\begin{figure}[p]
\centering
\begin{subfigure}[b]{0.475\textwidth}   
\centering 
\includegraphics[width=\textwidth]{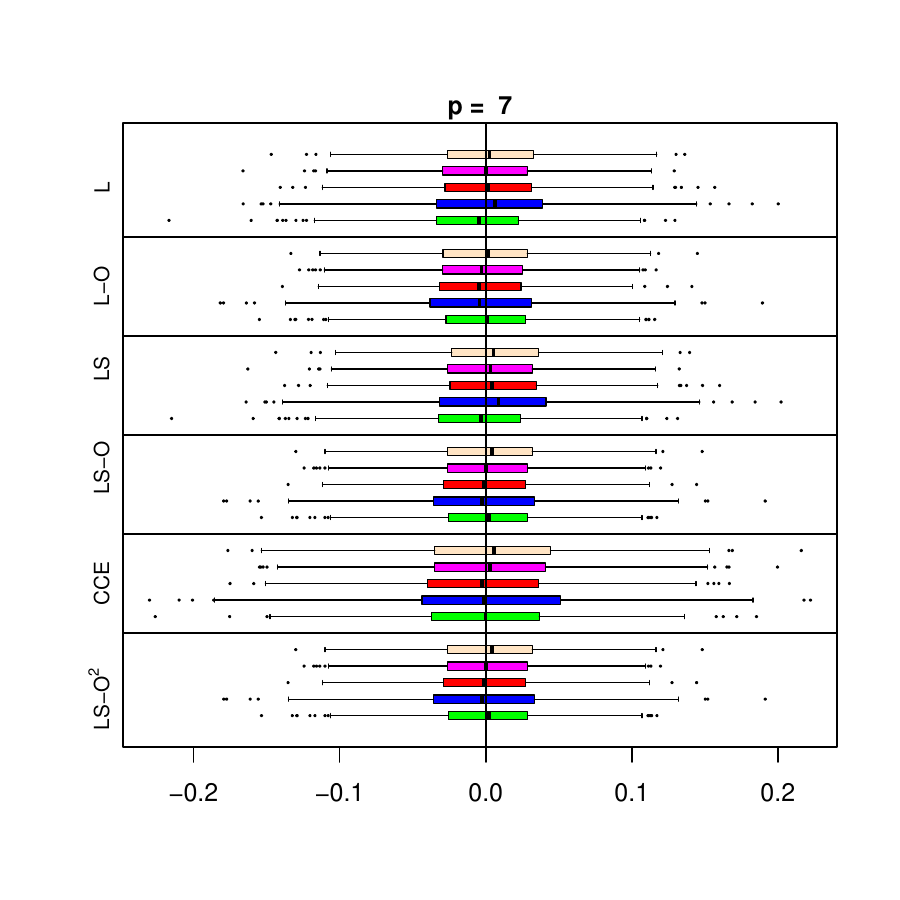} \\
\includegraphics[width=\textwidth]{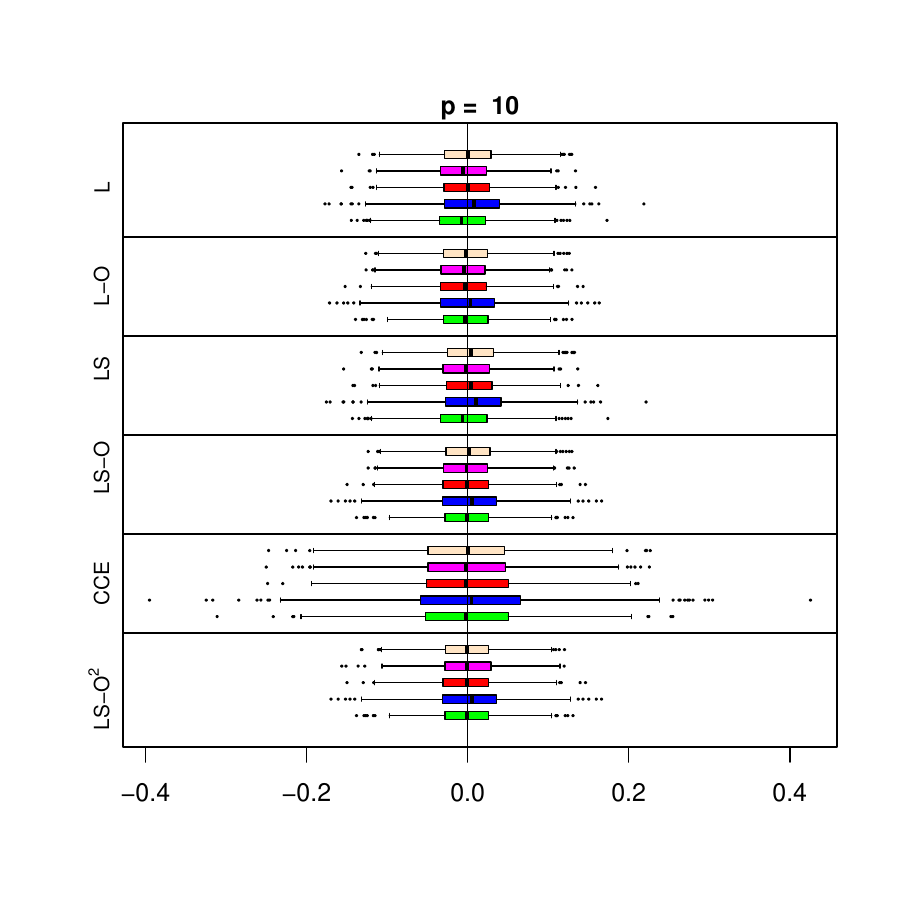} \\ 
\includegraphics[width=\textwidth]{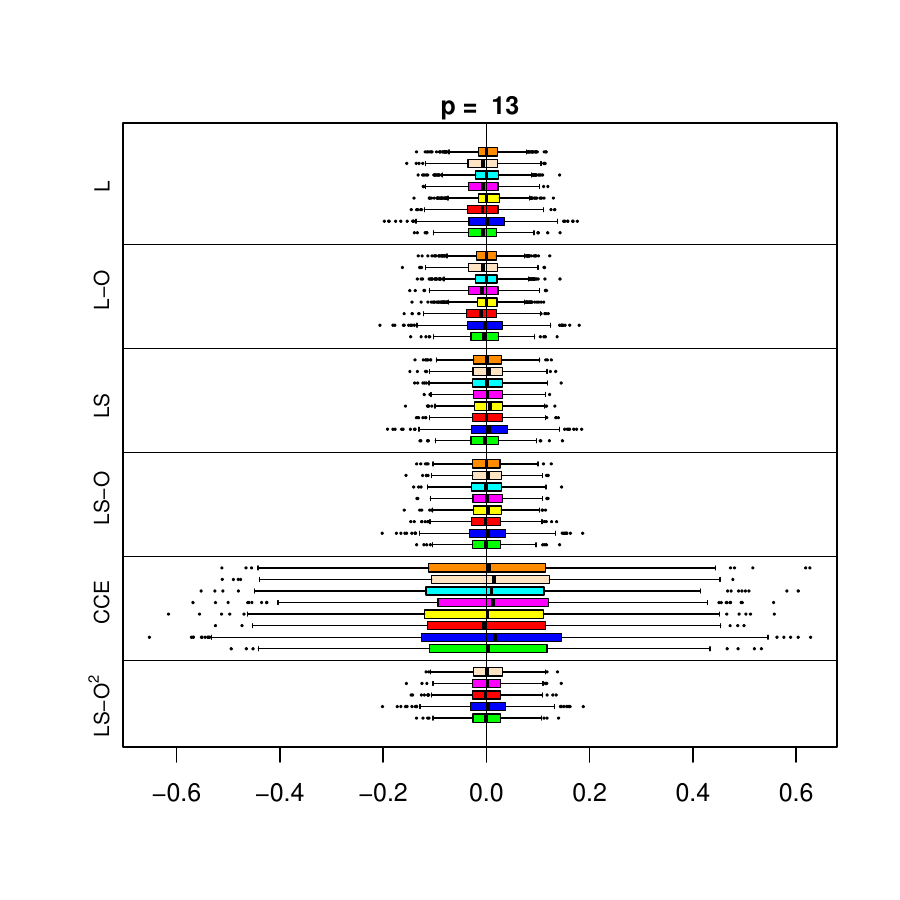}
\subcaption{$T=15$}\label{fig:results:A-smallT}
\end{subfigure}
\hspace{0.2cm}
\begin{subfigure}[b]{0.475\textwidth}   
\centering 
\includegraphics[width=\textwidth]{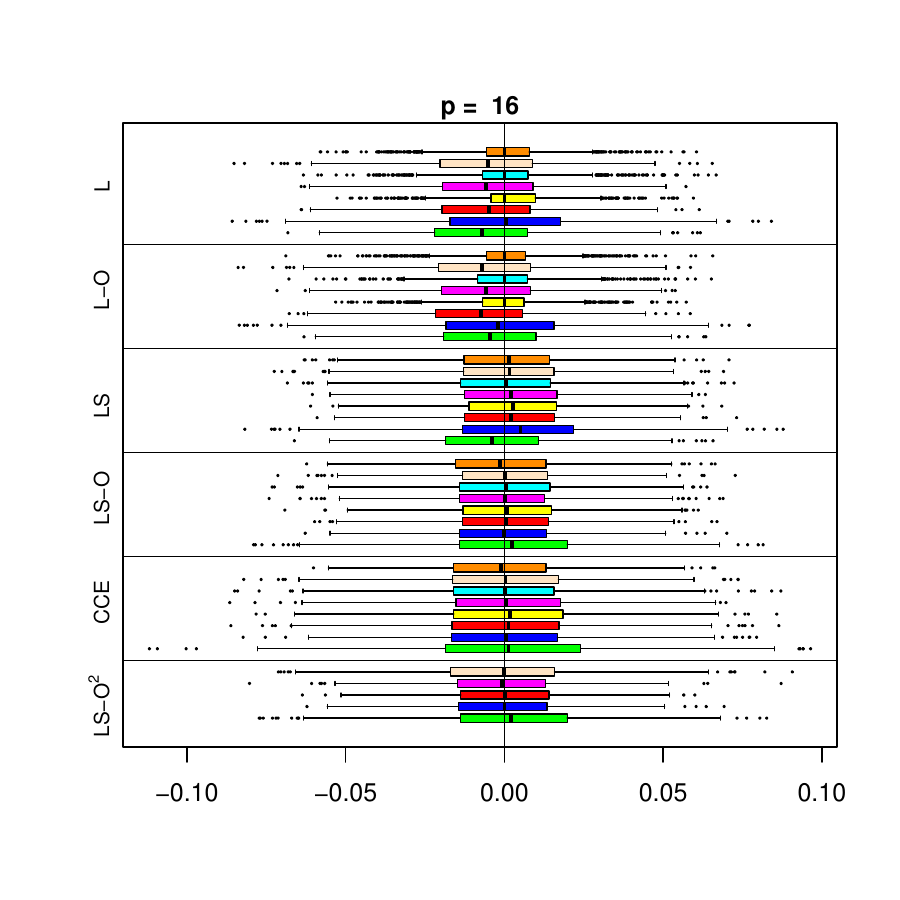} \\
\includegraphics[width=\textwidth]{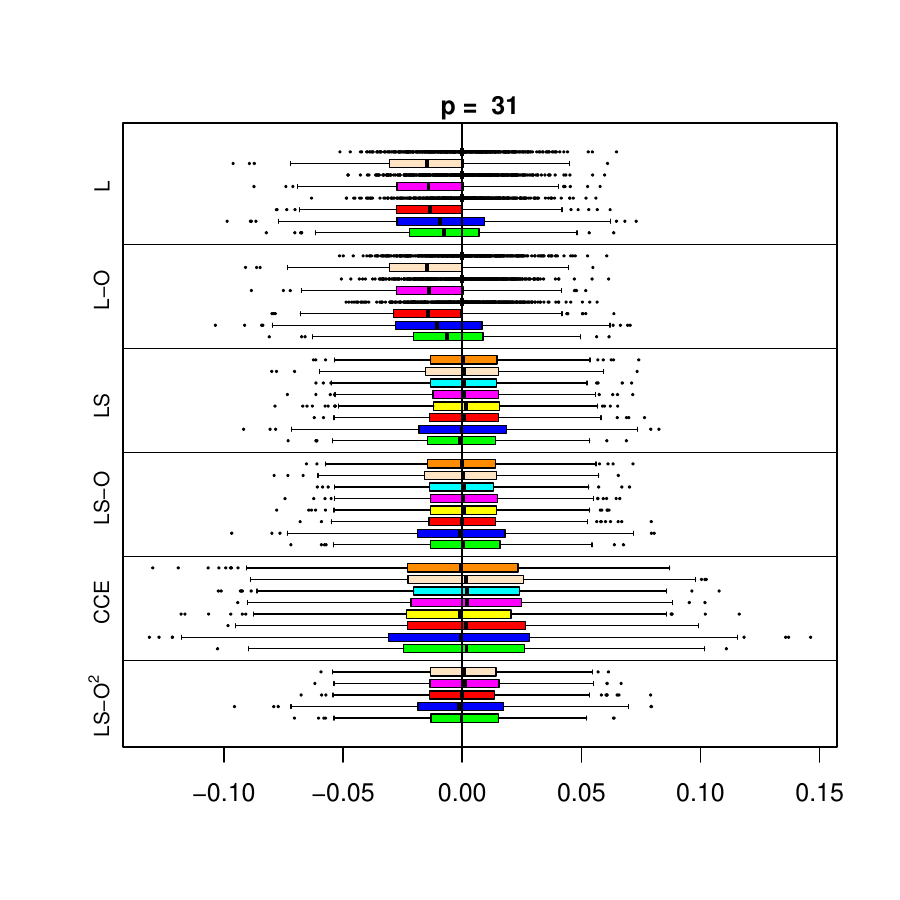} \\
\includegraphics[width=\textwidth]{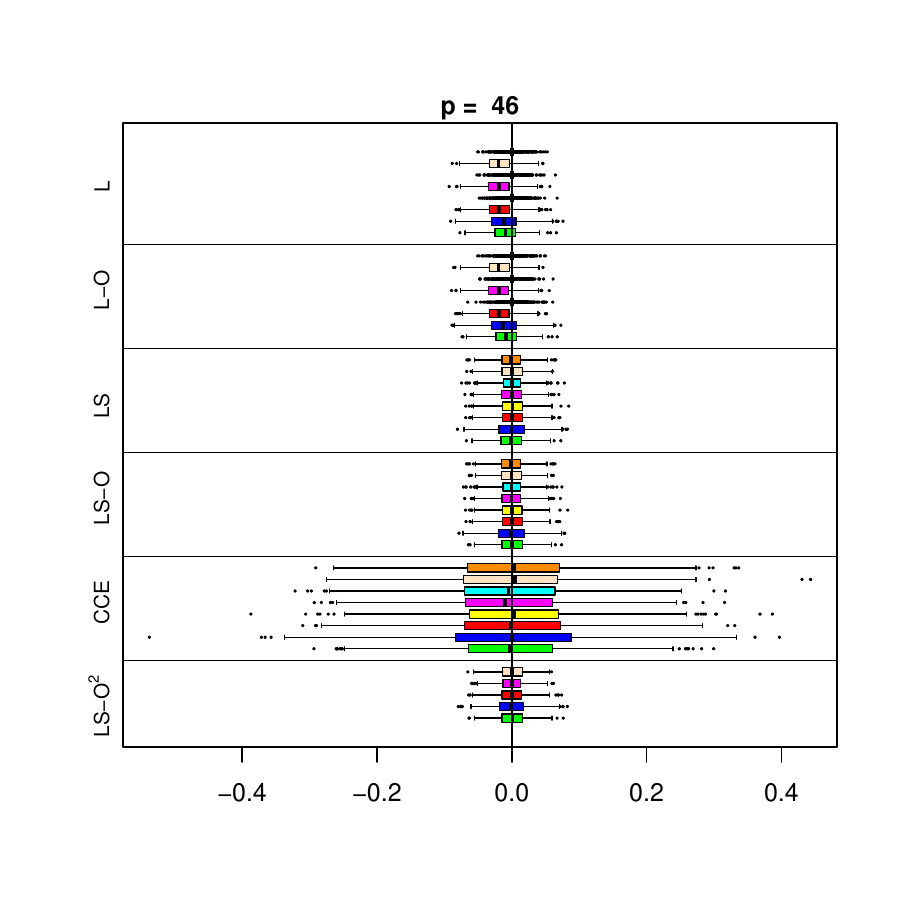}
\subcaption{$T=50$}\label{fig:results:A-largeT}
\end{subfigure}
\caption{Simulation results in Scenario A.}\label{fig:results:A}
\end{figure}

\begin{figure}[p]
\centering
\begin{subfigure}[b]{0.475\textwidth}   
\centering 
\includegraphics[width=\textwidth]{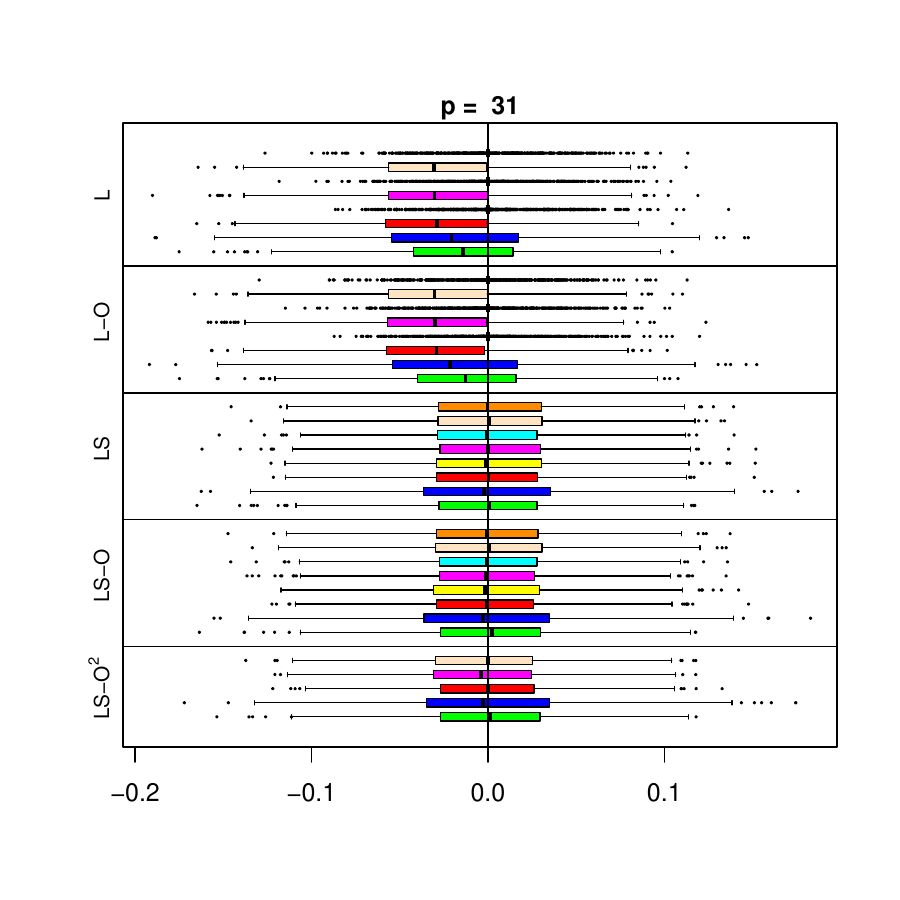} \\
\includegraphics[width=\textwidth]{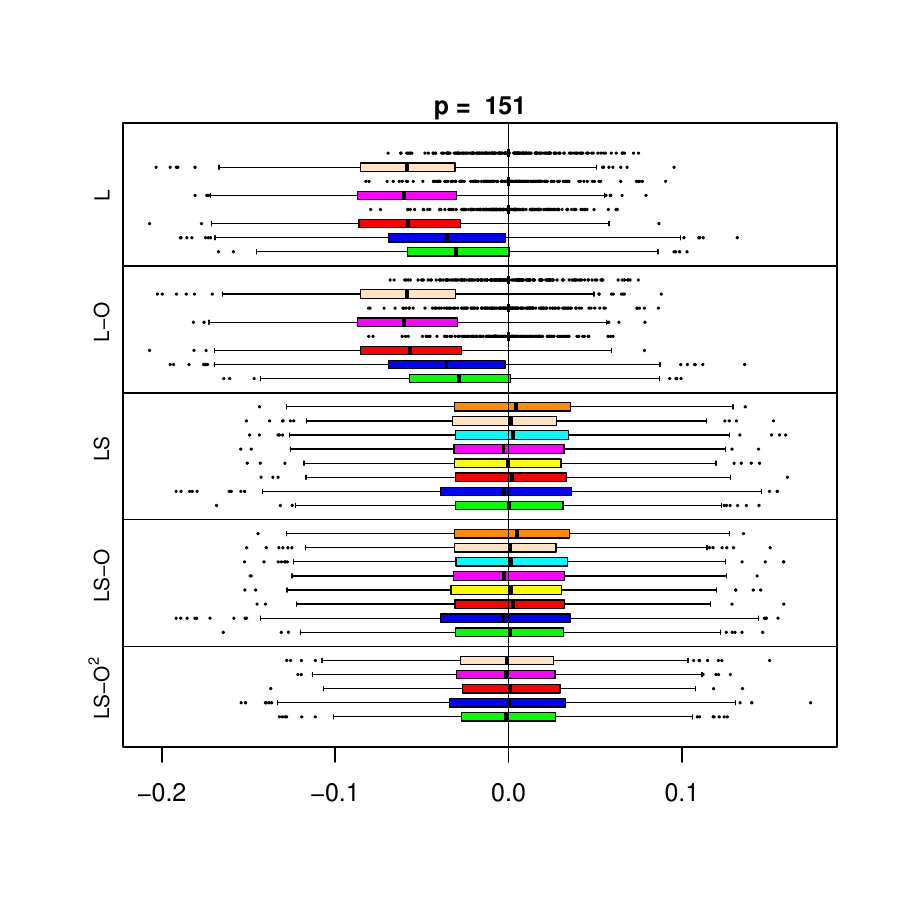} \\ 
\includegraphics[width=\textwidth]{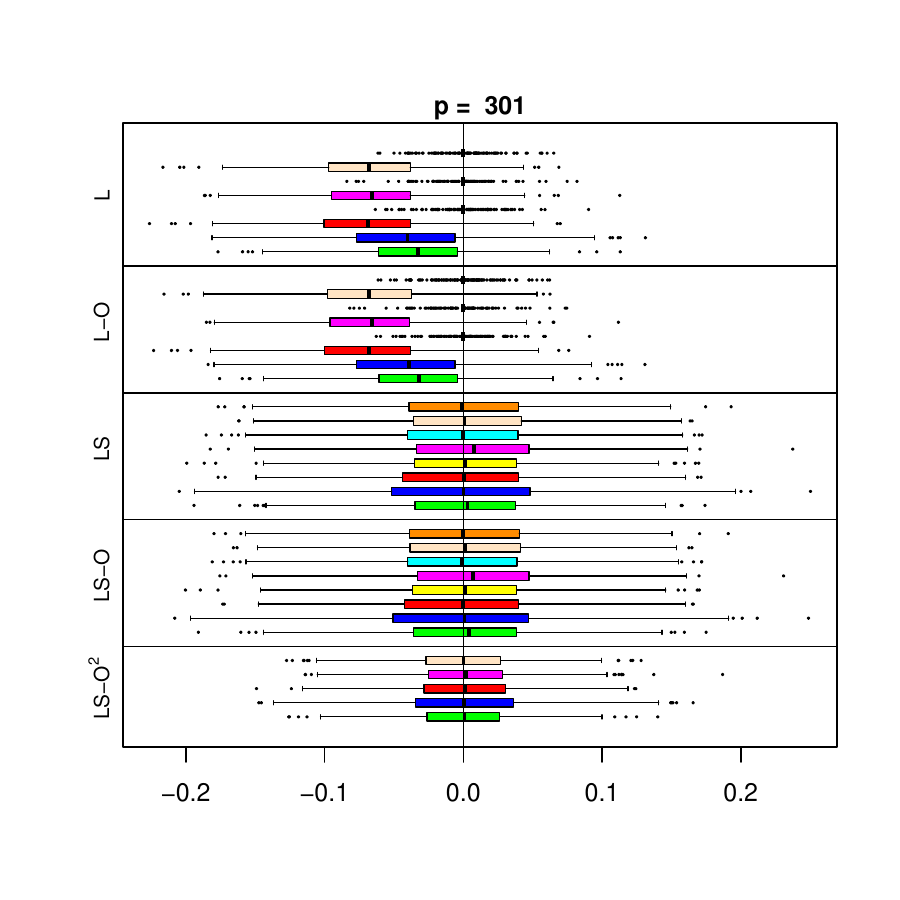}
\subcaption{$T=15$}\label{fig:results:B-smallT}
\end{subfigure}
\hspace{0.2cm}
\begin{subfigure}[b]{0.475\textwidth}   
\centering 
\includegraphics[width=\textwidth]{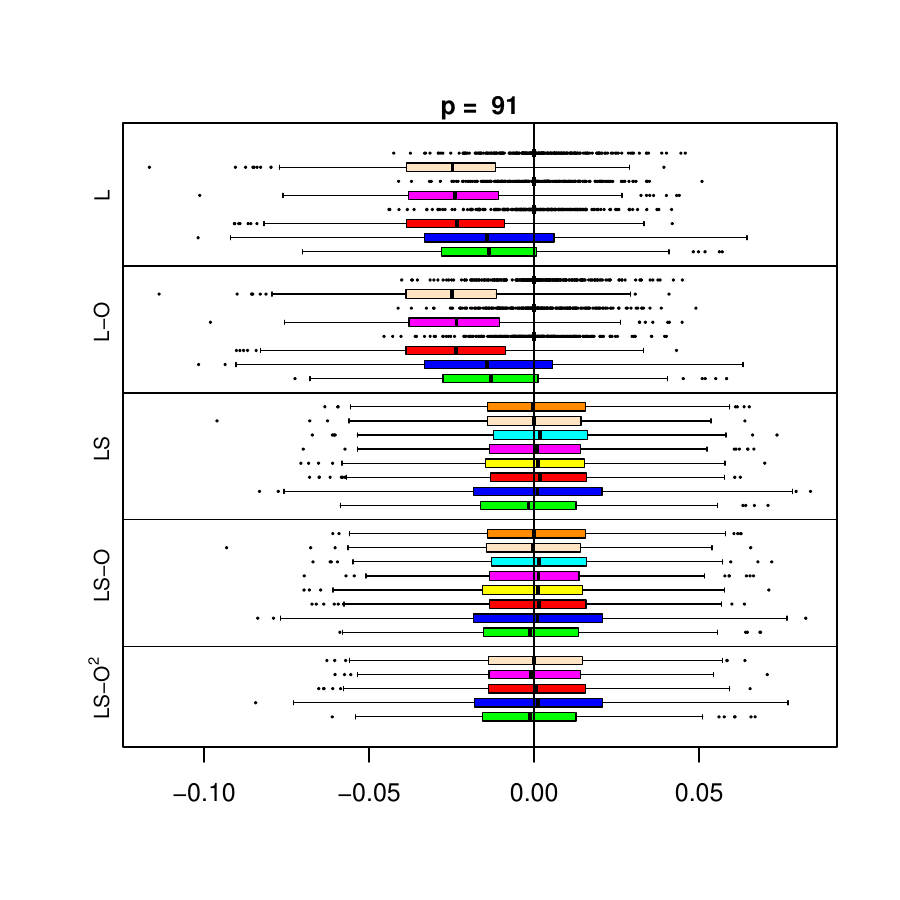} \\
\includegraphics[width=\textwidth]{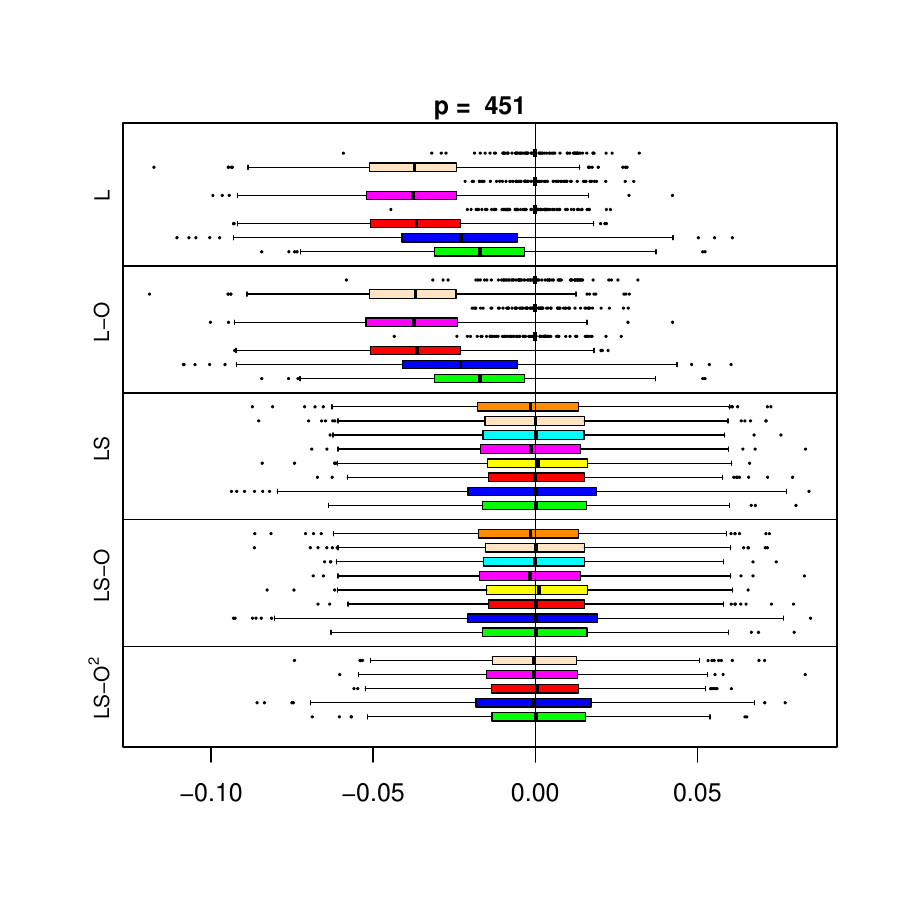} \\
\includegraphics[width=\textwidth]{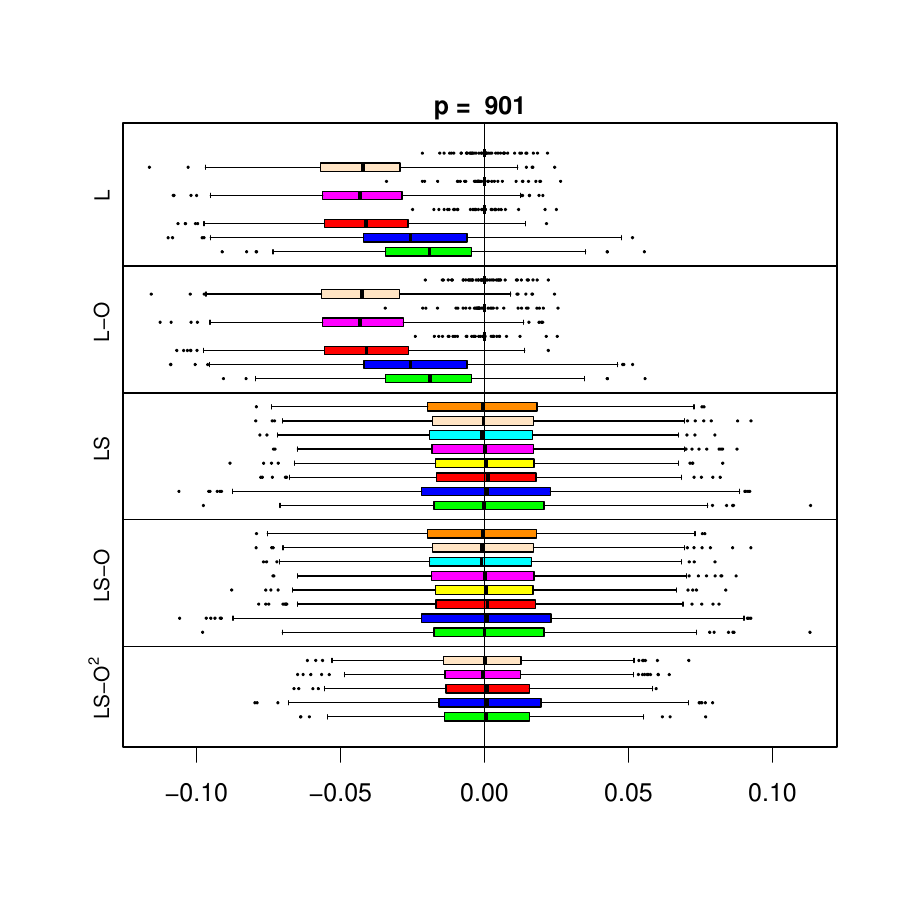}
\subcaption{$T=50$}\label{fig:results:B-largeT}
\end{subfigure}
\caption{Simulation results in Scenario B.}\label{fig:results:B}
\end{figure}

\begin{figure}[t]
\centering
\begin{subfigure}[b]{0.475\textwidth}   
\centering 
\includegraphics[width=\textwidth]{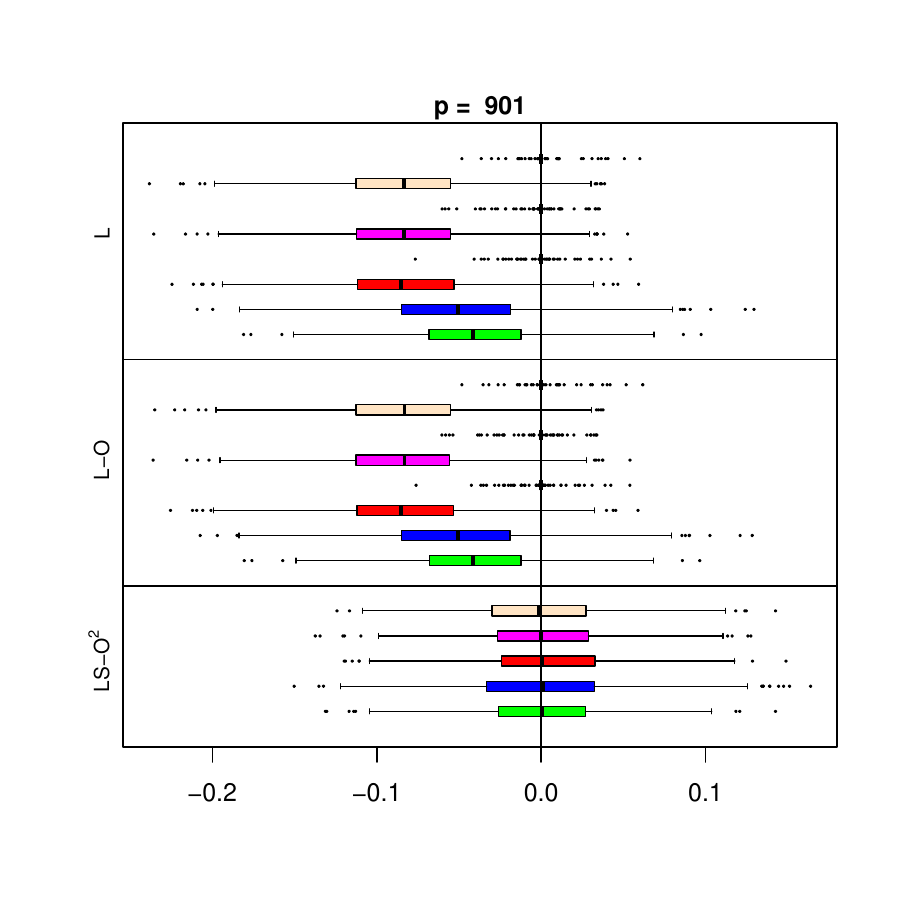} 
\subcaption{$T=15$}\label{fig:results:C-smallT}
\end{subfigure}
\hspace{0.2cm}
\begin{subfigure}[b]{0.475\textwidth}   
\centering 
\includegraphics[width=\textwidth]{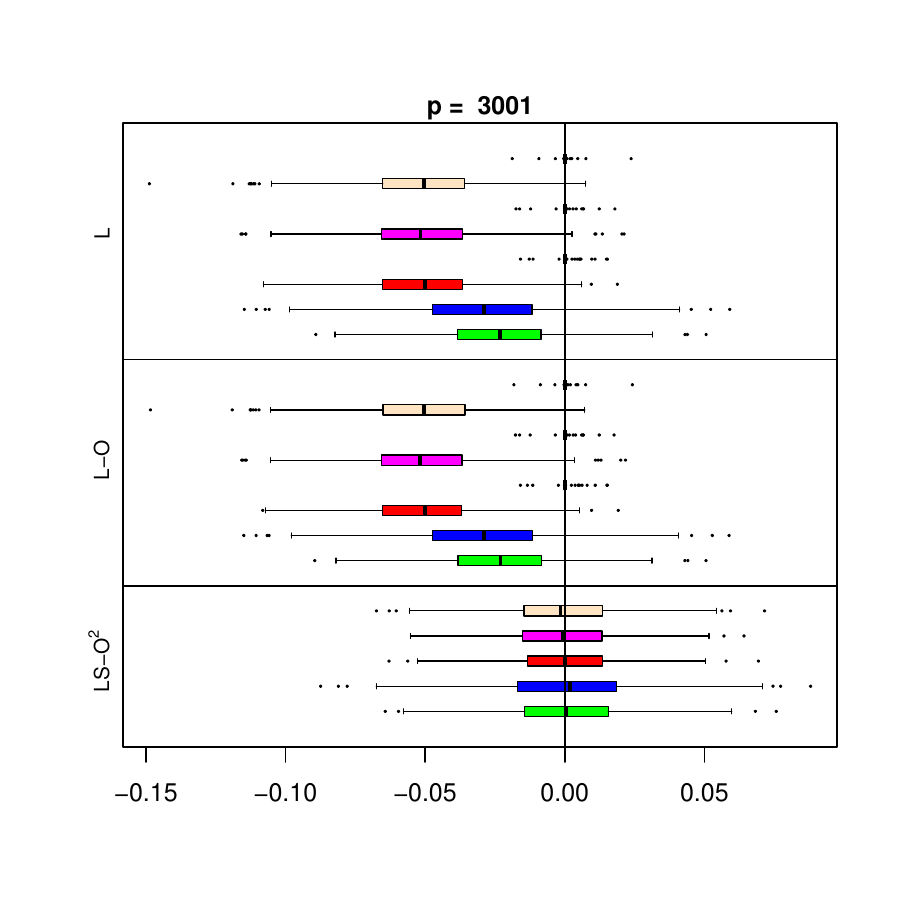}
\subcaption{$T=50$}
\end{subfigure}\label{fig:results:C-largeT}
\caption{Simulation results in Scenario C.}\label{fig:results:C}
\end{figure}

The simulation results are produced as follows: For each choice of $n$, $T$ and $p$, we compute the available estimators from Table \ref{table:estimators} over $1000$ simulation runs. In each run, we calculate the deviation $\Delta_j^{(\ell)} = \widehat{\beta}_j^{(\ell)} - \beta_j$ for each available estimator $\ell \in \{$L, L-O, LS, LS-O, LS-O$^2$, CCE$\}$ and for the eight representative regressors $j$. This leaves us with $1000$ values for each deviation $\Delta_j^{(\ell)}$, which are presented by means of box plots in Figures \ref{fig:results:A}--\ref{fig:results:C}.

Figure \ref{fig:results:A} depicts the results for Scenario A, each panel corresponding to a different choice of $n$, $T$ and $p$. In each panel, the first block (L) shows the box plots for the deviations $\Delta_j^{(\textnormal{L})}$ produced by our HD-CCE estimator, where $j$ runs over the representatives of the different regressor types. Analogously, the second block (L-O) shows the box plots for the deviations $\Delta_j^{(\textnormal{L-O})}$ produced by the oracle version of the HD-CCE estimator, and so on. The representative regressors $j$ in each block are sorted from bottom to top in increasing order (i.e., $j=1$ is at the bottom, followed by $j=2$, $j=3$, $j=5$ and so on) and they are colour-coded (i.e., the boxes for a given $j$ all have the same colour).
In all settings considered in Figure \ref{fig:results:A}, the box plots produced by our HD-CCE estimator (L) and its least squares version (LS) are very similar to those produced by the corresponding oracle versions (L-O and LS-O), which are based on the unobserved projection matrix $\bs{\Pi}$ rather than $\widehat{\bs{\Pi}}$. Moreover, the box plots of the least squares oracle (LS-O) are almost identical to those of the double oracle (LS-O$^2$). We thus find that our procedures work well in Scenario A and that the projection matrix $\widehat{\bs{\Pi}}$ is a quite accurate proxy of the unknown $\bs{\Pi}$. Whereas the CCE estimator shows a performance comparable to the other estimators for $p=7$ in the $T=15$ case and for $p=16$ in the $T=50$ case, its performance deteriorates considerably as the number of regressors $p$ gets larger and comes closer to the critical threshold $T$. The behaviour of our estimators, in contrast, is very stable across $p$.

The results for Scenario B are reported in Figure \ref{fig:results:B}. The box plots produced by our HD-CCE estimator (L) and its least squares version (LS) are again very close to those of the corresponding oracle estimators (L-O and LS-O). Moreover, quite unsurprisingly, the double oracle estimator (LS-O$^2$) which knows the true support $S$ of $\beta$ performs a bit better than our least squares estimator (LS) and its oracle version (LS-O). However, even for the largest $p$ under consideration, the difference in performance is quite moderate. Whereas the box plots of our least squares estimator (LS) and its oracle version (LS-O) are approximately centred around $0$, the box plots of our lasso estimator (L) and its oracle version (L-O) are biased downwards for those $j$ with a non-zero coefficient $\beta_j$. This is not surprising because by construction, the lasso shrinks the parameter values towards zero. The box plots of the lasso and its oracle for the components $j$ with $\beta_j=0$ may look a bit strange on first sight: one can only see the set of outliers, whereas the whole region between the whiskers is collapsed to zero. The reason for this is as follows: since $\beta_j = 0$, the lasso $\widehat{\beta}_{\pen,j}$ often takes exactly the value $0$. Only in a small fraction of the simulation runs, it takes a non-zero value. These non-zero values are visible as outliers in the box plots. 

We finally turn to the results for Scenario C which are provided in Figure \ref{fig:results:C}. As in Scenario B, the box plots of our HD-CCE estimator (L) are almost indistinguishable from those of the oracle (L-O). Moreover, one can again see clearly that both the HD-CCE estimator and its oracle have a downward bias. The box plots also indicate that the precision of our estimator and its oracle (measured in terms of variance) is comparable to that of the double oracle (LS-O$^2$).


In summary, the simulation results show that the HD-CCE estimator (L) and its least squares version (LS) exhibit a performance comparable to the oracle versions (L-O and LS-O) in all the considered settings of Scenarios A--C. 
Moreover, the results implicitly suggest that the cross-validation procedure for choosing the penalty $\pen$ of the lasso-based estimators (L and L-O) performs reasonably well. This is in particular suggested by the fact that the precision of the lasso-based estimators (measured in terms of variance) is quite close to that of the double oracle (LS-O$^2$) in all considered settings.  
All in all, the simulation exercises demonstrate that our estimation approach works well in both low and high dimensions and in particular allows to deal with the case $p \ge T$ where the original CCE approach is not available.

\section{Inference}\label{sec:sim-inf}
\setcounter{table}{2}
\setcounter{figure}{3}

In the second part of the simulation study, we use the desparsified HD-CCE estimator $\widetilde{b}_j$ to perform inference on the coefficient $\beta_j$. Specifically, we set $j=1$ and consider the problem of testing the null $H_0: \beta_1 = 0$ against the alternative $H_1: \beta_1 \ne 0$. We run the following test: reject $H_0$ at significance level $\alpha \in (0,1)$ if 
\[ |\widetilde{b}_1| > \frac{\widetilde{\scaling} \, q_{1-\frac{\alpha}{2}}}{\widetilde{\resnw}^\top \widetilde{X}_{(1)}}, \]
where $\widetilde{\scaling} = \widetilde{\scaling}^{\hspace{1pt} \text{IID}}$ and $q_\alpha$ is the $\alpha$-quantile of the standard normal distribution. We simulate data under the null by setting $c^{**} = 0$ in the parameter vector $\beta = (c^{**},\beta^{(1)},\beta^{(1)},\beta^{(1)})^\top$ and consider two different alternatives by setting $c^{**} = 0.1$ and $c^{**} = 0.2$.

Table \ref{table:size} reports the empirical size of the test under the null (defined as the number of rejections divided by the total number of simulation runs) for different values of the nominal size $\alpha$. The table shows that the test has good size properties. In particular, the empirical size is fairly close to the target $\alpha$ in all considered settings. Tables \ref{table:power0.1} and \ref{table:power0.2} report the empirical power of the test (which is defined exactly as the empirical size) against the two alternatives with $c^{**} = 0.1$ and $c^{**} = 0.2$. In the setting with $T=15$, the test has considerable power against the alternative with $c^{**} =0.1$, however, still substantially below $1$. The power increases quickly as we move further away from the null. Specifically, for $c^{**} = 0.2$, the power already reaches values close to $1$. As expected, the power numbers for $T=50$ are even better than those for $T=15$, being close to $1$ already in the case with $c^{**}=0.1$.

\begin{table}[hb!]
\caption{Empirical size of the test under $H_0$.}\label{table:size}
\begin{subtable}[t]{0.45\textwidth}
\subcaption{$T=15$}
{\small 
\begin{tabular}{@{\extracolsep{2pt}} lccc} 
\toprule\\[-0.45cm]
 & $\alpha = 0.01$ & $\alpha = 0.05$ & $\alpha = 0.1\phantom{0}$ \\ [0.5ex] 
\hline\\[-0.45cm]
$p=7$ & $0.013$ & $0.057$ & $0.113$ \\ 
$p=10$ & $0.018$ & $0.059$ & $0.114$ \\ 
$p=13$ & $0.008$ & $0.026$ & $0.072$ \\ 
$p=31$ & $0.014$ & $0.064$ & $0.121$ \\ 
$p=151$ & $0.015$ & $0.066$ & $0.117$ \\ 
$p=301$ & $0.006$ & $0.040$ & $0.089$ \\ 
$p=901$ & $0.014$ & $0.045$ & $0.091$ \\  
\bottomrule 
\end{tabular}}
\end{subtable} \hspace{0.75cm}
\begin{subtable}[t]{0.45\textwidth}
\subcaption{$T=50$}
{\small 
\begin{tabular}{@{\extracolsep{2pt}} lccc} 
\toprule\\[-0.45cm]
 & $\alpha = 0.01$ & $\alpha = 0.05$ & $\alpha = 0.1\phantom{0}$ \\ [0.5ex] 
\hline\\[-0.45cm]
$p=16$ & $0.009$ & $0.043$ & $0.086$ \\ 
$p=31$ & $0.009$ & $0.049$ & $0.109$ \\ 
$p=46$ & $0.008$ & $0.055$ & $0.107$ \\ 
$p=91$ & $0.008$ & $0.044$ & $0.094$ \\ 
$p=451$ & $0.010$ & $0.057$ & $0.111$ \\ 
$p=901$ & $0.014$ & $0.060$ & $0.105$ \\ 
$p=3001$ & $0.013$ & $0.050$ & $0.097$ \\ 
\bottomrule 
\end{tabular}}
\end{subtable}
\end{table}

\begin{table}[t]
\caption{Power of the test against the alternative with $c^{**}=0.1$.}\label{table:power0.1}
\begin{subtable}[t]{0.45\textwidth}
\subcaption{$T=15$}
{\small 
\begin{tabular}{@{\extracolsep{2pt}} lccc} 
\toprule\\[-0.45cm]
 & $\alpha = 0.01$ & $\alpha = 0.05$ & $\alpha = 0.1\phantom{0}$ \\ [0.5ex] 
\hline\\[-0.45cm]
$p=7$ & $0.425$ & $0.660$ & $0.765$ \\ 
$p=10$ & $0.405$ & $0.646$ & $0.752$ \\ 
$p=13$ & $0.431$ & $0.679$ & $0.778$ \\ 
$p=31$ & $0.456$ & $0.685$ & $0.787$ \\ 
$p=151$ & $0.440$ & $0.669$ & $0.769$ \\ 
$p=301$ & $0.461$ & $0.670$ & $0.782$ \\ 
$p=901$ & $0.455$ & $0.691$ & $0.786$ \\ 
\bottomrule 
\end{tabular}}
\end{subtable} \hspace{0.75cm}
\begin{subtable}[t]{0.45\textwidth}
\subcaption{$T=50$}
{\small 
\begin{tabular}{@{\extracolsep{2pt}} lccc} 
\toprule\\[-0.45cm]
 & $\alpha = 0.01$ & $\alpha = 0.05$ & $\alpha = 0.1\phantom{0}$ \\ [0.5ex] 
\hline\\[-0.45cm]
$p=16$ & $0.991$ & $0.999$ & $1$ \\ 
$p=31$ & $0.989$ & $0.996$ & $0.999$ \\ 
$p=46$ & $0.989$ & $0.998$ & $0.999$ \\ 
$p=91$ & $0.988$ & $1$ & $1$ \\ 
$p=451$ & $0.976$ & $0.995$ & $1$ \\ 
$p=901$ & $0.970$ & $0.995$ & $0.997$ \\ 
$p=3001$ & $0.976$ & $0.994$ & $0.996$ \\ 
\bottomrule 
\end{tabular}}
\end{subtable}
\vspace{0.75cm}

\caption{Power of the test against the alternative with $c^{**}=0.2$.}\label{table:power0.2}
\begin{subtable}[t]{0.45\textwidth}
\subcaption{$T=15$}
{\small 
\begin{tabular}{@{\extracolsep{2pt}} lccc} 
\toprule\\[-0.45cm]
 & $\alpha = 0.01$ & $\alpha = 0.05$ & $\alpha = 0.1\phantom{0}$ \\ [0.5ex] 
\hline\\[-0.45cm]
$p=7$ & $0.979$ & $0.992$ & $0.994$ \\ 
$p=10$ & $0.976$ & $0.993$ & $0.998$ \\ 
$p=13$ & $0.992$ & $0.998$ & $1$ \\ 
$p=31$ & $0.981$ & $0.995$ & $0.998$ \\ 
$p=151$ & $0.981$ & $0.996$ & $0.999$ \\ 
$p=301$ & $0.986$ & $0.996$ & $0.998$ \\ 
$p=901$ & $0.982$ & $0.997$ & $1$ \\  
\bottomrule 
\end{tabular}}
\end{subtable} \hspace{0.75cm}
\begin{subtable}[t]{0.45\textwidth}
\subcaption{$T=50$}
{\small 
\begin{tabular}{@{\extracolsep{2pt}} lccc} 
\toprule\\[-0.45cm]
 & $\alpha = 0.01$ & $\alpha = 0.05$ & $\alpha = 0.1\phantom{0}$ \\ [0.5ex] 
\hline\\[-0.45cm]
$p=16$ & $1$ & $1$ & $1$ \\ 
$p=31$ & $1$ & $1$ & $1$ \\ 
$p=46$ & $1$ & $1$ & $1$ \\ 
$p=91$ & $1$ & $1$ & $1$\\ 
$p=451$ & $1$ & $1$ & $1$ \\ 
$p=901$ & $1$ & $1$ & $1$\\ 
$p=3001$ & $1$ & $1$ & $1$ \\ 
\bottomrule 
\end{tabular}}
\end{subtable}
\vspace{0.25cm}
\end{table}

\section{Robustness checks}\label{sec:sim-robustness}
\setcounter{table}{5}
\setcounter{figure}{3}

\subsubsection*{Overestimation of $\bs{K}$}

\begin{figure}[p]
\centering
\begin{subfigure}[b]{0.475\textwidth}   
\centering 
\includegraphics[width=\textwidth]{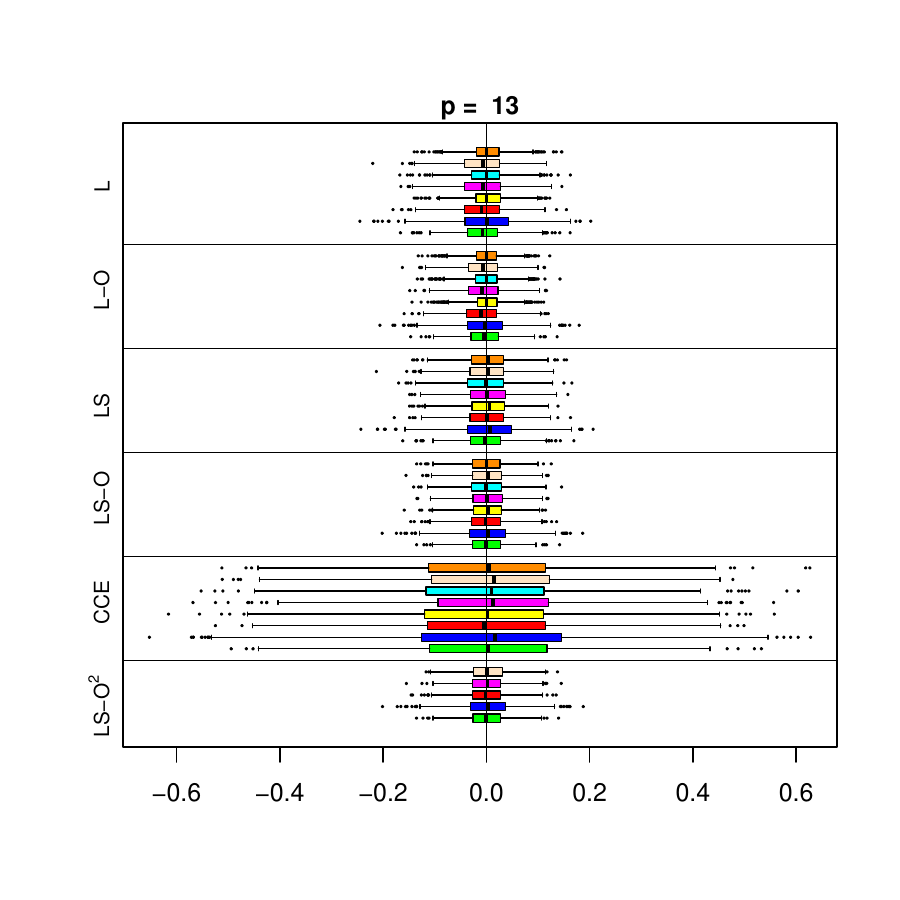} \\
\includegraphics[width=\textwidth]{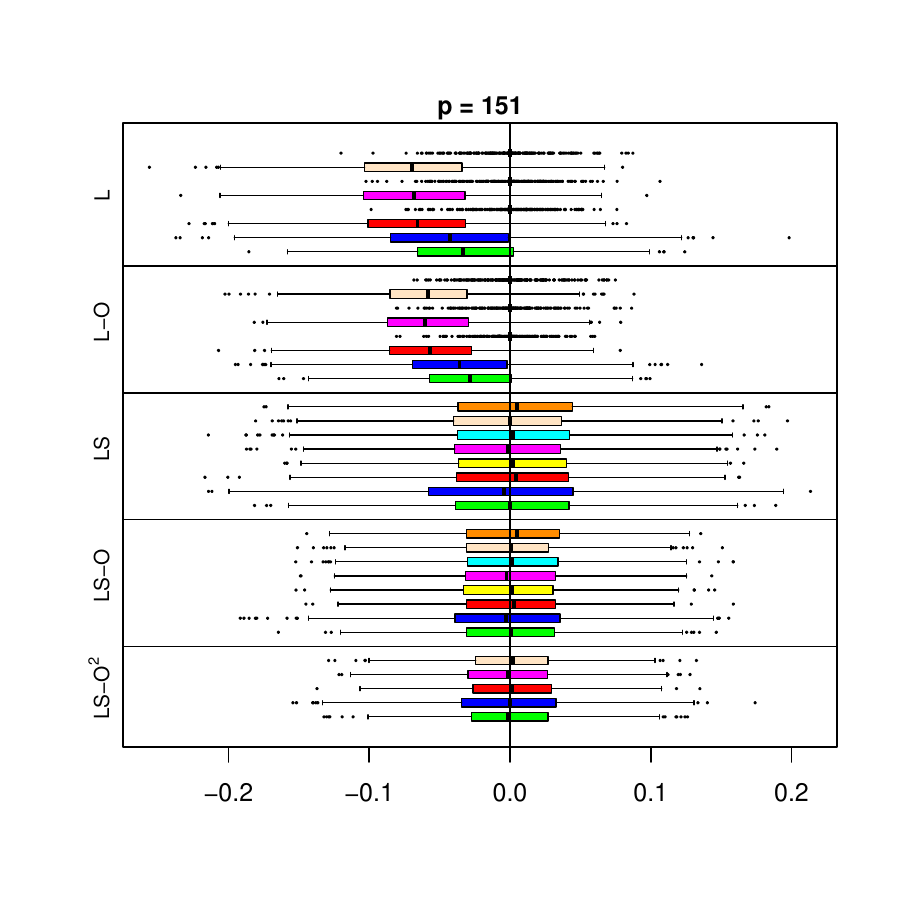} \\
\includegraphics[width=\textwidth]{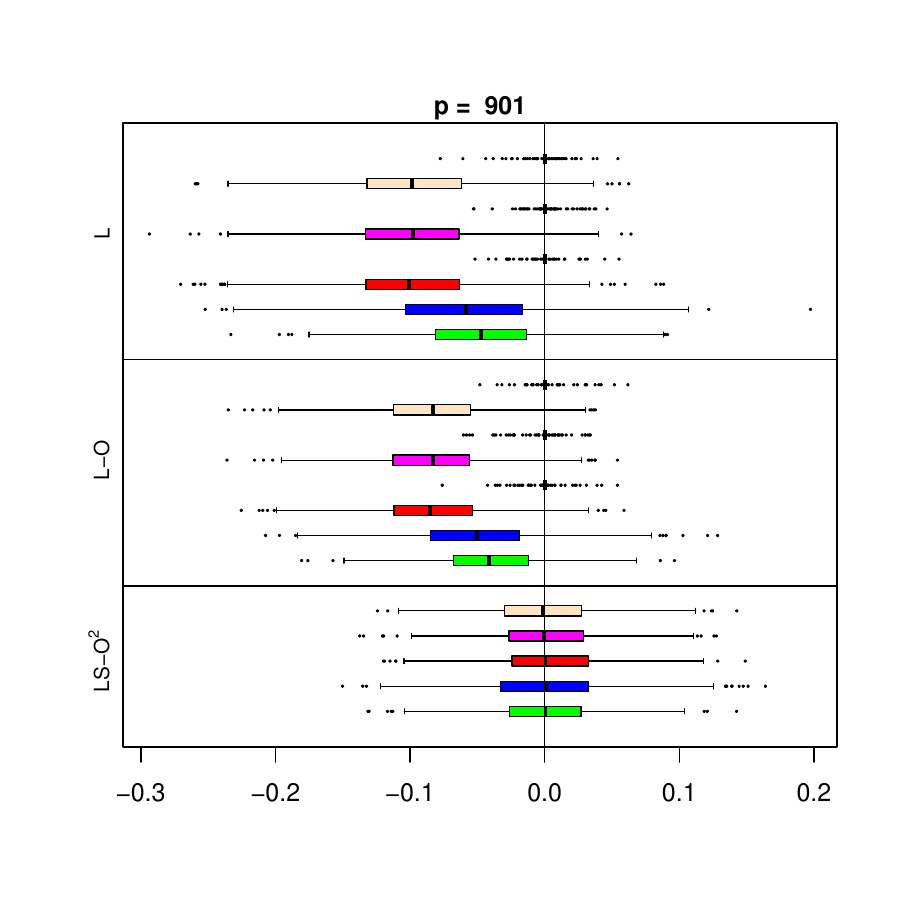}
\subcaption{$T=15$}
\end{subfigure}
\hspace{0.2cm}
\begin{subfigure}[b]{0.475\textwidth}   
\centering 
\includegraphics[width=\textwidth]{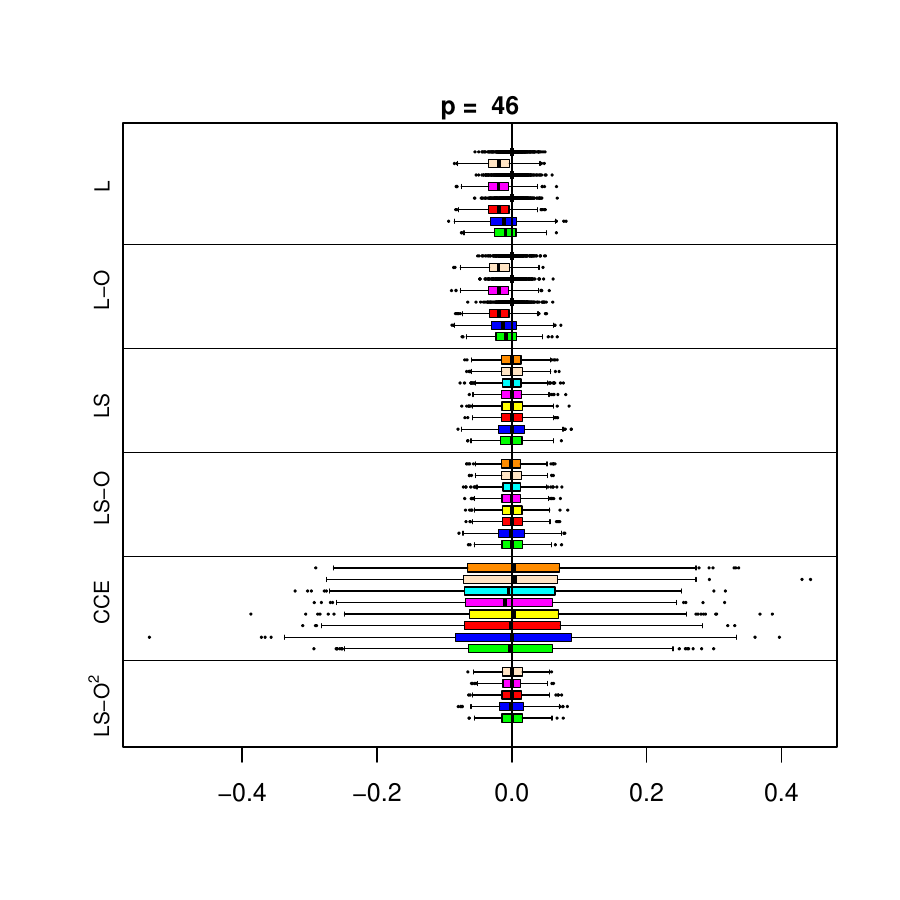} \\
\includegraphics[width=\textwidth]{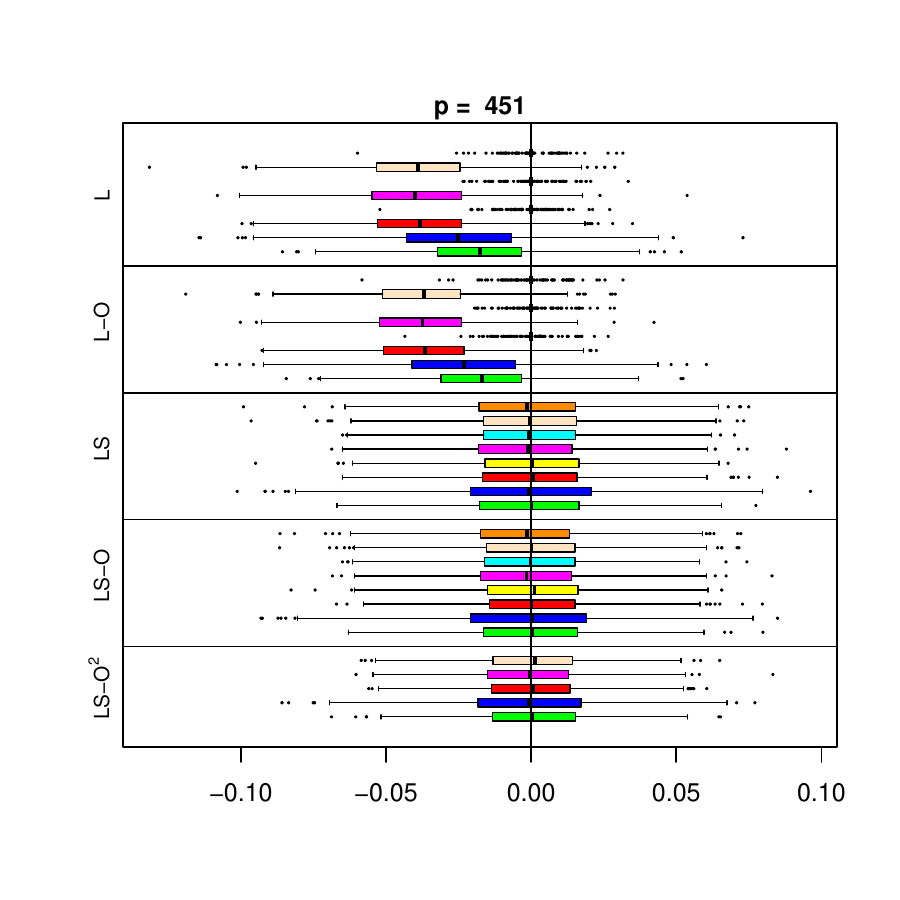} \\
\includegraphics[width=\textwidth]{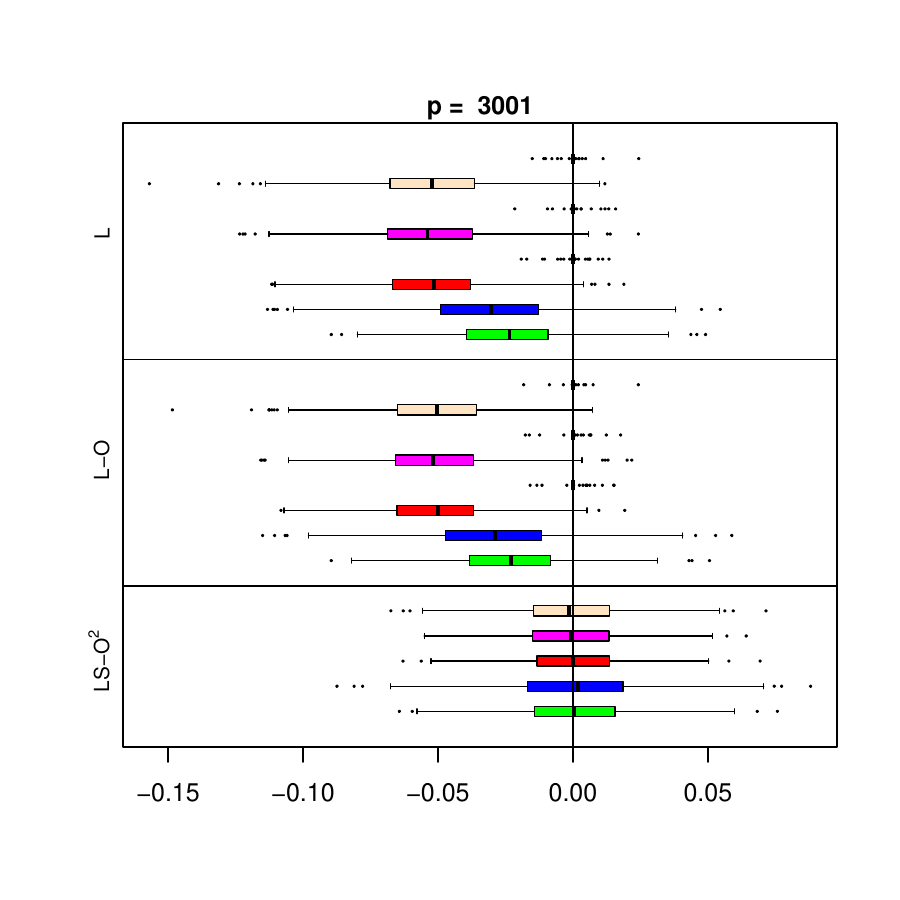}
\subcaption{$T=50$}
\end{subfigure}
\caption{Simulation results when the number of factors $K=3$ is overestimated by $6$.}\label{fig:results:overest}
\end{figure}

So far, we have not discussed the performance of the estimator $\widehat{K}$. Table \ref{table:K} makes up for this: it reports the number of simulations (out of a total of $1000$) in which $\widehat{K}$ takes a certain value. As can be seen, $\widehat{K}$ never underestimates the true number of factors $K=3$. However, for some values of $p$, it (moderately) overestimates $K$ in a considerable number of simulation runs. 
Notably, this appears not to have a strong negative effect on the quality of our estimators. In particular, Figures \ref{fig:results:A}--\ref{fig:results:C} reveal that in all cases under consideration, our HD-CCE estimator (L) and its least squares version (LS) perform very similar to their oracle versions (L-O and LS-O) which presuppose knowledge of the factors and their number $K$. Hence, (moderately) overestimating $K$ appears not to do much harm, which makes sense intuitively: if we overestimate $K$, we project away ``too much'', i.e., we do not only (approximately) project away the space spanned by the $K=3$ factors but a larger dimensional space. This results in a loss of efficiency, which can however be expected to be moderate as long as $K$ is not massively overestimated. We run some simulation exercises to support this conjecture. Specifically, we re-run the simulations from Section \ref{sec:sim-est} with the estimator $\widehat{K}$ replaced by the fixed number $6$. We thus overestimate the true $K=3$ by $6$ in all simulation runs. The results are reported in Figure \ref{fig:results:overest}. 
In the case with $T=50$, our HD-CCE estimator (L) and its least squares version (LS) exhibit a performance very similar to the respective oracle versions (L-O and LS-O, which are identical to the oracle procedures in Figures \ref{fig:results:A}--\ref{fig:results:C} as they are based on the true projection matrix $\bs{\Pi}$ and thus do not rely on our choice of the number of factors). In the case with $T=15$, our estimators also perform well even though they are a bit less precise than the oracle versions. 
Overall, these findings support our conjecture that (moderate) overestimation of $K$ is rather unproblematic: our estimators still produce good results even though we loose a bit in terms of estimation precision. (Note: To save space, we only report the results for $p \in \{13,151,901\}$ in the $T=15$ case and for $p \in \{46,451,3001\}$ in the $T=50$ case in Figure \ref{fig:results:overest}.)

\begin{table}[b] 
\caption{Performance of the estimator $\widehat{K}$ in the simulation exercises from Section \ref{sec:sim-est}. The numbers in the tables specify in how many simulation runs (out of a total of $1000$) $\widehat{K}$ takes a certain value.}\label{table:K}
\begin{subtable}[t]{0.5\textwidth}
\centering
\subcaption{$T=15$}
{\small 
\begin{tabular}{@{\extracolsep{2pt}}lcccccc} 
\toprule\\[-0.45cm]
$\widehat{K}$ & 1 & 2 & 3 & 4 & 5 & 6--7 \\ [0.5ex] 
\hline\\[-0.45cm]
$p=7$ & $0$ & $0$ & $646$ & $302$ & $52$ & $0$  \\ 
$p=10$ & $0$ & $0$ & $722$ & $224$ & $50$ & $4$  \\ 
$p=13$ & $0$ & $0$ & $781$ & $173$ & $43$ & $3$  \\ 
$p=31$ & $0$ & $0$ & $938$ & $48$ & $9$ & $5$ \\ 
$p=151$ & $0$ & $0$ & $995$ & $3$ & $1$  & $1$ \\ 
$p=301$ & $0$ & $0$ & $1000$ & $0$ & $0$ & $0$ \\ 
$p=901$ & $0$ & $0$ & $1000$ & $0$ & $0$  & $0$ \\
\bottomrule 
\end{tabular}}
\end{subtable} 
\begin{subtable}[t]{0.5\textwidth}
\centering
\subcaption{$T=50$}
{\small 
\begin{tabular}{@{\extracolsep{2pt}}lccccc} 
\toprule\\[-0.45cm]
$\widehat{K}$ & 1 & 2 & 3 & 4 & 5 \\ [0.5ex] 
\hline\\[-0.45cm]
$p=16$   & $0$ & $0$ & $966$ & $34$ & $0$  \\ 
$p=31$   &$0$ & $0$ & $1000$ & $0$ & $0$ \\ 
$p=46$   & $0$ & $0$ & $1000$ & $0$ & $0$ \\ 
$p=91$  & $0$ & $0$ & $1000$ & $0$ & $0$ \\ 
$p=451$ & $0$ & $0$ & $1000$ & $0$ & $0$ \\ 
$p=901$   &$0$ & $0$ & $1000$ & $0$ & $0$ \\ 
$p=3001$ & $0$ & $0$ & $1000$ & $0$ & $0$ \\ 
\bottomrule 
\end{tabular}}
\end{subtable}
\end{table}

\subsubsection*{Imbalanced eigenvalues}

\begin{figure}[p]
\centering
\begin{subfigure}[b]{0.475\textwidth}   
\centering 
\includegraphics[width=\textwidth]{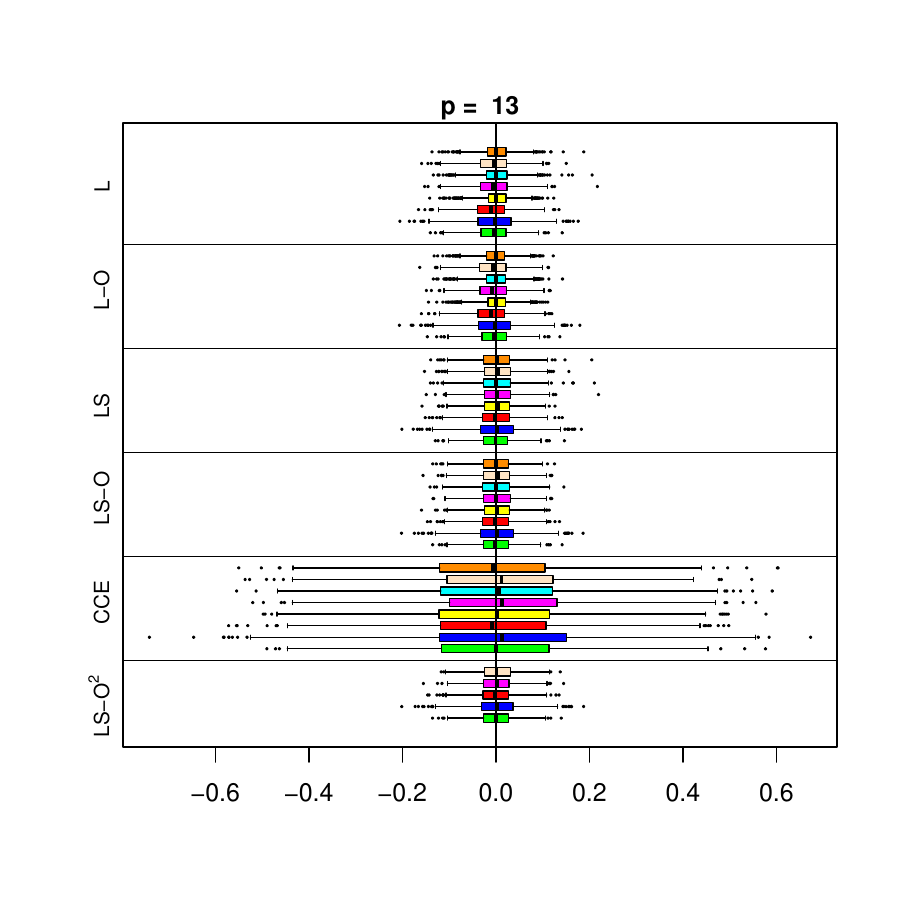} \\
\includegraphics[width=\textwidth]{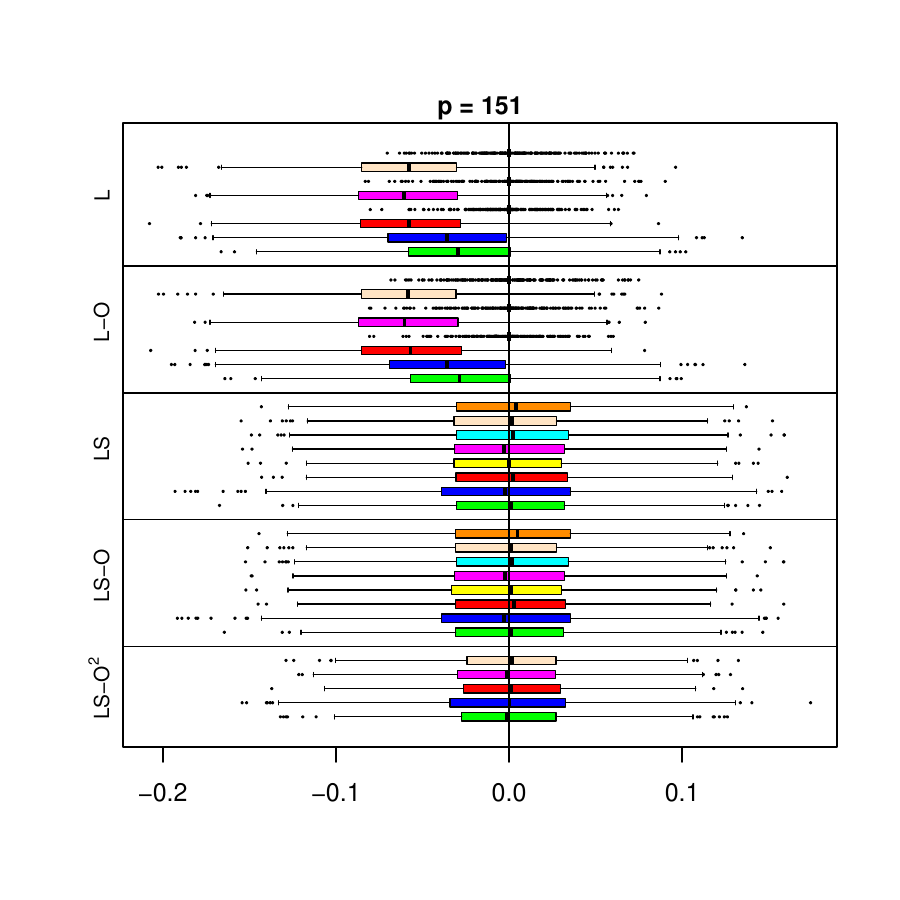} \\
\includegraphics[width=\textwidth]{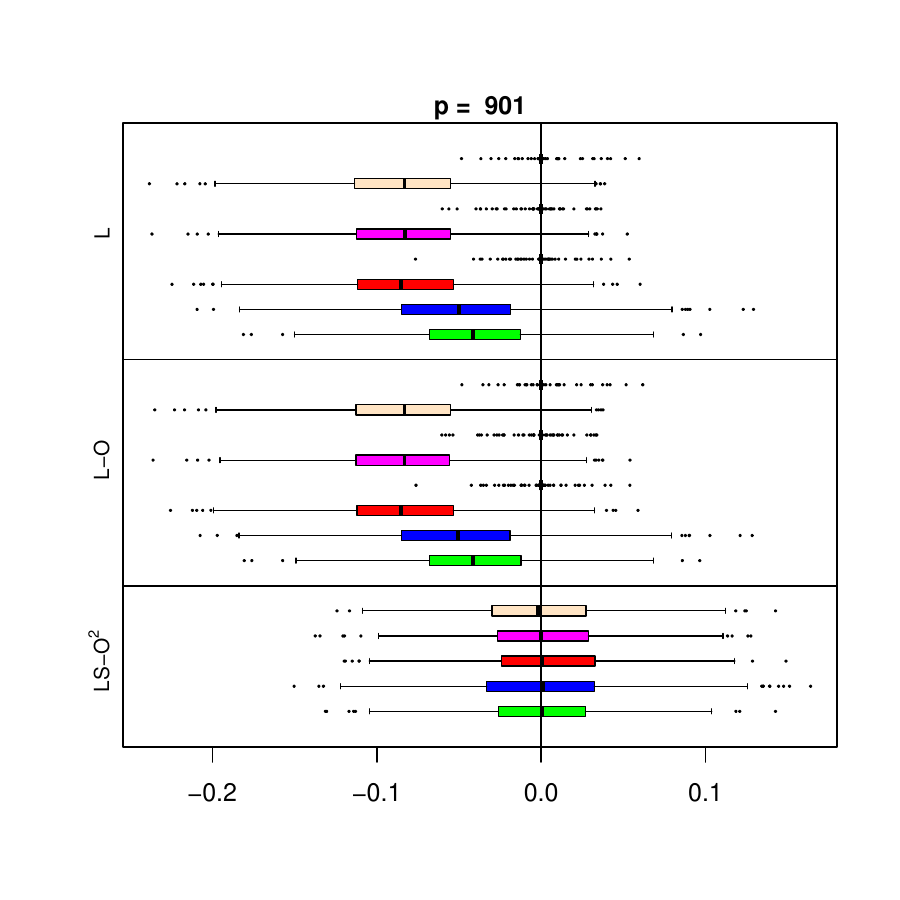}
\subcaption{$T=15$}
\end{subfigure}
\hspace{0.2cm}
\begin{subfigure}[b]{0.475\textwidth}   
\centering 
\includegraphics[width=\textwidth]{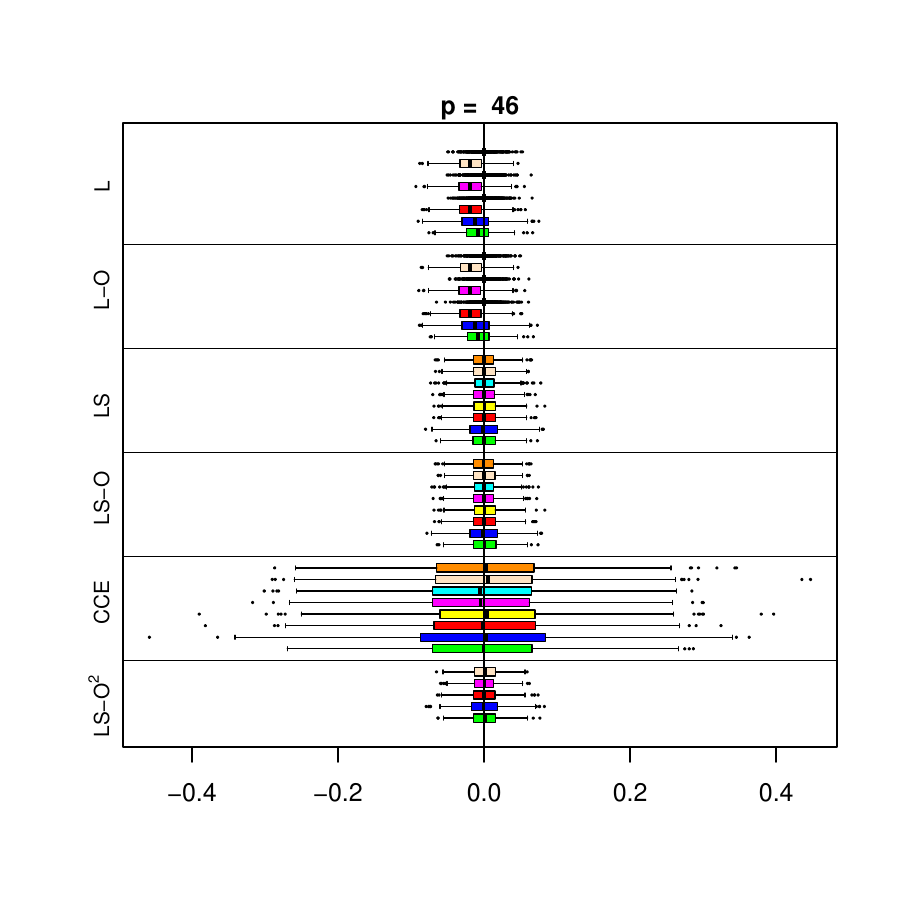} \\
\includegraphics[width=\textwidth]{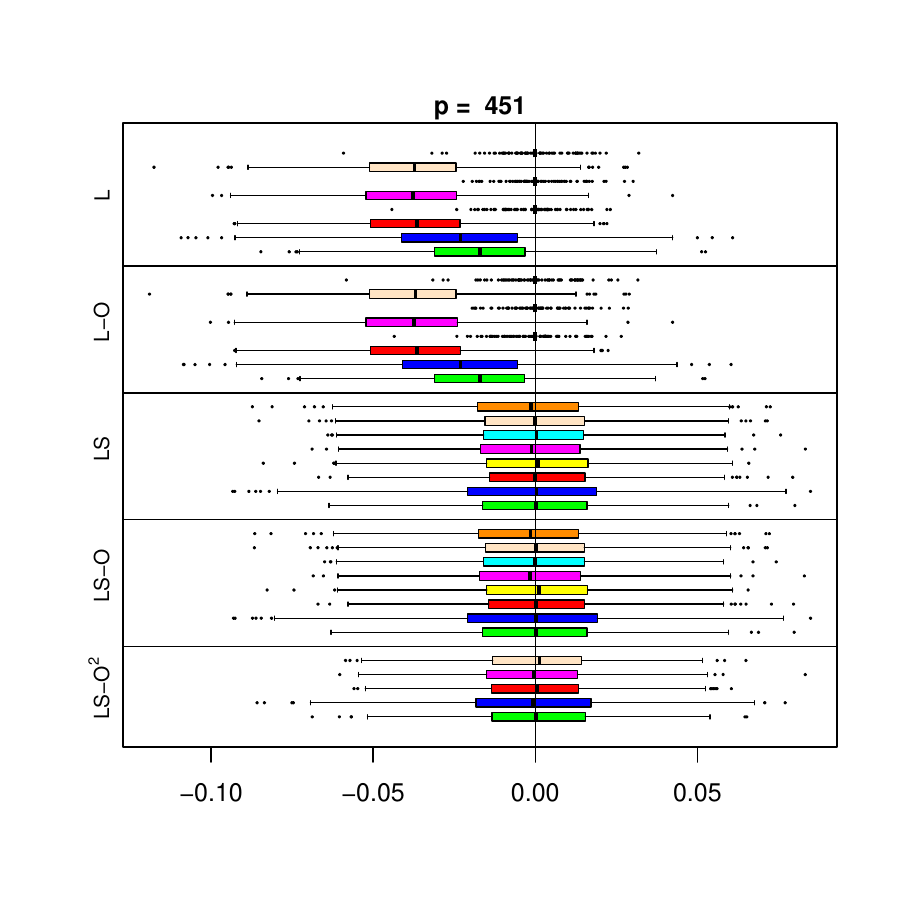} \\
\includegraphics[width=\textwidth]{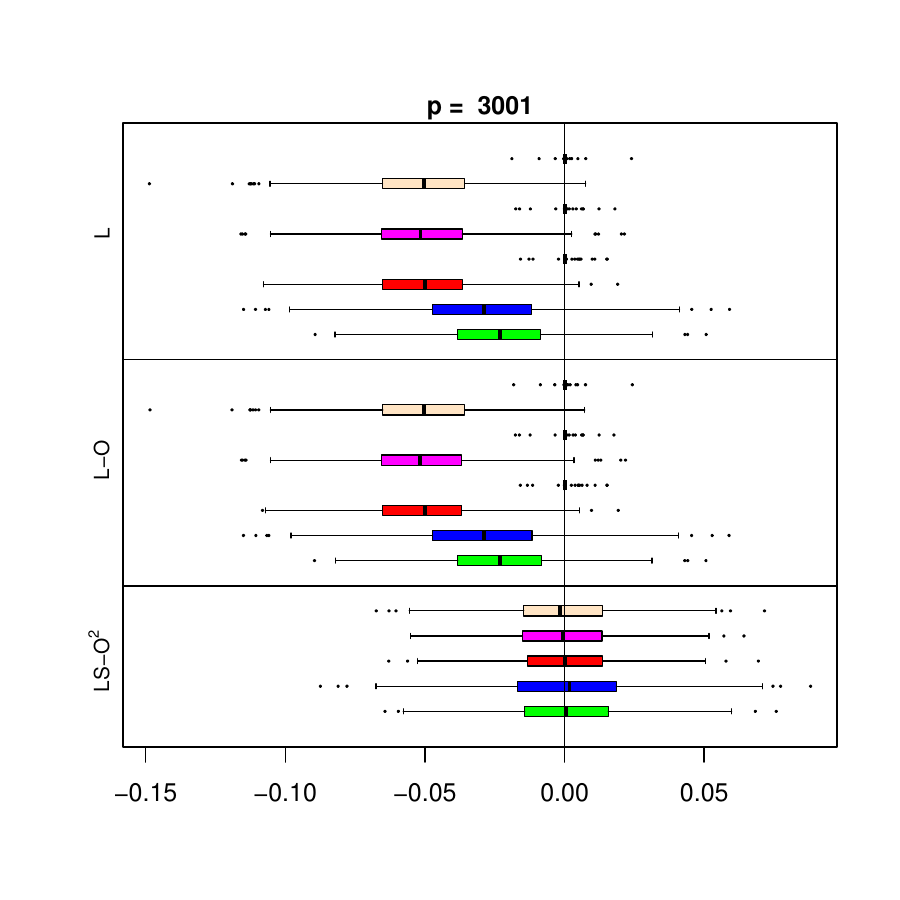}
\subcaption{$T=50$}
\end{subfigure}
 \caption{Estimation results in the imbalanced setting  where the entries of the mean vector $\ex[\Gamma_i^{(1)}]$ are all equal to $\sqrt{10}$.}\label{fig:results:imbalanced}
\end{figure}

The simulation design considered in Sections \ref{sec:sim-est} and \ref{sec:sim-inf} includes a factor loading matrix $\bs{\Gamma} = \ex [\bs{\Gamma}_i]$ which is well balanced. In particular, the $K=3$ eigenvalues of the $K \times K$ matrix $\bs{\Gamma}^\top \bs{\Gamma}/p$ are all approximately the same. We now consider a more imbalanced setting where the first eigenvalue is approximately $10$ times larger than the second and third one. To obtain such an imbalanced situation, we increase the entries of the mean vector $\ex[\Gamma_i^{(1)}]$ in the simulation setting from $1$ to $\sqrt{10}$. We then re-run the simulation exercises from Section \ref{sec:sim-est} with this modification. The results (reported in Table \ref{table:K-imbalanced} and Figure \ref{fig:results:imbalanced}) demonstrate that our methods perform reasonably well in this imbalanced setting. In particular, the box plots in Figure \ref{fig:results:imbalanced} are very similar to those in Figures \ref{fig:results:A}--\ref{fig:results:C}. The performance of our methods in the imbalanced setting is thus comparable to that in the balanced case. (Note: To save space, we again only report the results for $p \in \{13,151,901\}$ in the $T=15$ case and for $p \in \{46,451,3001\}$ in the $T=50$ case in Figure \ref{fig:results:imbalanced}.)

What happens if we make the simulation setting more and more imbalanced by driving up the values in the mean vector $\ex[\Gamma_i^{(1)}]$? At some point, the estimator $\widehat{K}$ will break down: it will only detect one factor ($\widehat{K} = 1$), thus underestimating the true $K=3$. Unlike (moderate) overestimation of $K$, underestimation is not harmless at all: if we underestimate $K$, we do not properly eliminate the factor structure because we project away only part of the space spanned by the factors. This will most likely result in poor performance of our estimators. For practical purposes, we thus recommend to corroborate the computed estimate $\widehat{K}$ by a look at the scree plot. Suppose e.g.\ that we have a similar situation as in the imbalanced simulation scenario under consideration: the screeplot shows a very large leading eigenvalue followed by two moderately large ones and then a drop to smaller eigenvalues.  In such a case, we suggest to set the number of factors to $3$ rather than to $1$. In the worst case scenario, this will result in rather harmless overestimation of $K$, in particular, in a moderate loss of precision. Generally speaking, we recommend to choose the number of factors a bit too liberally (i.e., a bit too large) in order to avoid underestimation of $K$.

\begin{table}[b] 
\caption{Performance of the estimator $\widehat{K}$ in the imbalanced simulation setting where the entries of the mean vector $\ex[\Gamma_i^{(1)}]$ are all equal to $\sqrt{10}$. The numbers in the tables specify in how many simulation runs (out of a total of $1000$) $\widehat{K}$ takes a certain value.}\label{table:K-imbalanced}
\begin{subtable}[t]{0.5\textwidth}
\centering
\subcaption{$T=15$}
{\small 
\begin{tabular}{@{\extracolsep{2pt}} lccccc} 
\toprule\\[-0.45cm]
$\widehat{K}$ & 1 & 2 & 3 & 4  \\ [0.5ex] 
\hline\\[-0.45cm]
$p=7$ & $1$ & $9$ & $989$ & $1$ \\ 
$p=10$ & $1$ & $9$ & $990$ & $0$ \\ 
$p=13$ & $0$ & $10$ & $990$ & $0$ \\ 
$p=31$ & $0$ & $4$ & $996$ & $0$ \\ 
$p=151$ & $0$ & $6$ & $994$ & $0$ \\ 
$p=301$ & $0$ & $8$ & $992$ & $0$ \\ 
$p=901$ & $0$ & $6$ & $994$ & $0$ \\ 
\bottomrule 
\end{tabular}}
\end{subtable}
\begin{subtable}[t]{0.5\textwidth}
\centering
\subcaption{$T=50$}
{\small 
\begin{tabular}{@{\extracolsep{2pt}} lccccc} 
\toprule\\[-0.45cm]
$\widehat{K}$ & 1 & 2 & 3 & 4 \\ [0.5ex] 
\hline\\[-0.45cm]
$p=16$ & $0$ & $0$ & $1000$ & $0$ \\ 
$p=31$ & $0$ & $0$ & $1000$ & $0$ \\ 
$p=46$ & $0$ & $0$ & $1000$ & $0$ \\ 
$p=91$ & $0$ & $0$ & $1000$ & $0$ \\ 
$p=451$ & $0$ & $0$ & $1000$ & $0$ \\ 
$p=901$ & $0$ & $0$ & $1000$ & $0$ \\ 
$p=3001$ & $0$ & $0$ & $1000$ & $0$ \\
\bottomrule 
\end{tabular}}
\end{subtable}
\end{table}

\subsubsection*{Heteroskedasticity and autocorrelation in the idiosyncratic errors}

\begin{table}[p!]
\caption{Empirical size of the test under $H_0$ in the setting with heteroskedastic but uncorrelated idiosyncratic errors.}\label{table:sizeHETEROSCEDASTIC}
\begin{subtable}[t]{0.45\textwidth}
\subcaption{$T=15$}
{\small 
\begin{tabular}{@{\extracolsep{2pt}} lccc} 
\toprule\\[-0.45cm]
 & $\alpha = 0.01$ & $\alpha = 0.05$ & $\alpha = 0.1\phantom{0}$ \\ [0.5ex] 
\hline\\[-0.45cm]
$p=7$ & $0.016$ & $0.054$ & $0.114$ \\ 
$p=10$ & $0.015$ & $0.054$ & $0.113$ \\ 
$p=13$ & $0.010$ & $0.040$ & $0.099$ \\ 
$p=31$ & $0.020$ & $0.059$ & $0.102$ \\ 
$p=151$ & $0.015$ & $0.055$ & $0.116$ \\ 
$p=301$ & $0.009$ & $0.044$ & $0.093$ \\ 
$p=901$ & $0.012$ & $0.059$ & $0.112$ \\ 
\bottomrule 
\end{tabular}}
\end{subtable} \hspace{0.75cm}
\begin{subtable}[t]{0.45\textwidth}
\subcaption{$T=50$}
{\small 
\begin{tabular}{@{\extracolsep{2pt}} lccc} 
\toprule\\[-0.45cm]
 & $\alpha = 0.01$ & $\alpha = 0.05$ & $\alpha = 0.1\phantom{0}$ \\ [0.5ex] 
\hline\\[-0.45cm]
$p=16$ & $0.012$ & $0.048$ & $0.109$ \\ 
$p=31$ & $0.014$ & $0.060$ & $0.117$ \\ 
$p=46$ & $0.006$ & $0.049$ & $0.112$ \\ 
$p=91$ & $0.011$ & $0.049$ & $0.107$ \\ 
$p=451$ & $0.015$ & $0.048$ & $0.114$ \\ 
$p=901$ & $0.011$ & $0.057$ & $0.108$ \\ 
$p=3001$ & $0.011$ & $0.050$ & $0.112$ \\ 
\bottomrule 
\end{tabular}}
\end{subtable}
\vspace{0.75cm}
\caption{Power of the test against the alternative with $c^{**}=0.1$.}\label{table:power0.1HETEROSCEDASTIC}
\begin{subtable}[t]{0.45\textwidth}
\subcaption{$T=15$}
{\small 
\begin{tabular}{@{\extracolsep{2pt}} lccc} 
\toprule\\[-0.45cm]
 & $\alpha = 0.01$ & $\alpha = 0.05$ & $\alpha = 0.1\phantom{0}$ \\ [0.5ex] 
\hline\\[-0.45cm]
$p=7$ & $0.432$ & $0.644$ & $0.756$ \\ 
$p=10$ & $0.408$ & $0.660$ & $0.776$ \\ 
$p=13$ & $0.420$ & $0.670$ & $0.774$ \\ 
$p=31$ & $0.460$ & $0.675$ & $0.768$ \\ 
$p=151$ & $0.421$ & $0.645$ & $0.749$ \\ 
$p=301$ & $0.450$ & $0.667$ & $0.781$ \\ 
$p=901$ & $0.453$ & $0.684$ & $0.773$ \\ 
\bottomrule 
\end{tabular}}
\end{subtable} \hspace{0.75cm}
\begin{subtable}[t]{0.45\textwidth}
\subcaption{$T=50$}
{\small 
\begin{tabular}{@{\extracolsep{2pt}} lccc} 
\toprule\\[-0.45cm]
 & $\alpha = 0.01$ & $\alpha = 0.05$ & $\alpha = 0.1\phantom{0}$ \\ [0.5ex] 
\hline\\[-0.45cm]
$p=16$ & $0.985$ & $0.997$ & $0.999$ \\ 
$p=31$ & $0.984$ & $0.996$ & $0.998$ \\ 
$p=46$ & $0.993$ & $1$ & $1$ \\ 
$p=91$ & $0.983$ & $0.996$ & $0.998$ \\ 
$p=451$ & $0.970$ & $0.992$ & $0.997$ \\ 
$p=901$ & $0.977$ & $0.998$ & $1$ \\ 
$p=3001$ & $0.982$ & $0.997$ & $0.999$ \\ 
\bottomrule 
\end{tabular}}
\end{subtable}
\vspace{0.75cm}
\caption{Power of the test against the alternative with $c^{**}=0.2$.}\label{table:power0.2HETEROSCEDASTIC}
\begin{subtable}[t]{0.45\textwidth}
\subcaption{$T=15$}
{\small 
\begin{tabular}{@{\extracolsep{2pt}} lccc} 
\toprule\\[-0.45cm]
 & $\alpha = 0.01$ & $\alpha = 0.05$ & $\alpha = 0.1\phantom{0}$ \\ [0.5ex] 
\hline\\[-0.45cm]
$p=7$ & $0.972$ & $0.995$ & $0.997$ \\ 
$p=10$ & $0.977$ & $0.997$ & $0.998$ \\ 
$p=13$ & $0.986$ & $0.998$ & $1$ \\ 
$p=31$ & $0.983$ & $0.994$ & $0.996$ \\ 
$p=151$ & $0.977$ & $0.995$ & $0.999$ \\ 
$p=301$ & $0.988$ & $0.998$ & $1$ \\ 
$p=901$ & $0.979$ & $0.997$ & $0.999$ \\ 
\bottomrule 
\end{tabular}}
\end{subtable} \hspace{0.75cm}
\begin{subtable}[t]{0.45\textwidth}
\subcaption{$T=50$}
{\small 
\begin{tabular}{@{\extracolsep{2pt}} lccc} 
\toprule\\[-0.45cm]
 & $\alpha = 0.01$ & $\alpha = 0.05$ & $\alpha = 0.1\phantom{0}$ \\ [0.5ex] 
\hline\\[-0.45cm]
$p=16$ & $1$ & $1$ & $1$ \\ 
$p=31$ & $1$ & $1$ & $1$ \\ 
$p=46$ & $1$ & $1$ & $1$ \\ 
$p=91$ & $1$ & $1$ & $1$\\ 
$p=451$ & $1$ & $1$ & $1$ \\ 
$p=901$ & $1$ & $1$ & $1$\\ 
$p=3001$ & $1$ & $1$ & $1$ \\ 
\bottomrule 
\end{tabular}}
\end{subtable}
\end{table}

\begin{table}[p!]
\caption{Empirical size of the test under $H_0$ in the setting with heteroskedastic and autocorrelated idiosyncratic errors.}\label{table:sizeHAC}
\begin{subtable}[t]{0.45\textwidth}
\subcaption{$T=15$}
{\small 
\begin{tabular}{@{\extracolsep{2pt}} lccc} 
\toprule\\[-0.45cm]
 & $\alpha = 0.01$ & $\alpha = 0.05$ & $\alpha = 0.1\phantom{0}$ \\ [0.5ex] 
\hline\\[-0.45cm]
$p=7$ & $0.012$ & $0.048$ & $0.112$ \\ 
$p=10$ & $0.012$ & $0.050$ & $0.110$ \\ 
$p=13$ & $0.021$ & $0.060$ & $0.116$ \\ 
$p=31$ & $0.010$ & $0.048$ & $0.106$ \\ 
$p=151$ & $0.012$ & $0.057$ & $0.103$ \\ 
$p=301$ & $0.013$ & $0.056$ & $0.101$ \\ 
$p=901$ & $0.015$ & $0.061$ & $0.102$ \\ 
\bottomrule 
\end{tabular}}
\end{subtable} \hspace{0.75cm}
\begin{subtable}[t]{0.45\textwidth}
\subcaption{$T=50$}
{\small 
\begin{tabular}{@{\extracolsep{2pt}} lccc} 
\toprule\\[-0.45cm]
 & $\alpha = 0.01$ & $\alpha = 0.05$ & $\alpha = 0.1\phantom{0}$ \\ [0.5ex] 
\hline\\[-0.45cm]
$p=16$ & $0.014$ & $0.045$ & $0.108$ \\ 
$p=31$ & $0.009$ & $0.064$ & $0.122$ \\ 
$p=46$ & $0.012$ & $0.065$ & $0.112$ \\ 
$p=91$ & $0.011$ & $0.043$ & $0.091$ \\ 
$p=451$ & $0.005$ & $0.063$ & $0.126$ \\ 
$p=901$ & $0.011$ & $0.052$ & $0.116$ \\ 
$p=3001$ & $0.014$ & $0.054$ & $0.108$ \\
\bottomrule 
\end{tabular}}
\end{subtable}
\vspace{0.75cm}
\caption{Power of the test against the alternative with $c^{**}=0.1$.}\label{table:power0.1HAC}
\begin{subtable}[t]{0.45\textwidth}
\subcaption{$T=15$}
{\small 
\begin{tabular}{@{\extracolsep{2pt}} lccc} 
\toprule\\[-0.45cm]
 & $\alpha = 0.01$ & $\alpha = 0.05$ & $\alpha = 0.1\phantom{0}$ \\ [0.5ex] 
\hline\\[-0.45cm]
$p=7$ & $0.287$ & $0.526$ & $0.645$ \\ 
$p=10$ & $0.310$ & $0.546$ & $0.670$ \\ 
$p=13$ & $0.317$ & $0.526$ & $0.652$ \\ 
$p=31$ & $0.309$ & $0.555$ & $0.657$ \\ 
$p=151$ & $0.320$ & $0.538$ & $0.652$ \\ 
$p=301$ & $0.335$ & $0.528$ & $0.639$ \\ 
$p=901$ & $0.348$ & $0.542$ & $0.655$ \\ 
\bottomrule 
\end{tabular}}
\end{subtable} \hspace{0.75cm}
\begin{subtable}[t]{0.45\textwidth}
\subcaption{$T=50$}
{\small 
\begin{tabular}{@{\extracolsep{2pt}} lccc} 
\toprule\\[-0.45cm]
 & $\alpha = 0.01$ & $\alpha = 0.05$ & $\alpha = 0.1\phantom{0}$ \\ [0.5ex] 
\hline\\[-0.45cm]
$p=16$ & $0.885$ & $0.959$ & $0.979$ \\ 
$p=31$ & $0.891$ & $0.961$ & $0.982$ \\ 
$p=46$ & $0.898$ & $0.969$ & $0.983$ \\ 
$p=91$ & $0.890$ & $0.959$ & $0.979$ \\ 
$p=451$ & $0.860$ & $0.953$ & $0.972$ \\ 
$p=901$ & $0.857$ & $0.954$ & $0.974$ \\ 
$p=3001$ & $0.867$ & $0.955$ & $0.975$ \\
\bottomrule 
\end{tabular}}
\end{subtable}
\vspace{0.75cm}
\caption{Power of the test against the alternative with $c^{**}=0.2$.}\label{table:power0.2HAC}
\begin{subtable}[t]{0.45\textwidth}
\subcaption{$T=15$}
{\small 
\begin{tabular}{@{\extracolsep{2pt}} lccc} 
\toprule\\[-0.45cm]
 & $\alpha = 0.01$ & $\alpha = 0.05$ & $\alpha = 0.1\phantom{0}$ \\ [0.5ex] 
\hline\\[-0.45cm]
$p=7$ & $0.911$ & $0.972$ & $0.989$ \\ 
$p=10$ & $0.915$ & $0.977$ & $0.987$ \\ 
$p=13$ & $0.910$ & $0.980$ & $0.988$ \\ 
$p=31$ & $0.929$ & $0.982$ & $0.991$ \\ 
$p=151$ & $0.907$ & $0.975$ & $0.985$ \\ 
$p=301$ & $0.904$ & $0.970$ & $0.990$ \\ 
$p=901$ & $0.916$ & $0.967$ & $0.988$ \\
\bottomrule 
\end{tabular}}
\end{subtable} \hspace{0.75cm}
\begin{subtable}[t]{0.45\textwidth}
\subcaption{$T=50$}
{\small 
\begin{tabular}{@{\extracolsep{2pt}} lccc} 
\toprule\\[-0.45cm]
 & $\alpha = 0.01$ & $\alpha = 0.05$ & $\alpha = 0.1\phantom{0}$ \\ [0.5ex] 
\hline\\[-0.45cm]
$p=16$ & $1$ & $1$ & $1$ \\ 
$p=31$ & $1$ & $1$ & $1$ \\ 
$p=46$ & $1$ & $1$ & $1$ \\ 
$p=91$ & $1$ & $1$ & $1$\\ 
$p=451$ & $1$ & $1$ & $1$ \\ 
$p=901$ & $1$ & $1$ & $1$\\ 
$p=3001$ & $1$ & $1$ & $1$ \\ 
\bottomrule 
\end{tabular}}
\end{subtable}
\end{table}

We next demonstrate that our inference methods are robust to heteroskedasticity and autocorrelation in the idiosyncratic error structure. To do so, we run the test from Section \ref{sec:sim-inf} in a modified simulation design, where the error terms $\varepsilon_{it}$ and $u_{it}$ are chosen as follows (and everything else is kept as in Section \ref{sec:sim-inf}).
\begin{enumerate}[label = \textbf{(\alph*)},leftmargin=0.725cm]

\item \textbf{Heteroskedastic (but uncorrelated) errors.} In this scenario, we generate the error variances $\{\sigma_{\varepsilon,i}^2: 1 \le i \le n\}$ as independent uniform draws from the interval $[0.5,1.5]$ in each simulation run and let $\varepsilon_{it}$ be i.i.d.\ $N(0,\sigma_{\varepsilon,i}^2)$-distributed. In contrast to $\varepsilon_{it}$, the nodewise error terms $u_{it}$ are i.i.d.\  $N(0,1)$-distributed and thus homoskedastic. (Reason: we want the covariates $X_{it,j}$ to have the same mean and variance for all $j$ as in the original simulation design. To ensure this, $\sigma_{u,i}^2  = \ex[u_{it}^2]$ must be independent of $i$.)
Tables \ref{table:sizeHETEROSCEDASTIC}--\ref{table:power0.2HETEROSCEDASTIC} report the empirical size and power of the test implemented with the normalization $\widetilde{\mathcal{N}} = \widetilde{\mathcal{N}}^{\textnormal{HET}}$. In all considered simulation settings, we find accurate size and power numbers comparable to those from Section \ref{sec:sim-inf}, suggesting that the normalization $\widetilde{\mathcal{N}}^{\textnormal{HET}}$ is robust with respect to heteroskedasticity. 

\item \textbf{Heteroskedastic and autocorrelated errors.} In this scenario, the error processes $\{ \varepsilon_{it}: 1 \le t \le T \}$ and $\{ u_{it}: 1 \le t \le T \}$ are generated as stationary AR(1) processes (that are independent across $i$) of the form 
\begin{align*}
 & \varepsilon_{it} = 0.5 \, \varepsilon_{it-1} + v_{it} \quad \, \text{with} \quad v_{it} \sim \normal(0, 0.75 \, \sigma_{\varepsilon,i}^2) \\
 & u_{it} = 0.5 \, u_{it-1} + v'_{it} \quad \text{with} \quad v'_{it}\sim \normal(0,0.75),
\end{align*} 
where $v_{it}$ and $v'_{it}$ are independent across $t$. With this choice, we in particular get that $\var(\varepsilon_{it}) = \sigma_{\varepsilon,i}^2$ depends on $i$, that is, we do not only have autocorrelation in the errors $\varepsilon_{it}$ across $t$ but also unconditional heteroskedasticity across $i$. As before, we choose the constants $\sigma_{\varepsilon,i}^2$ as independent uniform draws from the interval $[0.5,1.5]$ in each simulation run.
We run our test with the normalization $\widetilde{\mathcal{N}} = \widetilde{\mathcal{N}}^{\textnormal{HAC}}$. Tables \ref{table:sizeHAC}--\ref{table:power0.2HAC} show that the test performs reasonably well, even though 
the power numbers are a bit lower than in the previous case without autocorrelation. 

\end{enumerate}

\FloatBarrier

\newpage
\allowdisplaybreaks[3]

\begin{center}
{\LARGE \textbf{Supplement: Technical Details}}
\end{center}

\def\theequation{A.\arabic{equation}}
\setcounter{equation}{4}
\setcounter{page}{1}
\section*{A \hspace{0.1cm} Details on the proof of Theorem \ref{theo:rate}(a)}

\subsection*{Proof of Lemma \ref{lemma:aux1}}

We start with the proof of \ref{lemma:aux1:Z}. It suffices to show that
\begin{equation}\label{proof:aux1:Z:claim}
\pr \bigg( \max_{1 \le j \le p} \max_{1 \le t \le T} \Big| \frac{1}{n} \sum_{i=1}^n Z_{it,j} \Big| > C_0 \sqrt{\frac{\log(pT)}{n}} \bigg) = o(1)
\end{equation}
for some sufficiently large constant $C_0 > 0$. Let
\begin{align*}
Z_{it,j}^{\le} & = Z_{it,j} \, \ind\big(Z_{it,j} \le \{npT\}^{\frac{1}{\moments-\delta}}\big) \\
Z_{it,j}^{>} & = Z_{it,j} \, \ind\big(Z_{it,j} > \{npT\}^{\frac{1}{\moments-\delta}}\big), 
\end{align*} 
where $\delta > 0$ is an absolute constant that can be chosen as small as desired, and write 
\[ \frac{1}{n} \sum_{i=1}^n Z_{it,j} = \frac{1}{n} \sum_{i=1}^n (Z_{it,j}^\le - \ex Z_{it,j}^\le) + \frac{1}{n} \sum_{i=1}^n (Z_{it,j}^> - \ex Z_{it,j}^>). \]
With this notation, we get that 
\[ \pr \bigg( \max_{1 \le j \le p} \max_{1 \le t \le T} \Big| \frac{1}{n} \sum_{i=1}^n Z_{it,j} \Big| > C_0 \sqrt{\frac{\log(pT)}{n}} \bigg) \le P^{\le} + P^{>}, \]
where
\begin{align*} 
P^{\le} & = \pr \bigg( \max_{1 \le j \le p} \max_{1 \le t \le T} \Big| \frac{1}{n} \sum_{i=1}^n (Z_{it,j}^\le - \ex Z_{it,j}^\le) \Big| > \frac{C_0}{2} \sqrt{\frac{\log(pT)}{n}} \bigg) \\
P^{>} & = \pr \bigg( \max_{1 \le j \le p} \max_{1 \le t \le T} \Big| \frac{1}{n} \sum_{i=1}^n (Z_{it,j}^> - \ex Z_{it,j}^>) \Big| > \frac{C_0}{2} \sqrt{\frac{\log(pT)}{n}} \bigg).
\end{align*}
In what follows, we show that $P^{\le} = o(1)$ and $P^{>} = o(1)$ for some sufficiently large constant $C_0$, which implies \eqref{proof:aux1:Z:claim}.

We first have a closer look at $P^{>}$. It holds that $P^{>} \le P^{>}_1 + P^{>}_2$, where
\begin{align*}
P_1^{>} & = \pr \bigg( \max_{1 \le j \le p} \max_{1 \le t \le T} \Big| \frac{1}{n} \sum_{i=1}^n Z_{it,j}^> \Big| > \frac{C_0}{4} \sqrt{\frac{\log(pT)}{n}} \bigg) \\
P_2^{>} & = \pr \bigg( \max_{1 \le j \le p} \max_{1 \le t \le T} \Big| \frac{1}{n} \sum_{i=1}^n \ex Z_{it,j}^> \Big| > \frac{C_0}{4} \sqrt{\frac{\log(pT)}{n}} \bigg).
\end{align*} 
Since $\ex|Z_{it,j}|^\moments \le C < \infty$, 
\begin{align} 
P_1^{>} & \le \pr \Big( |Z_{it,j}| > \{npT\}^{\frac{1}{\moments-\delta}} \text{ for some indices } i,j \text{ and } t \Big) \nonumber \\
 & \le \sum\limits_{i=1}^n \sum_{j=1}^p \sum\limits_{t=1}^{T} \pr \Big( |Z_{it,j}| > \{npT\}^{\frac{1}{\moments-\delta}} \Big) 
   \le \sum\limits_{i=1}^n \sum_{j=1}^p \sum\limits_{t=1}^{T} \ex \bigg[ \frac{|Z_{it,j}|^\moments}{\{npT\}^{\frac{\moments}{\moments-\delta}}} \bigg] \nonumber \\[0.2cm] 
 & \le C \{npT\} \big/ \{npT\}^{\frac{\moments}{\moments-\delta}} = o(1). \label{eq:bound-P1>}
\end{align}
Moreover, since $|\ex Z_{it,T}^>| \le C / \{npT\}^{(\moments-1)/(\moments-\delta)}$ and $C / \{npT\}^{(\moments-1)/(\moments-\delta)} < (C_0 / 4) \linebreak \sqrt{\log(pT)/n}$ for sufficiently large $n$, it holds that $P_2^{>} = 0$ for $n$ large enough. Putting everything together, we obtain that $P^{>} = o(1)$ as desired.

We now turn to the analysis of $P^{\le}$. To make the notation more compact, we introduce the shorthand $B_{it,j} = (Z_{it,j}^\le - \ex Z_{it,j}^\le)/\sqrt{n}$. Since
\[ P^{\le} \le \sum_{j=1}^p \sum_{t=1}^T \pr \bigg( \Big| \sum_{i=1}^n B_{it,j} \Big| > \frac{C_0}{2} \sqrt{\log(pT)} \bigg), \]
it suffices to show that 
\begin{equation}\label{proof:aux1:Z:exp-bound} 
\pr \bigg( \Big| \sum_{i=1}^n B_{it,j} \Big| > \frac{C_0}{2} \sqrt{\log(pT)} \bigg) \le \frac{C}{(pT)^r} 
\end{equation}
uniformly over $j$ and $t$ with some constant $r > 1$. For the proof, we make use of the following two facts:
\begin{enumerate}[label=(\alph*),leftmargin=0.7cm]
\item For a real-valued random variable $B$ and $\gamma > 0$, Markov's inequality yields that $\pr( \pm B > \delta ) \le \ex \exp(\pm \gamma B) / \exp(\gamma \delta)$.  
\item Since $|B_{it,j}| \le 2 (npT)^{1/(\moments-\delta)} / \sqrt{n}$ and $(npT)^{1/(\moments-\delta)} / \sqrt{n} = o(1/\sqrt{\log(pT)})$ under assumption \ref{C:nTp-large}, we obtain that $\gamma |B_{it,j}| \le 1/2$ if we set $\gamma = c_\gamma \sqrt{\log(pT)}$ with some sufficiently small constant $c_\gamma$. As $\exp(x) \le 1 + x + x^2$ for $|x| \le 1/2$, it follows that
\[ \ex \Big[ \exp \big( \pm \gamma B_{it,j} \big) \Big] \le 1 + \gamma^2 \ex \big[ B_{it,j}^2 \big] \le \exp \big( \gamma^2 \ex \big[ B_{it,j}^2 \big] \big). \]
\item $\ex[B_{it,j}^2] \le C_V/n$ with some sufficiently large constant $C_V > 0$. 
\end{enumerate}
Using (a)--(c), we obtain that
\begin{align*}
\pr \bigg( \Big| \sum_{i=1}^n B_{it,j} \Big| > \frac{C_0}{2} \sqrt{\log(pT)} \bigg)
 & \le \pr \bigg( \sum_{i=1}^n B_{it,j} > \frac{C_0}{2} \sqrt{\log(pT)} \bigg) \\*
 & \quad + \pr \bigg( -\sum_{i=1}^n B_{it,j} > \frac{C_0}{2} \sqrt{\log(pT)} \bigg),
\end{align*}
where
\begin{align*} 
 & \pr \bigg( \pm \sum_{i=1}^n B_{it,j} > \frac{C_0}{2} \sqrt{\log(pT)} \bigg) \\
 & \le \exp \Big( - \frac{C_0 \gamma \sqrt{\log(pT)}}{2} \Big) \, \ex \bigg[ \exp \Big( \pm \gamma \sum\limits_{i=1}^n B_{it,j} \Big) \bigg] \\
 & \le \exp \Big( - \frac{C_0 \gamma \sqrt{\log(pT)}}{2} \Big) \, \prod\limits_{i=1}^n \ex \Big[ \exp \big( \pm \gamma B_{it,j} \big) \Big] \\
 & \le \exp \Big( - \frac{C_0 \gamma \sqrt{\log(pT)}}{2} \Big) \, \prod\limits_{i=1}^n \exp \big( \gamma^2 \ex \big[ B_{it,j}^2 \big] \big) \\
 & = \exp \Big( - \frac{C_0 \gamma \sqrt{\log(pT)}}{2} \Big) \exp \Big( \gamma^2 \sum_{i=1}^n \ex \big[ B_{it,j}^2 \big] \Big) \\
 & \le \exp \Big( - c_\gamma \Big[\frac{C_0}{2} - c_\gamma C_V \Big] \log(pT) \Big). 
\end{align*}
 Hence, 
\begin{align*}
\pr \bigg( \Big| \sum_{i=1}^n B_{it,j} \Big| > \frac{C_0}{2} \sqrt{\log(pT)} \bigg) & \le 2 \exp \Big( - c_\gamma \Big[\frac{C_0}{2} - c_\gamma C_V \Big] \log(pT) \Big) \le C (pT)^{-r}, 
\end{align*}
where the constant $r > 0$ can be made arbitrarily large by picking $C_0$ large enough. This completes the proof of statement \ref{lemma:aux1:Z} of the lemma. The proofs of statements \ref{lemma:aux1:Feps} and \ref{lemma:aux1:FZ} are omitted as they are similar (the only notable difference being that the dependencies in the data across $t$ need to be taken into account, which can be done by the same blocking argument as used in the proof of Lemma \ref{lemma:aux2}).

\subsection*{Proof of Lemma \ref{lemma:aux2}}

We only give the proof of \ref{lemma:aux2:ZZ} as \ref{lemma:aux2:Zeps} can be shown by analogous but somewhat simpler arguments. It holds that 
\[ \max_{i,j,j^\prime} \Big| \frac{1}{T} \sum_{t=1}^T \Big\{ \frac{1}{n} \sum_{i^\prime=1}^n Z_{i^\prime t,j^\prime} \Big\} Z_{it,j} \Big| \le Q_A + Q_B \]
with 
\begin{align*} 
Q_A & = \max_{i,j,j^\prime} \Big| \frac{1}{nT} \sum_{t=1}^T Z_{it,j^\prime} Z_{it,j} \Big| \\
Q_B & = \max_{i,j,j^\prime} \Big| \frac{1}{T} \sum_{t=1}^T \Big\{ \frac{1}{n} \sum_{i^\prime \ne i} Z_{i^\prime t,j^\prime} \Big\} Z_{it,j} \Big|.
\end{align*}
By arguments similar to those for Lemma \ref{lemma:aux1}, we obtain that 
\begin{align*}
Q_A 
 & \le \max_{i,j,j^\prime} \Big| \frac{1}{nT} \sum_{t=1}^T (Z_{it,j^\prime} Z_{it,j} - \ex Z_{it,j^\prime} Z_{it,j}) \Big| + \max_{i,j,j^\prime} \Big| \frac{1}{nT} \sum_{t=1}^T \ex Z_{it,j^\prime} Z_{it,j} \Big| \\
 & = O_p\Big(\frac{\sqrt{\log(p^2 n)}}{n\sqrt{T}} \Big) + O\Big(\frac{1}{n}\Big) = O_p\Big(\frac{1}{n}\Big). 
\end{align*}
To deal with the term $Q_B$, we rewrite it as
\[ Q_B = \max_{i,j,j^\prime} \Big| \frac{1}{T} \sum_{t=1}^T w_{it,j^\prime} Z_{it,j} \Big| \qquad \text{with} \qquad w_{it,j^\prime} = \frac{1}{n} \sum_{i^\prime \ne i} Z_{i^\prime t,j^\prime}, \]
where the random weights $w_{it,j^\prime}$ have the following properties: 
\begin{enumerate}[label=(P\arabic*),leftmargin=1cm]
\item \label{prop1} By essentially the same arguments as for Lemma \ref{lemma:aux1}\ref{lemma:aux1:Z}, 
\[ \pr \bigg( \max_{i,t,j^\prime} |w_{it,j^\prime}| > C_w \sqrt{\frac{\log (n p T)}{n}} \bigg) = o(1), \]
where $C_w$ is a sufficiently large absolute constant. 
\item \label{prop2} For each $i$, $j$ and $j^\prime$, the collections of random variables $\{ w_{it,j^\prime}: 1 \le t \le T \}$ and $\{Z_{it,j}: 1 \le t \le T \}$ are independent from each other. 
\end{enumerate}
In the sequel, we prove that 
\begin{equation}\label{proof:aux2:ZZ:Q2}
\pr \bigg( \max_{i,j,j^\prime} \Big| \frac{1}{T} \sum_{t=1}^T w_{it,j^\prime} Z_{it,j} \Big| > C_0 r_{n,p,T} \bigg) = o(1) 
\end{equation}
with $r_{n,p,T} = \sqrt{\log(npT)\log(np^2)}/\sqrt{nT}$ and some sufficiently large constant $C_0 > 0$, which implies that $Q_B = O_p(r_{n,p,T})$. For the proof of \eqref{proof:aux2:ZZ:Q2}, we define the truncated variables 
\begin{align*}
Z_{it,j}^{\le} & = Z_{it,j} \, \ind\big(Z_{it,j} \le \{npT\}^{\frac{1}{\moments-\delta}}\big) \\
Z_{it,j}^{>} & = Z_{it,j} \, \ind\big(Z_{it,j} > \{npT\}^{\frac{1}{\moments-\delta}}\big), 
\end{align*} 
where $\delta > 0$ is an absolute constant that can be chosen as small as desired. Since 
\[ \frac{1}{T} \sum_{t=1}^T w_{it,j^\prime} Z_{it,j} = \frac{1}{T} \sum_{t=1}^T w_{it,j^\prime} (Z_{it,j}^{\le} - \ex Z_{it,j}^{\le}) + \frac{1}{T} \sum_{t=1}^T w_{it,j^\prime} (Z_{it,j}^{>} - \ex Z_{it,j}^{>}), \] 
we obtain that 
\[ \pr \bigg( \max_{i,j,j^\prime} \Big| \frac{1}{T} \sum_{t=1}^T w_{it,j^\prime} Z_{it,j} \Big| > C_0 r_{n,p,T} \bigg) \le P^{\le} + P^{>} \]
with 
\begin{align*} 
P^{\le} & = \pr \bigg( \max_{i,j,j^\prime} \Big|  \frac{1}{T} \sum_{t=1}^T w_{it,j^\prime} (Z_{it,j}^{\le} - \ex Z_{it,j}^{\le}) \Big| > \frac{C_0 r_{n,p,T}}{2} \bigg) \\*
P^{>} & = \pr \bigg( \max_{i,j,j^\prime} \Big| \frac{1}{T} \sum_{t=1}^T w_{it,j^\prime} (Z_{it,j}^{>} - \ex Z_{it,j}^{>}) \Big| > \frac{C_0 r_{n,p,T}}{2} \bigg).
\end{align*} 
We now show that $P^{\le} = o(1)$ and $P^{>} = o(1)$ for some sufficiently large constant $C_0$, which implies \eqref{proof:aux2:ZZ:Q2}.

We first have a closer look at $P^{>}$. It holds that $P^{>} \le P^{>}_1 + P^{>}_2$, where
\begin{align*}
P_1^{>} & = \pr \Big( \max_{i,j,j^\prime} \Big| \frac{1}{T} \sum\limits_{t=1}^{T} w_{it,j^\prime} Z_{it,j}^{>} \Big| > \frac{C_0 r_{n,p,T}}{4} \Big) \\
P_2^{>} & = \pr \Big( \max_{i,j,j^\prime} \Big| \frac{1}{T} \sum\limits_{t=1}^{T} w_{it,j^\prime} \ex Z_{it,j}^{>} \Big| > \frac{C_0 r_{n,p,T}}{4} \Big). 
\end{align*} 
By the same arguments as for \eqref{eq:bound-P1>}, we obtain that
\begin{align*} 
P_1^{>} 
 & \le \pr \Big( |Z_{it,j}| > \{npT\}^{\frac{1}{\moments-\delta}} \text{ for some } i,j \text{ and } t \Big) 
   \le C \{npT\} \big/ \{npT\}^{\frac{\moments}{\moments-\delta}} = o(1). 
\end{align*}
Moreover, as $|\ex Z_{it,T}^>| \le C / \{npT\}^{(\moments-1)/(\moments-\delta)}$ and $\max_{i,t,j^\prime} |w_{it,j^\prime}| \le C_w \sqrt{\log (n p T) / n}$ with probability tending to $1$ by \ref{prop1}, we get that 
\begin{align*}
P_2^{>} & = \pr \bigg( \max_{i,j,j^\prime} \Big| \frac{1}{T} \sum\limits_{t=1}^{T} w_{it,j^\prime} \ex Z_{it,j}^{>} \Big| > \frac{C_0 r_{n,p,T}}{4}, \, \max_{i,t,j^\prime} |w_{it,j^\prime}| \le C_w \sqrt{\frac{\log (n p T)}{n}} \bigg) + o(1) \\
 & \le \pr \bigg( C_w \max_{i,j,t} |\ex Z_{it,j}^{>}| > \frac{C_0}{4} \sqrt{\frac{\log(np^2)}{T}} \bigg) + o(1) = o(1).
\end{align*}
As a result, we arrive at $P^{>} = o(1)$.

We next turn to the analysis of $P^{\le}$. Let $\mathcal{E}$ be the event that $\max_{i,t,j^\prime} |w_{it,j^\prime}| \le C_w \sqrt{\log(npT)/n}$ and $\mathcal{E}_{ij^\prime}$ the event that $\max_{t} |w_{it,j^\prime}| \le C_w \sqrt{\log(npT)/n}$. Using (P1) and noting that $\mathcal{E} \subseteq \mathcal{E}_{ij^\prime}$, we obtain that 
\begin{align}
P^{\le} 
 & = \pr \bigg( \max_{i,j,j^\prime} \Big|  \frac{1}{T} \sum_{t=1}^T w_{it,j^\prime} (Z_{it,j}^{\le} - \ex Z_{it,j}^{\le}) \Big| > \frac{C_0 r_{n,p,T}}{2}, \mathcal{E} \bigg) + o(1) \nonumber \\
 & = \pr \bigg( \ind(\mathcal{E}) \cdot \max_{i,j,j^\prime} \Big|  \frac{1}{T} \sum_{t=1}^T w_{it,j^\prime} (Z_{it,j}^{\le} - \ex Z_{it,j}^{\le}) \Big| > \frac{C_0 r_{n,p,T}}{2} \bigg) + o(1) \nonumber \\
 & \le \pr \bigg( \max_{i,j,j^\prime} \Big|  \frac{1}{T} \sum_{t=1}^T w_{it,j^\prime}^* (Z_{it,j}^{\le} - \ex Z_{it,j}^{\le}) \Big| > \frac{C_0 r_{n,p,T}}{2} \bigg) + o(1) \nonumber \\
 & \le \sum\limits_{i=1}^n \sum_{j=1}^p \sum_{j^\prime=1}^p \pr \bigg( \Big| \frac{1}{T} \sum_{t=1}^T w_{it,j^\prime}^* (Z_{it,j}^{\le} - \ex Z_{it,j}^{\le}) \Big| > \frac{C_0 r_{n,p,T}}{2} \bigg) + o(1) \label{proof:aux2:ZZ:exp-bound-prelim}
\end{align}
with $w_{it,j^\prime}^* = \ind(\mathcal{E}_{ij^\prime}) w_{it,j^\prime} = \ind (\max_{t} |w_{it,j^\prime}| \le C_w \sqrt{\log(npT)/n}) \, w_{it,j^\prime}$. In the following, we show that 
\begin{equation}\label{proof:aux2:ZZ:exp-bound} 
\pr \bigg( \Big| \frac{1}{T} \sum_{t=1}^T w_{it,j^\prime}^* \big\{ Z_{it,j}^{\le} - \ex  Z_{it,j}^{\le} \big\} \Big| > \frac{C_0 r_{n,p,T}}{2} \bigg) \le \frac{C}{(np)^{r}} 
\end{equation}
uniformly over $i$, $j$ and $j^\prime$ with $r > 0$ as large as desired. Together with \eqref{proof:aux2:ZZ:exp-bound-prelim}, this immediately implies that $P^{\le} = o(1)$. Since 
\begin{align*} 
\pr \bigg( \Big| & \frac{1}{T} \sum_{t=1}^T w_{it,j^\prime}^* \big\{ Z_{it,j}^{\le} - \ex  Z_{it,j}^{\le} \big\} \Big| > \frac{C_0 r_{n,p,T}}{2} \bigg) \\ & = \ex \bigg[ \pr \bigg( \Big| \frac{1}{T} \sum_{t=1}^T w_{it,j^\prime}^* \big\{ Z_{it,j}^{\le} - \ex  Z_{it,j}^{\le} \big\} \Big| > \frac{C_0 r_{n,p,T}}{2} \, \bigg| \, w_{1:T} \bigg) \bigg] 
\end{align*}
with $w_{1:T} = \{w_{it,j^\prime}: 1 \le t \le T\}$, it suffices to prove that 
\begin{equation}\label{proof:aux2:ZZ:conditional-exp-bound} 
\pr \bigg( \Big| \frac{1}{\sqrt{T}} \sum_{t=1}^T w_{it,j^\prime}^* \big\{ Z_{it,j}^{\le} - \ex  Z_{it,j}^{\le} \big\} \Big| > \frac{C_0 \sqrt{T} \, r_{n,p,T}}{2} \, \bigg| \, w_{1:T}  \bigg) \le \frac{C}{(np)^{r}}.  
\end{equation}
To do so, we split the term $T^{-1/2} \sum_{t=1}^T w_{it,j^\prime}^* \{ Z_{it,j}^{\le} - \ex  Z_{it,j}^{\le} \}$ into blocks as follows: 
\[ \frac{1}{\sqrt{T}} \sum_{t=1}^T w_{it,j^\prime}^* \big\{ Z_{it,j}^{\le} - \ex  Z_{it,j}^{\le} \big\} = \sum\limits_{m=1}^{\lceil M \rceil} B_{2m-1}(w_{1:T}) + \sum\limits_{m=1}^{\lfloor M \rfloor} B_{2m}(w_{1:T}) \] 
with
\[ B_{m}(w_{1:T}) = B_{m,ijj^\prime}(w_{1:T}) = \frac{1}{\sqrt{T}} \sum\limits_{t = (m-1) L + 1}^{\min\{m L,T\}} w_{it,j^\prime}^* \{ Z_{it,j}^{\le} - \ex  Z_{it,j}^{\le} \}, \]
where $L = \sqrt{T} / ( \{npT\}^{1/(\moments-\delta)} \sqrt{\log (np^2)})$ is the block length 
and $2 M$ with $M = \lceil T/L \rceil / 2$ is the number of blocks. Note that under assumption \ref{C:nTp-large}, it holds that $cT^\xi \le L \le CT^{1-\xi}$ with some sufficiently small $\xi > 0$. With this notation at hand, we obtain that
\begin{align}
 & \pr \bigg( \Big| \frac{1}{\sqrt{T}} \sum_{t=1}^T w_{it,j^\prime}^* \big\{ Z_{it,j}^{\le} - \ex  Z_{it,j}^{\le} \big\} \Big| > \frac{C_0 \sqrt{T} \, r_{n,p,T}}{2} \, \bigg| \, w_{1:T} \bigg) \nonumber \\
 & \qquad \qquad \le \pr \bigg( \Big| \sum\limits_{m=1}^{\lceil M \rceil} B_{2m-1}(w_{1:T}) \Big| > \frac{C_0 \sqrt{T} \, r_{n,p,T}}{4}  \, \bigg| \, w_{1:T} \bigg) \nonumber \\ & \qquad \qquad \qquad \qquad + \pr \bigg( \Big| \sum\limits_{m=1}^{\lfloor M \rfloor} B_{2m}(w_{1:T}) \Big| > \frac{C_0 \sqrt{T} \, r_{n,p,T}}{4} \, \bigg| \, w_{1:T} \bigg). \label{proof:aux2:ZZ:conditional-exp-bound-1}
\end{align}
As the two terms on the right-hand side of \eqref{proof:aux2:ZZ:conditional-exp-bound-1} can be treated analogously, we focus attention to the first one. By applying Bradley's strong approximation theorem \citep[see Theorem 3 in][]{Bradley1983} conditionally on $w_{1:T}$, we can construct a sequence of random variables $B_{1}^*(w_{1:T}), B_{3}^*(w_{1:T}), \ldots$ such that (I) $B_{1}^*(w_{1:T}), B_{3}^*(w_{1:T}), \ldots$ are independent, (II) $B_{2m-1}(w_{1:T})$ and $B_{2m-1}^*(w_{1:T})$ have the same distribution for each $m$, and (III) $\pr (|B_{2m-1}^*(w_{1:T}) - B_{2m-1}(w_{1:T})| > \mu \, | \, w_{1:T}) \le 18 (\| B_{2m-1}(w_{1:T}) \|_{\infty}/\mu)^{1/2}$ $\alpha(L)$ for $0 < \mu \le \| B_{2m-1}(w_{1:T}) \|_{\infty}$, where we use the symbol $\| \cdot \|_\infty$ to denote the $L_\infty$-norm of a real-valued random variable. With the variables $B_{2m-1}^*(w_{1:T})$, we can construct the bound
\begin{equation}\label{eq-exp-bound-2}
\pr \bigg( \Big| \sum\limits_{m=1}^{\lceil M \rceil} B_{2m-1}(w_{1:T}) \Big| > \frac{C_0 \sqrt{T} \, r_{n,p,T}}{4} \, \bigg| \, w_{1:T} \bigg) \le P_1^* + P_2^*, 
\end{equation}
where
\begin{align*}
P_1^* & = \pr \bigg( \Big| \sum\limits_{m=1}^{\lceil M \rceil} B_{2m-1}^*(w_{1:T}) \Big| > \frac{C_0 \sqrt{T} \, r_{n,p,T}}{8} \, \bigg| \, w_{1:T} \bigg) \\
P_2^* & = \pr \bigg( \Big| \sum\limits_{m=1}^{\lceil M \rceil} \big\{ B_{2m-1}(w_{1:T}) - B_{2m-1}^*(w_{1:T}) \big\} \Big| > \frac{C_0 \sqrt{T} \, r_{n,p,T}}{8} \, \bigg| \, w_{1:T} \bigg). 
\end{align*}
Using (III) together with the fact that the mixing coefficients $\alpha(\cdot)$ decay to $0$ exponentially fast, it is not difficult to see that $P_2^* \le C (np)^{-r}$, where the constant $r > 0$ can be picked as large as desired. To deal with $P_1^*$, we make use of the following three facts:
\begin{enumerate}[label=(\alph*),leftmargin=0.7cm]
\item For a real-valued random variable $B$ and $\gamma > 0$, Markov's inequality yields that $\pr( \pm B > \delta ) \le \ex \exp(\pm \gamma B) / \exp(\gamma \delta)$.  
\item Since $|B_{2m-1}(w_{1:T})| \le \{ 2 C_w L \sqrt{\log(npT)/n} \, (npT)^{1/(\moments-\delta)} \} / \sqrt{T}$, we can choose $\gamma = c_\gamma \sqrt{\log(np^2)} \sqrt{n/\log(npT)}$ with $c_\gamma > 0$ so small that $\gamma |B_{2m-1}(w_{1:T})| \le 1/2$. As $\exp(x) \le 1 + x + x^2$ for $|x| \le 1/2$, we get that
\begin{align*}
\ex \Big[ \exp \big( \pm \gamma B_{2m-1}(w_{1:T}) \big) \, \Big| \, w_{1:T} \Big] 
 & \le 1 + \gamma^2 \ex \big[ \{ B_{2m-1}(w_{1:T}) \}^2 \, \big| \, w_{1:T} \big] \\
 & \le \exp \big( \gamma^2 \ex \big[ \{B_{2m-1}(w_{1:T})\}^2 \, \big| \, w_{1:T} \big] \big)
\end{align*}
along with
\[ \ex \Big[ \exp \big( \pm \gamma B_{2m-1}^*(w_{1:T}) \big) \, \Big| \, w_{1:T} \Big] \le \exp \big( \gamma^2 \ex \big[ \{B_{2m-1}^*(w_{1:T})\}^2 \, \big| \, w_{1:T} \big] \big). \]
\item Standard calculations yield that 
\[ \sum\limits_{m=1}^{\lceil M \rceil} \ex \big[ \{B_{2m-1}(w_{1:T})\}^2 \big| w_{1:T} \big] \le \frac{C_V \log(npT)}{n} \]
with some sufficiently large constant $C_V$.   
\end{enumerate}
Using (a)--(c), we obtain that
\begin{align*}
P_1^* 
 & \le \pr \bigg( \sum\limits_{m=1}^{\lceil M \rceil} B_{2m-1}^*(w_{1:T}) > \frac{C_0 \sqrt{T} \, r_{n,p,T}}{8} \, \bigg| \, w_{1:T} \bigg) \\*
 & \quad + \pr \bigg( -\sum\limits_{m=1}^{\lceil M \rceil} B_{2m-1}^*(w_{1:T}) > \frac{C_0 \sqrt{T} \, r_{n,p,T}}{8} \, \bigg| \, w_{1:T} \bigg),
\end{align*}
where
\begin{align*} 
 & \pr \bigg( \pm \sum\limits_{m=1}^{\lceil M \rceil} B_{2m-1}^*(w_{1:T}) > \frac{C_0 \sqrt{T} \, r_{n,p,T}}{8} \, \bigg| \, w_{1:T} \bigg) \\
 & \le \exp \Big( - \frac{C_0 \gamma \sqrt{T} \, r_{n,p,T}}{8} \Big) \, \ex \bigg[ \exp \Big( \pm \gamma \sum\limits_{m=1}^{\lceil M \rceil} B_{2m-1}^*(w_{1:T}) \Big) \, \bigg| \, w_{1:T} \bigg] \\
 & = \exp \Big( - \frac{C_0 \gamma \sqrt{T} \, r_{n,p,T}}{8} \Big) \, \prod\limits_{m=1}^{\lceil M \rceil} \ex \Big[ \exp \big( \pm \gamma B_{2m-1}^*(w_{1:T}) \big) \, \Big| \, w_{1:T} \Big] \\
 & \le \exp \Big( - \frac{C_0 \gamma \sqrt{T} \, r_{n,p,T}}{8} \Big) \, \prod\limits_{m=1}^{\lceil M \rceil} \exp \Big( \gamma^2 \ex \big[ \{B_{2m-1}^*(w_{1:T})\}^2 \, \big| \, w_{1:T} \big] \Big) \\
 & =  \exp \Big( - \frac{C_0 \gamma \sqrt{T} \, r_{n,p,T}}{8} \Big) \exp \bigg( \gamma^2 \sum\limits_{m=1}^{\lceil M \rceil} \ex \big[ \{B_{2m-1}^*(w_{1:T})\}^2 \, \big| \, w_{1:T} \big] \bigg) \\
 & \le \exp \Big( - c_\gamma \Big[\frac{C_0}{8} - c_\gamma C_V \Big] \log(np^2) \Big). 
\end{align*}
Hence, 
\begin{align*}
P_1^* & \le 2 \exp \Big( - c_\gamma \Big[\frac{C_0}{8} - c_\gamma C_V \Big] \log(np^2) \Big) \le \frac{C}{(np)^r}, 
\end{align*}
where the constant $r > 0$ can be made arbitrarily large by picking $C_0$ large enough. To summarize, we have shown that $P_1^* \le C (np)^{-r}$ and $P_2^* \le C (np)^{-r}$ with some arbitrarily large $r > 0$. From this, it follows that $P^{\le} = o(1)$, which completes the proof.

\subsection*{Proof of Lemma \ref{lemma:aux5}}

The $K \times K$ matrix $(\overline{\bs{\Gamma}} - \bs{\Gamma})^\top (\overline{\bs{\Gamma}} - \bs{\Gamma})$ has the entries 
\[ \sum_{j=1}^p \Big\{ \frac{1}{n} \sum_{i=1}^n (\Gamma_{i,jk} - \Gamma_{jk}) \Big\} \Big\{ \frac{1}{n} \sum_{i=1}^n (\Gamma_{i,jk^\prime} - \Gamma_{jk^\prime}) \Big\} \]
for $1 \le k,k^\prime \le K$, where $\Gamma_{i,jk}$ and $\Gamma_{jk}$ denote the elements of $\bs{\Gamma}_i$ and $\bs{\Gamma}$, respectively. By arguments analogous to those for Lemma \ref{lemma:aux1}, it holds that
\begin{equation}\label{eq:lemma:aux5:conv-Gamma}
\max_{j,k} \Big | \frac{1}{n} \sum_{i=1}^n (\Gamma_{i,jk} - \Gamma_{jk}) \Big| = O_p\Big( \sqrt{\frac{\log p}{n}} \Big). 
\end{equation}
Hence, we obtain that 
\begin{align*} 
\big\| (\overline{\bs{\Gamma}} - \bs{\Gamma})^\top (\overline{\bs{\Gamma}} - \bs{\Gamma}) \big\|_{\max}
 & \le p \bigg\{ \max_{j,k} \Big | \frac{1}{n} \sum_{i=1}^n (\Gamma_{i,jk} - \Gamma_{jk}) \Big| \bigg\}^2 \\ & = O_p\Big( \frac{p \log p}{n} \Big),
\end{align*}
which implies that $\| (\overline{\bs{\Gamma}} - \bs{\Gamma})^\top (\overline{\bs{\Gamma}} - \bs{\Gamma}) \| = O_p(p \log p / n)$. This in turn yields that $\| \overline{\bs{\Gamma}} - \bs{\Gamma} \| = O_p( \sqrt{p \log p / n})$.

Next, the $p \times p$ matrix $\overline{\bs{\Gamma}}  \overline{\bs{\Gamma}}^\top - \ex \overline{\bs{\Gamma}} \overline{\bs{\Gamma}}^\top$ has the entries $\sum_{k=1}^K \big\{ D_{jk}  D_{j^\prime k} - \ex D_{jk}  D_{j^\prime k} \big\}$ for $1 \le j,j^\prime \le p$, where we use the notation $D_{jk} = n^{-1} \sum_{i=1}^n \Gamma_{i,jk}$. According to \eqref{eq:lemma:aux5:conv-Gamma}, it holds that 
\[ \max_{j,k} \big| D_{jk} - \ex D_{jk} \big| = O_p\Big(\sqrt{\frac{\log p}{n}}\Big). \]
Moreover, $\max_{j,k} |\ex D_{jk}| = O(1)$ and 
\begin{align*} 
\max_{j,j^\prime,k} \big| & \ex D_{jk} \ex D_{j^\prime k} - \ex D_{jk} D_{j^\prime k} \big| 
 = \max_{j,j^\prime,k} \Big| \frac{1}{n^2} \sum_{i=1}^n \big\{ \ex \Gamma_{i,jk} \ex \Gamma_{i,j^\prime k} - \ex \Gamma_{i,jk} \Gamma_{i,j^\prime k} \big\} \Big| = O\Big(\frac{1}{n}\Big). 
\end{align*}
From these observations, it follows that 
\begin{align*} 
 & \max_{j,j^\prime,k} \big| D_{jk} D_{j^\prime k} - \ex D_{jk} D_{j^\prime k} \big| \\*
 & = \max_{j,j^\prime,k} \big| \big\{ (D_{jk} - \ex D_{jk}) + \ex D_{jk} \big\} \big\{ (D_{j^\prime k} - \ex D_{j^\prime k}) + \ex D_{j^\prime k} \big\} - \ex D_{jk} D_{j^\prime k} \big| \\
 & \le \Big\{ \max_{j,k} \big| D_{jk} - \ex D_{jk} \big| \Big\}^2 + 2 \Big\{ \max_{j,k}  \big| \ex D_{jk} \big| \Big\} \Big\{ \max_{j,k} \big| D_{jk} - \ex D_{jk} \big| \Big\} \\*
 & \quad + \max_{j,j^\prime,k} \big|  \ex D_{jk} \ex D_{j^\prime k} -  \ex D_{jk} D_{j^\prime k} \big| \phantom{\Big\}} \\*
 & = O_p\Big(\sqrt{\frac{\log p}{n}}\Big).
\end{align*}
We can thus conclude that 
\begin{align*} 
\big\| \overline{\bs{\Gamma}} \overline{\bs{\Gamma}}^\top - \ex \overline{\bs{\Gamma}} \overline{\bs{\Gamma}}^\top \big\| 
 & \le p \big\| \overline{\bs{\Gamma}}  \overline{\bs{\Gamma}}^\top - \ex \overline{\bs{\Gamma}}  \overline{\bs{\Gamma}}^\top \big\|_{\max} \\
 & = p \max_{j,j^\prime} \Big| \sum_{k=1}^K \big\{ D_{jk} D_{j^\prime k} - \ex D_{jk} D_{j^\prime k} \big\} \Big| \\ 
 & \le pK\max_{j,j^\prime,k} \big| D_{jk} D_{j^\prime k} - \ex D_{jk} D_{j^\prime k} \big| \\
 & = O_p\Big(p \sqrt{\frac{\log p}{n}}\Big). \qedhere
\end{align*}

\subsection*{Proof of Lemma \ref{lemma:aux3}}

Statement \ref{lemma:aux3:Feps} is a direct consequence of Lemma \ref{lemma:aux1}\ref{lemma:aux1:Feps} since
\begin{align*} 
\max_{i} \Big\| \frac{\bs{F}^\top \varepsilon_i}{T} \Big\| 
 & = \max_i \sqrt{\sum_{k=1}^K \Big\{ \frac{1}{T} \sum_{t=1}^T F_{t,k} \varepsilon_{it} \Big\}^2} \\
 & \le \sqrt{K} \max_{i,k} \Big| \frac{1}{T} \sum_{t=1}^T F_{t,k} \varepsilon_{it} \Big| = O_p\Big( \sqrt{\frac{\log n}{T}} \Big). 
\end{align*}
Analogously, statement \ref{lemma:aux3:FZ} directly follows from Lemma \ref{lemma:aux1}\ref{lemma:aux1:FZ} as
\begin{align*}
\max_{i,j} \Big\| \frac{\bs{F}^\top Z_{i(j)}}{T} \Big\|
 & = \max_{i,j} \sqrt{\sum_{k=1}^K \Big\{ \frac{1}{T} \sum_{t=1}^T F_{t,k} Z_{it,j} \Big\}^2} \\
 & \le \sqrt{K} \max_{i,j,k} \Big| \frac{1}{T} \sum_{t=1}^T F_{t,k} Z_{it,j} \Big| = O_p\Big(\sqrt{\frac{\log(np)}{T}}\Big). \qedhere
\end{align*}

\subsection*{Proof of Lemma \ref{lemma:aux4}}

With Lemma \ref{lemma:aux1}\ref{lemma:aux1:Z}, we obtain that 
\begin{align*}
\| \overline{\bs{Z}} \| = \sqrt{\eig_{\max}( \overline{\bs{Z}}^\top \overline{\bs{Z}} )} 
 & \le \sqrt{p \max_{j,j^\prime} \Big| \sum_{t=1}^T \Big\{ \frac{1}{n} \sum_{i=1}^n Z_{it,j} \Big\} \Big\{ \frac{1}{n} \sum_{i=1}^n Z_{it,j^\prime} \Big\} \Big| } \\
 & \le \sqrt{pT \Big\{ \max_{j,t} \Big| \frac{1}{n} \sum_{i=1}^n Z_{it,j} \Big| \Big\}^2} = O_p\Big(\sqrt{\frac{pT\log(pT)}{n}}\Big),
\end{align*}
which gives \ref{lemma:aux4:barZ}. Moreover, 
\begin{align*}
\Big\| \frac{\overline{\bs{Z}}^\top \bs{F}}{T} \Big\| = \Big\| \frac{1}{T} \sum_{t=1}^T \overline{Z}_t F_t^\top \Big\| 
 & = \sqrt{\eig_{\max} \bigg( \Big\{ \frac{1}{T} \sum_{t=1}^T \overline{Z}_t F_t^\top \Big\}^\top \Big\{ \frac{1}{T} \sum_{t=1}^T \overline{Z}_t F_t^\top \Big\} \bigg) } \\*
 & \le \sqrt{ K \Big\| \Big\{ \frac{1}{T} \sum_{t=1}^T \overline{Z}_t F_t^\top \Big\}^\top \Big\{ \frac{1}{T} \sum_{t=1}^T \overline{Z}_t F_t^\top \Big\} \Big\|_{\max} }
\end{align*}
and 
\begin{align*} 
 & \Big\| \Big\{ \frac{1}{T} \sum_{t=1}^T \overline{Z}_t F_t^\top \Big\}^\top \Big\{ \frac{1}{T} \sum_{t=1}^T \overline{Z}_t F_t^\top \Big\} \Big\|_{\max} \\
 & = \max_{k,k^\prime} \Big| \sum_{j=1}^p \Big\{ \frac{1}{T} \sum_{t=1}^T \Big(\frac{1}{n} \sum_{i=1}^n Z_{it,j} \Big) F_{t,k} \Big\} \Big\{ \frac{1}{T} \sum_{t=1}^T \Big(\frac{1}{n} \sum_{i=1}^n Z_{it,j} \Big) F_{t,k^\prime} \Big\} \Big| \\
 & \le p \bigg\{ \max_{k,j} \Big| \frac{1}{T} \sum_{t=1}^T \Big(\frac{1}{n} \sum_{i=1}^n Z_{it,j} \Big) F_{t,k} \Big| \bigg\}^2 \\
 & = p \cdot O_p\Big( \sqrt{\frac{\log(npT) \log p}{nT}}\Big)^2,
\end{align*}
where the last equality is by Lemma \ref{lemma:aux2}\ref{lemma:aux2:ZF}. We thus obtain that 
\[ \Big\| \frac{\overline{\bs{Z}}^\top \bs{F}}{T} \Big\| = O_p\Big( \sqrt{\frac{p \log(npT) \log p}{nT}}\Big), \]
which is statement \ref{lemma:aux4:barZF}. Next, statement \ref{lemma:aux4:barZeps} is an immediate consequence of Lemma \ref{lemma:aux2}\ref{lemma:aux2:Zeps} since 
\begin{align*}
\max_{i} \Big\| \frac{\overline{\bs{Z}}^\top \varepsilon_i}{T} \Big\| 
 & = \max_i \sqrt{ \sum_{j=1}^p \Big\{ \frac{1}{T} \sum_{t=1}^T \Big( \frac{1}{n} \sum_{i^\prime = 1}^n Z_{i^\prime t,j} \Big) \varepsilon_{it} \Big\}^2 } \\
 & \le \sqrt{p} \max_{i,j} \Big| \frac{1}{T} \sum_{t=1}^T \Big( \frac{1}{n} \sum_{i^\prime = 1}^n Z_{i^\prime t,j} \Big) \varepsilon_{it} \Big| = O_p\Big( \sqrt{\frac{p \log(npT)\log(np)}{nT}}\Big).
\end{align*}
Finally, with Lemma \ref{lemma:aux2}\ref{lemma:aux2:ZZ}, we obtain that 
\begin{align*}
\max_{i,j} \Big\| \frac{\overline{\bs{Z}}^\top Z_{i(j)}}{T} \Big\| 
 & = \max_{i,j} \sqrt{ \sum_{j^\prime=1}^p \bigg( \frac{1}{T} \sum_{t=1}^T \Big\{ \frac{1}{n} \sum_{i^\prime=1}^n Z_{i^\prime t,j^\prime} \Big\} Z_{it,j} \bigg)^2 } \\
 & \le \sqrt{p} \max_{i,j,j^\prime} \Big| \frac{1}{T} \sum_{t=1}^T \Big\{ \frac{1}{n} \sum_{i^\prime=1}^n Z_{i^\prime t,j^\prime} \Big\} Z_{it,j} \Big| \\
 & = O_p\Big( \sqrt{\frac{p \log(npT)\log(np^2)}{nT}} + \frac{\sqrt{p}}{n}\Big). \qedhere
\end{align*}

\subsection*{Proof of Lemma \ref{lemma3:projection}}

By definition, 
\[ \widehat{\bs{\Eig}} = \frac{\widehat{\bs{W}}^\top \widehat{\bs{W}}}{T} = \frac{1}{T} (\bs{F} \overline{\bs{\Gamma}}^\top \widehat{\bs{U}} + \overline{\bs{Z}} \widehat{\bs{U}})^\top (\bs{F} \overline{\bs{\Gamma}}^\top \widehat{\bs{U}} + \overline{\bs{Z}} \widehat{\bs{U}}). \]
Hence, 
\begin{align*} 
\Big\| \widehat{\bs{\Eig}} - \frac{1}{T} (\bs{F} \overline{\bs{\Gamma}}^\top \widehat{\bs{U}})^\top (\bs{F} \overline{\bs{\Gamma}}^\top \widehat{\bs{U}}) \Big\| 
 & \le 2 \Big\| \frac{1}{T} (\bs{F} \overline{\bs{\Gamma}}^\top \widehat{\bs{U}})^\top (\overline{\bs{Z}} \widehat{\bs{U}}) \Big\| + \Big\| \frac{1}{T} (\overline{\bs{Z}} \widehat{\bs{U}})^\top (\overline{\bs{Z}} \widehat{\bs{U}}) \Big\| \\
 & \le 2 \big\| \overline{\bs{\Gamma}} \big\| \Big\| \frac{\bs{F}^\top \overline{\bs{Z}}}{T} \Big\| + \Big\| \frac{\overline{\bs{Z}}^\top \overline{\bs{Z}}}{T} \Big\|,  
\end{align*} 
where we have used that $\| \widehat{\bs{U}} \| = 1$. Since $\| \overline{\bs{\Gamma}} \| \le \| \overline{\bs{\Gamma}} - \bs{\Gamma} \| + \| \bs{\Gamma} \| = O_p(\sqrt{p})$ by Lemma \ref{lemma:aux5}\ref{lemma:aux5:1} and the fact that $\| \bs{\Gamma} \| = \{ \eig_{\max}(\bs{\Gamma}^\top \bs{\Gamma}) \}^{1/2} = O(\sqrt{p})$ by assumption \ref{C:id2}, we can use Lemma \ref{lemma:aux4}\ref{lemma:aux4:barZ} and \ref{lemma:aux4:barZF} to obtain that 
\[ \Big\| \widehat{\bs{\Eig}} - \frac{1}{T} (\bs{F} \overline{\bs{\Gamma}}^\top \widehat{\bs{U}})^\top (\bs{F} \overline{\bs{\Gamma}}^\top \widehat{\bs{U}}) \Big\| = O_p\bigg( p \Big \{ \frac{\log(pT)}{n} + \sqrt{\frac{\log(npT) \log p}{nT}} \Big\} \bigg), \]
which is statement (i) of the lemma.

To prove (ii), we make use of the following bound for invertible matrices $\bs{A}$ and $\bs{B}$: Since $\bs{A}^{-1} - \bs{B}^{-1} = (\bs{A}^{-1} - \bs{B}^{-1} + \bs{B}^{-1}) (\bs{B} - \bs{A}) \bs{B}^{-1}$, it holds that  $\| \bs{A}^{-1} - \bs{B}^{-1} \| \le ( \|\bs{A}^{-1} - \bs{B}^{-1} \| + \| \bs{B}^{-1} \|) \| \bs{B} - \bs{A} \| \| \bs{B}^{-1} \|$ and thus
\[ \| \bs{A}^{-1} - \bs{B}^{-1} \| \le \frac{\| \bs{B}^{-1} \|^2 \|\bs{B} - \bs{A}\|}{1 - \| \bs{B}^{-1} \| \| \bs{B} - \bs{A} \|}, \] 
provided that $\| \bs{B}^{-1} \| \| \bs{B} - \bs{A} \| < 1$. With this bound, we obtain that 
\begin{align*}
\Big\| \widehat{\bs{\Eig}}^{-1} - \Big[ \frac{1}{T} (\bs{F} \overline{\bs{\Gamma}}^\top \widehat{\bs{U}})^\top (\bs{F} \overline{\bs{\Gamma}}^\top \widehat{\bs{U}}) \Big]^{-1} \Big\| 
 & \le \frac{\| \widehat{\bs{\Eig}}^{-1} \|^2 \|\widehat{\bs{\Eig}} - \frac{1}{T} (\bs{F} \overline{\bs{\Gamma}}^\top \widehat{\bs{U}})^\top (\bs{F} \overline{\bs{\Gamma}}^\top \widehat{\bs{U}}) \|}{1 - \| \widehat{\bs{\Eig}}^{-1} \| \|\widehat{\bs{\Eig}} - \frac{1}{T} (\bs{F} \overline{\bs{\Gamma}}^\top \widehat{\bs{U}})^\top (\bs{F} \overline{\bs{\Gamma}}^\top \widehat{\bs{U}}) \|}.
\end{align*}
Since $\widehat{\bs{\Eig}}^{-1} = \text{diag}(\widehat{\eig}_1^{-1},\ldots,\widehat{\eig}_{\widehat{K}}^{-1})$ and $\widehat{\eig}_1^{-1} \le \ldots \le \widehat{\eig}_{\widehat{K}}^{-1} \le C/p$ with probability tending to $1$ by Proposition \ref{lemma4:eigenstructure} (whose proof only makes use of Lemmas \ref{lemma:aux1}--\ref{lemma:aux4}), \KOM{[order of proofs a bit unfortunate here, maybe shift this lemma to the propositions]} we can use statement (i) to infer that 
\[ \Big\| \widehat{\bs{\Eig}}^{-1} - \Big[ \frac{1}{T} (\bs{F} \overline{\bs{\Gamma}}^\top \widehat{\bs{U}})^\top (\bs{F} \overline{\bs{\Gamma}}^\top \widehat{\bs{U}}) \Big]^{-1} \Big\| = O_p\bigg( \frac{1}{p} \Big \{ \frac{\log(pT)}{n} + \sqrt{\frac{\log(npT) \log p}{nT}} \Big\} \bigg). \]
This completes the proof of (ii).

\subsection*{Proof of Proposition \ref{lemma1:eigenstructure}}

We first verify the bound on $\| \widehat{\bs{\Sigma}} - \overline{\bs{\Sigma}} \|$. Since 
\begin{align*} 
\| \widehat{\bs{\Sigma}} - \overline{\bs{\Sigma}} \|
 & \le \Big\| \frac{1}{T} \sum_{t=1}^T \big\{ \overline{\bs{\Gamma}} F_t F_t^\top \overline{\bs{\Gamma}}^\top - \ex \overline{\bs{\Gamma}} F_t F_t^\top \overline{\bs{\Gamma}}^\top \big\} \Big\| \\
 & \quad + 2 \Big\| \overline{\bs{\Gamma}} \Big\{ \frac{1}{T} \sum_{t=1}^T F_t \overline{Z}_t^\top \Big\} \Big\| + \Big\| \frac{1}{T} \sum_{t=1}^T \big( \overline{Z}_t \overline{Z}_t^\top - \ex \overline{Z}_t \overline{Z}_t^\top \big) \Big\|,
\end{align*} 
it suffices to bound the three terms on the right-hand side. As $\| \bs{\Gamma} \| = O(\sqrt{p})$ by \ref{C:id2} and $\| \overline{\bs{\Gamma}} - \bs{\Gamma} \|= O_p(\sqrt{p \log p / n})$ by Lemma \ref{lemma:aux5}\ref{lemma:aux5:1}, we obtain that $\| \overline{\bs{\Gamma}} \| \le \| \overline{\bs{\Gamma}} - \bs{\Gamma} \| + \| \bs{\Gamma} \| = O_p(\sqrt{p})$. From this, the law of large numbers and Lemma \ref{lemma:aux5}\ref{lemma:aux5:2}, it follows that 
\begin{align*} 
 & \Big\| \frac{1}{T} \sum_{t=1}^T \big\{ \overline{\bs{\Gamma}} F_t F_t^\top \overline{\bs{\Gamma}}^\top - \ex \overline{\bs{\Gamma}} F_t F_t^\top \overline{\bs{\Gamma}}^\top \big\} \Big\|   =   \Big\| \overline{\bs{\Gamma}} \overline{\bs{\Gamma}}^\top - \ex \overline{\bs{\Gamma}} \overline{\bs{\Gamma}}^\top \Big\| = O_p \Big(  p \sqrt{\frac{\log p}{n}} \Big).
\end{align*}
 Moreover, Lemma \ref{lemma:aux4}\ref{lemma:aux4:barZF} yields that
\begin{align*} 
\Big\| \overline{\bs{\Gamma}} \Big\{ \frac{1}{T} \sum_{t=1}^T F_t \overline{Z}_t^\top \Big\} \Big\| & \le \big\| \overline{\bs{\Gamma}} \big \| \Big\| \frac{1}{T} \sum_{t=1}^T F_t \overline{Z}_t^\top \Big\| = O_p\Big( p \sqrt{\frac{\log(npT) \log p}{nT}}\Big).
\end{align*}
Finally, since $|\ex ( n^{-1} \sum_{i=1}^n Z_{it,j} ) ( n^{-1} \sum_{i=1}^n Z_{it,j^\prime} )| \le C/n$ and $\max_{j,t} | n^{-1} \sum_{i=1}^n Z_{it,j} | = O_p(\sqrt{\log(pT)/n})$ by Lem\-ma \ref{lemma:aux1}\ref{lemma:aux1:Z}, we obtain that 
\begin{align*}
\Big\| & \frac{1}{T} \sum_{t=1}^T \big( \overline{Z}_t \overline{Z}_t^\top - \ex \overline{Z}_t \overline{Z}_t^\top \big) \Big\|
 \le p \Big\| \frac{1}{T} \sum_{t=1}^T \big( \overline{Z}_t \overline{Z}_t^\top - \ex \overline{Z}_t \overline{Z}_t^\top \big) \Big\|_{\max} \\
 & = p \max_{j,j^\prime} \Big| \frac{1}{T} \sum_{t=1}^T \bigg\{ \Big( \frac{1}{n} \sum_{i=1}^n Z_{it,j} \Big) \Big( \frac{1}{n} \sum_{i=1}^n Z_{it,j^\prime} \Big) - \ex \Big( \frac{1}{n} \sum_{i=1}^n Z_{it,j} \Big) \Big( \frac{1}{n} \sum_{i=1}^n Z_{it,j^\prime} \Big) \bigg\} \Big| \\
 & \le p \bigg\{ \Big( \max_{j,t} \Big| \frac{1}{n} \sum_{i=1}^n Z_{it,j} \Big| \Big)^2 + \frac{C}{n} \bigg\} = O_p\Big( \frac{p \log(pT)}{n} \Big). 
\end{align*} 
Putting everything together, we arrive at the bound $\| \widehat{\bs{\Sigma}} - \overline{\bs{\Sigma}} \| = O_p( p \sqrt{\log p/n})$, which is the first statement of the proposition.

We next turn to the bound on $\| \overline{\bs{\Sigma}} - \bs{\Sigma} \|$. Since $\overline{Z}_t$ and $\overline{\bs{\Gamma}}$ are independent from each other, it holds that 
\begin{align*}
\overline{\bs{\Sigma}} - \bs{\Sigma} 
 & = \ex \big[ (\overline{\bs{\Gamma}} - \bs{\Gamma})  (\overline{\bs{\Gamma}} - \bs{\Gamma})^\top \big] +  \frac{1}{T} \sum_{t=1}^T \ex \big[ \overline{Z}_t \overline{Z}_t^\top \big].
\end{align*}
The entries of the $p \times p$ matrix $(\overline{\bs{\Gamma}} - \bs{\Gamma}) (\overline{\bs{\Gamma}} - \bs{\Gamma})^\top$ are given by
\[ \sum_{k=1}^K \Big\{ \frac{1}{n} \sum_{i=1}^n \big( \Gamma_{i,jk} - \Gamma_{jk} \big) \Big\}  \Big\{ \frac{1}{n} \sum_{i=1}^n \big( \Gamma_{i,j^\prime k} - \Gamma_{j^\prime k} \big) \Big\}, \]
where $\Gamma_{i,jk}$ and $\Gamma_{jk}$ denote the elements of the matrices $\bs{\Gamma}_i$ and $\bs{\Gamma}$, respectively. Since the variables $\Gamma_{i,jk} - \Gamma_{jk}$ have mean zero and are independent across $i$, we can infer that 
\begin{align*} 
\Big\| & \ex \big[ (\overline{\bs{\Gamma}} - \bs{\Gamma})  (\overline{\bs{\Gamma}} - \bs{\Gamma})^\top \big] \Big\| 
 \le p \Big\| \ex \big[ (\overline{\bs{\Gamma}} - \bs{\Gamma}) (\overline{\bs{\Gamma}} - \bs{\Gamma})^\top \big] \Big\|_{\max} \\
 & = p \max_{j,j^\prime} \Big| \ex \sum_{k=1}^K \Big\{ \frac{1}{n} \sum_{i=1}^n \big( \Gamma_{i,jk} - \Gamma_{jk} \big) \Big\}  \Big\{ \frac{1}{n} \sum_{i=1}^n \big( \Gamma_{i,j^\prime k} - \Gamma_{j^\prime k} \big) \Big\} \Big| \\
 & \le p K \max_{j,j^\prime,k} \Big| \ex \Big\{ \frac{1}{n} \sum_{i=1}^n \big( \Gamma_{i,jk} - \Gamma_{jk} \big) \Big\} \Big\{ \frac{1}{n} \sum_{i=1}^n \big( \Gamma_{i,j^\prime k} - \Gamma_{j^\prime k} \big) \Big\} \Big| \le \frac{C p}{n}.
\end{align*} 
Similarly, we obtain that  
\begin{align*}
\Big\| \frac{1}{T} \sum_{t=1}^T \ex \big[ \overline{Z}_t \overline{Z}_t^\top \big] \Big\| 
 & \le p \Big\| \frac{1}{T} \sum_{t=1}^T \ex \big[ \overline{Z}_t \overline{Z}_t^\top \big] \Big\|_{\max} \\
 & \le p \max_{j,j^\prime,t} \Big| \ex \Big(\frac{1}{n} \sum_{i=1}^n Z_{it,j} \Big) \Big(\frac{1}{n} \sum_{i=1}^n Z_{it,j^\prime} \Big) \Big| \le  \frac{C p}{n}. 
\end{align*}
Taken together, these computations show that $\| \overline{\bs{\Sigma}} - \bs{\Sigma} \| = O_p( p/n)$, which is the second statement of the proposition.

\subsection*{Proof of Proposition \ref{lemma2:projection}}

Let $\widehat{\bs{U}}_{1:K} = (\widehat{U}_1 \ldots \widehat{U}_K)$. As a first preliminary step, we prove that 
\begin{equation}\label{eq:lemma1:projection:prelim}
\big\| (\overline{\bs{\Gamma}}^\top \widehat{\bs{U}}_{1:K})^\top (\overline{\bs{\Gamma}}^\top \widehat{\bs{U}}_{1:K}) - \widehat{\bs{U}}_{1:K}^\top \widehat{\bs{\Sigma}} \widehat{\bs{U}}_{1:K} \big\| = O_p\Big(p \sqrt{\frac{\log p}{n}}\Big).
\end{equation}
To do so, we use the following facts: 
\begin{enumerate}[label=(\alph*),leftmargin=0.7cm]

\item Since $\widehat{\bs{U}}_{1:K}^\top \widehat{\bs{U}}_{1:K} = \bs{I}_K$, it holds that $\| \widehat{\bs{U}}_{1:K} \| = 1$. 

\item From Lemma \ref{lemma:aux5}\ref{lemma:aux5:1} and the fact that $\| \bs{\Gamma} \| = O(\sqrt{p})$, it follows that   
\begin{align*} 
\big\| \overline{\bs{\Gamma}} \overline{\bs{\Gamma}}^\top - \bs{\Gamma} \bs{\Gamma}^\top \big\|
 & \le \big\| (\overline{\bs{\Gamma}} - \bs{\Gamma}) (\overline{\bs{\Gamma}} - \bs{\Gamma})^\top \big\| + 2 \big\| (\overline{\bs{\Gamma}} - \bs{\Gamma}) \bs{\Gamma}^\top \big\| \\
 & \le \big\| \overline{\bs{\Gamma}} - \bs{\Gamma} \big\|^2 + 2 \big\| \overline{\bs{\Gamma}} - \bs{\Gamma} \big\| \big\| \bs{\Gamma} \big\| = O_p\Big(p \sqrt{\frac{\log p}{n}}\Big).
\end{align*} 

\item By Proposition \ref{lemma1:eigenstructure},
\[ \big\| \bs{\Sigma} - \widehat{\bs{\Sigma}} \big\| = O_p\Big( p \sqrt{\frac{\log p}{n}}\Big). \]

\end{enumerate}
Using (a)--(c) along with the identity $\bs{\Sigma} = \bs{\Gamma} \bs{\Gamma}^\top$, we can conclude that 
\begin{align*}
\big\| (\overline{\bs{\Gamma}}^\top \widehat{\bs{U}}_{1:K})^\top (\overline{\bs{\Gamma}}^\top \widehat{\bs{U}}_{1:K}) - \widehat{\bs{U}}_{1:K}^\top \widehat{\bs{\Sigma}} \widehat{\bs{U}}_{1:K} \big\| 
 & \le \big\| \widehat{\bs{U}}_{1:K} \big\|^2 \big\| \overline{\bs{\Gamma}} \overline{\bs{\Gamma}}^\top - \widehat{\bs{\Sigma}} \big\| \\
 & \le \big\| \overline{\bs{\Gamma}} \overline{\bs{\Gamma}}^\top - \bs{\Gamma} \bs{\Gamma}^\top \big\| + \big\| \bs{\Sigma} - \widehat{\bs{\Sigma}} \big\| \\
 & = O_p\Big( p \sqrt{\frac{\log p}{n}}\Big),
\end{align*} 
which is the statement of \eqref{eq:lemma1:projection:prelim}.

Now let $\widetilde{\eig}_1 \ge \ldots \ge \widetilde{\eig}_K$ be the eigenvalues of $(\overline{\bs{\Gamma}}^\top \widehat{\bs{U}}_{1:K})^\top (\overline{\bs{\Gamma}}^\top \widehat{\bs{U}}_{1:K})$. The eigenvalues of $\widehat{\bs{U}}_{1:K}^\top \widehat{\bs{\Sigma}} \widehat{\bs{U}}_{1:K}$ are identical to the $K$ largest eigenvalues $\widehat{\eig}_1 \ge \ldots \ge \widehat{\eig}_K$ of the matrix $\widehat{\bs{\Sigma}}$, since the columns of $\widehat{\bs{U}}_{1:K}$ are the first $K$ eigenvectors of $\widehat{\bs{\Sigma}}$ and thus
$\widehat{\bs{U}}_{1:K}^\top \widehat{\bs{\Sigma}} \widehat{\bs{U}}_{1:K} = \text{diag}(\widehat{\eig}_1,\ldots,\widehat{\eig}_K)$. 
According to Proposition \ref{lemma4:eigenstructure}, it holds that 
\begin{equation}\label{eq:lemma1:projection:intermediate1}
\widehat{\eig}_1 \ge \ldots \ge \widehat{\eig}_K \ge c_0 p 
\end{equation}
with probability tending to $1$. Moreover, by Weyl's theorem and \eqref{eq:lemma1:projection:prelim},  
\begin{align}
|\widetilde{\eig}_k - \widehat{\eig}_k| 
 & \le \big\| (\overline{\bs{\Gamma}}^\top \widehat{\bs{U}}_{1:K})^\top (\overline{\bs{\Gamma}}^\top \widehat{\bs{U}}_{1:K}) - \widehat{\bs{U}}_{1:K}^\top \widehat{\bs{\Sigma}} \widehat{\bs{U}}_{1:K} \big\| \nonumber \\
 & = O_p\Big( p \sqrt{\frac{\log p}{n}}\Big) = o_p(p) \label{eq:lemma1:projection:intermediate2}
\end{align}
for any $k$. Taken together, \eqref{eq:lemma1:projection:intermediate1} and \eqref{eq:lemma1:projection:intermediate2} immediately yield that  
\[ \widetilde{\eig}_1 \ge \ldots \ge \widetilde{\eig}_K \ge c p \]
for some sufficiently small constant $c > 0$ with probability tending to $1$. This in particular implies that the matrix $(\overline{\bs{\Gamma}}^\top \widehat{\bs{U}}_{1:K})^\top (\overline{\bs{\Gamma}}^\top \widehat{\bs{U}}_{1:K})$, and thus the matrix $\overline{\bs{\Gamma}}^\top \widehat{\bs{U}}_{1:K}$, is invertible with probability approaching $1$.

Finally, since $\widehat{K} = K$ with probability tending to $1$, it holds that $\overline{\bs{\Gamma}}^\top \widehat{\bs{U}}_{1:K} = \overline{\bs{\Gamma}}^\top \widehat{\bs{U}}$ with probability tending to $1$. We can thus conclude that $\overline{\bs{\Gamma}}^\top \widehat{\bs{U}}$ is invertible with probability approaching $1$, which immediately implies that $\widehat{\bs{\Pi}} = \bs{\Pi} -\bs{R}$ with probability approaching $1$.

\subsection*{Proof of Proposition \ref{lemma1:Lasso}}

Suppose we are on the event $\mathcal{T}_\pen$. By the basic inequality for the lasso, it holds that 
\[ \frac{1}{nT} \| \widehat{\bs{X}} (\widehat{\beta}_\pen - \beta) \|^2 \le \frac{2 \| \widehat{\bs{X}}^\top e \|_\infty}{nT} \| \widehat{\beta}_\pen - \beta \|_1 + \pen \| \beta \|_1 - \pen \| \widehat{\beta}_\pen \|_1. \]
From this, it follows that
\begin{equation}\label{eq:Lasso-ineq-RE}
\frac{2}{nT} \| \widehat{\bs{X}} (\widehat{\beta}_\pen - \beta) \|^2 \le 3 \pen \| \widehat{\beta}_{\pen,S} - \beta_S \|_1 - \pen \| \widehat{\beta}_{\pen,S^c} \|_1, 
\end{equation}
which in turn implies that the approximation error $\delta = \widehat{\beta}_{\pen} - \beta$ of the lasso is such that $3 \| \delta_S \|_1 \ge \| \delta_{S^c} \|_1$. 
With \eqref{eq:Lasso-ineq-RE}, we obtain that  
\begin{align}
 & \frac{2}{nT} \| \widehat{\bs{X}} (\widehat{\beta}_\pen - \beta) \|^2 + \pen \| \widehat{\beta}_\pen - \beta \|_1 \nonumber \\*
 & = \frac{2}{nT} \| \widehat{\bs{X}} (\widehat{\beta}_\pen - \beta) \|^2 + \pen \| \widehat{\beta}_{\pen,S} - \beta_S \|_1 + \pen \| \widehat{\beta}_{\pen,S^c} \|_1 \nonumber \\
 & \le 3 \pen \| \widehat{\beta}_{\pen,S} - \beta_S \|_1 - \pen \| \widehat{\beta}_{\pen,S^c} \|_1 + \pen \| \widehat{\beta}_{\pen,S} - \beta_S \|_1 + \pen \| \widehat{\beta}_{\pen,S^c} \|_1 \nonumber \\*
 & \le 4 \pen \| \widehat{\beta}_{\pen,S} - \beta_S \|_1. \label{eq:lemma1:Lasso:star1}
\end{align}
Moreover, since   
\begin{align*}
\| \widehat{\beta}_{\pen,S} - \beta_S \|_1^2 
 & \le \frac{\| \widehat{\bs{X}} (\widehat{\beta}_{\pen} - \beta) \|^2}{nT} \frac{s}{\phi^2}
\end{align*}
on the event $\mathcal{T}_{\text{RE}}$, it holds that 
\begin{align}
4 \pen \| \widehat{\beta}_{\pen,S} - \beta_S \|_1 
 & \le \frac{4 \pen}{\phi} \sqrt{\frac{s}{nT}} \| \widehat{\bs{X}} (\widehat{\beta}_{\pen} - \beta) \| 
   \le \frac{1}{nT} \| \widehat{\bs{X}} (\widehat{\beta}_{\pen} - \beta) \|^2 + \frac{4 \pen^2s}{\phi^2}, \label{eq:lemma1:Lasso:star2} 
\end{align}
where the last inequality uses that $4ab \le b^2 + 4a^2$. Plugging \eqref{eq:lemma1:Lasso:star2} into \eqref{eq:lemma1:Lasso:star1}, we arrive at  
\[  \frac{1}{nT} \| \widehat{\bs{X}} (\widehat{\beta}_\pen - \beta) \|^2 + \pen \| \widehat{\beta}_\pen - \beta \|_1 \le \frac{4}{\phi^2} \pen^2s, \]
which immediately implies the claim.

\subsection*{Proof of Proposition \ref{lemma2:Lasso}}

It holds that 
\[ \frac{\| \widehat{\bs{X}}^\top e \|_\infty}{nT} = \frac{1}{nT} \max_{1 \le j \le p} \Big| \sum_{i=1}^n \widehat{X}_{i(j)}^\top e_i \Big| \]
with $\widehat{X}_{i(j)} = \widehat{\bs{\Pi}} X_{i(j)}$ and $X_{i(j)} = \bs{F} \Gamma_{i,j} + Z_{i(j)}$, where $\Gamma_{i,j}$ is the $j$-th row of the matrix $\bs{\Gamma}_i$ and $X_{i(j)}$ is the $j$-th column of the matrix $\bs{X}_i$. Moreover, since 
\[ \sum_{i=1}^n \widehat{X}_{i(j)}^\top e_i = \sum_{i=1}^n \big\{ \widehat{\bs{\Pi}} X_{i(j)} \big\}^\top \big\{ \widehat{\bs{\Pi}} e_i \big\} = \sum_{i=1}^n \big\{ \widehat{\bs{\Pi}} (\bs{F} \Gamma_{i,j} + Z_{i(j)}) \big\}^\top \big\{ \widehat{\bs{\Pi}} (\bs{F} \gamma_i + \varepsilon_i) \big\}, \]
we have that
\begin{align*}
\frac{\| \widehat{\bs{X}}^\top e \|_\infty}{nT} \nonumber 
 & \le \frac{1}{nT} \max_{1 \le j \le p} \Big| \sum_{i=1}^n \big\{ \widehat{\bs{\Pi}} \bs{F} \Gamma_{i,j} \big\}^\top \big\{ \widehat{\bs{\Pi}} \bs{F} \gamma_i \big\} \Big| \\
 & \quad + \frac{1}{nT} \max_{1 \le j \le p} \Big| \sum_{i=1}^n \big\{ \widehat{\bs{\Pi}} \bs{F} \Gamma_{i,j} \big\}^\top \big\{ \widehat{\bs{\Pi}} \varepsilon_i \big\} \Big| \\
 & \quad + \frac{1}{nT} \max_{1 \le j \le p} \Big| \sum_{i=1}^n \big\{ \widehat{\bs{\Pi}} Z_{i(j)} \big\}^\top \big\{ \widehat{\bs{\Pi}} \bs{F} \gamma_i \big\} \Big| \\
 & \quad + \frac{1}{nT} \max_{1 \le j \le p} \Big| \sum_{i=1}^n \big\{ \widehat{\bs{\Pi}} Z_{i(j)} \big\}^\top \big\{ \widehat{\bs{\Pi}} \varepsilon_i \big\} \Big|. 
\end{align*}
We now bound the four terms on the right-hand side one after the other. In particular, we prove that
\begin{align}
\frac{1}{nT} \max_{1 \le j \le p} \Big| \sum_{i=1}^n \big\{ \widehat{\bs{\Pi}} \bs{F} \Gamma_{i,j} \big\}^\top \big\{ \widehat{\bs{\Pi}} \bs{F} \gamma_i \big\} \Big| & = O_p\Big( \frac{\log(pT)}{n} \Big) \label{lemma2:Lasso:claim1} \\
\frac{1}{nT} \max_{1 \le j \le p} \Big| \sum_{i=1}^n \big\{ \widehat{\bs{\Pi}} \bs{F} \Gamma_{i,j} \big\}^\top \big\{ \widehat{\bs{\Pi}} \varepsilon_i \big\} \Big| & = O_p\Big( \sqrt{\frac{\log(npT)\log(np)}{nT}}\Big) \label{lemma2:Lasso:claim2} \\
\frac{1}{nT} \max_{1 \le j \le p} \Big| \sum_{i=1}^n \big\{ \widehat{\bs{\Pi}} Z_{i(j)} \big\}^\top \big\{ \widehat{\bs{\Pi}} \bs{F} \gamma_i \big\} \Big| & = O_p\Big( \sqrt{\frac{\log(npT)\log(np^2)}{nT}} + \frac{1}{n}\Big) \label{lemma2:Lasso:claim3} \\
\frac{1}{nT} \max_{1 \le j \le p} \Big| \sum_{i=1}^n \big\{ \widehat{\bs{\Pi}} Z_{i(j)} \big\}^\top \big\{ \widehat{\bs{\Pi}} \varepsilon_i \big\} \Big| & = O_p\Big( \sqrt{\frac{\log p}{nT}} \Big). \label{lemma2:Lasso:claim4} 
\end{align}
Proposition \ref{lemma2:Lasso} is a direct consequence of these four statements.

\begin{proof}[Proof of \eqref{lemma2:Lasso:claim1}]
In this and the following proofs, we repeatedly use that $\| \widehat{\bs{U}} \| = 1$, $\| \bs{\Gamma} \| = O(\sqrt{p})$ by \ref{C:id2} and $\| \widehat{\bs{\Eig}}^{-1} \| = O_p(p^{-1})$ by Proposition \ref{lemma4:eigenstructure}.
As a first preliminary step, we derive a bound on the term $\| \widehat{\bs{\Pi}} \bs{F} \|$. Since $\widehat{\bs{\Pi}} = \bs{\Pi} - \widehat{\bs{R}}$ with probability tending to $1$ by Proposition \ref{lemma2:projection}, it holds that 
\begin{align*} 
\| \widehat{\bs{\Pi}} \bs{F} \|
 & \le \Big\| \frac{1}{T} (\bs{F} \overline{\bs{\Gamma}}^\top \widehat{\bs{U}}) \bigg\{ \widehat{\bs{\Eig}}^{-1} - \Big[ \frac{1}{T} (\bs{F} \overline{\bs{\Gamma}}^\top \widehat{\bs{U}})^\top (\bs{F} \overline{\bs{\Gamma}}^\top \widehat{\bs{U}}) \Big]^{-1} \bigg\} (\bs{F} \overline{\bs{\Gamma}}^\top \widehat{\bs{U}})^\top \bs{F} \Big\| \\*
 & \quad + \Big\| \frac{1}{T} (\bs{F} \overline{\bs{\Gamma}}^\top \widehat{\bs{U}}) \widehat{\bs{\Eig}}^{-1} (\overline{\bs{Z}} \widehat{\bs{U}})^\top \bs{F} \Big\| 
   + \Big\| \frac{1}{T} (\overline{\bs{Z}} \widehat{\bs{U}}) \widehat{\bs{\Eig}}^{-1} (\bs{F} \overline{\bs{\Gamma}}^\top \widehat{\bs{U}})^\top \bs{F} \Big\| \\
 & \quad + \Big\| \frac{1}{T} (\overline{\bs{Z}} \widehat{\bs{U}}) \widehat{\bs{\Eig}}^{-1} (\overline{\bs{Z}} \widehat{\bs{U}})^\top \bs{F} \Big\| 
\end{align*} 
with probability tending to $1$, where
\begin{align*}
 & \Big\| \frac{1}{T} (\bs{F} \overline{\bs{\Gamma}}^\top \widehat{\bs{U}}) \bigg\{ \widehat{\bs{\Eig}}^{-1} - \Big[ \frac{1}{T} (\bs{F} \overline{\bs{\Gamma}}^\top \widehat{\bs{U}})^\top (\bs{F} \overline{\bs{\Gamma}}^\top \widehat{\bs{U}}) \Big]^{-1} \bigg\} (\bs{F} \overline{\bs{\Gamma}}^\top \widehat{\bs{U}})^\top \bs{F} \Big\| \\
 & \le \| \bs{F} \| \| \overline{\bs{\Gamma}} \|^2 \Big\| \widehat{\bs{\Eig}}^{-1} - \Big[ \frac{1}{T} (\bs{F} \overline{\bs{\Gamma}}^\top \widehat{\bs{U}})^\top (\bs{F} \overline{\bs{\Gamma}}^\top \widehat{\bs{U}}) \Big]^{-1} \Big\| \Big\| \frac{\bs{F}^\top \bs{F}}{T} \Big\| \\
 & = O_p \Big( \frac{\sqrt{T} \log(pT)}{n} + \sqrt{\frac{\log(npT) \log p}{n}} \Big)
\end{align*} 
by Lemmas \ref{lemma:aux5}\ref{lemma:aux5:1} and \ref{lemma3:projection}\ref{lemma3:projection:ii},
\begin{align*}
\Big\| \frac{1}{T} (\bs{F} \overline{\bs{\Gamma}}^\top \widehat{\bs{U}}) \widehat{\bs{\Eig}}^{-1} (\overline{\bs{Z}} \widehat{\bs{U}})^\top \bs{F} \Big\| 
 & \le \| \bs{F} \| \| \overline{\bs{\Gamma}} \| \big\| \widehat{\bs{\Eig}}^{-1} \big\| \Big\| \frac{\overline{\bs{Z}}^\top \bs{F}}{T} \Big\| \\
 & = O_p\Big( \sqrt{\frac{\log(npT) \log p}{n}}\Big)
\end{align*}
by Lemmas \ref{lemma:aux5}\ref{lemma:aux5:1} and \ref{lemma:aux4}\ref{lemma:aux4:barZF}, 
\begin{align*}
\Big\| \frac{1}{T} (\overline{\bs{Z}} \widehat{\bs{U}}) \widehat{\bs{\Eig}}^{-1} (\bs{F} \overline{\bs{\Gamma}}^\top \widehat{\bs{U}})^\top \bs{F} \Big\|
 & \le \| \overline{\bs{Z}} \| \big\| \widehat{\bs{\Eig}}^{-1} \big\| \| \overline{\bs{\Gamma}} \| \Big\| \frac{\bs{F}^\top \bs{F}}{T} \Big\| 
   = O_p\Big(\sqrt{\frac{T\log(pT)}{n}}\Big)
\end{align*}
by Lemmas \ref{lemma:aux5}\ref{lemma:aux5:1} and \ref{lemma:aux4}\ref{lemma:aux4:barZ}, and
\begin{align*}
\Big\| \frac{1}{T} (\overline{\bs{Z}} \widehat{\bs{U}}) \widehat{\bs{\Eig}}^{-1} (\overline{\bs{Z}} \widehat{\bs{U}})^\top \bs{F} \Big\|
 & \le \| \overline{\bs{Z}} \| \big \| \widehat{\bs{\Eig}}^{-1} \big \| \Big\| \frac{\overline{\bs{Z}}^\top \bs{F}}{T} \Big\| \\
 & = O_p\Big( \frac{\sqrt{\log(npT) \log(pT) \log p}}{n}\Big) 
\end{align*}
by Lemma \ref{lemma:aux4}\ref{lemma:aux4:barZ} and \ref{lemma:aux4:barZF}. As a result, we obtain that 
\begin{equation}\label{eq:lemma2:Lasso:claim1:1}
\| \widehat{\bs{\Pi}} \bs{F} \| = O_p\Big( \sqrt{\frac{T\log(pT)}{n}} \Big). 
\end{equation} 
Moreover, by arguments analogous to those for Lemma \ref{lemma:aux1}\ref{lemma:aux1:Z}, 
\[ \max_{1 \le j \le p} \Big\{ \frac{1}{n} \sum_{i=1}^n \| \Gamma_{i,j} \| \| \gamma_i \| - \ex \| \Gamma_{i,j} \| \| \gamma_i \| \Big\} = O_p\Big( \sqrt{\frac{\log p}{n}} \Big), \]
which implies that 
\begin{equation}\label{eq:lemma2:Lasso:claim1:2}
\max_{1 \le j \le p} \Big\{ \frac{1}{n} \sum_{i=1}^n \| \Gamma_{i,j} \| \| \gamma_i \| \Big\} = O_p(1) 
\end{equation}
under the conditions of \ref{C:loadings}. With \eqref{eq:lemma2:Lasso:claim1:1} and \eqref{eq:lemma2:Lasso:claim1:2}, we can conclude that 
\begin{align*} 
\frac{1}{nT} \max_{1 \le j \le p} \Big| \sum_{i=1}^n \big\{ \widehat{\bs{\Pi}} \bs{F} \Gamma_{i,j} \big\}^\top \big\{ \widehat{\bs{\Pi}} \bs{F} \gamma_i \big\} \Big|
 & \le \frac{1}{nT} \max_{1 \le j \le p} \sum_{i=1}^n \| \Gamma_{i,j} \| \| \widehat{\bs{\Pi}} \bs{F} \|^2 \| \gamma_i \| \\
 & = \frac{\| \widehat{\bs{\Pi}} \bs{F} \|^2}{T} \max_{1 \le j \le p} \Big\{ \frac{1}{n} \sum_{i=1}^n \| \Gamma_{i,j} \| \| \gamma_i \| \Big\} \\
 & = O_p\Big( \frac{\log(pT)}{n} \Big). \qedhere
\end{align*}
\end{proof}

\begin{proof}[Proof of \eqref{lemma2:Lasso:claim2}]
It holds that 
\begin{align*} 
 & \max_{1 \le i \le n} \| \bs{F}^\top \widehat{\bs{R}}^\top \varepsilon_i \| \\*
 & \le \max_{1 \le i \le n} \Big\| \frac{1}{T} \bs{F}^\top (\bs{F} \overline{\bs{\Gamma}}^\top \widehat{\bs{U}}) \bigg\{ \widehat{\bs{\Eig}}^{-1} - \Big[ \frac{1}{T} (\bs{F} \overline{\bs{\Gamma}}^\top \widehat{\bs{U}})^\top (\bs{F} \overline{\bs{\Gamma}}^\top \widehat{\bs{U}}) \Big]^{-1} \bigg\} (\bs{F} \overline{\bs{\Gamma}}^\top \widehat{\bs{U}})^\top \varepsilon_i \Big\| \\*
 & \quad + \max_{1 \le i \le n} \Big\| \frac{1}{T} \bs{F}^\top (\bs{F} \overline{\bs{\Gamma}}^\top \widehat{\bs{U}}) \widehat{\bs{\Eig}}^{-1} (\overline{\bs{Z}} \widehat{\bs{U}})^\top \varepsilon_i \Big\| 
         + \max_{1 \le i \le n} \Big\| \frac{1}{T} \bs{F}^\top (\overline{\bs{Z}} \widehat{\bs{U}}) \widehat{\bs{\Eig}}^{-1} (\bs{F} \overline{\bs{\Gamma}}^\top \widehat{\bs{U}})^\top \varepsilon_i \Big\| \\
 & \quad + \max_{1 \le i \le n} \Big\| \frac{1}{T} \bs{F}^\top (\overline{\bs{Z}} \widehat{\bs{U}}) \widehat{\bs{\Eig}}^{-1} (\overline{\bs{Z}} \widehat{\bs{U}})^\top \varepsilon_i \Big\|, 
\end{align*} 
where
\begin{align*}
 & \max_{1 \le i \le n} \Big\| \frac{1}{T} \bs{F}^\top (\bs{F} \overline{\bs{\Gamma}}^\top \widehat{\bs{U}}) \bigg\{ \widehat{\bs{\Eig}}^{-1} - \Big[ \frac{1}{T} (\bs{F} \overline{\bs{\Gamma}}^\top \widehat{\bs{U}})^\top (\bs{F} \overline{\bs{\Gamma}}^\top \widehat{\bs{U}}) \Big]^{-1} \bigg\} (\bs{F} \overline{\bs{\Gamma}}^\top \widehat{\bs{U}})^\top \varepsilon_i \Big\| \\
 & \le \Big\| \frac{1}{T} \bs{F}^\top \bs{F} \Big\| \| \overline{\bs{\Gamma}} \|^2 \Big\| \widehat{\bs{\Eig}}^{-1} - \Big[ \frac{1}{T} (\bs{F} \overline{\bs{\Gamma}}^\top \widehat{\bs{U}})^\top (\bs{F} \overline{\bs{\Gamma}}^\top \widehat{\bs{U}}) \Big]^{-1} \Big\| \max_{1 \le i \le n} \| \bs{F}^\top \varepsilon_i \| \\
 & = O_p\Big( \frac{\sqrt{T} \sqrt{\log n} \log(pT)}{n} + \sqrt{\frac{\log(npT) \log n \log p}{n}} \Big)
\end{align*}
by Lemmas \ref{lemma:aux5}\ref{lemma:aux5:1},  \ref{lemma:aux3}\ref{lemma:aux3:Feps} and \ref{lemma3:projection}\ref{lemma3:projection:ii},
\begin{align*}
\max_{1 \le i \le n} \Big\| \frac{1}{T} \bs{F}^\top (\bs{F} \overline{\bs{\Gamma}}^\top \widehat{\bs{U}}) \widehat{\bs{\Eig}}^{-1} (\overline{\bs{Z}} \widehat{\bs{U}})^\top \varepsilon_i \Big\|
 & \le \Big\| \frac{\bs{F}^\top \bs{F}}{T} \Big\| \| \overline{\bs{\Gamma}} \| \big\| \widehat{\bs{\Eig}}^{-1} \big\| \max_{1 \le i \le n} \| \overline{\bs{Z}}^\top \varepsilon_i \| \\
 & = O_p\Big( \sqrt{\frac{T \log(npT)\log(np)}{n}}\Big)
\end{align*}
by Lemmas \ref{lemma:aux5}\ref{lemma:aux5:1} and \ref{lemma:aux4}\ref{lemma:aux4:barZeps}, 
\begin{align*}
\max_{1 \le i \le n} \Big\| \frac{1}{T} \bs{F}^\top (\overline{\bs{Z}} \widehat{\bs{U}}) \widehat{\bs{\Eig}}^{-1} (\bs{F} \overline{\bs{\Gamma}}^\top \widehat{\bs{U}})^\top \varepsilon_i \Big\|
 & \le \Big\| \frac{\bs{F}^\top \overline{\bs{Z}}}{T} \Big\| \big\| \widehat{\bs{\Eig}}^{-1} \big\| \| \overline{\bs{\Gamma}} \| \max_{1 \le i \le n} \| \bs{F}^\top \varepsilon_i \| \\
 & = O_p\Big( \sqrt{\frac{\log(npT) \log n \log p}{n}}\Big)
\end{align*}
by Lemmas \ref{lemma:aux5}\ref{lemma:aux5:1}, \ref{lemma:aux3}\ref{lemma:aux3:Feps} and \ref{lemma:aux4}\ref{lemma:aux4:barZF}, and
\begin{align*}
\max_{1 \le i \le n} \Big\| \frac{1}{T} \bs{F}^\top (\overline{\bs{Z}} \widehat{\bs{U}}) \widehat{\bs{\Eig}}^{-1} (\overline{\bs{Z}} \widehat{\bs{U}})^\top \varepsilon_i \Big\|
 & \le \Big\| \frac{\bs{F}^\top \overline{\bs{Z}}}{T} \Big\| \big\| \widehat{\bs{\Eig}}^{-1} \big\| \max_{1 \le i \le n} \| \overline{\bs{Z}}^\top \varepsilon_i \| \\
 & = O_p\Big( \frac{\log(npT)\sqrt{\log p \log(np)}}{n}\Big)
\end{align*}
by Lemma \ref{lemma:aux4}\ref{lemma:aux4:barZF} and \ref{lemma:aux4:barZeps}. Consequently, we obtain that 
\[ \max_{1 \le i \le n} \| \bs{F}^\top \widehat{\bs{R}}^\top \varepsilon_i \| = O_p\Big( \sqrt{\frac{T \log(npT)\log(np)}{n}}\Big). \]
With this and the fact that $\max_{j} \{n^{-1} \sum_{i=1}^n \| \Gamma_{i,j} \|\} = O_p(1)$, which can be verified analogously as \eqref{eq:lemma2:Lasso:claim1:2}, we can conclude that 
\begin{align*}
 & \frac{1}{nT} \max_{1 \le j \le p} \Big| \sum_{i=1}^n \big\{ \widehat{\bs{\Pi}} \bs{F} \Gamma_{i,j} \big\}^\top \big\{ \widehat{\bs{\Pi}} \varepsilon_i \big\} \Big| \\
 & = \frac{1}{nT} \max_{1 \le j \le p} \Big| \sum_{i=1}^n \Gamma_{i,j}^\top \big\{ \widehat{\bs{\Pi}} \bs{F} \big\}^\top \varepsilon_i \Big| \\
 & = \frac{1}{nT} \max_{1 \le j \le p} \Big| \sum_{i=1}^n \Gamma_{i,j}^\top \big\{ -\widehat{\bs{R}} \bs{F} \big\}^\top \varepsilon_i \Big| \quad \text{(w.p.} \to 1) \\
 & \le \max_{1 \le j \le p} \Big\{ \frac{1}{n} \sum_{i=1}^n \| \Gamma_{i,j} \| \Big\} \Big\{ \frac{1}{T} \max_{1 \le i \le n} \| \bs{F}^\top \widehat{\bs{R}}^\top \varepsilon_i \| \Big\} \\*
 & = O_p\Big( \sqrt{\frac{\log(npT)\log(np)}{nT}}\Big). \qedhere
\end{align*}
\end{proof}

\begin{proof}[Proof of \eqref{lemma2:Lasso:claim3}]
It holds that 
\begin{align*} 
 & \max_{i,j} \| Z_{i(j)}^\top \widehat{\bs{R}} \bs{F} \| \\
 & \le \max_{i,j} \Big\| \frac{1}{T} Z_{i(j)}^\top (\bs{F} \overline{\bs{\Gamma}}^\top \widehat{\bs{U}}) \bigg\{ \widehat{\bs{\Eig}}^{-1} - \Big[ \frac{1}{T} (\bs{F} \overline{\bs{\Gamma}}^\top \widehat{\bs{U}})^\top (\bs{F} \overline{\bs{\Gamma}}^\top \widehat{\bs{U}}) \Big]^{-1} \bigg\} (\bs{F} \overline{\bs{\Gamma}}^\top \widehat{\bs{U}})^\top \bs{F} \Big\| \\*
 & \quad + \max_{i,j} \Big\| \frac{1}{T} Z_{i(j)}^\top (\bs{F} \overline{\bs{\Gamma}}^\top \widehat{\bs{U}}) \widehat{\bs{\Eig}}^{-1} (\overline{\bs{Z}} \widehat{\bs{U}})^\top \bs{F} \Big\| 
 + \max_{i,j} \Big\| \frac{1}{T} Z_{i(j)}^\top (\overline{\bs{Z}} \widehat{\bs{U}}) \widehat{\bs{\Eig}}^{-1} (\bs{F} \overline{\bs{\Gamma}}^\top \widehat{\bs{U}})^\top \bs{F} \Big\| \\
 & \quad + \max_{i,j} \Big\| \frac{1}{T} Z_{i(j)}^\top (\overline{\bs{Z}} \widehat{\bs{U}}) \widehat{\bs{\Eig}}^{-1} (\overline{\bs{Z}} \widehat{\bs{U}})^\top \bs{F} \Big\|, 
\end{align*}
where
\begin{align*}
 & \max_{i,j} \Big\| \frac{1}{T} Z_{i(j)}^\top (\bs{F} \overline{\bs{\Gamma}}^\top \widehat{\bs{U}}) \bigg\{ \widehat{\bs{\Eig}}^{-1} - \Big[ \frac{1}{T} (\bs{F} \overline{\bs{\Gamma}}^\top \widehat{\bs{U}})^\top (\bs{F} \overline{\bs{\Gamma}}^\top \widehat{\bs{U}}) \Big]^{-1} \bigg\} (\bs{F} \overline{\bs{\Gamma}}^\top \widehat{\bs{U}})^\top \bs{F} \Big\| \\
 & \le \max_{i,j} \Big\| \frac{Z_{i(j)}^\top \bs{F}}{T} \Big\| \| \overline{\bs{\Gamma}} \|^2 \Big\| \widehat{\bs{\Eig}}^{-1} - \Big[ \frac{1}{T} (\bs{F} \overline{\bs{\Gamma}}^\top \widehat{\bs{U}})^\top (\bs{F} \overline{\bs{\Gamma}}^\top \widehat{\bs{U}}) \Big]^{-1} \Big\| \| \bs{F}^\top \bs{F} \| \\
 & = O_p\bigg( \frac{\sqrt{T} \sqrt{\log(np)}\log(pT)}{n} + \sqrt{\frac{\log(npT) \log(np) \log p}{n}} \bigg)
\end{align*} 
by Lemmas \ref{lemma:aux5}\ref{lemma:aux5:1}, \ref{lemma:aux3}\ref{lemma:aux3:FZ} and \ref{lemma3:projection}\ref{lemma3:projection:ii},
\begin{align*}
\max_{i,j} \Big\| \frac{1}{T} Z_{i(j)}^\top (\bs{F} \overline{\bs{\Gamma}}^\top \widehat{\bs{U}}) \widehat{\bs{\Eig}}^{-1} (\overline{\bs{Z}} \widehat{\bs{U}})^\top \bs{F} \Big\| 
 & \le \max_{i,j} \Big\| \frac{Z_{i(j)}^\top \bs{F}}{T} \Big\| \| \overline{\bs{\Gamma}} \| \big\| \widehat{\bs{\Eig}}^{-1} \big\| \| \overline{\bs{Z}}^\top \bs{F} \| \\
 & = O_p\Big( \sqrt{\frac{\log(npT) \log(np) \log p}{n}}\Big)
\end{align*}
by Lemmas \ref{lemma:aux5}\ref{lemma:aux5:1}, \ref{lemma:aux3}\ref{lemma:aux3:FZ} and \ref{lemma:aux4}\ref{lemma:aux4:barZF}, 
\begin{align*}
\max_{i,j} \Big\| \frac{1}{T} Z_{i(j)}^\top (\overline{\bs{Z}} \widehat{\bs{U}}) \widehat{\bs{\Eig}}^{-1} (\bs{F} \overline{\bs{\Gamma}}^\top \widehat{\bs{U}})^\top \bs{F} \Big\|
 & \le \max_{i,j} \Big\| \frac{Z_{i(j)}^\top \overline{\bs{Z}}}{T} \Big\| \big\| \widehat{\bs{\Eig}}^{-1} \big\| \| \overline{\bs{\Gamma}} \| \| \bs{F}^\top \bs{F} \| \\
 & = O_p\Big( \sqrt{\frac{T \log(npT)\log(np^2)}{n}} + \frac{T}{n}\Big)
\end{align*}
by Lemmas \ref{lemma:aux5}\ref{lemma:aux5:1} and \ref{lemma:aux4}\ref{lemma:aux4:barZZ}, and
\begin{align*}
 & \max_{i,j} \Big\| \frac{1}{T} Z_{i(j)}^\top (\overline{\bs{Z}} \widehat{\bs{U}}) \widehat{\bs{\Eig}}^{-1} (\overline{\bs{Z}} \widehat{\bs{U}})^\top \bs{F} \Big\| \\
 & \le \max_{i,j} \Big\| \frac{Z_{i(j)}^\top \overline{\bs{Z}}}{T} \Big\| \big\| \widehat{\bs{\Eig}}^{-1} \big\| \| \overline{\bs{Z}}^\top \bs{F} \| \\
 & = O_p\Big( \frac{\log(npT)\sqrt{\log(np^2) \log p}}{n} + \frac{\sqrt{T} \sqrt{\log(npT) \log p}}{n^{3/2}}\Big) 
\end{align*} 
by Lemma \ref{lemma:aux4}\ref{lemma:aux4:barZF} and \ref{lemma:aux4:barZZ}. Hence, we arrive at 
\[ \max_{i,j} \| Z_{i(j)}^\top \widehat{\bs{R}} \bs{F} \| = O_p\Big( \sqrt{\frac{T \log(npT)\log(np^2)}{n}} + \frac{T}{n}\Big). \]
With this and the fact that $n^{-1} \sum_{i=1}^n \| \gamma_i \| = O_p(1)$, we can conclude that 
\begin{align*}
\frac{1}{nT} \max_{1 \le j \le p} \Big| \sum_{i=1}^n \big\{ \widehat{\bs{\Pi}} Z_{i(j)} \big\}^\top \big\{ \widehat{\bs{\Pi}} \bs{F} \gamma_i \big\} \Big| 
 & = \frac{1}{nT} \max_{1 \le j \le p} \Big| \sum_{i=1}^n Z_{i(j)}^\top \widehat{\bs{R}} \bs{F} \gamma_i \Big| \quad \text{(w.p.} \to 1) \\
 & \le \Big\{ \frac{1}{n} \sum_{i=1}^n \| \gamma_i \| \Big\} \Big\{ \frac{1}{T} \max_{i,j} \| Z_{i(j)}^\top \widehat{\bs{R}} \bs{F} \| \Big\} \\
 & = O_p\Big( \sqrt{\frac{\log(npT)\log(np^2)}{nT}} + \frac{1}{n}\Big). \qedhere 
\end{align*}
\end{proof}

\begin{proof}[Proof of \eqref{lemma2:Lasso:claim4}]
Since $\bs{\Pi} = \bs{I} - \bs{F} (\bs{F}^\top \bs{F})^{-1} \bs{F}^\top$ and $\widehat{\bs{\Pi}} = \bs{\Pi} - \widehat{\bs{R}}$ with probability tending to $1$, we obtain the bound  
\begin{align*}
\frac{1}{nT} \max_{1 \le j \le p} \Big| \sum_{i=1}^n \big\{ \widehat{\bs{\Pi}} Z_{i(j)} \big\}^\top \big\{ \widehat{\bs{\Pi}} \varepsilon_i \big\} \Big|
 & \le \frac{1}{nT} \max_{1 \le j \le p} \Big| \sum_{i=1}^n Z_{i(j)}^\top \varepsilon_i \Big| \\
 & \quad + \frac{1}{nT} \max_{1 \le j \le p} \Big| \sum_{i=1}^n Z_{i(j)}^\top \bs{F} (\bs{F}^\top \bs{F})^{-1} \bs{F}^\top \varepsilon_i \Big| \\
 & \quad + \frac{1}{nT} \max_{1 \le j \le p} \Big| \sum_{i=1}^n Z_{i(j)}^\top \widehat{\bs{R}} \varepsilon_i \Big|
\end{align*}
with probability tending to $1$. Arguments analogous to those for Lemma \ref{lemma:aux1}\ref{lemma:aux1:Z} yield that
\begin{equation}\label{lemma2:Lasso:claim4a}
\frac{1}{nT} \max_{1 \le j \le p} \Big| \sum_{i=1}^n Z_{i(j)}^\top \varepsilon_i \Big| = O_p\Big( \sqrt{\frac{\log p}{nT}} \Big). 
\end{equation}
Moreover, we show below that
\begin{align}
\frac{1}{nT} \max_{1 \le j \le p} \Big| \sum_{i=1}^n Z_{i(j)}^\top \bs{F} (\bs{F}^\top \bs{F})^{-1} \bs{F}^\top \varepsilon_i \Big| & = o_p\Big( \sqrt{\frac{1}{nT}} \Big) \label{lemma2:Lasso:claim4b} \\
\frac{1}{nT} \max_{1 \le j \le p} \Big| \sum_{i=1}^n Z_{i(j)}^\top \widehat{\bs{R}} \varepsilon_i \Big| & = o_p\Big( \sqrt{\frac{1}{nT}} \Big). \label{lemma2:Lasso:claim4c}
\end{align} 
Statement \eqref{lemma2:Lasso:claim4} follows upon combining \eqref{lemma2:Lasso:claim4a}--\eqref{lemma2:Lasso:claim4c}.

We first prove \eqref{lemma2:Lasso:claim4c}. It holds that 
\begin{align*} 
 & \max_{i,j} | Z_{i(j)}^\top \widehat{\bs{R}} \varepsilon_i | \\
 & \le \max_{i,j} \Big| \frac{1}{T} Z_{i(j)}^\top (\bs{F} \overline{\bs{\Gamma}}^\top \widehat{\bs{U}}) \bigg\{ \widehat{\bs{\Eig}}^{-1} - \Big[ \frac{1}{T} (\bs{F} \overline{\bs{\Gamma}}^\top \widehat{\bs{U}})^\top (\bs{F} \overline{\bs{\Gamma}}^\top \widehat{\bs{U}}) \Big]^{-1} \bigg\} (\bs{F} \overline{\bs{\Gamma}}^\top \widehat{\bs{U}})^\top \varepsilon_i \Big| \\*
 & \quad + \max_{i,j} \Big| \frac{1}{T} Z_{i(j)}^\top (\bs{F} \overline{\bs{\Gamma}}^\top \widehat{\bs{U}}) \widehat{\bs{\Eig}}^{-1} (\overline{\bs{Z}} \widehat{\bs{U}})^\top \varepsilon_i \Big|
         + \max_{i,j} \Big| \frac{1}{T} Z_{i(j)}^\top (\overline{\bs{Z}} \widehat{\bs{U}}) \widehat{\bs{\Eig}}^{-1} (\bs{F} \overline{\bs{\Gamma}}^\top \widehat{\bs{U}})^\top \varepsilon_i \Big| \\
 & \quad + \max_{i,j} \Big| \frac{1}{T} Z_{i(j)}^\top (\overline{\bs{Z}} \widehat{\bs{U}}) \widehat{\bs{\Eig}}^{-1} (\overline{\bs{Z}} \widehat{\bs{U}})^\top \varepsilon_i \Big|, 
\end{align*}
where
\begin{align*}
 & \max_{i,j} \Big| \frac{1}{T} Z_{i(j)}^\top (\bs{F} \overline{\bs{\Gamma}}^\top \widehat{\bs{U}}) \bigg\{ \widehat{\bs{\Eig}}^{-1} - \Big[ \frac{1}{T} (\bs{F} \overline{\bs{\Gamma}}^\top \widehat{\bs{U}})^\top (\bs{F} \overline{\bs{\Gamma}}^\top \widehat{\bs{U}}) \Big]^{-1} \bigg\} (\bs{F} \overline{\bs{\Gamma}}^\top \widehat{\bs{U}})^\top \varepsilon_i \Big| \\
 & \le \max_{i,j} \Big\| \frac{Z_{i(j)}^\top \bs{F}}{T} \Big\| \| \overline{\bs{\Gamma}} \|^2 \Big\| \widehat{\bs{\Eig}}^{-1} - \Big[ \frac{1}{T} (\bs{F} \overline{\bs{\Gamma}}^\top \widehat{\bs{U}})^\top (\bs{F} \overline{\bs{\Gamma}}^\top \widehat{\bs{U}}) \Big]^{-1} \Big\| \max_i \| \bs{F}^\top \varepsilon_i \| \\
 & = O_p\bigg( \frac{\log(pT)\sqrt{\log n \log(np)}}{n} + \sqrt{\frac{\log(npT) \log(np) \log p \log n}{nT}} \bigg) 
\end{align*}
by Lemmas \ref{lemma:aux5}\ref{lemma:aux5:1}, \ref{lemma:aux3}\ref{lemma:aux3:Feps}, \ref{lemma:aux3}\ref{lemma:aux3:FZ} and \ref{lemma3:projection}\ref{lemma3:projection:ii},
\begin{align*}
\max_{i,j} \Big| \frac{1}{T} Z_{i(j)}^\top (\bs{F} \overline{\bs{\Gamma}}^\top \widehat{\bs{U}}) \widehat{\bs{\Eig}}^{-1} (\overline{\bs{Z}} \widehat{\bs{U}})^\top \varepsilon_i \Big| 
 & \le \max_{i,j} \Big\| \frac{Z_{i(j)}^\top \bs{F}}{T} \Big\| \| \overline{\bs{\Gamma}} \| \big\| \widehat{\bs{\Eig}}^{-1} \big\| \| \overline{\bs{Z}}^\top \varepsilon_i \| \\
 & = O_p\Big( \frac{\sqrt{\log(npT)}\log(np)}{\sqrt{n}}\Big)
\end{align*}
by Lemmas \ref{lemma:aux5}\ref{lemma:aux5:1}, \ref{lemma:aux3}\ref{lemma:aux3:FZ} and \ref{lemma:aux4}\ref{lemma:aux4:barZeps},
\begin{align*}
\max_{i,j} \Big| \frac{1}{T} Z_{i(j)}^\top (\overline{\bs{Z}} \widehat{\bs{U}}) \widehat{\bs{\Eig}}^{-1} (\bs{F} \overline{\bs{\Gamma}}^\top \widehat{\bs{U}})^\top \varepsilon_i \Big|
 & \le \max_{i,j} \Big\| \frac{Z_{i(j)}^\top \overline{\bs{Z}}}{T} \Big\| \big\| \widehat{\bs{\Eig}}^{-1} \big\| \| \overline{\bs{\Gamma}} \| \| \bs{F}^\top \varepsilon_i \| \\
 & = O_p\Big( \sqrt{\frac{\log(npT)\log(np^2) \log n}{n}} + \frac{\sqrt{T \log n}}{n}\Big) 
\end{align*}
by Lemmas \ref{lemma:aux5}\ref{lemma:aux5:1}, \ref{lemma:aux3}\ref{lemma:aux3:Feps} and \ref{lemma:aux4}\ref{lemma:aux4:barZZ}, and
\begin{align*}
 & \max_{i,j} \Big| \frac{1}{T} Z_{i(j)}^\top (\overline{\bs{Z}} \widehat{\bs{U}}) \widehat{\bs{\Eig}}^{-1} (\overline{\bs{Z}} \widehat{\bs{U}})^\top \varepsilon_i \Big| \\*
 & \le \max_{i,j} \Big\| \frac{Z_{i(j)}^\top \overline{\bs{Z}}}{T} \Big\| \big\| \widehat{\bs{\Eig}}^{-1} \big\| \| \overline{\bs{Z}}^\top \varepsilon_i \| \\
 & = O_p\Big( \frac{\log(npT) \sqrt{\log(np^2)\log(np)}}{n} + \frac{\sqrt{T \log(npT)\log(np)}}{n^{3/2}}\Big)
\end{align*}
by Lemma \ref{lemma:aux4}\ref{lemma:aux4:barZeps} and \ref{lemma:aux4:barZZ}. As a result, we obtain that 
\begin{align*}
\frac{1}{nT} \max_{1 \le j \le p} \Big| \sum_{i=1}^n Z_{i(j)}^\top \widehat{\bs{R}} \varepsilon_i \Big| 
 & \le \frac{1}{T} \max_{i,j} |  Z_{i(j)}^\top \widehat{\bs{R}} \varepsilon_i | \\
 & = O_p \Big( \frac{\sqrt{\log(npT)\log^2(np) \log n}}{\sqrt{n}T} + \frac{\sqrt{\log n}}{n \sqrt{T}} \Big) = o_p\Big( \sqrt{\frac{1}{nT}} \Big),
\end{align*}
which completes the proof of \eqref{lemma2:Lasso:claim4c}.

We next turn to the proof of \eqref{lemma2:Lasso:claim4b}. The quantity of interest can be expressed as 
\[ \frac{1}{nT} \sum_{i=1}^n Z_{i(j)}^\top \bs{F} (\bs{F}^\top \bs{F})^{-1} \bs{F}^\top \varepsilon_i = \frac{1}{nT} \sum_{i=1}^n \sum_{t=1}^T  w_{it,j}\varepsilon_{it}, \]
where the weight vectors $w_{i,j} = (w_{i1,j},\ldots,w_{iT,j})^\top$ are defined as $w_{i,j} = \bs{F} (\bs{F}^\top \bs{F})^{-1} \linebreak \bs{F}^\top Z_{i(j)} = T^{-1} \bs{F} \bs{F}^\top Z_{i(j)}$. Importantly, the weight vectors $w_{i,j}$ have the following properties:
\begin{enumerate}[label=(\alph*),leftmargin=0.7cm]
\item Since $w_{i,j}$ depends only on $\bs{F}$ and $Z_{i(j)}$, it is independent from $\varepsilon_i$.
\item It holds that 
\[ \max_{i,j,t} |w_{it,j}| = O_p(T^{-\xi}) \]
for some small $\xi > 0$, since 
\begin{align*} 
\max_{i,j,t} |w_{it,j}| = \max_{i,j} \Big\| \bs{F} \Big(\frac{\bs{F}^\top Z_{i(j)}}{T}\Big) \Big\|_\infty
 & = \max_{i,j,s} \Big| \sum_{k=1}^K F_{s,k} \Big( \frac{1}{T} \sum_{t=1}^T F_{t,k} Z_{it,j} \Big) \Big| \\
 & = \max_{i,j,s} \Big|\frac{1}{T} \sum_{t=1}^T \Big\{ \sum_{k=1}^K F_{s,k} F_{t,k} \Big\} Z_{it,j} \Big| \\
 & = O_p\Big( \frac{\{ \log(np) \}^{\frac{1}{2}}}{T^{\frac{1}{2}-\frac{2}{\moments}}} \Big),
\end{align*}
where the last line follows by arguments analogous to those for the proof of Lemma \ref{lemma:aux1}, taking into account that $\max_{1 \le k \le K} \max_{1 \le t \le T} |F_{t,k}| \le C T^{1/\moments}$ (which is a direct consequence of the last condition in \ref{C:F}) and thus $|\sum_k F_{s,k} F_{t,k}| \le C T^{2/\moments}$.   
\end{enumerate}
With properties (a) and (b), we can proceed analogously as in the proof of Lemma \ref{lemma:aux2} in order to obtain that 
\[ \max_{1 \le j \le p} \Big| \frac{1}{nT} \sum_{i=1}^n \sum_{t=1}^T w_{it,j} \varepsilon_{it} \Big| = o_p\Big( \sqrt{\frac{1}{nT}} \Big), \]
which proves \eqref{lemma2:Lasso:claim4b}. 
\end{proof}

\subsection*{Proof of Proposition \ref{lemma3:Lasso}}

It holds that $\widehat{\bs{X}}^\top \widehat{\bs{X}} = \sum_{i=1}^n \widehat{\bs{X}}_i^\top \widehat{\bs{X}}_i = \sum_{i=1}^n \{\widehat{\bs{\Pi}} \bs{X}_i\}^\top \{\widehat{\bs{\Pi}} \bs{X}_i\}$, where
\begin{align*}
\{\widehat{\bs{\Pi}} \bs{X}_i\}^\top \{\widehat{\bs{\Pi}} \bs{X}_i\} 
 & = \{\widehat{\bs{\Pi}} \bs{F} \bs{\Gamma}_i^\top\}^\top \{\widehat{\bs{\Pi}} \bs{F} \bs{\Gamma}_i^\top\} + \{\widehat{\bs{\Pi}} \bs{Z}_i\}^\top \{\widehat{\bs{\Pi}} \bs{F} \bs{\Gamma}_i^\top\} \\
 & \quad + \{\widehat{\bs{\Pi}} \bs{F} \bs{\Gamma}_i^\top\}^\top \{\widehat{\bs{\Pi}} \bs{Z}_i\} + \{\widehat{\bs{\Pi}} \bs{Z}_i\}^\top \{\widehat{\bs{\Pi}} \bs{Z}_i\}
\end{align*} 
and 
\begin{align*} 
\{\widehat{\bs{\Pi}} \bs{Z}_i\}^\top \{\widehat{\bs{\Pi}} \bs{Z}_i\} = \bs{Z}_i^\top \widehat{\bs{\Pi}} \bs{Z}_i 
 & = \bs{Z}_i^\top \{\bs{\Pi} - \widehat{\bs{R}}\} \bs{Z}_i \\
 & = \bs{Z}_i^\top \{\bs{I} - \bs{F}(\bs{F}^\top \bs{F})^{-1} \bs{F}^\top - \widehat{\bs{R}}\} \bs{Z}_i 
\end{align*}
with probability tending to $1$. From this, it follows that  
\begin{align*}
\Big\| \frac{\widehat{\bs{X}}^\top \widehat{\bs{X}}}{nT} - \frac{\bs{Z}^\top \bs{Z}}{nT} \Big\|_{\max} 
 & \le \max_{j,j^\prime} \Big| \frac{1}{nT} \sum_{i=1}^n \{\widehat{\bs{\Pi}} \bs{F} \Gamma_{i,j}\}^\top \{\widehat{\bs{\Pi}} \bs{F} \Gamma_{i,j^\prime}\} \Big| \\
 & \quad + 2 \max_{j,j^\prime} \Big| \frac{1}{nT} \sum_{i=1}^n \{\widehat{\bs{\Pi}} Z_{i(j)}\}^\top \{\widehat{\bs{\Pi}} \bs{F} \Gamma_{i,j^\prime}\} \Big| \\
 & \quad + \max_{j,j^\prime} \Big| \frac{1}{nT} \sum_{i=1}^n Z_{i(j)}^\top \bs{F}(\bs{F}^\top \bs{F})^{-1} \bs{F}^\top Z_{i(j^\prime)} \Big| \\
& \quad + \max_{j,j^\prime} \Big| \frac{1}{nT} \sum_{i=1}^n Z_{i(j)}^\top \widehat{\bs{R}} Z_{i(j^\prime)} \Big|
\end{align*}
with probability approaching $1$. In the remainder of the proof, we bound the four terms on the right-hand side in the above display. Analogous calculations as in the proof of Proposition \ref{lemma2:Lasso} yield that 
\begin{align}
\max_{j,j^\prime} \Big| \frac{1}{nT} \sum_{i=1}^n \{\widehat{\bs{\Pi}} \bs{F} \Gamma_{i,j}\}^\top \{\widehat{\bs{\Pi}} \bs{F} \Gamma_{i,j^\prime}\} \Big| & = O_p\Big( \frac{\log(pT)}{n} \Big) \label{lemma3:Lasso:claim1} \\
\max_{j,j^\prime} \Big| \frac{1}{nT} \sum_{i=1}^n \{\widehat{\bs{\Pi}} Z_{i(j)}\}^\top \{\widehat{\bs{\Pi}} \bs{F} \Gamma_{i,j^\prime}\} \Big| & = O_p\Big( \sqrt{\frac{\log(npT)\log(np^2)}{nT}} + \frac{1}{n}\Big) \label{lemma3:Lasso:claim2} \\
\max_{j,j^\prime} \Big| \frac{1}{nT} \sum_{i=1}^n Z_{i(j)}^\top \widehat{\bs{R}} Z_{i(j^\prime)} \Big| & = O_p \Big( \frac{\sqrt{\log(npT)}\log(np)}{\sqrt{n}T} \nonumber \\* & \phantom{ = O_p \Big( } + \frac{\sqrt{\log (np)}}{n \sqrt{T}} + \frac{1}{n^2} \Big). \label{lemma3:Lasso:claim3}
\end{align}
In particular, the proofs of \eqref{lemma3:Lasso:claim1}, \eqref{lemma3:Lasso:claim2} and \eqref{lemma3:Lasso:claim3} parallel those of \eqref{lemma2:Lasso:claim1}, \eqref{lemma2:Lasso:claim3} and \eqref{lemma2:Lasso:claim4c}, respectively. Moreover, 
\begin{align*}
\max_{j,j^\prime} \Big| \frac{1}{nT} \sum_{i=1}^n Z_{i(j)}^\top \bs{F}(\bs{F}^\top \bs{F})^{-1} \bs{F}^\top Z_{i(j^\prime)} \Big| 
 & = \max_{j,j^\prime} \Big| \frac{1}{n} \sum_{i=1}^n \Big(\frac{Z_{i(j)}^\top \bs{F}}{T}\Big) \Big(\frac{\bs{F}^\top Z_{i(j^\prime)}}{T}\Big) \Big| \\
 & \le \bigg( \max_{i,j} \Big\| \frac{\bs{F}^\top Z_{i(j)}}{T} \Big\| \bigg)^2 \\
 & = O_p\Big(\frac{\log(np)}{T}\Big)
\end{align*}
by Lemma \ref{lemma:aux3}\ref{lemma:aux3:FZ}. To summarize, we obtain that 
\[ \Big\| \frac{\widehat{\bs{X}}^\top \widehat{\bs{X}}}{nT} - \frac{\bs{Z}^\top \bs{Z}}{nT} \Big\|_{\max} = O_p\Big(\frac{\log(npT)}{\min\{n,T\}}\Big). \]

\def\theequation{A'.\arabic{equation}}
\setcounter{equation}{0}
\section*{A' \hspace{0.1cm} Proof of Theorem \ref{theo:rate}(b)}

The proof strategy is the same as for the large-$T$-case in Appendix A. The various propositions and auxiliary lemmas, however, that are derived in the course of the proof must be adapted. As they can be adapted in a quite straightforward way, we do not give full proofs but only comment on noteworthy differences.

\subsection*{Step 1: Analysis of the eigenstructure of $\bs{\widehat{\Sigma}}$}

Propositions \ref{lemma1:eigenstructure}--\ref{lemma5:eigenstructure} remain unchanged in the small-$T$-case and can be proven by completely analogous arguments.

\subsection*{Step 2: Analysis of the projection matrix $\bs{\widehat{\Pi}}$}

We decompose $\bs{\widehat{\Pi}}$ exactly as in the large-$T$-case. In particular, we write
\begin{align*} 
\widehat{\bs{\Pi}} 
 & = \bs{I} - \widehat{\bs{W}} (\widehat{\bs{W}}^\top \widehat{\bs{W}})^{-1} \widehat{\bs{W}}^\top \\
 & = \bigg\{ \bs{I} - \frac{1}{T} (\bs{F} \overline{\bs{\Gamma}}^\top \widehat{\bs{U}}) \Big[ \frac{1}{T} (\bs{F} \overline{\bs{\Gamma}}^\top \widehat{\bs{U}})^\top (\bs{F} \overline{\bs{\Gamma}}^\top \widehat{\bs{U}}) \Big]^{-1} (\bs{F} \overline{\bs{\Gamma}}^\top \widehat{\bs{U}})^\top \bigg\} - \widehat{\bs{R}},
\end{align*}
where $\widehat{\bs{W}} = \overline{\bs{X}} \widehat{\bs{U}}$ and $\widehat{\bs{R}}$ is defined as before. This decomposition allows us to link the proxy $\widehat{\bs{\Pi}}$ to the unknown projection matrix $\bs{\Pi}$ in the same way as in the large-$T$-case. Specifically, by arguments completely analogous to those for Proposition \ref{lemma2:projection}, we can prove that 
\[ \widehat{\bs{\Pi}} = \bs{\Pi} - \widehat{\bs{R}} \]
with probability tending to $1$.

\subsection*{Step 3: Analysis of the lasso $\bs{\widehat{\beta}_\pen}$}

As in the large-$T$-case, let $\mathcal{T}_{\text{RE}}$ be the event that the design matrix $\widehat{\bs{X}}$ fulfills the $\text{RE}(S,\phi)$ condition with some constant $\phi > 0$ and define the event $\mathcal{T}_\pen$ as 
\[ \mathcal{T}_\pen = \Big\{ \frac{4 \| \widehat{\bs{X}}^\top e \|_\infty}{nT} \le \pen \Big\}. \]
Proposition \ref{lemma1:Lasso} and its proof remain completely unchanged, yielding that 
\begin{equation}\label{statement:lemma1prime:Lasso}
\| \widehat{\beta}_\pen - \beta \|_1 \le \frac{4}{\phi^2} \pen s \quad \text{on the event } \mathcal{T}_\pen \cap \mathcal{T}_{\textnormal{RE}}. 
\end{equation}
We next show that the event $\mathcal{T}_\pen \cap \mathcal{T}_{\textnormal{RE}}$ occurs with probability tending to $1$. To do so, we derive the following two results which parallel Propositions \ref{lemma2:Lasso} and \ref{lemma3:Lasso}.

\setcounter{propAprime}{7}
\begin{propAprime}\label{lemma2prime:Lasso} 
It holds that 
\[ \frac{\| \widehat{\bs{X}}^\top e \|_\infty}{nT} = O_p\Big((n^2 p)^{1/\moments} \sqrt{\frac{\log p}{n}} \Big). \]
\end{propAprime}

\begin{propAprime}\label{lemma3prime:Lasso}
It holds that
\[ \Big\| \frac{\widehat{\bs{X}}^\top \widehat{\bs{X}}}{nT} - \frac{\bs{Z}^\top \bs{Z}}{nT} \Big\|_{\max} = O_p \Big( (np)^{2/\moments} \sqrt{\frac{\log p}{n}} \Big). \]
\end{propAprime}

\noindent These two results can be proven by a fairly straightforward adaption of the arguments for Propositions \ref{lemma2:Lasso} and \ref{lemma3:Lasso}. Some more details are provided below. From Proposition \ref{lemma2prime:Lasso}, it immediately follows that 
\begin{equation}\label{eq:T-lambda-small}
\pr(\mathcal{T}_\pen) \to 1 \quad \text{for any choice} \quad \pen = h_n (n^2 p)^{1/\moments} \sqrt{\frac{\log p}{n}}, 
\end{equation}
where $h_n$ slowly diverges to infinity. Moreover, since $s = o((np)^{-2/\moments} \sqrt{n/\log p})$ by \ref{C:s-small}, Proposition \ref{lemma3prime:Lasso} implies that 
\begin{equation*}
\frac{32 s}{\varphi^2} \Big\| \frac{\widehat{\bs{X}}^\top \widehat{\bs{X}}}{nT} - \frac{\bs{Z}^\top \bs{Z}}{nT} \Big\|_{\max} \le 1
\end{equation*}
with probability tending to $1$ for any given constant $\varphi > 0$. As in the large-$T$-case, we can now use Corollary 6.8 in \cite{BuehlmannvandeGeer2011} to get that 
\begin{equation}\label{eq:T-RE-small}
\pr(\mathcal{T}_{\text{RE}}) \to 1. 
\end{equation}
Taken together, \eqref{eq:T-lambda-small} and \eqref{eq:T-RE-small} imply that the event $\mathcal{T}_\pen \cap \mathcal{T}_{\textnormal{RE}}$ occurs with probability tending to $1$. Combining this with \eqref{statement:lemma1prime:Lasso}, we finally arrive at the following statement: 
\[ \| \widehat{\beta}_\pen - \beta \|_1 \le \frac{4}{\phi^2} \pen s \]
for any $\pen = h_n (n^2 p)^{1/\theta} \sqrt{\log p /n}$ with probability tending to $1$.

\subsection*{Step 4: Proof of intermediate results}

We now adapt the convergence rates derived in Lemmas \ref{lemma:aux1}--\ref{lemma3:projection} to the small-$T$-case. We demonstrate how to modify the proof of Lemma \ref{lemma:aux1} while omitting the proofs of the other lemmas, which are either straightforward to modify or can be modified in the same way as Lemma \ref{lemma:aux1}. In addition, we give some details on the proofs of Propositions \ref{lemma2prime:Lasso} and \ref{lemma3prime:Lasso}.

\begin{lemmaAprime}\label{lemmaprime:aux1}
It holds that 
\begin{enumerate}[label=(\roman*),leftmargin=0.975cm,topsep=0.5cm]
\item \label{lemmaprime:aux1:Z} $\displaystyle{\max_{1 \le j \le p} \max_{1 \le t \le T} \Big| \frac{1}{n} \sum_{i=1}^n Z_{it,j} \Big| = O_p\Big(\sqrt{\frac{\log p}{n}} \Big)}$.
\item \label{lemmaprime:aux1:Feps} $\displaystyle{\max_{1 \le k \le K} \max_{1 \le i \le n} \Big| \frac{1}{T} \sum_{t=1}^T F_{t,k} \varepsilon_{it} \Big| = O_p\big(n^{1/\moments}\big)}$.
\item \label{lemmaprime:aux1:FZ} $\displaystyle{\max_{1 \le k \le K} \max_{1 \le i \le n} \max_{1 \le j \le p} \Big| \frac{1}{T} \sum_{t=1}^T F_{t,k} Z_{it,j} \Big| = O_p\big( \{np\}^{1/\moments}\big)}$.
\end{enumerate}
\end{lemmaAprime}

\begin{lemmaAprime}\label{lemmaprime:aux2}
It holds that 
\begin{enumerate}[label=(\roman*),leftmargin=0.975cm,topsep=0.5cm]
\item \label{lemmaprime:aux2:ZF} $\displaystyle{\max_{1 \le k \le K} \max_{1 \le j \le p} \Big| \frac{1}{T} \sum_{t=1}^T \Big\{ \frac{1}{n} \sum_{i=1}^n Z_{it,j} \Big\} F_{t,k} \Big| = O_p\Big(  \sqrt{\frac{\log p}{n}} \Big)}$.
\item \label{lemmaprime:aux2:Zeps} $\displaystyle{\max_{1 \le i \le n} \max_{1 \le j \le p} \Big| \frac{1}{T} \sum_{t=1}^T \Big\{ \frac{1}{n} \sum_{i^\prime=1}^n Z_{i^\prime t,j} \Big\} \varepsilon_{it} \Big| = O_p\Big( n^{1/\moments} \sqrt{\frac{\log p}{n}} \Big)}$.
\item \label{lemmaprime:aux2:ZZ} $\displaystyle{\max_{1 \le i \le n} \max_{1 \le j \le p} \max_{1 \le j^\prime \le p} \Big| \frac{1}{T} \sum_{t=1}^T \Big\{ \frac{1}{n} \sum_{i^\prime=1}^n Z_{i^\prime t,j^\prime} \Big\} Z_{it,j} \Big| = O_p\Big(  \{np\}^{1/\moments} \sqrt{\frac{\log p}{n}} \Big)}$.
\end{enumerate}
\end{lemmaAprime}

\pagebreak
\begin{lemmaAprime}
\label{lemmaprime:aux5}
It holds that 
\begin{enumerate}[label=(\roman*),leftmargin=0.975cm,topsep=0.5cm]
\item \label{lemmaprime:aux5:1} $\displaystyle{\big\| \overline{\bs{\Gamma}} - \bs{\Gamma} \big\| = O_p \Big( \sqrt{\frac{p \log p}{n}} \Big)}$.
\item \label{lemmaprime:aux5:2} $\displaystyle{\big\| \overline{\bs{\Gamma}} \overline{\bs{\Gamma}}^\top - \ex \, \overline{\bs{\Gamma}}  \overline{\bs{\Gamma}}^\top \big\| = O_p \Big( p \sqrt{\frac{\log p}{n}} \Big)}$.
\end{enumerate}
\end{lemmaAprime}

\begin{lemmaAprime}\label{lemmaprime:aux3}
It holds that
\begin{enumerate}[label=(\roman*),leftmargin=0.975cm,topsep=0.5cm]
\item \label{lemmaprime:aux3:Feps} $\displaystyle{\max_{1 \le i \le n} \Big\| \frac{\bs{F}^\top \varepsilon_i}{T} \Big\| = O_p\big( n^{1/\moments} \big)}$.
\item \label{lemmaprime:aux3:FZ} $\displaystyle{\max_{1 \le i \le n} \max_{1 \le j \le p} \Big\| \frac{\bs{F}^\top Z_{i(j)}}{T} \Big\| = O_p\big( \{ np \}^{1/\moments}\big)}$.
\end{enumerate}
\end{lemmaAprime}

\begin{lemmaAprime}\label{lemmaprime:aux4}
It holds that 
\begin{enumerate}[label=(\roman*),leftmargin=0.975cm,topsep=0.5cm]
\item \label{lemmaprime:aux4:barZ} $\displaystyle{\big\| \overline{\bs{Z}} \big\| = O_p\Big(\sqrt{\frac{p\log p}{n}}\Big)}$.
\item \label{lemmaprime:aux4:barZF} $\displaystyle{\Big\| \frac{\overline{\bs{Z}}^\top \bs{F}}{T} \Big\| = O_p\Big(\sqrt{\frac{p \log p}{n}}\Big)}$.
\item \label{lemmaprime:aux4:barZeps} $\displaystyle{\max_{1 \le i \le n} \Big\| \frac{\overline{\bs{Z}}^\top \varepsilon_i}{T} \Big\| = O_p\Big(n^{1/\moments} \sqrt{\frac{p \log p}{n}} \Big)}$.
\item \label{lemmaprime:aux4:barZZ} $\displaystyle{\max_{1 \le i \le n} \max_{1 \le j \le p} \Big\| \frac{\overline{\bs{Z}}^\top Z_{i(j)}}{T} \Big\| = O_p\Big(  \{np\}^{1/\moments} \sqrt{\frac{p \log p}{n}} \Big)}$. 
\end{enumerate}
\end{lemmaAprime}

\begin{lemmaAprime}\label{lemma3prime:projection}
It holds that 
\begin{enumerate}[label=(\roman*),leftmargin=0.975cm,topsep=0.5cm]
\item \label{lemma3prime:projection:i}
$\displaystyle{\Big\| \widehat{\bs{\Eig}} - \frac{1}{T} (\bs{F} \overline{\bs{\Gamma}}^\top \widehat{\bs{U}})^\top (\bs{F} \overline{\bs{\Gamma}}^\top \widehat{\bs{U}}) \Big\| = O_p\bigg( p \sqrt{\frac{\log p}{n}} \bigg)}$.
\item \label{lemma3prime:projection:ii}
$\displaystyle{\Big\| \widehat{\bs{\Eig}}^{-1} - \Big[ \frac{1}{T} (\bs{F} \overline{\bs{\Gamma}}^\top \widehat{\bs{U}})^\top (\bs{F} \overline{\bs{\Gamma}}^\top \widehat{\bs{U}}) \Big]^{-1} \Big\|  = O_p\bigg( \frac{1}{p} \sqrt{\frac{\log p}{n}}  \bigg)}$. 
\end{enumerate}
\end{lemmaAprime}

\subsection*{Proof of Lemma \ref{lemmaprime:aux1}}

The proof of \ref{lemmaprime:aux1:Z} is essentially identical to that of \ref{lemma:aux1:Z} in Lemma \ref{lemma:aux1}. Moreover, as the proofs of \ref{lemmaprime:aux1:Feps} and \ref{lemmaprime:aux1:FZ} are completely analogous, we only verify \ref{lemmaprime:aux1:FZ}. It holds that 
\begin{align*}
\max_{i,j,k} \Big| \frac{1}{T} \sum_{t=1}^T F_{t,k} Z_{it,j} \Big| 
 & \le \Big\{ \max_k \frac{1}{T} \sum_{t=1}^T |F_{t,k}| \Big\} \max_{i,t,j} |Z_{it,j}| \\
 & \le C \max_{i,t,j} |Z_{it,j}|,
\end{align*}
where $C = C(F_1,\ldots,F_T) := \max_k T^{-1} \sum_{t=1}^T |F_{t,k}|$ is a fixed number for all $t$. Since $(\ex \max_{i,j,t} |Z_{it,j}|^\moments)^{1/\moments} \le C (npT)^{1/\moments}$, we further have that
\begin{align*} 
\pr \Big( \max_{i,t,j} |Z_{it,j}| > C_0 \{npT\}^{1/\moments} \Big) \le \frac{\ex \max_{i,j,t} |Z_{it,j}|^\moments}{C_0^\moments npT} \le \Big(\frac{C}{C_0}\Big)^\moments
\end{align*}
for any $C_0 > 0$, which implies that $\max_{i,t,j} |Z_{it,j}| = O_p(\{npT\}^{1/\moments}) = O_p(\{np\}^{1/\moments})$. Therefore, we obtain that
\[ \max_{i,j,k} \Big| \frac{1}{T} \sum_{t=1}^T F_{t,k} Z_{it,j} \Big| = O_p\big(\{np\}^{1/\moments}\big) \] for all $t$.

\subsection*{Proof of Proposition \ref{lemma2prime:Lasso}}

It holds that 
\begin{align*}
\frac{\| \widehat{\bs{X}}^\top e \|_\infty}{nT} \nonumber 
 & \le \frac{1}{nT} \max_{1 \le j \le p} \Big| \sum_{i=1}^n \big\{ \widehat{\bs{\Pi}} \bs{F} \Gamma_{i,j} \big\}^\top \big\{ \widehat{\bs{\Pi}} \bs{F} \gamma_i \big\} \Big| \\
 & \quad + \frac{1}{nT} \max_{1 \le j \le p} \Big| \sum_{i=1}^n \big\{ \widehat{\bs{\Pi}} \bs{F} \Gamma_{i,j} \big\}^\top \big\{ \widehat{\bs{\Pi}} \varepsilon_i \big\} \Big| \\
 & \quad + \frac{1}{nT} \max_{1 \le j \le p} \Big| \sum_{i=1}^n \big\{ \widehat{\bs{\Pi}} Z_{i(j)} \big\}^\top \big\{ \widehat{\bs{\Pi}} \bs{F} \gamma_i \big\} \Big| \\
 & \quad + \frac{1}{nT} \max_{1 \le j \le p} \Big| \sum_{i=1}^n \big\{ \widehat{\bs{\Pi}} Z_{i(j)} \big\}^\top \big\{ \widehat{\bs{\Pi}} \varepsilon_i \big\} \Big|. 
\end{align*}
The same proof strategy as for Proposition \ref{lemma2:Lasso} yields that 
\begin{align*}
\frac{1}{nT} \max_{1 \le j \le p} \Big| \sum_{i=1}^n \big\{ \widehat{\bs{\Pi}} \bs{F} \Gamma_{i,j} \big\}^\top \big\{ \widehat{\bs{\Pi}} \bs{F} \gamma_i \big\} \Big| & = O_p\Big( \frac{\log p}{n} \Big) \\
\frac{1}{nT} \max_{1 \le j \le p} \Big| \sum_{i=1}^n \big\{ \widehat{\bs{\Pi}} \bs{F} \Gamma_{i,j} \big\}^\top \big\{ \widehat{\bs{\Pi}} \varepsilon_i \big\} \Big| & = O_p\Big( n^{1/\moments} \sqrt{\frac{\log p}{n}} \Big) \\
\frac{1}{nT} \max_{1 \le j \le p} \Big| \sum_{i=1}^n \big\{ \widehat{\bs{\Pi}} Z_{i(j)} \big\}^\top \big\{ \widehat{\bs{\Pi}} \bs{F} \gamma_i \big\} \Big| & = O_p\Big( \{np\}^{1/\moments} \sqrt{\frac{\log p}{n}} \Big) \\
\frac{1}{nT} \max_{1 \le j \le p} \Big| \sum_{i=1}^n \big\{ \widehat{\bs{\Pi}} Z_{i(j)} \big\}^\top \big\{ \widehat{\bs{\Pi}} \varepsilon_i \big\} \Big| & = O_p\Big( \{n^2p\}^{1/\moments} \sqrt{\frac{\log p}{n}} \Big). 
\end{align*}
Proposition \ref{lemma2prime:Lasso} is a direct consequence of these four statements.

\subsection*{Proof of Proposition \ref{lemma3prime:Lasso}}

Following the same line of argument as in the proof of Proposition \ref{lemma3:Lasso} yields
\begin{align*}
\Big\| \frac{\widehat{\bs{X}}^\top \widehat{\bs{X}}}{nT} - \frac{(\bs{X}^\perp)^\top (\bs{X}^\perp)}{nT} \Big\|_{\max} 
 & \le \max_{j,j^\prime} \Big| \frac{1}{nT} \sum_{i=1}^n \{\widehat{\bs{\Pi}} \bs{F} \Gamma_{i,j}\}^\top \{\widehat{\bs{\Pi}} \bs{F} \Gamma_{i,j^\prime}\} \Big| \\
 & \quad + 2 \max_{j,j^\prime} \Big| \frac{1}{nT} \sum_{i=1}^n \{\widehat{\bs{\Pi}} Z_{i(j)}\}^\top \{\widehat{\bs{\Pi}} \bs{F} \Gamma_{i,j^\prime}\} \Big| \\
& \quad + \max_{j,j^\prime} \Big| \frac{1}{nT} \sum_{i=1}^n Z_{i(j)}^\top \widehat{\bs{R}} Z_{i(j^\prime)} \Big|
\end{align*}
with probability approaching $1$, where
\begin{align*}
\max_{j,j^\prime} \Big| \frac{1}{nT} \sum_{i=1}^n \{\widehat{\bs{\Pi}} \bs{F} \Gamma_{i,j}\}^\top \{\widehat{\bs{\Pi}} \bs{F} \Gamma_{i,j^\prime}\} \Big| & = O_p\Big( \frac{\log p}{n} \Big) \\
\max_{j,j^\prime} \Big| \frac{1}{nT} \sum_{i=1}^n \{\widehat{\bs{\Pi}} Z_{i(j)}\}^\top \{\widehat{\bs{\Pi}} \bs{F} \Gamma_{i,j^\prime}\} \Big| & = O_p\Big( \{np\}^{1/\moments} \sqrt{\frac{\log p}{n}} \Big) \\
\max_{j,j^\prime} \Big| \frac{1}{nT} \sum_{i=1}^n Z_{i(j)}^\top \widehat{\bs{R}} Z_{i(j^\prime)} \Big| & = O_p \Big( \{np\}^{2/\moments} \sqrt{\frac{\log p}{n}} \Big). 
\end{align*}
This immediately implies Proposition \ref{lemma3prime:Lasso}.

\def\theequation{B.\arabic{equation}}
\setcounter{equation}{3}
\section*{B \hspace{0.1cm} Details on the proof of Theorem \ref{theo:normality}(a)}

\subsection*{Proof of Lemma \ref{lemmaC:aux1}}

\begin{proof}[Proof of \ref{lemmaC:aux1:normbarZ-j},  \ref{lemmaC:aux1:maxFui} and \ref{lemmaC:aux1:maxnormZijF}]
Slightly adapting the arguments in the proof of Lemma \ref{lemma:aux1},
we obtain that for sufficiently large $\Summable_5$, $\Summable_6$ and $\Summable_7$,
\begin{align*}
\pr\left( \max_{j,t} \Big| \frac{1}{n} \sum_{i=1}^n Z_{it,j} \Big| > \Summable_5 \sqrt{\frac{\log(pT)}{n}} \right) & \le \summable_n^{(5)} \\
\pr\left(\sqrt{K}\max_{i,k} \left| \frac{1}{T} \sum_{t=1}^T F_{t,k}u_{it} \right| > \Summable_6 \sqrt{\frac{\log n}{T}} \right) & \leq \summable_n^{(6)} \\
\pr\left( \sqrt{K}\max_{i,j,k} \Big| \frac{1}{T}\sum_{t=1}^T Z_{it,j} F_{t,k} \Big| > \Summable_7 \sqrt{\frac{\log(np)}{T}}\right) & = \summable_n^{(7)},
\end{align*}
where $\{\summable_n^{(\ell)}\}$ is summable for $5 \le \ell \le 7$. Statements \ref{lemmaC:aux1:normbarZ-j},  \ref{lemmaC:aux1:maxFui} and \ref{lemmaC:aux1:maxnormZijF} follow from these three bounds together with some straightforward algebra. 
\end{proof}

\pagebreak
\begin{proof}[Proof of \ref{lemmaC:aux1:normbarZF}, \ref{lemmaC:aux1:normbarZui} and \ref{lemmaC:aux1:maxnormZijbarZ}]
Minor modifications of the arguments in the proof of Lemma \ref{lemma:aux2}
yield that for sufficiently large $\Summable_8$, $\Summable_9$ and $\Summable_{10}$,
\begin{align*}
\pr\left( \sqrt{K} \max_{j,k} \Big |\frac{1}{T}\sum_{t=1}^T \Big\{ \frac{1}{n} \sum_{i=1}^n Z_{it,j} \Big\} F_{t,k} \Big|   > \Summable_8 \sqrt{\frac{\log(npT) \log p}{nT}} \right) & \leq \summable_n^{(8)} \\
\pr\left( \max_{i,j} \left\lvert \frac{1}{T}\sum_{t=1}^T\Big\{ \frac{1}{n} \sum_{i'=1}^n Z_{i't,j} \Big\} u_{it} \right\rvert > \Summable_9 \sqrt{\frac{\log(npT)\log(np)}{nT}}\right) & \leq \summable_n^{(9)} \\
\pr\left( \max_{i,j,j^\prime} \Big| \frac{1}{T} \sum_{t=1}^T \Big\{ \frac{1}{n} \sum_{i^\prime=1}^n Z_{i^\prime t,j^\prime} \Big\} Z_{it,j} \Big| > \Summable_{10} \bigg[ \sqrt{\frac{\log(npT)\log(np^2)}{nT}} + \frac{1}{n} \bigg] \right) & \le \summable_n^{(10)}, 
\end{align*}
where $\{\summable_n^{(\ell)}\}$ is summable for $8 \le \ell \le 10$. Combining these bounds with simple algebra yields statements \ref{lemmaC:aux1:normbarZF}, \ref{lemmaC:aux1:normbarZui} and \ref{lemmaC:aux1:maxnormZijbarZ}. 
\end{proof}

\subsection*{Proof of Lemma \ref{lemmaC:aux2}}

\begin{proof}[Proof of \ref{lemmaC:aux2:normbargamma}]
By \ref{C:Gamma-INF}, $\| \bs{\Gamma}_{-j} \| \le C_\Gamma \sqrt{p}$ for some $C_\Gamma > 0$. Hence, choosing $\Summable_{11} > 2 C_\Gamma$, we obtain that
\begin{align*}
\pr\left(\|\overline{\bs{\Gamma}}_{-j}\| > \Summable_{11} \sqrt{p} \right) 
 & \leq \pr\left(\|\overline{\bs{\Gamma}}_{-j} - \bs{\Gamma}_{-j}\| > \frac{\Summable_{11} \sqrt{p}}{2} \right) + \pr\left(\|\bs{\Gamma}_{-j}\| > \frac{\Summable_{11}\sqrt{p}}{2} \right) \\
 & = \pr\left(\|\overline{\bs{\Gamma}}_{-j} - \bs{\Gamma}_{-j}\| > \frac{\Summable_{11}\sqrt{p}}{2} \right).
\end{align*}
Letting $\{\bs{A}\}_{jk}$ be the element in the $j$-th row and the $k$-th column of a generic matrix $\bs{A}$ and numbering the columns of $\bs{\Gamma}_{-j}$ and $\overline{\bs{\Gamma}}_{-j}$ by $1,\ldots,j-1,j+1,\ldots,p$ (thus leaving out the index $j$), we further get that 
\begin{align*}
 & \pr\left(\|\overline{\bs{\Gamma}}_{-j} - \bs{\Gamma}_{-j}\| > \frac{\Summable_{11}\sqrt{p}}{2} \right) \\
 & \leq \pr\left(\sqrt{K} \sqrt{\max_{k,k'}\big| \{ (\overline{\bs{\Gamma}}_{-j} - \bs{\Gamma}_{-j})^\top(\overline{\bs{\Gamma}}_{-j} - \bs{\Gamma}_{-j}) \}_{kk'} \big| } > \frac{\Summable_{11}\sqrt{p}}{2} \right) \\
 & = \pr\left(\sqrt{K} \sqrt{ \max_{k,k'} \left\lvert\sum_{j'=1, j'\neq j}^p \{\overline{\bs{\Gamma}}_{-j} - \bs{\Gamma}_{-j}\}_{j'k} \{\overline{\bs{\Gamma}}_{-j} - \bs{\Gamma}_{-j}\}_{j'k'} \right\rvert}> \frac{\Summable_{11}\sqrt{p}}{2} \right) \\
 & \leq \pr\left(\sqrt{Kp} \sqrt{ \max_{j,k,k'}  \big|\{\overline{\bs{\Gamma}}-\bs{\Gamma}\}_{jk}\{\overline{\bs{\Gamma}}-\bs{\Gamma}\}_{jk'}\big| }> \frac{\Summable_{11}\sqrt{p}}{2} \right) \\
 & \leq \pr\left(\sqrt{K} \max_{j,k} \big|\{\overline{\bs{\Gamma}}-\bs{\Gamma}\}_{jk}\big| > \frac{\Summable_{11}}{2} \right).
\end{align*}
Finally, slightly adapting the proof of Lemma \ref{lemma:aux1} yields that  
\begin{equation}\label{lemmaC:aux2:normbargamma:proof1}
\pr\left(\max_{j,k} \big|\{\overline{\bs{\Gamma}}-\bs{\Gamma}\}_{jk}\big| > \Summable_{11} \sqrt{\frac{\log p}{n}}\right) \leq \summable_n^{(11)} 
\end{equation}
with some summable sequence $\{\summable_n^{(11)}\}$.
\end{proof}

\begin{proof}[Proof of \ref{lemmaC:aux2:sigmasigmabar}] 
It holds that
\begin{align}
\left\|\bs{\Gamma}_{-j} \bs{\Gamma}_{-j}^\top - \ex\left[\overline{\bs{\Gamma}}_{-j}\overline{\bs{\Gamma}}_{-j}^\top\right]  \right\| & = O\left(\frac{p}{n}\right) \label{lemmaC:aux2:sigmasigmabar:proof1} \\
\left\| \ex\left[\frac{1}{T}\sum_{t=1}^T\overline{Z}_{t,-j}\overline{Z}_{t,-j}^\top\right] \right\| & = O\left(\frac{p}{n}\right), \label{lemmaC:aux2:sigmasigmabar:proof2}
\end{align}
since
\begin{align*}
\left \| \bs{\Gamma}_{-j}\bs{\Gamma}_{-j}^\top - \ex\left[\overline{\bs{\Gamma}}_{-j}\overline{\bs{\Gamma}}_{-j}^\top\right]\right\| 
 &= \left \| \ex\left[(\bs{\Gamma}_{-j}-\overline{\bs{\Gamma}}_{-j})(\bs{\Gamma}_{-j}-\overline{\bs{\Gamma}}_{-j})^\top\right] \right\| \\
 &\leq p \max_{j,j'} \bigg| \ex\bigg[ \sum_{k=1}^K (\Gamma_{jk}-\overline{\Gamma}_{jk})  (\Gamma_{j'k} -\overline{\Gamma}_{j'k} ) \bigg] \bigg| \\
 &\leq K p\max_{j,j',k} \left| \ex\left[  (\Gamma_{jk}-\overline{\Gamma}_{jk})  (\Gamma_{j'k} -\overline{\Gamma}_{j'k} )\right] \right| \\
 &\leq K p\max_{j,j',k} \sqrt{\ex\left(\Gamma_{jk}-\overline{\Gamma}_{jk}\right)^2 \ex\left(\Gamma_{j'k} -\overline{\Gamma}_{j'k}\right)^2} \\
 & \leq K p \max_{j,k} \ex\left(\Gamma_{jk}-\overline{\Gamma}_{jk}\right)^2 \\
 & = K p \max_{j,k} \var\left(\frac{1}{n}\sum_{i=1}^n \Gamma_{i,jk}\right) \leq \frac{CKp}{n}
\end{align*}
and 
\begin{align*}
\bigg\| \ex\bigg[ \frac{1}{T}\sum_{t=1}^T \overline{Z}_{t,-j}\overline{Z}_{t,-j}^\top\bigg] \bigg\| 
 & = \bigg\| \ex\bigg[ \frac{1}{T}\sum_{t=1}^T \bigg(\frac{1}{n}\sum_{i=1}^n Z_{it,-j}\bigg)\bigg(\frac{1}{n}\sum_{i=1}^n Z_{it,-j}\bigg)^\top  \bigg] \bigg\| \\
 & = \bigg\| \ex\bigg[ \frac{1}{T}\sum_{t=1}^T \frac{1}{n^2}\sum_{i,l=1}^n Z_{it,-j}Z_{lt,-j}^\top \bigg] \bigg\| \\
 & = \bigg\| \frac{1}{T} \sum_{t=1}^T\frac{1}{n^2}\sum_{i=1}^n \ex\big[Z_{it,-j}Z_{it,-j}^\top  \big] \bigg\| \\
 & \leq p \max_{j,j'} \bigg|  \frac{1}{T} \sum_{t=1}^T\frac{1}{n^2}\sum_{i=1}^n \ex\big[Z_{it,j}Z_{it,j'}  \big] \bigg| \\
 & \leq p \max_{j,j'} \bigg|  \frac{1}{T} \sum_{t=1}^T\frac{1}{n^2}\sum_{i=1}^n \sqrt{\ex[Z_{it,j}^2] \ex[Z_{it,j'}^2]} \bigg| \\
 & \leq \frac{p}{n} \max_{i,t,j} \ex[Z_{it,j}^2] \le \frac{Cp}{n}. 
\end{align*}
With \eqref{lemmaC:aux2:sigmasigmabar:proof1} and \eqref{lemmaC:aux2:sigmasigmabar:proof2}, we can directly conclude that
\begin{align*}
\|\bs{\Sigma}^{[-j]}-\overline{\bs{\Sigma}}^{[-j]}\| 
 & \leq \Big\|\bs{\Gamma}_{-j} \bs{\Gamma}_{-j}^\top - \ex\left[\overline{\bs{\Gamma}}_{-j}\overline{\bs{\Gamma}}_{-j}^\top\right] \Big\| + \left\| \ex\left[\frac{1}{T}\sum_{t=1}^T\overline{Z}_{t,-j}\overline{Z}_{t,-j}^\top\right] \right\| \le \frac{\Summable_{12}p}{n}
\end{align*}
for $\Summable_{12}$ chosen sufficiently large. 
\end{proof}

\begin{proof}[Proof of \ref{lemmaC:aux2:sigmahatsigmabar}]
We have
\[ \pr\left(\|\overline{\bs{\Sigma}}^{[-j]}-\widetilde{\bs{\Sigma}}\| > \Summable_{13} p\sqrt{\frac{\log p}{n}} \right) \le P_1 + P_2 + P_3 \]
with
\begin{align*}
P_1 & = \pr \Bigg(\left\|\ex\left[\overline{\bs{\Gamma}}_{-j}\overline{\bs{\Gamma}}_{-j}^\top\right] - \overline{\bs{\Gamma}}_{-j}\overline{\bs{\Gamma}}_{-j}^\top  \right\| > \frac{\Summable_{13}p}{3} \sqrt{\frac{\log p}{n}} \Bigg) \\
P_2 & = \pr\left(\left\|\frac{1}{T}\sum_{t=1}^T \ex\left[ \overline{Z}_{t,-j}\overline{Z}_{t,-j}^\top\right] - \overline{Z}_{t,-j}\overline{Z}_{t,-j}^\top \right\| > \frac{\Summable_{13}p}{3} \sqrt{\frac{\log p}{n}} \right) \\
P_3 & = \pr\left( 2\left\| \overline{\bs{\Gamma}}_{-j} \right\| \left\|\frac{1}{T}\sum_{t=1}^T F_t\overline{Z}_{t,-j}^\top \right\| > \frac{\Summable_{13}p}{3} \sqrt{\frac{\log p}{n}} \right). 
\end{align*}
Combining the  arguments from Lemma \ref{lemma:aux5} with \eqref{lemmaC:aux2:normbargamma:proof1} yields that $P_1$ is bounded by a summable sequence $\{\summable_n\}$. Moreover, since 
\begin{align*}
P_2 & \leq \pr\left(\left\|\frac{1}{T}\sum_{t=1}^T \ex\left[ \overline{Z}_{t,-j}\overline{Z}_{t,-j}^\top\right] \right\| > \frac{\Summable_{13}p}{6} \sqrt{\frac{\log p}{n}} \right) \\
    & \quad + \pr\left(\left\|\frac{1}{T}\sum_{t=1}^T  \overline{Z}_{t,-j}\overline{Z}_{t,-j}^\top   \right\| > \frac{\Summable_{13}p}{6} \sqrt{\frac{\log p}{n}} \right) \\
    & = \pr\left(\left\|\frac{1}{T}\sum_{t=1}^T \ex\left[ \overline{Z}_{t,-j}\overline{Z}_{t,-j}^\top\right] \right\| > \frac{\Summable_{13}p}{6} p\sqrt{\frac{\log p}{n}} \right) \\
    & \quad +\pr\left( \|\overline{\bs{Z}}_{(-j)}\|^2> \frac{\Summable_{13}pT}{6} \sqrt{\frac{\log p}{n}} \right),
\end{align*}
we can use \eqref{lemmaC:aux2:sigmasigmabar:proof2} and Lemma \ref{lemmaC:aux1}\ref{lemmaC:aux1:normbarZ-j} to infer that $P_2$ is bounded by a summable sequence as well. Finally, with Lemma \ref{lemmaC:aux1}\ref{lemmaC:aux1:normbarZF} and Lemma \ref{lemmaC:aux2}\ref{lemmaC:aux2:normbargamma},
\begin{align*}
P_3 & \leq \pr\left( 2 \Summable_{11}\sqrt{p}  \left\|\frac{1}{T}\sum_{t=1}^T F_t\overline{Z}_{t,-j}^\top \right\| > \frac{\Summable_{13}p}{3} \sqrt{\frac{\log p}{n}} \right) + a_n^{(11)} \leq a_n^{(8)} + a_n^{(11)} 
\end{align*}
for $\Summable_{13}$ sufficiently large.
\end{proof}

\pagebreak
\begin{proof}[Proof of \ref{lemmaC:aux2:normpsi-1}]
Let $\widetilde{\eig}_{1}, \ldots, \widetilde{\eig}_K$, $\overline{\eig}_1^{[-j]}, \ldots, \overline{\eig}_K^{[-j]}$ and $\eig_1^{[-j]},\ldots,\eig_K^{[-j]}$ denote the $K$ largest eigenvalues of the matrices $\widetilde{\bs{\Sigma}} = \overline{\bs{X}}_{(-j)}^\top\overline{\bs{X}}_{(-j)}/T$, $\overline{\bs{\Sigma}}^{[-j]} = \ex[\overline{\bs{X}}_{(-j)}^\top\overline{\bs{X}}_{(-j)}]/T$ and $\bs{\Sigma}^{[-j]} = \bs{\Gamma}_{-j}\bs{\Gamma}_{-j}^\top$, respectively. Then
\begin{align*}
 & \pr\left( \big\|\widetilde{\bs{\Eig}}^{-1}\big\| > \frac{\Summable_{14}}{p}\right) = \pr\left( \lambda_{\textnormal{min}}(\widetilde{\bs{\Eig}}) < \frac{p}{\Summable_{14}}\right) = \pr\left( \min\{\widetilde{\eig}_1, \ldots, \widetilde{\eig}_K\} < \frac{p}{\Summable_{14}}\right) \\
 & \leq \pr\left( \min\{\widetilde{\eig}_1, \ldots, \widetilde{\eig}_K\} < \frac{p}{\Summable_{14}}, \|\overline{\bs{\Sigma}}^{[-j]}-\widetilde{\bs{\Sigma}}\| \leq \Summable_{13} p\sqrt{\frac{\log(p)}{n}}\right) + \summable_n^{(13)} \\
 & = \pr\bigg( \min\{\widetilde{\eig}_1, \ldots, \widetilde{\eig}_K\} < \frac{p}{\Summable_{14}}, \\
 & \phantom{\leq \pr\bigg( \ } |\overline{\eig}_k^{[-j]}-\widetilde{\eig}_k| \leq \|\overline{\bs{\Sigma}}^{[-j]}-\widetilde{\bs{\Sigma}}\| \leq \Summable_{13}p\sqrt{\frac{\log(p)}{n}} \text{ for } k=1,\ldots,K  \bigg) + \summable_n^{(13)} \\
 & \leq \pr\bigg( \min\{\widetilde{\eig}_1, \ldots, \widetilde{\eig}_K\} < \frac{p}{\Summable_{14}}, \\
 & \phantom{\leq \pr\bigg( \ } |\overline{\eig}_{k}^{[-j]}-\widetilde{\eig}_k| \leq \|\overline{\bs{\Sigma}}^{[-j]}-\widetilde{\bs{\Sigma}}\| \leq \Summable_{13}p\sqrt{\frac{\log(p)}{n}} \text{ for } k=1,\ldots,K, \\
 & \phantom{\leq \pr\bigg( \ } |\eig_{k}^{[-j]} - \overline{\eig}_{k}^{[-j]}| \leq \|\bs{\Sigma}^{[-j]} - \overline{\bs{\Sigma}}^{[-j]} \| \leq \frac{\Summable_{12} p}{n} \text{ for } k=1,\ldots,K \bigg) + \summable_n^{(13)} \\
 & = \summable_n^{(13)} =: \summable_n^{(14)} 
\end{align*}
for $\Summable_n^{(14)}$ chosen sufficiently large, where we have used assumption \ref{C:Gamma-INF} in the final line.  
\end{proof}

\begin{proof}[Proof of \ref{lemmaC:aux2:psibar1gammaUF1} and \ref{lemmaC:aux2:psibar-1gammaUF-1}]
The two statements follow by extending the arguments for Lemma \ref{lemma3:projection} in the same way as outlined in the proof of Lemma \ref{lemmaC:aux1} above. 
\end{proof}

\subsection*{Proof of Lemma \ref{lemmaC:propertiesofPiu}}

\begin{proof}[Proof of \ref{lemmaC:propertiesofPiu_itemi}]
By straightforward calculations,
\begin{align*}
\ex\big[\|\bs{\Pi} u_i\|^2\big]
 & = \ex\bigg [ \bigg\| \Big(I-\frac{\bs{F}\bs{F}^\top}{T}\Big) u_i \bigg\|^2 \bigg] \\
 & =  \ex\big[\| u_i\|^2\big] - \ex\bigg[2u_i^\top \Big(\frac{\bs{F}\bs{F}^\top}{T}\Big) u_i\bigg] + \ex\bigg[\bigg\| \Big(\frac{\bs{F} \bs{F}^\top}{T}\Big) u_i \bigg\|^2 \bigg] \\
 & = \ex\big[\| u_i\|^2\big]  - \ex\bigg[\bigg\| \Big( \frac{\bs{F} \bs{F}^\top}{T}\Big) u_i \bigg\|^2\bigg] \\
 & = \ex\big[\| u_i\|^2\big] - \frac{\ex[\| \bs{F}^\top u_i\|^2]}{T}.
\end{align*}
Moreover, since $\ex[\| u_i\|^2] = \sum_{t=1}^T \ex[u_{it}^2] = T \ex[u_{11}^2]$ and  
\begin{align*}
\frac{\ex[\| \bs{F}^\top u_i\|^2]}{T}
 & = \frac{1}{T} \sum_{k=1}^K \ex\left[\left(\sum_{t=1}^T F_{t,k} u_{it}\right)^2\right]  \\
 & = \frac{1}{T} \sum_{k=1}^K \sum_{t,t'=1}^T F_{t,k}F_{t',k} \ex\left[ u_{it}u_{it'} \right] \\
 & = \frac{1}{T} \sum_{k=1}^K \sum_{t=1}^T F_{t,k}^2 \ex[ u_{it}^2] = K \ex[u_{11}^2],  
\end{align*}
we obtain that $\ex[\|\bs{\Pi} u_i\|^2] = (T-K) \ex[u_{11}^2]$.
\end{proof}

\begin{proof}[Proof of \ref{lemmaC:propertiesofPiu-itemii}]
Since $\bs{\Pi}$ is deterministic and $u_i$ is independent and identically distributed across $i$ with sufficiently many moments, we can prove the claim by adapting the arguments for Lemma \ref{lemma:aux1} in the same way as outlined in the proof of Lemma \ref{lemmaC:aux1} above. 
\end{proof}

\subsection*{Proof of Proposition \ref{propC:tilde-pi-equals-pi-r}}

The proof extends the technical arguments for Proposition \ref{lemma2:projection}. 
Since the event $\{\overline{\bs{\Gamma}}_{-j}^\top\widetilde{\bs{U}} \ \ \text{is invertible}\}$ is a subset of the event $\{\widetilde{\bs{\Pi}} = \bs{\Pi} - \widetilde{\bs{R}}\}$, it holds that
\begin{align*}
\pr\left(\widetilde{\bs{\Pi}} \neq \bs{\Pi} -\widetilde{\bs{R}} \right) 
 & \leq \pr\left(\overline{\bs{\Gamma}}_{-j}^\top\widetilde{\bs{U}} \ \ \text{is not invertible} \right) \\
 & = \pr\left(\text{There exists an eigenvalue of } \overline{\bs{\Gamma}}_{-j}^\top\widetilde{\bs{U}} \text{ equal to } 0 \right) \\
 & = \pr\left(\text{There exists an eigenvalue of } \widetilde{\bs{U}}^\top\overline{\bs{\Gamma}}_{-j}\overline{\bs{\Gamma}}_{-j}^\top\widetilde{\bs{U}} \text{ equal to } 0\right) \\
 & \leq P_1 + P_2,
\end{align*}
where
\begin{align*}
P_1 & = \pr\bigg(\text{There exists an eigenvalue of } \widetilde{\bs{U}}^\top\overline{\bs{\Gamma}}_{-j}\overline{\bs{\Gamma}}_{-j}^\top\widetilde{\bs{U}} \text{ equal to } 0, \\
    & \phantom{\leq \pr\bigg( \ } \Big\|\widetilde{\bs{U}}^\top\overline{\bs{\Gamma}}_{-j}\overline{\bs{\Gamma}}_{-j}^\top\widetilde{\bs{U}} - \widetilde{\bs{U}}^\top\bigg(\frac{\overline{\bs{X}}_{(-j)}^\top\overline{\bs{X}}_{(-j)}}{T}\bigg)\widetilde{\bs{U}} \Big\| \leq C p\sqrt{\frac{\log(p)}{n}}\bigg) \\
P_2 & = \pr\left( \Big\| \widetilde{\bs{U}}^\top\overline{\bs{\Gamma}}_{-j}\overline{\bs{\Gamma}}_{-j}^\top\widetilde{\bs{U}} - \widetilde{\bs{U}}^\top\bigg(\frac{\overline{\bs{X}}_{(-j)}^\top\overline{\bs{X}}_{(-j)}}{T}\bigg)\widetilde{\bs{U}} \Big\| > Cp\sqrt{\frac{\log(p)}{n}}\right) 
\end{align*}
with $C$ chosen sufficiently large. To complete the proof, we show that $P_1$ and $P_2$ can both be bounded by a summable sequence $\{\summable_n\}$.

We first consider $P_1$. By \ref{C:Gamma-INF}, the eigenvalues of $\bs{\Sigma}^{[-j]} = \bs{\Gamma}_{-j}\bs{\Gamma}_{-j}^\top$ are lower bounded by $c_{\min} p$. Using Lemma \ref{lemmaC:aux2} \ref{lemmaC:aux2:sigmasigmabar} and \ref{lemmaC:aux2:sigmahatsigmabar} together with Weyl's theorem, we can infer from this that the eigenvalues $\widetilde{\eig}_1, \ldots, \widetilde{\eig}_K$ of the matrix $\widetilde{\bs{\Sigma}} = (\overline{\bs{X}}_{(-j)}^\top\overline{\bs{X}}_{(-j)})/T$ have the following property: there exists a constant $\Summable > 0$ and a summable sequence  $\{a_n\}$ such that 
\[ \pr \big( \min\{\widetilde{\eig}_1, \ldots, \widetilde{\eig}_K\} \leq \Summable p \big) \le \summable_n. \]
(Notably, this statement has already been verified in the proof of Lemma \ref{lemmaC:aux2}\ref{lemmaC:aux2:normpsi-1} above.) Since $\widetilde{\bs{\Sigma}}$ and $\widetilde{\bs{U}}^\top(\overline{\bs{X}}_{(-j)}^\top\overline{\bs{X}}_{(-j)} / T)\widetilde{\bs{U}} = \widetilde{\bs{U}}^\top \widetilde{\bs{\Sigma}} \widetilde{\bs{U}}$ have the same eigenvalues, this implies that 
\begin{align*}
P_1 
 & \le \pr\bigg(\text{There exists an eigenvalue of } \widetilde{\bs{U}}^\top\overline{\bs{\Gamma}}_{-j}\overline{\bs{\Gamma}}_{-j}^\top\widetilde{\bs{U}} \text{ equal to } 0, \\*
 & \phantom{\leq \pr\bigg( \ } \Big\|\widetilde{\bs{U}}^\top\overline{\bs{\Gamma}}_{-j}\overline{\bs{\Gamma}}_{-j}^\top\widetilde{\bs{U}} - \widetilde{\bs{U}}^\top\bigg(\frac{\overline{\bs{X}}_{(-j)}^\top\overline{\bs{X}}_{(-j)}}{T}\bigg)\widetilde{\bs{U}} \Big\| \leq C p\sqrt{\frac{\log(p)}{n}}, \\*
 & \phantom{\leq \pr\bigg( \ } \min\{\widetilde{\eig}_1, \ldots, \widetilde{\eig}_K\} > \Summable p\Bigg) 
   + \pr \big( \min\{\widetilde{\eig}_1, \ldots, \widetilde{\eig}_K\} \leq \Summable p \big) \\*
 & \le 0 + \summable_n
\end{align*}
for sufficiently large $n$, showing that $P_1$ can be bounded by a summable sequence. Turning to $P_2$, we have 
\begin{align*}
P_2   
 & \leq \pr\left( \Big\|\overline{\bs{\Gamma}}_{-j}\overline{\bs{\Gamma}}_{-j}^\top - \frac{\overline{\bs{X}}_{(-j)}^\top\overline{\bs{X}}_{(-j)}}{T} \Big\| > Cp\sqrt{\frac{\log(p)}{n}} \right) \\
 & \leq \pr\left(\|\overline{\bs{\Gamma}}_{-j}\overline{\bs{\Gamma}}_{-j}^\top - \bs{\Gamma}_{-j}\bs{\Gamma}_{-j}^\top \| > \frac{Cp}{2}\sqrt{\frac{\log(p)}{n}} \right) \\
 & \quad + \pr\left( \Big\|\bs{\Gamma}_{-j}\bs{\Gamma}_{-j}^\top - \frac{\overline{\bs{X}}_{(-j)}^\top\overline{\bs{X}}_{(-j)}}{T} \Big\| > \frac{Cp}{2}\sqrt{\frac{\log(p)}{n}} \right)\\
 & \leq \pr\left(\|(\overline{\bs{\Gamma}}_{-j}-\bs{\Gamma}_{-j} + \bs{\Gamma}_{-j})(\overline{\bs{\Gamma}}_{-j}-\bs{\Gamma}_{-j} + \bs{\Gamma}_{-j})^\top - \bs{\Gamma}_{-j}\bs{\Gamma}_{-j}^\top \|> \frac{Cp}{2} \sqrt{\frac{\log(p)}{n}} \right) \\
 & \quad + \pr\left( \Big\|\bs{\Gamma}_{-j}\bs{\Gamma}_{-j}^\top - \frac{\overline{\bs{X}}_{(-j)}^\top\overline{\bs{X}}_{(-j)}}{T} \Big\| > \frac{Cp}{2} \sqrt{\frac{\log(p)}{n}} \right)\\
 & \leq \pr\left(\|\overline{\bs{\Gamma}}_{-j}-\bs{\Gamma}_{-j}\|^2  > \frac{Cp}{4} \sqrt{\frac{\log(p)}{n}} \right) \\*
 & \quad +  \pr\left( 2\|\bs{\Gamma}_{-j}\|\|\overline{\bs{\Gamma}}_{-j}-\bs{\Gamma}_{-j}\| > \frac{Cp}{4} \sqrt{\frac{\log(p)}{n}}\right) \\
 & \quad + \pr\left( \Big\|\bs{\Gamma}_{-j}\bs{\Gamma}_{-j}^\top - \frac{\overline{\bs{X}}_{(-j)}^\top\overline{\bs{X}}_{(-j)}}{T} \Big\| > \frac{Cp}{2} \sqrt{\frac{\log(p)}{n}} \right) \\
 & \leq \pr\left(\|\overline{\bs{\Gamma}}_{-j}-\bs{\Gamma}_{-j}\|^2  > \frac{Cp}{4} \sqrt{\frac{\log(p)}{n}} \right) \\
 & \quad +  \pr\left( 2\|\bs{\Gamma}_{-j}\|\|\overline{\bs{\Gamma}}_{-j}-\bs{\Gamma}_{-j}\| > \frac{Cp}{4} \sqrt{\frac{\log(p)}{n}}\right) \\
 & \quad + \pr\left( \Big\| \bs{\Gamma}_{-j}\bs{\Gamma}_{-j}^\top - \ex\bigg[\frac{\overline{\bs{X}}_{(-j)}^\top\overline{\bs{X}}_{(-j)}}{T}\bigg] \Big\| > \frac{Cp}{4} \sqrt{\frac{\log(p)}{n}} \right)  \\
 & \quad + \pr\left( \Big\| \ex\bigg[\frac{\overline{\bs{X}}_{(-j)}^\top\overline{\bs{X}}_{(-j)}}{T}\bigg] -\frac{\overline{\bs{X}}_{(-j)}^\top\overline{\bs{X}}_{(-j)}}{T} \Big\| > \frac{Cp}{4} \sqrt{\frac{\log(p)}{n}} \right),
\end{align*}
the four terms in the upper bound being summable according to \ref{C:Gamma-INF}, Lemma \ref{lemmaC:aux2} \ref{lemmaC:aux2:sigmasigmabar} and \ref{lemmaC:aux2:sigmahatsigmabar} as well as the fact that $\pr(\|\overline{\bs{\Gamma}}_{-j} - \bs{\Gamma}_{-j}\| > C  \sqrt{p \log(p) / n} )$ is summable for $C$ large enough, which follows from \eqref{lemmaC:aux2:normbargamma:proof1}.

\subsection*{Proof of Proposition \ref{propC:max-tildeR-u}}

\begin{proof}[Proof of \ref{propC:max-tildeR-u-i}]
It holds that
\[ \pr\left(\max_{i}\| \widetilde{\bs{R}}u_i\| \geq \Summable_2 \frac{\sqrt{\log(npT)\log(np)}}{\sqrt{n}}  \right) \le P_1 + P_2 + P_3 + P_4 \]
with 
\begin{align*}
P_1 & = \pr\Bigg(\max_{i}\left\| \frac{1}{T}(\bs{F}\overline{\bs{\Gamma}}_{-j}^\top\widetilde{\bs{U}})\left[ \widetilde{\bs{\Eig}}^{ -1} - \left(\frac{1}{T}(\bs{F}\overline{\bs{\Gamma}}_{-j}^\top\widetilde{\bs{U}})^\top(\bs{F}\overline{\bs{\Gamma}}_{-j}^\top\widetilde{\bs{U}})\right)^{-1}\right](\bs{F}\overline{\bs{\Gamma}}_{-j}^\top\widetilde{\bs{U}})^\top u_i\right\| \\
 & \phantom{\leq \pr\bigg(} \geq \frac{\Summable_2 \sqrt{\log(npT)\log(np)}}{4 \sqrt{n}} \Bigg) \\
P_2 & = \pr\left(\max_{i}\left\|\frac{1}{T}(\bs{F}\overline{\bs{\Gamma}}_{-j}^\top\widetilde{\bs{U}}) \widetilde{\bs{\Eig}}^{-1} (\overline{\bs{Z}}_{(-j)}\widetilde{\bs{U}})^\top u_i\right\| \geq \frac{\Summable_2 \sqrt{\log(npT)\log(np)}}{4 \sqrt{n}}  \right) \\
P_3 & = \pr\left(\max_{i}\left\|\frac{1}{T}\overline{\bs{Z}}_{(-j)}\widetilde{\bs{U}} \widetilde{\bs{\Eig}}^{-1}  (\bs{F}\overline{\bs{\Gamma}}_{-j}^\top\widetilde{\bs{U}})^\top u_i\right\| \geq \frac{\Summable_2 \sqrt{\log(npT)\log(np)}}{4\sqrt{n}} \right) \\
P_4 & = \pr\left(\max_{i}\left\|\frac{1}{T}\overline{\bs{Z}}_{(-j)}\widetilde{\bs{U}} \widetilde{\bs{\Eig}}^{-1}  (\overline{\bs{Z}}_{(-j)}\widetilde{\bs{U}})^\top u_i\right\| \geq \frac{\Summable_2 \sqrt{\log(npT)\log(np)}}{4\sqrt{n}} \right).
\end{align*}
Using Lemmas \ref{lemmaC:aux1} and \ref{lemmaC:aux2} together with the fact that $\|\widetilde{\bs{U}}\| =1$ and $\| \bs{F}/T\| = T^{-1/2}$, we get that 
\begin{align*}
P_1 & \leq \pr\bigg(\left\| \frac{1}{T}\bs{F}\overline{\bs{\Gamma}}_{-j}^\top \right\|\left\| \widetilde{\bs{\Eig}}^{ -1} - \left(\frac{1}{T}(\bs{F}\overline{\bs{\Gamma}}_{-j}^\top\widetilde{\bs{U}})^\top\bs{F}\overline{\bs{\Gamma}}_{-j}^\top\widetilde{\bs{U}}\right)^{-1}\right\|\max_{i}\left\| \overline{\bs{\Gamma}}_{-j}\bs{F}^\top u_i\right\| \\*
    & \phantom{\leq \pr\bigg(} \geq \frac{\Summable_2\sqrt{\log(npT)\log(np)}}{4 \sqrt{n}} \bigg) \\
    & \leq \pr\bigg(\left\| \frac{\bs{F}}{T}\right\| \left\| \overline{\bs{\Gamma}}_{-j} \right\|^2\left\| \widetilde{\bs{\Eig}}^{ -1} - \left(\frac{1}{T}(\bs{F}\overline{\bs{\Gamma}}_{-j}^\top \widetilde{\bs{U}})^\top \bs{F}\overline{\bs{\Gamma}}_{-j}^\top \widetilde{\bs{U}}\right)^{-1}\right\|\max_{i}\left\| \bs{F}^\top  u_i\right\| \\
    & \phantom{\leq \pr\bigg(} \geq \frac{\Summable_2\sqrt{\log(npT)\log(np)}}{4 \sqrt{n}} \bigg) \\
    & \leq \pr\bigg(T^{-1/2} \left\| \overline{\bs{\Gamma}}_{-j} \right\|^2\left\| \widetilde{\bs{\Eig}}^{ -1} - \left(\frac{1}{T}(\bs{F}\overline{\bs{\Gamma}}_{-j}^\top \widetilde{\bs{U}})^\top \bs{F}\overline{\bs{\Gamma}}_{-j}^\top \widetilde{\bs{U}}\right)^{-1}\right\|\max_{i}\left\| \bs{F}^\top  u_i\right\| \\
    & \phantom{\leq \pr\bigg(} \geq \frac{\Summable_2 \sqrt{\log(npT)\log(np)}}{4 \sqrt{n}} \bigg) \\
    & \leq \pr\bigg(T^{-1/2} \Summable_{11}^2 p \, \frac{\Summable_{16}}{p} \left( \sqrt{\frac{\log(p)\log(npT)}{nT}} + \frac{\log(pT)}{n} \right) \Summable_6 \sqrt{ T \log(n)}  \\
    & \phantom{\leq \pr\bigg(} \geq \frac{\Summable_2 \sqrt{\log(npT)\log(np)}}{4\sqrt{n}} \bigg) + \summable_n^{(11)} + \summable_n^{(16)} + \summable_n^{(6)} \\
    & = 0 + \summable_n^{(11)} + \summable_n^{(16)} + \summable_n^{(6)}
\end{align*}
for sufficiently large $n$. Similarly,   
\begin{align*}
P_2 & \le \pr\left(\left\|\frac{1}{T}\bs{F}\overline{\bs{\Gamma}}_{-j}^\top\right\| \|\widetilde{\bs{\Eig}}^{-1}\| \max_{i}\left\|\overline{\bs{Z}}_{(-j)}^\top u_i\right\| \geq \frac{\Summable_2 \sqrt{\log(npT)\log(np)}}{4\sqrt{n}} \right) \\
    & \le \pr\left( T^{-1/2} \|\overline{\bs{\Gamma}}_{-j}\| \|\widetilde{\bs{\Eig}}^{-1}\| \max_{i}\left\|\overline{\bs{Z}}_{(-j)}^\top u_i\right\| \geq \frac{\Summable_2 \sqrt{\log(npT)\log(np)}}{4\sqrt{n}} \right)\\
    &\le \pr\left( T^{-1/2} \Summable_{11} \sqrt{p} \frac{\Summable_{14}}{p} \Summable_{9} \frac{\sqrt{pT\log(npT)\log(np)}}{\sqrt{n}} \geq \frac{\Summable_2 \sqrt{\log(npT)\log(np)}}{4\sqrt{n}} \right) \\
    & \quad + \summable_n^{(11)} + \summable_n^{(14)} + \summable_n^{(9)} \\
    & = 0 + \summable_n^{(11)} + \summable_n^{(14)} + \summable_n^{(9)}
\end{align*}
and 
\begin{align*}
P_3 & \le \pr\left(\left\|\frac{1}{T}\overline{\bs{Z}}_{(-j)}\right\| \|\widetilde{\bs{\Eig}}^{-1} \| \max_{i}\left\| \overline{\bs{\Gamma}}_{-j} \bs{F}^\top u_i\right\| \geq \frac{\Summable_2 \sqrt{\log(npT)\log(np)}}{4\sqrt{n}} \right) \\
    & \le \pr\left(\left\|\frac{1}{T}\overline{\bs{Z}}_{(-j)}\right\| \|\overline{\bs{\Gamma}}_{-j}\| \|\widetilde{\bs{\Eig}}^{-1} \| \max_{i}\left\|  \bs{F}^\top u_i\right\| \geq \frac{\Summable_2 \sqrt{\log(npT)\log(np)}}{4\sqrt{n}}  \right) \\
    & \le \pr\left(\left\|\frac{1}{T}\overline{\bs{Z}}_{(-j)}\right\| \|\overline{\bs{\Gamma}}_{-j}\| \|\widetilde{\bs{\Eig}}^{-1} \| \max_{i}\left\|  \bs{F}^\top u_i\right\| \geq \frac{\Summable_2 \sqrt{\log(npT)\log(np)}}{4\sqrt{n}} \right) \\
    & \le \pr\left(\frac{1}{T} \Summable_{5} \sqrt{\frac{pT \log(pT)}{n}} \Summable_{11} \sqrt{p} \frac{\Summable_{14}}{p} \Summable_{6} \sqrt{T \log(n)}  \geq \frac{\Summable_2 \sqrt{\log(npT)\log(np)}}{4\sqrt{n}} \right) \\
    & \quad + \summable_n^{(5)} + \summable_n^{(11)} + \summable_n^{(14)} + \summable_n^{(6)} \\
    & = 0 + \summable_n^{(5)} + \summable_n^{(11)} + \summable_n^{(14)} + \summable_n^{(6)}
\end{align*}
as well as
\begin{align*} 
P_4 & \le \pr\left( \|\overline{\bs{Z}}_{(-j)} \| \| \widetilde{\bs{\Eig}}^{-1}\| \max_{i}\left\|\frac{1}{T}\overline{\bs{Z}}_{(-j)}^\top u_i\right\| \geq \frac{\Summable_2 \sqrt{\log(npT)\log(np)}}{4\sqrt{n}} \right) \\
    & \le \pr\left( \Summable_{5} \sqrt{\frac{pT\log(pT)}{n}} \frac{\Summable_{14}}{p} \Summable_{9} \sqrt{\frac{p \log(npT)\log(np)}{nT}} \geq \frac{\Summable_2 \sqrt{\log(npT)\log(np)}}{4\sqrt{n}}  \right) \\
    & \quad + \summable_n^{(5)} + \summable_n^{(14)} + \summable_n^{(9)} \\
    & = 0 + \summable_n^{(5)} + \summable_n^{(14)} + \summable_n^{(9)}
\end{align*}
for sufficiently large $n$. From these bounds on $P_j$ ($j=1,\ldots,4$), the statement directly follows. 
\end{proof}

\begin{proof}[Proof of \ref{propC:max-tildeR-u-ii}]
We write
\[ \pr\left( \left|\frac{1}{nT}\sum_{i=1}^n \| \widetilde{\bs{\Pi}}u_i \|^2 - \frac{1}{nT}\sum_{i=1}^n \| \bs{\Pi}u_i \|^2 \right| > \varepsilon \right) \le P_< + P_> \]
with
\begin{align*}
P_< & = \pr\left( \frac{1}{nT}\sum_{i=1}^n \| \widetilde{\bs{\Pi}}u_i \|^2 < \frac{1}{nT}\sum_{i=1}^n \| \bs{\Pi}u_i \|^2 - \varepsilon \right) \\
P_> & = \pr\left( \frac{1}{nT}\sum_{i=1}^n \| \widetilde{\bs{\Pi}}u_i \|^2 > \frac{1}{nT}\sum_{i=1}^n \| \bs{\Pi}u_i \|^2 + \varepsilon \right).
\end{align*}
It holds that
\begin{align*}
P_<
 & \leq \pr\left( \frac{1}{nT}\sum_{i=1}^n \| \bs{\Pi}u_i  - \widetilde{\bs{R}}u_i \|^2 < \frac{1}{nT}\sum_{i=1}^n \| \bs{\Pi}u_i \|^2 - \varepsilon \right) + \summable_n^{(1)} \\
 & = \pr\left( \frac{1}{nT}\sum_{i=1}^n \big\{ \| \bs{\Pi}u_i \|^2  - 2(\bs{\Pi}u_i )^\top \widetilde{\bs{R}}u_i  + \| \widetilde{\bs{R}}u_i \|^2 \big\} < \frac{1}{nT}\sum_{i=1}^n \| \bs{\Pi} u_i \|^2 -\varepsilon \right) + \summable_n^{(1)} \\
 & \leq \pr\left( \frac{1}{nT}\sum_{i=1}^n 2\| u_i \| \| \widetilde{\bs{R}}u_i  \| > \varepsilon \right) + \summable_n^{(1)} \\
 & \leq \pr\left( 2\sqrt{\frac{1}{nT}\sum_{i=1}^n \| u_i \|^2}\sqrt{\frac{1}{nT}\sum_{i=1}^n \| \widetilde{\bs{R}}u_i  \|^2 } > \varepsilon \right) + \summable_n^{(1)} \\ 
 & \leq \pr\left( 2\sqrt{\frac{3}{2}\ex[u_{11}^2]}\sqrt{\frac{1}{T} \max_i  \| \widetilde{\bs{R}}u_i \|^2 } > \varepsilon \right) + \pr\left( \frac{1}{nT}\sum_{i=1}^n \| u_i \|^2 > \frac{3}{2}\ex[u_{11}^2]\right) + \summable_n^{(1)} 
\end{align*}
and
\begin{align*}
P_>
 & \le \pr\left( \frac{1}{nT}\sum_{i=1}^n \| \bs{\Pi}u_i - \widetilde{\bs{R}}u_i  \|^2 > \frac{1}{nT}\sum_{i=1}^n \| \bs{\Pi}u_i \|^2 + \varepsilon \right) + \summable_n^{(1)} \\
 & \le \pr\left( \frac{1}{nT}\sum_{i=1}^n \big\{ \| \bs{\Pi}u_i \|^2  - 2(\bs{\Pi}u_i )^\top \widetilde{\bs{R}}u_i  + \| \widetilde{\bs{R}}u_i \|^2 \big\} > \frac{1}{nT}\sum_{i=1}^n \| \bs{\Pi} u_i \|^2 +\varepsilon \right) + \summable_n^{(1)} \\
 & \le \pr\left( \frac{1}{nT}\sum_{i=1}^n  2\| \bs{\Pi}u_i \| \| \widetilde{\bs{R}}u_i \| + \frac{1}{nT}\sum_{i=1}^n \| \widetilde{\bs{R}}u_i \|^2 >  \varepsilon   \right) + \summable_n^{(1)} \\
 & \le \pr\left( 2 \sqrt{\frac{1}{nT}\sum_{i=1}^n  \| u_i \|^2} \sqrt{\frac{1}{nT}\sum_{i=1}^n \| \widetilde{\bs{R}}u_i \|^2} + \frac{1}{nT}\sum_{i=1}^n \| \widetilde{\bs{R}}u_i \|^2 > \varepsilon \right) + \summable_n^{(1)} \\
 & \le \pr\left( 2\sqrt{\frac{3}{2}\ex[u_{11}^2]}\sqrt{\frac{1}{T} \max_i  \| \widetilde{\bs{R}}u_i  \|^2 } + \frac{1}{T} \max_i  \| \widetilde{\bs{R}}u_i  \|^2 > \varepsilon \right) \\
&\quad + \pr\left( \frac{1}{nT}\sum_{i=1}^n \| u_i  \|^2 > \frac{3}{2}\ex[u_{11}^2]\right) + \summable_n^{(1)}. 
\end{align*}
These upper bounds on $P_<$ and $P_>$ are summable by part \ref{propC:max-tildeR-u-i} of Proposition \ref{propC:max-tildeR-u} and the fact that $\pr((nT)^{-1}\sum_{i=1}^n \| u_i  \|^2 > 3\ex[\sigma_{11}^2]/2)$ can be bounded by a summable sequence, which follows by slightly adapting the arguments for Lemma \ref{lemma:aux1}.
\end{proof}

\subsection*{Proof of Proposition \ref{propC:norm-tilde-pi-F}}

It holds that 
\[ \pr\left(\| \widetilde{\bs{\Pi}}\bs{F}\| > \Summable_3 \sqrt{\frac{T\log(pT)}{n}} \right) \leq \pr\left(\| \widetilde{\bs{R}} \bs{F}\| > \Summable_3 \sqrt{\frac{T\log(pT)}{n}} \right) + \summable_n^{(1)} \]
and
\[ \pr\left(\| \widetilde{\bs{R}} \bs{F}\| > \Summable_3 \sqrt{\frac{T\log(pT)}{n}} \right) \le P_1 + P_2 + P_3 + P_4 \]
with
\begin{align*}
P_1 & = \pr\bigg( \frac{\| \bs{F} \|}{T} \left\| \overline{\bs{\Gamma}}_{-j} \right\|^2\left\| \widetilde{\bs{\Eig}}^{ -1} - \left(\frac{1}{T}(\bs{F}\overline{\bs{\Gamma}}_{-j}^\top\widetilde{\bs{U}})^\top\bs{F}\overline{\bs{\Gamma}}_{-j}^\top\widetilde{\bs{U}}\right)^{-1}\right\|\left\| \bs{F}^\top \bs{F}\right\| \geq \frac{\Summable_3}{4}\sqrt{\frac{T\log(pT)}{n}} \bigg) \\
P_2 & = \pr\left(\frac{\| \bs{F} \|}{T} \|\overline{\bs{\Gamma}}_{-j}\| \|\widetilde{\bs{\Eig}}^{-1}\|\left\|\overline{\bs{Z}}_{(-j)}^\top\bs{F}\right\| \geq \frac{\Summable_3}{4} \sqrt{\frac{T\log(pT)}{n}} \right) \\ 
P_3 & = \pr\left(\left\|\frac{1}{T}\overline{\bs{Z}}_{(-j)}\right\| \|\widetilde{\bs{\Eig}}^{-1} \| \|\overline{\bs{\Gamma}}_{-j}\| \left\| \bs{F}^\top\bs{F}\right\| \geq \frac{\Summable_3}{4} \sqrt{\frac{T\log(pT)}{n}} \right) \\
P_4 & = \pr\left( \|\overline{\bs{Z}}_{(-j)} \| \| \widetilde{\bs{\Eig}}^{-1}\|   \left\|\frac{1}{T}\overline{\bs{Z}}_{(-j)}^\top \bs{F}\right\| \geq \frac{\Summable_3}{4} \sqrt{\frac{T\log(pT)}{n}}\right).
\end{align*}
Moreover, for sufficiently large $n$, 
\begin{align*}
P_1 & \leq \pr\bigg( T^{-1/2} \Summable_{11}^2 p \frac{\Summable_{16}}{p} \left( \sqrt{\frac{\log(p)\log(npT)}{nT}} + \frac{\log(pT)}{n}\right) T \geq \frac{\Summable_2}{4}\sqrt{\frac{T\log(pT)}{n}} \bigg) \\*
 & \quad + \summable_n^{(11)} + \summable_n^{(16)} \\
 & = 0 + \summable_n^{(11)} + \summable_n^{(16)}, \\[0.3cm]
P_2 & \le \pr\left( T^{-1/2} \Summable_{11}\sqrt{p} \frac{\Summable_{14}}{p} \Summable_{8} \sqrt{\frac{pT \log(npT)\log(p)}{n}} \geq \frac{\Summable_2}{4}\sqrt{\frac{T\log(pT)}{n}} \right) \\ 
 & \quad + \summable_n^{(11)} + \summable_n^{(14)} + \summable_n^{(8)} \\
 & = 0 + \summable_n^{(11)} + \summable_n^{(14)} + \summable_n^{(8)}, \\[0.3cm]
P_3 & \le \pr\left(\Summable_{5} \sqrt{\frac{p\log(pT)}{nT}} \frac{\Summable_{14}}{p} \Summable_{11}\sqrt{p} T \geq \frac{\Summable_2}{4}\sqrt{\frac{T\log(pT)}{n}} \right) \\
 &\quad + \summable_n^{(5)} + \summable_n^{(14)} + \summable_n^{(11)} \\
 & = 0 + \summable_n^{(5)} + \summable_n^{(14)} + \summable_n^{(11)}, \\[0.3cm]
P_4 & \le \pr\left( \Summable_{5} \sqrt{\frac{pT \log(pT)}{n}} \frac{\Summable_{14}}{p} \Summable_{8} \sqrt{\frac{p \log(npT)\log(p)}{nT}} \geq \frac{\Summable_2}{4}\sqrt{\frac{T\log(pT)}{n}}\right) \\
 & \quad + \summable_n^{(5)} + \summable_n^{(14)} + \summable_n^{(8)} \\
 & = 0 + \summable_n^{(5)} + \summable_n^{(14)} + \summable_n^{(8)}. \qedhere
\end{align*}

\subsection*{Proof of Proposition \ref{propC:eff-noise-nodewise}}

We consider the simple bound
\begin{align*}
\pr\left(\frac{4\|\widetilde{\bs{X}}_{(-j)}^\top w\|_\infty }{nT} > \pennw  \right)
 & \leq \pr\left( \frac{4}{nT}\max_{j'\neq j} \Bigg|\sum_{i=1}^n \widetilde{X}_{i(j')} ^\top  \bs{F}\nu_i \Bigg|  > \frac{\pennw}{2} \right) \\
 & \quad + \pr\left( \frac{4}{nT}\max_{j'\neq j} \Bigg|\sum_{i=1}^n \widetilde{X}_{i(j')} ^\top  u_i \Bigg| > \frac{\pennw}{2} \right) \\
 & \le P_1 + P_2 + P_3 + P_4,
\end{align*}
where
\begin{align*}
P_1 & = \pr\left( \frac{4}{nT}\max_{j'\neq j} \Bigg| \sum_{i=1}^n (\widetilde{\bs{\Pi}} \bs{F} \Gamma_{i,j'}^\top)^\top \bs{F}\nu_i \Bigg| > \frac{\pennw}{4} \right)\\
P_2 & = \pr\left( \frac{4}{nT}\max_{j'\neq j} \Bigg|\sum_{i=1}^n (\widetilde{\bs{\Pi}} Z_{i(j')}) ^\top  \bs{F}\nu_i \Bigg|> \frac{\pennw}{4} \right) \\
P_3 & = \pr\left( \frac{4}{nT}\max_{j'\neq j} \Bigg|\sum_{i=1}^n (\widetilde{\bs{\Pi}} \bs{F} \Gamma_{i,j'}^\top)^\top u_i \Bigg| > \frac{\pennw}{4} \right) \\
P_4 & = \pr\left( \frac{4}{nT}\max_{j'\neq j} \Bigg| \sum_{i=1}^n (\widetilde{\bs{\Pi}} Z_{i(j')}) ^\top   u_i \Bigg| > \frac{\pennw}{4} \right).
\end{align*}
To complete the proof, we show that the probabilities $P_j$ ($j=1,\ldots,4$) can be bounded by summable sequences:
\begin{enumerate}[label=(\roman*),leftmargin=0.95cm]

\item Slightly adapting the proof of Lemma \ref{lemma:aux1} shows that  
\begin{align}
 & \pr\left(\max_{1 \le j \le p} \Big| \frac{1}{n} \sum_{i=1}^n \big\{ \| \Gamma_{i,j} \|^2 - \ex \| \Gamma_{i,j} \|^2 \big\} \Big| > \Summable \sqrt{\frac{\log(p)}{n}}\right) \leq \summable_n \label{tail-bound-Gammaij2}
\end{align}
with a sufficiently large constant $\Summable > 0$ and a summable sequence $\{\summable_n\}$. Using this together with \ref{C:loadings}, \ref{C:nodewise-loadings} and Proposition \ref{propC:norm-tilde-pi-F}, we obtain that 
\begin{align*}
P_1 
 & = \pr\left( \frac{4}{nT}\max_{j'\neq j} \Bigg| \sum_{i=1}^n (\widetilde{\bs{\Pi}} \bs{F} \Gamma_{i,j'}^\top)^\top \widetilde{\bs{\Pi}}\bs{F}\nu_i \Bigg| > \frac{\pennw}{4} \right) \\
 & \leq \pr\left(\|\widetilde{\bs{\Pi}} \bs{F} \|^2 \frac{4}{nT}\max_{j'\neq j} \sum_{i=1}^n \| \Gamma_{i,j'} \|  \|\nu_i\| > \frac{\pennw}{4} \right) \\
 & \leq \pr\left(\frac{\|\widetilde{\bs{\Pi}} \bs{F} \|^2}{T} \sqrt{\frac{4}{n}\max_{j'\neq j} \sum_{i=1}^n \| \Gamma_{i,j'} \|^2} \sqrt{\frac{4}{n} \sum_{i=1}^n \|\nu_i\|^2} > \frac{\pennw}{4} \right) \leq \summable_n
\end{align*}
for a summable sequence $\{\summable_n\}$.

\item Combining the arguments used for the proof of \eqref{lemma2:Lasso:claim3} with Lemmas \ref{lemmaC:aux1} and \ref{lemmaC:aux2}, one can show that
\begin{align*}
\pr\left(\max_{i,j'} \|Z_{i(j')} ^\top \widetilde{\bs{\Pi}} \bs{F} \|> \Summable \left[ \sqrt{\frac{T\log(np^2) \log(npT)}{n}} + \frac{T}{n} \right] \right) \leq \summable_n
\end{align*}
with a summable sequence $\{\summable_n\}$ and sufficiently large $\Summable$. From this and assumption \ref{C:nodewise-loadings}, it follows that
\begin{align*}
P_2 
 & = \pr\left( \frac{4}{nT}\max_{j'\neq j} \Bigg| \sum_{i=1}^n Z_{i(j')}^\top \widetilde{\bs{\Pi}} \bs{F} \nu_i\Bigg| > \frac{\pennw}{4} \right) \\
 & \leq \pr\left(\max_{i,j'} \|Z_{i(j')}^\top \widetilde{\bs{\Pi}} \bs{F} \| \frac{4}{nT}\sum_{i=1}^n \|\nu_i\| > \frac{\pennw}{4} \right) \leq \summable_n
\end{align*}
for some summable sequence $\{\summable_n\}$.

\item Combining the arguments from the proof of \eqref{lemma2:Lasso:claim2} with Lemmas \ref{lemmaC:aux1} and \ref{lemmaC:aux2}, one can show that 
\begin{align*}
\pr\left(\max_{i} \| (\widetilde{\bs{\Pi}}\bs{F})^\top u_i \|> \Summable \sqrt{\frac{T\log(np)\log(npT)}{n}} \right) \leq \summable_n
\end{align*}
with a summable sequence $\{\summable_n\}$ and sufficiently large $\Summable$. Using this together with \eqref{tail-bound-Gammaij2}, we can conclude that 
\begin{align*}
P_3
 & \leq \pr\left(\max_{i} \|(\widetilde{\bs{\Pi}} \bs{F})^\top u_i \| \frac{4}{nT}\max_{j'\neq j} \sum_{i=1}^n \| \Gamma_{i(j')} \| > \frac{\pennw}{4} \right) \leq \summable_n
\end{align*}
for some summable sequence $\{\summable_n\}$.

\item With Proposition \ref{propC:tilde-pi-equals-pi-r}, we obtain that
\begin{align*}
P_4  
 & \leq \pr\left( \frac{4}{nT}\max_{j'\neq j} \Bigg|\sum_{i=1}^n  Z_{i(j')} ^\top (\bs{\Pi}- \widetilde{\bs{R}})  u_i\Bigg|  > \frac{\pennw}{4} \right) + \summable_n^{(1)} \\
 & \leq \pr\left( \frac{4}{nT}\max_{j'\neq j} \Bigg| \sum_{i=1}^n  Z_{i(j')} ^\top \Big\{\bs{I} - \Big(\frac{\bs{F}\bs{F}^\top}{T}\Big) - \widetilde{\bs{R}}\Big\} u_i \Bigg|  > \frac{\pennw}{4} \right) + \summable_n^{(1)} \\
 & \le P_{4a} + P_{4b} + P_{4c} + \summable_n^{(1)}
\end{align*}
with
\begin{align*}
P_{4a} & = \pr\left( \frac{4}{nT}\max_{j'\neq j} \Bigg| \sum_{i=1}^n  Z_{i(j')} ^\top u_i \Bigg| > \frac{\pennw}{12} \right) \\
P_{4b} & = \pr\left( \frac{4}{nT}\max_{j'\neq j} \Bigg| \sum_{i=1}^n  Z_{i(j')} ^\top  \Big(\frac{\bs{F}\bs{F}^\top}{T}\Big) u_i \Bigg| > \frac{\pennw}{12} \right) \\
P_{4c} & = \pr\left( \frac{4}{nT}\max_{j'\neq j} \Bigg| \sum_{i=1}^n  Z_{i(j')} ^\top  \widetilde{\bs{R}} u_i  \Bigg| > \frac{\pennw}{12} \right). 
\end{align*}
The first term $P_{4a}$ can be shown to be summable by adapting the arguments for Lemma \ref{lemma:aux1} in the same way as outlined in the proof of Lemma \ref{lemmaC:aux1}.
In order to deal with the second term $P_{4b}$, we reformulate it as 
\begin{align*}
P_{4b} = \pr\left( \frac{4}{nT}\max_{j'\neq j} \bigg|\sum_{i=1}^n \sum_{t=1}^T  w_{it} Z_{it,j'} \bigg|> \frac{\pennw}{12} \right),
\end{align*}
where $w_{it} = \{T^{-1} (\bs{F}\bs{F}^\top) u_{i}\}_t$ is the $t$-th entry of the vector $T^{-1} (\bs{F}\bs{F}^\top) u_{i}$. To prove that $P_{4b}$ is summable, one may now follow the proof strategy of Lemma \ref{lemma:aux2} with the weights $w_{it}$ and adapt this strategy analogously as in Lemma \ref{lemmaC:aux1}.
Finally, the third term $P_{4c}$ can be shown to be summable by employing the arguments from the proof of \eqref{lemma2:Lasso:claim4c} together with Lemmas \ref{lemmaC:aux1} and \ref{lemmaC:aux2}.

\end{enumerate}

\subsection*{Proof of Proposition \ref{propC:residuals-unif-lower-bounded}}

In what follows, we show that
\begin{equation}\label{propC:residuals-unif-lower-bounded-claim}
\pr\big(\{\mathcal{E}_n^{>}\}^c\big) = \pr\left( \frac{\| \widetilde{\resnw} \|^2}{nT} \leq c_\resnw \right) \le \summable_n
\end{equation}
with some summable sequence $\{\summable_n\}$. The proposition then follows from the Borel-Cantelli lemma. In order to prove \eqref{propC:residuals-unif-lower-bounded-claim}, we make use of the following facts:
\begin{enumerate}[label=(\roman*),leftmargin=0.95cm]
\item By Proposition \ref{propC:eff-noise-nodewise}, $\pr( \{\widetilde{\mathcal{T}}_\pennw^{\textnormal{node}}\}^c)$ is bounded by a summable sequence.
\item By assumption \ref{C:RE-nodewise-INF}, $\pr\left( \{\widetilde{\mathcal{T}}_{\textnormal{RE}}^{\textnormal{node}}\}^c\right)$ is bounded by a summable sequence. 
\item By Proposition \ref{propC:max-tildeR-u}\ref{propC:max-tildeR-u-ii} and Lemma \ref{lemmaC:propertiesofPiu}, there exists a summable sequence $\{\summable_n^{(\textnormal{iii})}\}$ such that for any $\varepsilon > 0$,
\begin{align*} 
\pr\left ( \bigg | \frac{1}{nT}\sum_{i=1}^n\|\widetilde{\bs{\Pi}}u_i\|^2  - \frac{T-K}{T} \ex[u_{11}^2] \bigg| \geq \varepsilon \right) \leq \summable_n^{(\textnormal{iii})}. 
\end{align*} 
\item By Proposition \ref{propC:norm-tilde-pi-F} and assumption \ref{C:loadings}, there exists a constant $\Summable_{\textnormal{iv}}$ and a summable sequence $\{\summable_n^{(\textnormal{iv})}\}$ such that
\begin{align*}
\pr\left( \frac{1}{nT}\sum_{i=1}^n \|\widetilde{\bs{\Pi}}\bs{F}\nu_i\|^2 > \frac{\Summable_{\textnormal{iv}} \log(pT)}{n} \right) \leq \summable_n^{(\textnormal{iv})}.
\end{align*}
\end{enumerate}
With these facts at hand, we obtain that
\begin{align*}
 & \pr\left( \frac{\| \widetilde{\resnw} \|^2}{nT} \leq c_\resnw \right) \\*
 & = \pr\left( \frac{1}{nT}\| \widetilde{\bs{X}}_{(-j)}(\betanw-\widetilde{\betanw}_\pennw) + \widetilde{w} \|^2 \leq c_\resnw \right) \\
 & = \pr\left( \frac{1}{nT}\| \widetilde{\bs{X}}_{(-j)}(\betanw-\widetilde{\betanw}_\pennw)\|^2 + \frac{2}{nT}(\widetilde{\bs{X}}_{(-j)}(\betanw-\widetilde{\betanw}_\pennw))^\top \widetilde{w} + \frac{1}{nT}\|\widetilde{w} \|^2 \leq c_\resnw \right) \\
 & \leq \pr\left( - \frac{2}{nT}\|\widetilde{\bs{X}}_{(-j)}(\betanw-\widetilde{\betanw}_\pennw)\| \|\widetilde{w}\| +  \frac{1}{nT}\|\widetilde{w}\|^2 \leq c_\resnw \right) \\
 & \leq \pr\left(-  2 \sqrt{\frac{4}{\phi^2} \pennw^2 \|\betanw\|_0} \frac{1}{\sqrt{nT}} \|\widetilde{w}\| +  \frac{1}{nT}\|\widetilde{w} \|^2 \leq c_\resnw \right) + \pr\left( \{\widetilde{\mathcal{T}}_\pennw ^{\textnormal{node}}\}^c\right) + \pr\left( \{\widetilde{\mathcal{T}}_{\textnormal{RE}} ^{\textnormal{node}}\}^c\right)
\end{align*}
(the last line using Proposition \ref{propC:nodewise-prediction-l1}) and
\begin{align*}
 & \pr\left(-  2 \sqrt{\frac{4}{\phi^2} \pennw^2 \|\betanw\|_0} \frac{1}{\sqrt{nT}} \|\widetilde{w}\| +  \frac{1}{nT}\|\widetilde{w} \|^2 \leq c_\resnw \right) \\
 & = \pr\bigg( -\frac{4}{\phi} \pennw \sqrt{\|\betanw\|_0} \sqrt{ \frac{1}{nT} \sum_{i=1}^n\|\widetilde{\bs{\Pi}}u_i + \widetilde{\bs{\Pi}}\bs{F}\nu_i\|^2 } + \frac{1}{nT}\sum_{i=1}^n\|\widetilde{\bs{\Pi}}u_i + \widetilde{\bs{\Pi}}\bs{F}\nu_i\|^2 \leq c_\resnw \bigg) \\
 & \leq \pr\bigg( -\frac{4}{\phi} \pennw \sqrt{\|\betanw\|_0} \sqrt{\frac{4}{nT}\sum_{i=1}^n \big\{ \|\widetilde{\bs{\Pi}}u_i \|^2 + \|\widetilde{\bs{\Pi}}\bs{F}\nu_i\|^2 \big\} } \\
 & \phantom{\leq \pr\bigg(} + \frac{1}{nT}\sum_{i=1}^n \big\{ \|\widetilde{\bs{\Pi}}u_i \|^2 - 2\| \widetilde{\bs{\Pi}}u_i \| \| \widetilde{\bs{\Pi}}\bs{F}\nu_i \| \big\} \leq c_\resnw \bigg) \\
 & \leq \pr\bigg( -\frac{4}{\phi} \pennw \sqrt{\|\betanw\|_0} \sqrt{\frac{4}{nT}\sum_{i=1}^n  \|\widetilde{\bs{\Pi}}u_i \|^2 + \frac{4}{nT}\sum_{i=1}^n  \|\widetilde{\bs{\Pi}}\bs{F}\nu_i\|^2 } \\
 & \phantom{\leq \pr\bigg(} + \frac{1}{nT}\sum_{i=1}^n\|\widetilde{\bs{\Pi}}u_i \|^2 - 2\sqrt{\frac{1}{nT}\sum_{i=1}^n \| \widetilde{\bs{\Pi}}u_i\|^2} \sqrt{\frac{1}{nT}\sum_{i=1}^n \|\widetilde{\bs{\Pi}}\bs{F}\nu_i\|^2} \leq c_\resnw \bigg) \\
 & \leq \pr\bigg( -\frac{8}{\phi} \pennw \sqrt{\|\betanw\|_0} \sqrt{ \Big\{\frac{T-K}{T} \ex[u_{11}^2] + \varepsilon \Big\} + \frac{\Summable_{\textnormal{iv}} \log(pT)}{n} } \\
 & \phantom{\leq \pr\bigg(} + \Big\{ \frac{T-K}{T} \ex[u_{11}^2] - \varepsilon \Big\} - 2\sqrt{\Big\{\frac{T-K}{T} \ex[u_{11}^2] + \varepsilon\Big\}} \sqrt{\frac{\Summable_{\textnormal{iv}} \log(pT)}{n}} \leq c_\resnw \bigg) \\
 & \quad + \summable_n^{(\textnormal{iii})} + \summable_n^{(\textnormal{iv})}. 
\end{align*}
Choosing $c_\resnw \le \ex[u_{11}^2]/2$ and noting that $\pennw \sqrt{\|\betanw\|_0} = o(1)$ by assumption \ref{C:nTp-large-INF}, the probability in the final upper bound will be exactly equal to $0$ for sufficiently large $n$. This completes the proof.



\subsection*{Proof of Proposition \ref{propC:residuals-growth}}

The proposition follows from the Borel-Cantelli lemma and the fact that 
\begin{equation}\label{propC:residuals-growth-claim}
\pr\left(\{\mathcal{E}_n^{\le}\}^c\right) \le \summable_n
\end{equation}
for some summable sequence $\{\summable_n\}$. It thus remains to show \eqref{propC:residuals-growth-claim}. By definition, 
\[ \pr(\{\mathcal{E}_n^{\le}\}^c) = \pr \left(  \| \widetilde{\resnw} \|_\infty > C_\resnw \sqrt{T} (npT)^{\frac{2+\xi}{\moments}} \right) \le P_1 + P_2 \]
with
\begin{align*}
P_1 & = \pr\left( \max_{i,t}\|\widetilde{X}_{it}\|_\infty \|\betanw - \widetilde{\betanw}_\pennw \|_1 > \frac{C_\resnw \sqrt{T} (npT)^{\frac{2+\xi}{\moments}}}{2} \right) \\ 
P_2 & = \pr\left( \max_{i}\|\widetilde{\bs{\Pi}}u_i +  \widetilde{\bs{\Pi}}\bs{F}\nu_i\|_\infty > \frac{C_\resnw \sqrt{T} (npT)^{\frac{2+\xi}{\moments}}}{2} \right). 
\end{align*}
To complete the proof, we verify that $P_1$ and $P_2$ can be bounded by summable sequences: 
\begin{enumerate}[label=(\roman*),leftmargin=0.95cm]

\item By \eqref{eq:conv-Tlambda-Tre-nodewise}, $\pr( \{\widetilde{\mathcal{T}}_\pennw^{\textnormal{node}} \cap \widetilde{\mathcal{T}}_{\textnormal{RE}}^{\textnormal{node}}\}^c) \le \summable_n^{(4)}$, where $\{\summable_n^{(4)}\}$ is summable. Using this together with Proposition \ref{propC:eff-noise-nodewise}, we obtain that
\begin{align*}
P_1
 & \le \pr\left( \max_{i,t}\|\widetilde{X}_{it}\|_\infty \|\betanw - \widetilde{\betanw}_\pennw \|_1 > \frac{C_\resnw \sqrt{T} (npT)^{\frac{2+\xi}{\moments}}}{2},\widetilde{\mathcal{T}}_\pennw^{\textnormal{node}} \cap \widetilde{\mathcal{T}}_{\textnormal{RE}}^{\textnormal{node}} \right) \\
 & \quad + \pr\left( \{\widetilde{\mathcal{T}}_\pennw^{\textnormal{node}} \cap \widetilde{\mathcal{T}}_{\textnormal{RE}}^{\textnormal{node}}\}^c \right) \\
 & \le \pr\left( \max_{i,t}\|\widetilde{X}_{it}\|_\infty \Big\{ \frac{4}{\phi^2} \pennw \|\betanw\|_0 \Big\} > \frac{C_\resnw \sqrt{T} (npT)^{\frac{2+\xi}{\moments}}}{2}, \widetilde{\mathcal{T}}_\pennw^{\textnormal{node}} \cap \widetilde{\mathcal{T}}_{\textnormal{RE}}^{\textnormal{node}} \right) + \summable_n^{(4)} \\
 & \le \pr\left( \max_{i,t}\|\widetilde{X}_{it}\|_\infty > \frac{C_\resnw \sqrt{T} (npT)^{\frac{2+\xi}{\moments}}}{2}, \widetilde{\mathcal{T}}_\pennw^{\textnormal{node}} \cap \widetilde{\mathcal{T}}_{\textnormal{RE}}^{\textnormal{node}} \right) + \summable_n^{(4)} \\
 & \le \pr\left( \max_{i,t}\|\widetilde{X}_{it}\|_\infty > \frac{C_\resnw \sqrt{T} (npT)^{\frac{2+\xi}{\moments}}}{2} \right) + \summable_n^{(4)}
\end{align*}
for sufficiently large $n$, where the third inequality uses that $\pennw \|\betanw\|_0 = o(1)$ by assumption \ref{C:nTp-large-INF}. Moreover, since $\| \widetilde{\bs{\Pi}}\|=1$ and $\| x \|_\infty \leq \| x \| \leq  \sqrt{T }\| x \|_\infty$ for any $x \in \reals^T$,  
\begin{align*}
 & \pr\left( \max_{i,t}\|\widetilde{X}_{it}\|_\infty > \frac{C_\resnw \sqrt{T}  (npT)^{\frac{2+\xi}{\moments}}}{2} \right) \\
 & = \pr\left(\max_{i,j}\| \widetilde{X}_{i(j)}\|_\infty> \frac{C_\resnw \sqrt{T} (npT)^{\frac{2+\xi}{\moments}}}{2} \right) \\
 & \leq \pr\left(\| \widetilde{\bs{\Pi}}\bs{F} \| \max_{i,j}\| \Gamma_{i,j} \|> \frac{C_\resnw \sqrt{T} (npT)^{\frac{2+\xi}{\moments}}}{4} \right) \\
 & \quad + \pr\left(\max_{i,j}\| \widetilde{\bs{\Pi}}Z_{i(j)} \|_\infty > \frac{C_\resnw \sqrt{T} (npT)^{\frac{2+\xi}{\moments}}}{4} \right)\\
 & \leq \pr\left(\| \widetilde{\bs{\Pi}}\bs{F} \| \sqrt{K}\max_{i,j,k}|\Gamma_{i,jk}| > \frac{C_\resnw \sqrt{T} (npT)^{\frac{2+\xi}{\moments}}}{4} \right) \\*
 & \quad + \pr\left(\max_{i,j,t}|Z_{it,j}| > \frac{C_\resnw (npT)^{\frac{2+\xi}{\moments}}}{4} \right). 
\end{align*}
As $\ex[\max_{i,j,t}|Z_{it,j}|^{\moments}] \le \sum_{i,j,t} \ex[|Z_{it,j}|^{\moments}] \le C (npT)$ by \ref{C:Z},  
\begin{align}
\pr\left( \max_{i,j,t}|Z_{it,j}|> \frac{C_\resnw (npT)^{\frac{2+\xi}{\moments}}}{4} \right) \nonumber 
 & \leq \frac{4^\moments}{C^{\moments}_\resnw(npT)^{2+\xi}}\ex[\max_{i,j,t}|Z_{it,j}|^{\moments}] \nonumber \\
 & \leq \frac{4^\moments C}{C^{\moments}_\resnw}(npT)^{-(1+\xi)} \le \summable_n \label{eq:simple-maximal-ineq}
\end{align}
with some summable sequence $\{\summable_n\}$ under our conditions on $n$, $T$ and $p$. Using Proposition \ref{propC:norm-tilde-pi-F}, we further get that
\begin{align*}
 & \pr\left(\| \widetilde{\bs{\Pi}}\bs{F} \| \sqrt{K}\max_{i,j,k}|\Gamma_{i,jk}| > \frac{C_\resnw \sqrt{T} (npT)^{\frac{2+\xi}{\moments}}}{4} \right) \\
 & \le \pr\left( \Summable_3 \sqrt{\frac{T\log(pT)}{n}} \sqrt{K}\max_{i,j,k}|\Gamma_{i,jk}| > \frac{C_\resnw \sqrt{T} (npT)^{\frac{2+\xi}{\moments}}}{4} \right) + \summable_n^{(3)} \\
 & = \pr\left( \max_{i,j,k}|\Gamma_{i,jk}| > \frac{C_\resnw}{4\Summable_3\sqrt{K}} \sqrt{\frac{n}{\log(pT)}} (npT)^{\frac{2+\xi}{\moments}} \right) + \summable_n^{(3)},
\end{align*}
where the probability in the last line is summable by \ref{C:loadings} and the same arguments as used in \eqref{eq:simple-maximal-ineq}. Putting everything together, we arrive at the desired result that $P_1$ is bounded by a summable sequence.

\item We have
\begin{align*}
P_2 
 & = \pr\left( \max_{i}\|\widetilde{\bs{\Pi}}u_i +  \widetilde{\bs{\Pi}}\bs{F}\nu_i\|_\infty > \frac{C_\resnw \sqrt{T} (npT)^{\frac{2+\xi}{\moments}}}{2} \right)\\
 & \leq \pr\left( \sqrt{T}\max_{i,t} |u_{it}|  > \frac{C_\resnw \sqrt{T} (npT)^{\frac{2+\xi}{\moments}}}{4} \right) \\
 & \quad + \pr\left(\| \widetilde{\bs{\Pi}}\bs{F} \| \sqrt{T}\max_{i,k} |\nu_{ik}| > \frac{C_\resnw \sqrt{T} (npT)^{\frac{2+\xi}{\moments}}}{4} \right). 
\end{align*}
Using Proposition \ref{propC:norm-tilde-pi-F} and proceeding analogously as in \eqref{eq:simple-maximal-ineq} shows that the two probabilities on the right-hand side are dominated by a summable sequence.

\end{enumerate}

\subsection*{Proof of Proposition \ref{propC:error-variance}}

Rewriting the estimator $\widetilde{\sigma}_\varepsilon^2$ leads to
\begin{align*}
\widetilde{\sigma}_\varepsilon^2 
 & = \frac{1}{n(T-K)} \sum_{i=1}^n\|\widetilde{\bs{\Pi}} \bs{X}_i(\beta - \widetilde{\beta}_\pen) \|^2 \\
 & \quad - \frac{2}{n(T-K)} \sum_{i=1}^n  (\widetilde{\bs{\Pi}} \bs{X}_i(\beta - \widetilde{\beta}_\pen))^\top (\widetilde{\bs{\Pi}}\bs{F}\gamma_i + \widetilde{\bs{\Pi}}\varepsilon_i) \\
 & \quad + \frac{1}{n(T-K)} \sum_{i=1}^n\| \widetilde{\bs{\Pi}}\bs{F}\gamma_i + \widetilde{\bs{\Pi}}\varepsilon_i \|^2. 
\end{align*}
In what follows, we show that 
\begin{enumerate}[label=(\roman*),leftmargin=0.95cm]
\item $\displaystyle{\frac{1}{n(T-K)} \sum_{i=1}^n\|\widetilde{\bs{\Pi}} \bs{X}_i(\beta - \widetilde{\beta}_\pen) \|^2  = o_p(1) }$  \label{Lemma:PROOF:Errorvarianceeq1}
\item $\displaystyle{\frac{1}{n(T-K)} \sum_{i=1}^n\| (\widetilde{\bs{\Pi}}\bs{F}\gamma_i + \widetilde{\bs{\Pi}}\varepsilon_i) \|^2 \convp \ex[\varepsilon_{11}^2] = \sigma_\varepsilon^2}$ \label{Lemma:PROOF:Errorvarianceeq3}
\item $\displaystyle{\frac{2}{n(T-K)} \sum_{i=1}^n (\widetilde{\bs{\Pi}} \bs{X}_i(\beta - \widetilde{\beta}_\pen))^\top (\widetilde{\bs{\Pi}}\bs{F}\gamma_i + \widetilde{\bs{\Pi}}\varepsilon_i)=o_p(1) }$. \label{Lemma:PROOF:Errorvarianceeq2}
\end{enumerate}
This completes the proof.

\begin{proof}[Proof of \ref{Lemma:PROOF:Errorvarianceeq1}]
The claim is a direct consequence of \eqref{eq:conv-Tlambda-Tre-nodewise} and Proposition \ref{propC:Lasso}.
\end{proof}

\begin{proof}[Proof of \ref{Lemma:PROOF:Errorvarianceeq3}]
Using Proposition \ref{propC:norm-tilde-pi-F}, we can show that 
\[\frac{1}{n(T-K)} \sum_{i=1}^n  \|\widetilde{\bs{\Pi}}\bs{F}\gamma_i\|^2 = O_p\left(\frac{\log(pT)}{n} \right)=o_p(1). \] 
Moreover, Proposition \ref{propC:max-tildeR-u}\ref{propC:max-tildeR-u-ii} and Lemma \ref{lemmaC:propertiesofPiu} (with $u_i$ replaced by $\varepsilon_i$) yield that 
\[\frac{1}{n(T-K)} \sum_{i=1}^n  \| \widetilde{\bs{\Pi}}\varepsilon_i\|^2  \convp \ex[\varepsilon_{11}^2].\] 
From this and straightforward calculations, it follows that 
\begin{align*}
\frac{1}{n(T-K)} \sum_{i=1}^n\| (\widetilde{\bs{\Pi}}\bs{F}\gamma_i + \widetilde{\bs{\Pi}}\varepsilon_i) \|^2
 & \leq \frac{1}{n(T-K)} \sum_{i=1}^n \|\widetilde{\bs{\Pi}}\bs{F}\gamma_i\|^2 \\*
 & \quad + \frac{2}{n(T-K)} \sum_{i=1}^n (\widetilde{\bs{\Pi}}\bs{F}\gamma_i)^\top \widetilde{\bs{\Pi}}\varepsilon_i  \\*
 & \quad +  \frac{1}{n(T-K)} \sum_{i=1}^n \| \widetilde{\bs{\Pi}}\varepsilon_i\|^2 \\*
 & = \ex[\varepsilon_{11}^2] + o_p(1). \qedhere
\end{align*}
\end{proof}

\begin{proof}[Proof of \ref{Lemma:PROOF:Errorvarianceeq2}]
Applying the Cauchy-Schwarz inequality, we obtain that
\begin{align*}
 & \frac{2}{n(T-K)} \sum_{i=1}^n  (\widetilde{\bs{\Pi}} \bs{X}_i(\beta - \widetilde{\beta}_\pen))^\top (\widetilde{\bs{\Pi}}\bs{F}\gamma_i + \widetilde{\bs{\Pi}}\varepsilon_i) \\
 & \leq \sqrt{\frac{2}{n(T-K)} \sum_{i=1}^n  \|\widetilde{\bs{\Pi}} \bs{X}_i(\beta - \widetilde{\beta}_\pen)\|^2}\sqrt{\frac{2}{n(T-K)} \sum_{i=1}^n  \|\widetilde{\bs{\Pi}}\bs{F}\gamma_i + \widetilde{\bs{\Pi}}\varepsilon_i\|^2} \\
 & = o_p(1) \, O_p(1) = o_p(1),
\end{align*}
using \ref{Lemma:PROOF:Errorvarianceeq1} and \ref{Lemma:PROOF:Errorvarianceeq3}.
\end{proof}

\def\theequation{B'.\arabic{equation}}
\setcounter{equation}{0}
\section*{B' \hspace{0.1cm} Proof of Theorem \ref{theo:normality}(b)}

The overall structure of the proof is the same as for the large-T-case in Appendix B. Part of the propositions and lemmas, however, that are derived along the way need to be modified. Nevertheless, their proofs are analogous to those in the large-$T$-case and thus omitted. In what follows, we give a short summary over the main steps of the proof, highlighting noteworthy differences.

\subsection*{Step 1: Analysis of $\widetilde{\bs{\Pi}}$}

We consider the same decomposition as in Appendix B, that is, 
\begin{align*} 
\widetilde{\bs{\Pi}} 
 & = \bigg\{ \bs{I} - \frac{1}{T} (\bs{F} \overline{\bs{\Gamma}}_{-j}^\top \widetilde{\bs{U}}) \Big[ \frac{1}{T} (\bs{F} \overline{\bs{\Gamma}}_{-j}^\top \widetilde{\bs{U}})^\top (\bs{F} \overline{\bs{\Gamma}}_{-j}^\top \widetilde{\bs{U}}) \Big]^{-1} (\bs{F} \overline{\bs{\Gamma}}_{-j}^\top \widetilde{\bs{U}})^\top \bigg\} - \widetilde{\bs{R}}.
\end{align*}
As in Proposition \ref{propC:tilde-pi-equals-pi-r}, we can show that 
\[ \pr\left( \widetilde{\bs{\Pi}} \neq \bs{\Pi} - \bs{\widetilde{R}} \right) \leq \summable_n \]
with some summable sequence $\{\summable_n\}$. Moreover, Propositions \ref{propC:max-tildeR-u} and \ref{propC:norm-tilde-pi-F} can be adapted as follows. 
\setcounter{propBprime}{1}
\begin{propBprime}\label{propCPRIME:max-tildeR-u}
There exists a non-negative summable sequence $\{\summable_n\}$ with the following properties:
\begin{enumerate}[label=(\roman*),leftmargin=0.975cm,topsep=0.5cm]
\item \label{propCPRIME:max-tildeR-u-i} For $\Summable > 0$ sufficiently large and $\xi > 0$ arbitrarily small but fixed,
\begin{align*}
\pr\left(\max_{i}\| \widetilde{\bs{R}}u_i\| \geq \Summable \sqrt{\frac{\log(p)}{n}}(np)^{\frac{2+\xi}{\moments}} \right) \leq \summable_n.
\end{align*}    
\item \label{propCPRIME:max-tildeR-u-ii} For any $\varepsilon>0$,
\begin{align*}
\pr\left(  \left|\frac{1}{nT}\sum_{i=1}^n \| \widetilde{\bs{\Pi}}u_i\|^2 - \frac{1}{nT}\sum_{i=1}^n \| \bs{\Pi}u_i\|^2 \right| >\varepsilon \right) \leq \summable_n.
\end{align*}
\end{enumerate}
\end{propBprime}
\begin{propBprime}\label{propCPRIME:norm-tilde-pi-F}
There exist a constant $\Summable > 0$ and a non-negative summable sequence $\{\summable_n\}$ such that
\begin{align*}
\pr\left( \|\widetilde{\bs{\Pi}}\bs{F}\| > \Summable \sqrt{\frac{\log(p)}{n}} \right) \leq \summable_n.
\end{align*}
\end{propBprime}

\subsection*{Step 2: Analysis of the lasso}

Proposition \ref{propC:Lasso} and its proof remain completely unchanged. Hence, on the event $\widetilde{\mathcal{T}}_\pen \cap \widetilde{\mathcal{T}}_{\textnormal{RE}}$, we have     
\begin{align*}
\| \widetilde{\beta}_\pen - \beta \|_1 & \le \frac{4}{\phi^2} \pen \|\beta\|_0 \\
\frac{1}{nT} \| \widetilde{\bs{X}} (\widetilde{\beta}_\pen - \beta) \|^2 & \le \frac{4}{\phi^2} \pen^2 \|\beta\|_0.    
\end{align*}  
Moreover, the same arguments as in Section A' above yield that $\pr(\widetilde{\mathcal{T}}_\pen \cap \widetilde{\mathcal{T}}_{\textnormal{RE}}) \to 1$ for our choice of $\lambda$.

\subsection*{Step 3: Analysis of the nodewise lasso}

Analogous arguments as in the proof of Proposition \ref{propC:eff-noise-nodewise} give the following: 
there exist a positive constant $C_\pennw$ and a summable sequence $\{\summable_n\}$ such that
\KOM{[Double-check!]}
\begin{align*}
\pr\left(\frac{4\|\widetilde{\bs{X}}_{(-j)}^\top w\|_\infty }{nT} > C_\pennw  \sqrt{\frac{\log(p)}{n}}(np)^{\frac{4+2\xi}{\moments}} \right) \leq \summable_n
\end{align*}
for sufficiently large $n$. Choosing $\pennw =  C_\pennw\sqrt{\log(p)/n}(np)^{(4+2\xi)/\moments}$, this implies that $\pr(\widetilde{\mathcal{T}}_\pennw^{\textnormal{node}}) \ge 1 - \summable_n$. As $\pr(\mathcal{T}_{\textnormal{RE}}^{\textnormal{node}}) \ge 1 - \summable_n$ with a summable sequence $\{ \summable_n \}$ by \ref{C:RE-nodewise-INF}, we obtain that
\begin{equation*}
\pr(\widetilde{\mathcal{T}}_\pen^{\textnormal{node}} \cap \widetilde{\mathcal{T}}_{\textnormal{RE}}^{\textnormal{node}}) \ge 1 - \summable_n
\end{equation*}
with $\{\summable_n\}$ summable. Moreover, exactly the same arguments as in the large-$T$-case show that on the event 
$\mathcal{T}_\pennw^{\textnormal{node}} \cap \mathcal{T}_{\textnormal{RE}}^{\textnormal{node}}$,
\begin{align*}
\| \widetilde{\betanw}_\pennw - \betanw \|_1 & \le \frac{4}{\phi^2} \pennw \|\betanw\|_0 \\
\frac{1}{nT} \| \widetilde{\bs{X}}_{(-j)} (\widetilde{\betanw}_\pennw - \betanw) \|^2 & \le \frac{4}{\phi^2} \pennw^2 \|\betanw\|_0.    
\end{align*}  
We next analyze the nodewise lasso residuals $\widetilde{\resnw}$. Propositions \ref{propC:residuals-unif-lower-bounded} and \ref{propC:residuals-growth} carry over to the small-$T$-case almost unchanged. In particular, we can show the following.
\setcounter{propBprime}{6}
\begin{propBprime}\label{propCPRIME:residuals-unif-lower-bounded}
The event 
\begin{align*}
\mathcal{E}_n^{>} = \left\{ \frac{\| \widetilde{\resnw} \|^2}{nT} > c_\resnw \right\}
\end{align*}
with $c_\resnw > 0$ sufficiently small has the property that
\begin{align*}
\pr\left(\bigcup_{m=1}^\infty \bigcap_{n\geq m} \mathcal{E}_n^{>} \right) = 1.
\end{align*}
\end{propBprime}
\begin{propBprime}\label{propCPRIME:residuals-growth}
For sufficiently large $C_\resnw$, the event 
\begin{align*}
\mathcal{E}_n^{\le} = \left\{ \| \widetilde{\resnw} \|_\infty \leq C_\resnw (np)^{\frac{2+\xi}{\moments}} \right\}
\end{align*}
has the property that 
\begin{align*}
\pr\left(\bigcup_{m=1}^\infty \bigcap_{n\geq m} \mathcal{E}_n^{\le} \right) =1.
\end{align*}
\end{propBprime}

\subsection*{Steps 4-6: Decomposition of the desparsified lasso $\bs{\widetilde{b}_j}$ and ana\-lysis of the different components}

In this central part of the proof, we proceed as in the large-$T$-case: we consider the decomposition 
\begin{align*}
\frac{\widetilde{\resnw}^\top \widetilde{X}_{(j)}}{\|\widetilde{\resnw}\|}(\widetilde{b}_{j} - \beta_j) = \widetilde{\Upsilon}_A + \widetilde{\Upsilon}_B + \widetilde{\Upsilon}_C
\end{align*}
with $\widetilde{\Upsilon}_\ell$ for $\ell=A,B,C$ as before and show that 
$\widetilde{\Upsilon}_A = o_p(1)$,
$\widetilde{\Upsilon}_B = o_p(1)$ and
$\widetilde{\Upsilon}_C \convd \normal(0,\sigma_\varepsilon^2)$. The proofs proceed analogously as in the large-$T$-case.

\subsection*{Step 7: Consistent estimation of $\bs{\sigma_\varepsilon^2}$}

We can use the same estimator $\widetilde{\sigma}_{\varepsilon}^2$ as in the large-$T$-case, which can be shown to be consistent by analogous arguments as before.

\pagebreak
\subsection*{Step 8: Proof of intermediate results}

We finally state versions of Lemmas \ref{lemmaC:aux1}--\ref{lemmaC:propertiesofPiu} for the small-$T$-case.

\begin{lemmaBprime}\label{lemmaCPRIME:aux1}
There exist a positive constant $\Summable$ and a non-negative summable sequence $\{\summable_n\}$ such that for sufficiently large $n$ and $\xi > 0$ arbitrarily small but fixed,
\begin{enumerate}[label=(\roman*),leftmargin=0.975cm,topsep=0.5cm]
\item \label{lemmaCPRIME:aux1:normbarZ-j} 
$\displaystyle{ \pr\left( \left\| \overline{\bs{Z}}_{(-j)} \right\| > \Summable \sqrt{\frac{p\log(p)}{n}} \right) \leq \summable_n}$

\item \KOM{[Doublecheck!]}\label{lemmaCPRIME:aux1:maxFui}
$\displaystyle{\pr\left(\max_{1 \le i \le n} \left\| \frac{\bs{F}^\top u_i}{T} \right\| > \Summable n^{(2+\xi)/\moments} \right) \leq \summable_n}$
\item \KOM{[Doublecheck!]} \label{lemmaCPRIME:aux1:maxnormZijF}
$\displaystyle{ \pr\Bigg( \max_{\substack{1 \le i \le n \\ 1 \le j' \le n, j'\neq j}} \bigg\| \frac{Z_{i(j')}^\top \bs{F}}{T} \bigg\| > \Summable (np)^{(2+\xi)/\moments} \Bigg) \leq \summable_n}$
\item \label{lemmaCPRIME:aux1:normbarZF}
$\displaystyle{\pr\left( \bigg\| \frac{\bs{F}^\top \overline{\bs{Z}}_{(-j)}}{T} \bigg\| > \Summable \sqrt{\frac{p\log(p)}{n}} \right) \leq \summable_n}$ 
\item \KOM{[Doublecheck!]} \label{lemmaCPRIME:aux1:normbarZui}
$\displaystyle{\pr\left(\max_{1 \le i \le n} \bigg\| \frac{\overline{\bs{Z}}_{(-j)}^\top u_i}{T} \bigg\| > \Summable n^{(2+\xi)/\moments}\sqrt{\frac{p\log(p)}{n}} \right) \leq \summable_n}$
\item \KOM{[Doublecheck!]} \label{lemmaCPRIME:aux1:maxnormZijbarZ}
$\displaystyle{\pr\Bigg( \max_{\substack{1 \le i \le n \\ 1 \le j' \le n, j'\neq j}} \bigg\|\frac{Z_{i(j')}^\top \overline{\bs{Z}}_{(-j)}}{T} \bigg\| > \Summable (np)^{(2+\xi)/\moments}\sqrt{\frac{p\log(p)}{n}} \Bigg) \leq \summable_n}$.            
\end{enumerate}
\end{lemmaBprime}

\begin{lemmaBprime}\label{lemmaCPRIME:aux2}
There exist a positive constant $\Summable$ and a non-negative summable sequence $\{\summable_n\}$ such that for sufficiently large $n$,
\begin{enumerate}[label=(\roman*),leftmargin=0.975cm,topsep=0.5cm]
\item \label{lemmaCPRIME:aux2:normbargamma}
$\displaystyle{\pr\left(\|\overline{\bs{\Gamma}}_{-j}\| > \Summable \sqrt{p} \right) \leq \summable_n}$
\item \label{lemmaCPRIME:aux2:sigmasigmabar}
$\displaystyle{\|\bs{\Sigma}^{[-j]}-\overline{\bs{\Sigma}}^{[-j]}\| \le \frac{\Summable p}{n}}$

\item \label{lemmaCPRIME:aux2:sigmahatsigmabar}
$\displaystyle{\pr\left(\|\overline{\bs{\Sigma}}^{[-j]}-\widetilde{\bs{\Sigma}}\| > \Summable p\sqrt{\frac{\log(p)}{n}} \right) \leq \summable_n}$ 
\item \label{lemmaCPRIME:aux2:normpsi-1}
$\displaystyle{\pr\left(\big\| \widetilde{\bs{\Eig}}^{-1} \big\| > \frac{\Summable}{p}\right) \leq \summable_n}$   
\item \label{lemmaCPRIME:aux2:psibar1gammaUF1}
$\displaystyle{\pr\Bigg(\bigg\| \widetilde{\bs{\Eig}} - \left(\frac{1}{T}(\bs{F}\overline{\bs{\Gamma}}_{-j}^\top\widetilde{\bs{U}})^\top\bs{F}\overline{\bs{\Gamma}}_{-j}^\top\widetilde{\bs{U}}\right)\bigg\| > \Summable p\sqrt{\frac{\log(p)}{n}} \Bigg) \leq \summable_n}$
\item \label{lemmaCPRIME:aux2:psibar-1gammaUF-1}
$\displaystyle{\pr\Bigg(\bigg\| \widetilde{\bs{\Eig}}^{ -1} - \left(\frac{1}{T}(\bs{F}\overline{\bs{\Gamma}}_{-j}^\top\widetilde{\bs{U}})^\top \bs{F}\overline{\bs{\Gamma}}_{-j}^\top\widetilde{\bs{U}}\right)^{-1} \bigg\| > \frac{\Summable}{p}\sqrt{\frac{\log(p)}{n}} \Bigg) \leq \summable_n}$.
\end{enumerate}
\end{lemmaBprime}

\pagebreak
\begin{lemmaBprime}\label{lemmaCPRIME:propertiesofPiu}
\hfill
\begin{enumerate}[label=(\roman*),leftmargin=0.975cm]
\item \label{lemmaCPRIME:propertiesofPiu_itemi}
It holds that $\ex\left[ \|\bs{\Pi} u_i \|^2 \right] = (T - K)\ex[u_{11}^2] $.
\item \label{lemmaCPRIME:propertiesofPiu-itemii} 
For any $\varepsilon>0$, there exists a non-negative summable sequence $\{\summable_n\}$ such that 
\begin{align*}
\pr\left( \left|\frac{1}{nT}\sum_{i=1}^n (\| \bs{\Pi}u_i \|^2 -\ex\left[ \|\bs{\Pi} u_i \|^2 \right]) \right| >\varepsilon   \right) \leq \summable_n.
\end{align*}
\end{enumerate}
\end{lemmaBprime}

\def\theequation{C.\arabic{equation}}
\setcounter{equation}{0}
\section*{C \hspace{0.1cm} Results on identification}

\begin{proof}[Proof of Theorem \ref{lemma:id-K-large}.]
We only give the proof for the large-$T$-case as the arguments for the small-$T$-case are completely analogous.
By Proposition \ref{lemma3:eigenstructure}, the eigenvalues $\overline{\eig}_1 \ge \ldots \ge \overline{\eig}_p \ge 0$ of the $p \times p$ matrix $\overline{\bs{\Sigma}} = \ex[ T^{-1} \sum_{t=1}^T \overline{X}_t \overline{X}_t^\top ]$ have the following property for sufficiently large $n$: there exists a constant $c_0 > 0$ such that 
\[ \overline{\eig}_k \ge c_0 \, p \quad \text{for all } k \le K, \]
whereas $\overline{\eig}_k = O(p/n)$ for all $k > K$. From this, it immediately follows that
\[ K = \sum_{j=1}^p \ind\Big( \overline{\eig}_j > \frac{c_0 \, p}{2} \Big) \]
for sufficiently large $n$. Hence, $K$ can be expressed as a function of the eigenvalues of the matrix $\ex[ T^{-1} \sum_{t=1}^T \overline{X}_t \overline{X}_t^\top ]$, which is uniquely determined by the data. As a result, $K$ is identified.
\end{proof}

\begin{proof}[Proof of Theorem \ref{lemma:id-beta-large}.]
Once again, we only give the proof for the large-$T$-case, the one for the small-$T$-case being essentially the same. 
The proof is by contradiction. Suppose there are two $s$-sparse vectors $\beta$ and $\beta^\prime$ with active sets $S$ and $S^\prime$, respectively, that satisfy model \eqref{eq:model-CCE}. Since 
\begin{align*}
\ex \big\| Y^\perp - \bs{X}^\perp b \big\|^2
 & = (\beta - b)^\top \ex \big[ (\bs{X}^\perp)^\top \bs{X}^\perp \big] (\beta - b) + \ex\big[ (\varepsilon^\perp)^\top \varepsilon^\perp \big]
\end{align*}
for any $b \in \reals^p$, $\beta$ minimizes the function $Q(b) := \ex \| Y^\perp - \bs{X}^\perp b \|^2$. By the same argument, $\beta^\prime$ must be a minimizer of $Q(b)$ as well, which implies that 
\begin{equation}\label{eq:proof-lemma-id-K-to-be-contradicted}
(\beta - \beta^\prime)^\top \ex \big[ (\bs{X}^\perp)^\top \bs{X}^\perp \big] (\beta - \beta^\prime) = 0. 
\end{equation}
Since $\beta - \beta^\prime$ is a $2s$-sparse vector whose active set is contained in $I:= S \cup S^\prime$, we get by \ref{C:id3} that 
\[ \| \bs{X}^\perp (\beta - \beta^\prime) \|^2 \ge \frac{nT \varphi^2}{2s} \| \beta_{I} - \beta_{I}^\prime \|_1^2 > 0 \]
with probability $\ge 1 - c_{n,T}$. Hence, for sufficiently large sample sizes, 
\[ \ex \| \bs{X}^\perp (\beta - \beta^\prime) \| ^2 = (\beta - \beta^\prime)^\top \ex \big[ (\bs{X}^\perp)^\top \bs{X}^\perp \big] (\beta - \beta^\prime) > 0, \]
which contradicts \eqref{eq:proof-lemma-id-K-to-be-contradicted}.
\end{proof}

\begin{lemmaC}\label{lemma:RE-Z-large-T}
Consider the large-$T$-case and let \ref{C:F}--\ref{C:mixing}, \ref{C:nTp-large}--\ref{C:K-large} and \ref{C:id1}--\ref{C:id2} be satisfied. Moreover, let $\varphi > 0$ be a fixed constant and $\{c_{n,T}\}$ a sequence of non-negative numbers with $c_{n,T} \to 0$. If the matrix $\bs{Z}$ has the property that 
\[ \pr\Big( \bs{Z} \text{ fulfills } \textnormal{RE}(I,\varphi) \text { for all } I \subseteq \{1,\ldots,p\} \text{ with } |I| \le 2s\Big) \ge 1 - c_{n,T}, \]
then the matrix $\bs{X}^\perp$ is such that 
\[ \pr\Big( \bs{X}^\perp \text{ fulfills } \textnormal{RE}(I,\phi) \text { for all } I \subseteq \{1,\ldots,p\} \text{ with } |I| \le 2s\Big) \ge 1 - c_{n,T} \]
with $\phi=\varphi/\sqrt{2}$. 
\end{lemmaC}

\begin{proof}
The proof of Proposition \ref{lemma3:Lasso} shows that 
\[ \Big\| \frac{(\bs{X}^\perp)^\top \bs{X}^\perp}{nT} - \frac{\bs{Z}^\top \bs{Z}}{nT} \Big\|_{\max} = O_p \Big( \frac{\log(npT)}{\min\{n,T\}} \Big). \]
Since $s = o(\min\{n,T\} / \log(npT))$ by \ref{C:s-large}, this implies that 
\begin{equation}\label{eq:Corollary6.8-vdGB-supp} 
\frac{32 (2s)}{\varphi^2} \Big\| \frac{(\bs{X}^\perp)^\top \bs{X}^\perp}{nT} - \frac{\bs{Z}^\top \bs{Z}}{nT} \Big\|_{\max} \le 1 
\end{equation}
with probability tending to $1$ for any given constant $\varphi > 0$. By Corollary 6.8 in \cite{BuehlmannvandeGeer2011}, the following holds: Whenever the matrix $\bs{Z}$ fulfills the $\text{RE}(I,\varphi)$ condition for all $I$ with $|I| \le 2s$ and \eqref{eq:Corollary6.8-vdGB-supp} is fulfilled, the matrix $\bs{X}^\perp$ satisfies the $\text{RE}(I,\phi)$ condition with $\phi = \varphi/\sqrt{2}$ for all $I$ with $|I| \le 2s$. Since $\bs{Z}$ obeys the $\text{RE}(I,\varphi)$ condition  for all $I$ with $|I| \le 2s$  with probability tending to $1$ by assumption, we can infer that $\bs{X}^\perp$ must satisfy the $\text{RE}(S,\phi)$ condition  for all $I$ with $|I| \le 2s$ with probability tending to $1$. 
\end{proof}

\begin{lemmaC} \label{lemma:RE-Z-small-T}
Consider the small $T$-case and let \ref{C:F}--\ref{C:Z}, \ref{C:nTp-small}--\ref{C:K-small} and \ref{C:id1}--\ref{C:id2} be satisfied. In addition, suppose that the variables $Z_{it}$ are i.i.d.\ across $i$ and $t$. Let $\varphi > 0$ be a fixed constant and $\{c_{n}\}$ a sequence of non-negative numbers with $c_{n} \to 0$. If the matrix $\bs{Z}$ has the property that 
\begin{align*}
\pr\Big( \bs{Z} & \text{ fulfills } \textnormal{RE}(I,\varphi) \text{ for all } I \subseteq \{1,\ldots,p\} \text{ with } |I| \le 2s \Big) \ge 1 - c_{n}, 
\end{align*}
then the matrix $\bs{X}^\perp$ is such that 
\begin{align*}
\pr\Big( \bs{X}^{\perp} & \text{ fulfills } \textnormal{RE}(I,\phi) \text{ for all } I \subseteq \{1,\ldots,p\} \text{ with } |I| \le 2s \Big) \ge 1 - c_{n} 
\end{align*}
with $\phi = \varphi \sqrt{(1-\frac{K}{T})/(1+\delta)}$ and $\delta > 0$ arbitrarily small but fixed.
\end{lemmaC}

\begin{proof}
We first show that 
\begin{align}
\frac{\| \bs{Z} b \|^2}{nT} & = \sum_{j,j^\prime=1}^p \nu_{jj^\prime} b_j b_{j^\prime} + \| b_I \|_1^2 \, O_p\Big( \sqrt{\frac{\log p}{n}} \Big) \label{lemma:RE-Z-small-T:claim1} \\
\frac{\| \bs{X}^\perp b \|^2}{nT} & = \Big(1-\frac{K}{T}\Big) \sum_{j,j^\prime=1}^p \nu_{jj^\prime} b_j b_{j^\prime} + \| b_I \|_1^2 \, O_p\Big( \sqrt{\frac{\log p}{n}} \Big) \label{lemma:RE-Z-small-T:claim2}
\end{align}
for any $I \subseteq \{1,\ldots,p\}$ with $|I| \le 2s$ and $b \ne 0$ with $\|b_{I^c}\|_1 \le 3 \|b_I\|_1$, where $\nu_{jj^\prime} = \ex[Z_{it,j} Z_{it,j^\prime}]$. Since $\|b_{I^c}\|_1 \le 3 \|b_I\|_1$ and 
\[ \max_{j,j^\prime,t} \Big| \frac{1}{n} \sum_{i=1}^n \big( Z_{it,j} Z_{it,j^\prime} - \ex[Z_{it,j} Z_{it,j^\prime}] \big) \Big| = O_p\Big( \sqrt{\frac{\log p}{n}} \Big) \]
by standard arguments to derive uniform convergence rates (cp.\ e.g.\ Lemma \ref{lemmaprime:aux1}\ref{lemmaprime:aux1:Z} in Section A' above), we obtain that 
\begin{align*}
\frac{\| \bs{Z} b \|^2}{nT} 
 & = \frac{1}{nT} \sum_{i=1}^n \| \bs{Z}_i b \|^2 = \frac{1}{nT} \sum_{i=1}^n \sum_{t=1}^T \Big\{ \sum_{j=1}^p Z_{it,j} b_j \Big\}^2 \\
 & = \sum_{j,j^\prime=1}^p \ex[Z_{it,j} Z_{it,j^\prime}] b_j b_{j^\prime} + \sum_{j,j^\prime=1}^p \Big\{ \frac{1}{nT} \sum_{t=1}^T \sum_{i=1}^n \big( Z_{it,j} Z_{it,j^\prime} - \ex[Z_{it,j} Z_{it,j^\prime}] \big) \Big\} b_j b_{j^\prime} \\
 & = \sum_{j,j^\prime=1}^p \nu_{jj^\prime} b_j b_{j^\prime} + \| b_I \|_1^2 \, O_p\Big( \sqrt{\frac{\log p}{n}} \Big), 
\end{align*}
which is the statement of \eqref{lemma:RE-Z-small-T:claim1}. Since $\bs{\Pi} = \bs{I} - (\bs{F} \bs{F}^\top)/T$ under \ref{C:id1} and $\| \bs{\Pi} \bs{Z}_i b \|^2 = b^\top \bs{Z}_i^\top \bs{\Pi} \bs{Z}_i b = \| \bs{Z}_i b \|^2 - \| \bs{F}^\top \bs{Z}_i b \|^2/T$, we similarly get that  
\begin{align*}
\frac{\| \bs{X}^\perp b \|^2}{nT} 
 & = \frac{1}{nT} \sum_{i=1}^n \| \bs{\Pi} \bs{Z}_i b \|^2 = \frac{\| \bs{Z} b \|^2}{nT} - \frac{1}{nT} \sum_{i=1}^n \frac{\| \bs{F}^\top \bs{Z}_i b \|^2}{T}
\end{align*}
and
\begin{align*}
 & \frac{1}{nT} \sum_{i=1}^n \frac{\| \bs{F}^\top \bs{Z}_i b \|^2}{T} \\*
 & = \frac{1}{nT^2} \sum_{i=1}^n \sum_{k=1}^K \Big\{ \sum_{t=1}^T F_{t,k} \sum_{j=1}^p Z_{it,j} b_j \Big\}^2 \\
 & = \frac{1}{T^2} \sum_{k=1}^K \sum_{t,t^\prime=1}^T F_{t,k} F_{t^\prime,k} \sum_{j,j^\prime=1}^p \Big\{ \frac{1}{n} \sum_{i=1}^n Z_{it,j} Z_{it^\prime,j^\prime} \Big\} b_j b_{j^\prime} \\
 & = \frac{1}{T} \sum_{k=1}^K \frac{1}{T} \sum_{t=1}^T F_{t,k}^2 \sum_{j,j^\prime=1}^p \ex[Z_{it,j} Z_{it,j^\prime}] b_j b_{j^\prime} \\
 & \quad + \frac{1}{T^2} \sum_{k=1}^K \sum_{t,t^\prime=1}^T F_{t,k} F_{t^\prime,k} \sum_{j,j^\prime=1}^p \Big\{ \frac{1}{n} \sum_{i=1}^n \big( Z_{it,j} Z_{it^\prime,j^\prime} - \ex[ Z_{it,j} Z_{it^\prime,j^\prime}] \big) \Big\} b_j b_{j^\prime} \\
 & = \frac{K}{T} \sum_{j,j^\prime=1}^p \nu_{jj^\prime} b_j b_{j^\prime} + \| b_I \|_1^2 \, O_p\Big( \sqrt{\frac{\log p}{n}} \Big), 
\end{align*}
which yields the statement of \eqref{lemma:RE-Z-small-T:claim2}.

By assumption, $\bs{Z}$ fulfills the $\text{RE}(I,\varphi)$ condition with probability tending to $1$, where without loss of generality we let $I \neq \emptyset$. Hence, $\varphi^2/|I| \le \| \bs{Z} b \|^2 / \{ nT \| b_I \|_1^2 \}$ for any $b \neq 0$ with $\| b_{I^c} \|_1 \le 3 \| b_I \|_1$ with probability tending to $1$. From this and \eqref{lemma:RE-Z-small-T:claim1}, it follows that 
\[ \frac{\varphi^2}{|I|} \le \frac{1}{\| b_I \|_1^2} \sum_{j,j^\prime=1}^p \nu_{jj^\prime} b_j b_{j^\prime} + O_p\Big( \sqrt{\frac{\log p}{n}} \Big) \]
with probability tending to $1$. Moreover, since $|I| \le 2s = o(\sqrt{n/\log p})$ by \ref{C:s-small}, $1/|I|$ is of larger order than $\sqrt{\log p/n}$ and thus $\varphi^2/\{(1+\delta)|I|\} + O(\sqrt{\log p / n}) \le \varphi^2/|I|$ for any fixed $\delta > 0$ and sufficiently large $n$. Consequently, we obtain that 
\[ \frac{\varphi^2}{(1+\delta)|I|} \le \frac{1}{\| b_I \|_1^2} \sum_{j,j^\prime=1}^p \nu_{jj^\prime} b_j b_{j^\prime} \]
with probability tending to $1$ for any fixed $\delta > 0$. Combining this with \eqref{lemma:RE-Z-small-T:claim2}, we can infer that 
\[ \frac{\varphi^2\{ (1-\frac{K}{T}) / (1+\delta) \}}{|I|} \le \frac{1}{nT} \frac{\| \bs{X}^\perp b \|^2}{\|b_I\|_1^2} \]
with probability tending to $1$. Hence, $\bs{X}^\perp$ fulfills the $\text{RE}(I,\varphi \sqrt{(1-\frac{K}{T})/(1+\delta)})$ condition with probability tending to $1$. 
\end{proof}

\end{document}